\documentclass[rmp,aps,prd,nofootinbib]{revtex4}
\usepackage{amsmath}
\usepackage{amssymb}
\usepackage{graphics}
\usepackage{epsfig}
\usepackage{verbatim}
\usepackage{textcomp}
\usepackage[colorlinks=true,urlcolor=blue,linkcolor=blue,citecolor=blue]{hyperref}
\usepackage{tabularx}
\usepackage{xcolor}
\usepackage{array}
\setcitestyle{round,unsortedaddress}
\newcommand{\chapquote}[3]{\begin{quotation} \textit{#1} \end{quotation} \begin{flushright}  #2 \textit{#3}\end{flushright} }

\newcolumntype{C}[1]{>{\centering\arraybackslash}m{#1}}
\newcolumntype{s}[1]{>{\centering\arraybackslash}X}
\makeatletter
\def\cat@comma@active{\catcode`\,12}%
\makeatother

\begin{document}
	
\title{Nonadiabatic Landau-Zener-St\"{u}ckelberg-Majorana transitions, dynamics, and interference}
\author{Oleh~V.~Ivakhnenko}
\affiliation{B.~Verkin Institute for Low Temperature Physics and Engineering,~Kharkiv 61103,~Ukraine}
\affiliation{Center for Quantum computing, Cluster for Pioneering Research, RIKEN, Wako-shi, Saitama 351-0198, Japan}
\author{Sergey~N.~Shevchenko}
\email[e-mail: ]{sshevchenko@ilt.kharkov.ua}
\affiliation{B.~Verkin Institute for Low Temperature Physics and Engineering, Kharkiv 61103, Ukraine}
\affiliation{V.~N.~Karazin Kharkiv National University, Kharkiv 61022, Ukraine}
\affiliation{Center for Quantum computing, Cluster for Pioneering Research, RIKEN, Wako-shi, Saitama 351-0198, Japan}
\author{Franco~Nori}
\affiliation{Center for Quantum computing, Cluster for Pioneering Research, RIKEN, Wako-shi, Saitama 351-0198, Japan\\
Physics Department, University of Michigan, Ann Arbor, MI 48109-1040, USA}
	
	\begin{abstract}
		Since the pioneering works by Landau, Zener, St\"{u}ckelberg, and Majorana
		(LZSM), it has been known that driving a quantum two-level system results in
		tunneling between its states. Even though the interference between these
		transitions is known to be important, it is only recently that it became
		both accessible, controllable, and useful for manipulating a growing number of quantum systems.
		Here, we study systematically various aspects of LZSM physics and review the
		relevant literature, significantly expanding the review article in Ref.~\cite%
		{Shevchenko2010}.
	\end{abstract}
	
	
	
	
	\maketitle
	\vspace*{-0.8cm}
	\tableofcontents
	\vspace*{0.5cm}
	\textbf{Abbreviations and most-often-used symbols:}
	
	For the readers’ convenience, below we list
	the main abbreviations used in this work. This list also includes some of the topics
	covered.

	AIM:\:\:\: adiabatic-impulse model;
	
	CDT: \:\:coherent destruction of tunneling;
	
	JJ: \:\:\:\:\:\:\:Josephson junction;
	
	KZM: \:Kibble-Zurek mechanism;
	
	LZSM: Landau-Zener-St\"{u}ckelberg-Majorana;
	
	PAT:  \:\:\:\:photon-assisted tunneling;
	
	RAP: \:\:rapid adiabatic passage;
	
	RWA: \:\:rotating-wave approximation;
	
	TLS: \:\:\:\:two-level system;
	
	TM:  \:\:\:\:\:transfer matrix.
	
	------------------------------
	
	$\mathcal{P}$: single-passage LZSM transition probability;
	
	$\Delta $: minimal energy-level splitting;
	
	$\varepsilon $: energy bias;
	
	$A,\omega, T_{\mathrm{d}}$: amplitude, frequency, and period of the driving
	field;
	
	$\delta $: adiabaticity parameter;
	
	$\Delta E$: qubit energy-level gap;
	
	$\Gamma _{1,2}=T_{1,2}^{-1}$: relaxation and
	decoherence rates;
	
	$T$: temperature.
	
	\chapquote{\textquotedblleft Without nonadiabatic transition, this world would have been dead, because no basic chemical and biological processes, such as electron and proton transfer, could have occurred. Nonadiabatic transition is certainly an origin of mutability of this world.\textquotedblright\ }{
	}{\cite{Zhu2007}} 
	
	\section{Introduction}
	
	\label{Sec:Introduction}
	
	The quantum  two-level system (TLS) is  one of the basic models in quantum physics and
	describes systems that are ubiquitous in nature. On the one hand, this is the
	\textquotedblleft simplest nonsimple quantum problem\textquotedblright, 
	quoting \cite{Berry1995}; and, on the other hand, this provides the basis for
	quantum technologies, in which a TLS refers to a qubit.
	
	If a quantum system is excited by a time-dependent drive, it displays a
	variety of interesting and important effects. Note that several Nobel Prizes
	in physics have been awarded to physicists who exploited time-dependent
	few-level quantum systems:
	\begin{itemize}
	\item 1944: Rabi on molecular beams and nuclear
	magnetic resonance; 
	\item 1952: Bloch and Purcell on magnetic fields in atomic
	nuclei and nuclear magnetic moments; 
	\item 1964: Townes, Basov, and Prochorov on
	masers, lasers, and quantum optics;
	\item 1966: Kastler on optical pumping;
	\item 1989: Ramsey, Dehmelt, and Paul on atomic spectroscopy, hydrogen maser, and atomic
	clocks; 
	\item 1997: Chu, Cohen-Tannoudji, and Philips on cooling and trapping
	atoms with laser light; 
	\item 2012: Haroche and Wineland on coupled atoms and
	photons;
	\item 2022: Aspect, Clauser, and  Zeilinger on entangled photons and quantum information science.
	\end{itemize}
	
	\subsection{Relation to previous work and structure of this paper}
	We could ask ourselves a question here, quoting Ref.~\cite{Benderskii2003}: 	\begin{quotation} \textquotedblleft \textit{The title of this paper might sound perplexing at first sight. What else can be said about the Landau--Zener (LZ) problem after the numerous descriptions in both research and textbook literature?\textquotedblright} \end{quotation}
	  Below we give several reasons, starting
	from the fact that this topic should be called LZSM, not only LZ, and
	ending with the point that this evergreen topic is nowadays important for many areas of
	physics and its applications are growing over time.
	
	LZSM transitions are ubiquitous and important and have been addressed in
	several review articles~(e.g., \citet{Kazantsev1985, Shimshoni1991, Grifoni1998,
		Zhu2007, Shevchenko2010, Dziarmaga2010, Silveri2017, Sen2021}) and books~\cite%
	{Nakamura2012, Shevchenko2019,Nakamura2019}. In particular, the central idea of a
	previous review Ref.~\cite{Shevchenko2010} was a detailed presentation of
	the theoretical description of periodically driven TLSs. Here, we briefly
	mention the key aspects where the present work significantly extends this
	previous one:
	
	\begin{itemize}
		\item We show how to derive the LZSM formula by following the original works
		and not only presenting the readers these often-cited and difficult-to-access
		works. We also convincingly demonstrate that what is known as Zener or
		Landau-Zener transition/formula should be attributed to the four
		physicists: Landau, Zener, St\"{u}ckelberg, and Majorana (LZSM).
		
		\item We address different important aspects of the nonadiabatic
		transition, such as transition time, nonlinearity, and dissipation.
		
		\item We relate the LZSM formalism for avoided-level crossing with the
		Kibble–Zurek mechanism (KZM), which has been widely used to describe second-order phase
		transitions.
		
		\item We review new results that have appeared in the decade following the
		previous review article \cite{Shevchenko2010} and also cover various
		physical realizations.
		
		\item We emphasize that the detailed understanding of LZSM\ dynamics and
		its aspects, such as multi-photon transitions, is important not only for
		spectroscopy or interferometry, but also for quantum control.
	\end{itemize}
	
	For example, the original LZSM problem was covered very briefly in~\cite%
	{DiGiacomo2005}. However, we must ask what was studied in the original works of LZSM. Many
	(probably, the vast majority of) researchers cite the original works by LZSM
	without seeing those papers, which are difficult to access and read.
	Moreover, out of those five papers \cite{Landau1932a, Landau1932b,
		Zener1932, Stueckelberg1932, Majorana1932}, only one \cite{Zener1932} was
	written in English. [See the translations in Refs.~\cite{Landau_EN,
		Stueckelberg_EN, Cifarelli2020}.] One of the tasks in the current review article is
	to present a pedagogical summary of these original works of LZSM. We believe
	that seeing all four approaches together is both instructive and
	pedagogical.
	
	The present review paper is organized as follows:\footnote{%
		Please note that we do not always cite the relevant papers in
		chronological order. We have mostly aimed to tell the story about LZSM
		physics, for which it is sometimes more illustrative to refer to later
		publications or review articles for the sake of the readers’ convenience.}
	First, in the rest of Sec.~\ref{Sec:Introduction}, we present diverse
	physical systems that can effectively be described as TLSs. A nonadiabatic
	transition between energy levels, known as the LZSM transition, is described
	in Sec.~\ref{Sec:Linear}, with details provided in Appendix~\ref%
	{Sec:AppendixA}. Various approaches to the description of a periodically
	driven TLS are the subject of Sec.~\ref{Sec:Repetitive} and Appendix~\ref%
	{Sec:AppendixB}. We devote Sec.~\ref{Sec:Interferometry} to the description
	of experimental studies, in which the LZSM interference is relevant. Quantum
	control with nonadiabatic transitions and periodic driving is outlined in
	Sec.~\ref{Sec:Quantum}. Related classical coherent phenomena are considered
	in Sec.~\ref{Sec:Classics}. Section~\ref{Sec:Conclusion} presents the
	conclusions.
	
	\subsection{Driven few-level systems}
	
\begin{figure}[h]
		\centering{\includegraphics[width=0.81\columnwidth]{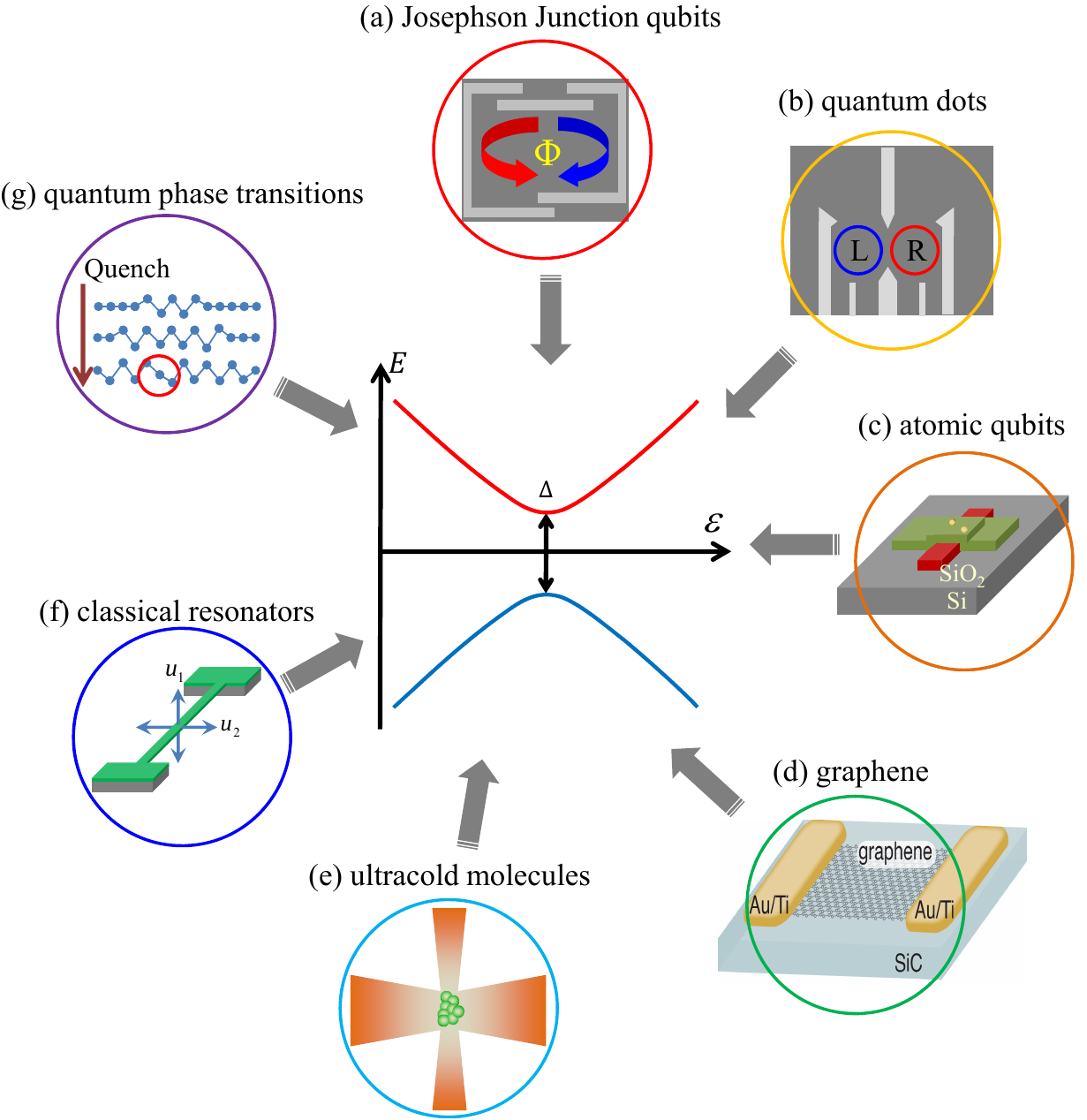}}
		\caption{\textbf{Various physical systems that can be described by the
				two-level model. }These can be driven by external fields and exhibit
			Landau-Zener-St\"{u}ckelberg-Majorana (LZSM) transitions. \textbf{(a)}%
			~Superconducting quantum circuits with Josephson junction (JJ) qubits are
			illustrated by the flux qubit, made of a superconducting ring with three
			tunnel JJs and the qubit states formed by the current
			direction \protect(e.g., \citet{Izmalkov2004}). \textbf{(b)}~Semiconductor artificial
			atoms are illustrated here by a gate-defined double quantum dot \protect\cite%
			{Cao2013}, where the charge qubit is formed by electron states localized in
			either the left (L) or right (R) dot. \textbf{(c)}~An impurity-based
			qubit, formed by phosphorus or arsenium donor atoms in a silicon nanowire
			transistor \protect\cite{Dupont-Ferrier2013}. \textbf{(d)}~Graphene strip
			contacted with gold electrodes, where the current displays
			nonadiabatic transitions between the valence and conduction bands around
			the Dirac point \protect\cite{Higuchi2017}. \textbf{(e)}~Ultracold Caesium
			Feshbach molecules in a laser trap, where ramping the magnetic field results
			in the transfer of molecular states \protect\cite{Mark2007}. \textbf{(f)}%
			~Classical nanomechanical resonator \protect\cite{Faust2012}, where the
			coherent superposition of the in-plane and out-of-plane modes behave as a
			driven TLS. \textbf{(g)}~Topological defect formation is schematically represented here by an
			ion Coulomb crystal; this is a trapped ion chain, where changing the
			confining potential results in the chain buckling; this breaking of the
			axial symmetry is a second-order phase transition, with the density of
			defects described by the Kibble-Zurek mechanism~\protect\cite{Pyka2013}. }
		\label{Fig:Fig1}
	\end{figure}

	First, we outline the systems in which the phenomena are taking
	place. These are very different in their physical origins because the objects can
	be microscopic (electron or nuclear spins, photons, atoms), mesoscopic
	(superconducting qubits, quantum dots, graphene structures), or macroscopic
	(mechanical or electrical resonators) \cite{Nakamura2019,Kjaergaard2020}. Our aim is to demonstrate
	that the description of all these can be reduced to a quantum two- or
	few-level system. Here, the key idea is to show that a basic notion in
	quantum mechanics—a TLS with avoided-level crossing—is ubiquitous and that for
	such systems, LZSM physics is relevant. We have chosen several
	illustrative examples and presented them in Fig.~\ref{Fig:Fig1}, with some
	details given in Table~\ref{Table:TLSs} and in the main text below.
	
	Note that neither Fig.~\ref{Fig:Fig1} nor Table~\ref{Table:TLSs} are
	comprehensive because neither gives a complete picture of the variety of
	the respective systems. The aim is to show the
	diversity of the systems and their characteristic parameters, including
	their typical sizes. A goal of this review article is to present different
	realizations of LZSM phenomena. In particular, the experimental realizations
	of the single- and multiple-passage transitions in quantum systems will be
	presented in Sec.~\ref{Sec:exprt_single} and Sec.~\ref{Sec:Interferometry},
	respectively, while their classical counterparts will be presented in Sec.~%
	\ref{Sec:Classics}. After making these references to subsequent sections,
	we briefly describe the quantum systems. Here, a note is in order. We
	describe classical realizations in a separate section and otherwise consider
	 quantum systems. Interestingly, nonadiabatic LZSM transitions are
	largely associated with quantum systems; and it is a rare example in physics
	when classical related phenomena are studied later than their quantum
	counterparts and not vice versa.
	
	It is difficult to give a complete picture of those processes where nonadiabatic
	transitions between potential energy curves matter because these are
	ubiquitous in natural sciences. Here, we can briefly consider different
	physical realizations.

	\begin{itemize}
		\item[*] As a recent example of this, in the review article \cite%
		{Koehler2006}, the role of LZSM nonadiabatic transitions is considered in
		the production of cold molecules and molecular association and dissociation;
		single and repeated nonadiabatic transitions were shown to transfer between
		molecular and atomic states \cite{Mark2007, Lang2008}.
		
		\item[*] LZSM transitions become important for describing atoms being scattered
		by a standing light wave \cite{Kazantsev1985}. The ground and excited states
		of an atom correspond to two effective potentials and two trajectories of
		motion. The possibility of nonadiabatic transitions between the two states
		(beams) results in changing an atom’s trajectory, which leads to
		interference of the translational motion states; this is similar to a two-channel
		optical interferometer.
		
		\item[*] Superconducting quantum circuits are based on Josephson junctions
		(JJs) (see, e.g., \citet{You2005,You2011,Xiang2013, Gu2017}). Depending on the system parameters, there
		are three basic types of JJ-based qubits: charge, phase, and flux ones. The
		energy-level spacing in these can be controlled by an external parameter:
		gate voltage, bias current, or magnetic flux, respectively. There are also
		newer subtypes, including a transmon, which is the capacitor-shunted charge
		qubit coupled with a transmission line; such layouts allow for better
		isolation from external noise, allowing for longer coherence times.
		
		\item[*] Quantum dots with controllable parameters are mainly based on
		electrons that are localized in gate-defined depleted regions of semiconductor
		heterostructures (typically a few tens of nanometers in size), such as
		GaAs/AlGaAs and Si/SiGe, or in nanowire structures \cite{Zwanenburg2013}.
		These show Coulomb blockade and display single-electron physics. Depending on
		which degree of freedom is relevant, we can have spin or charge qubits,
		which involve one or several electrons. The energy levels, including the
		minimal splitting, can be controlled by an external magnetic field and gate
		voltages.
		\begin{table}[t]
			\begin{center}
				\begin{tabularx}{500pt}{|C{4mm}|C{15mm}|C{18mm}|C{32.05mm}|C{19.5mm}|c|C{25.4mm}|C{21.2mm}|}
					\hline
					& \rule{0pt}{2ex} \textbf{System} & \textbf{Size} & \textbf{Basis} & \textbf{Variable} & \textbf{$\Delta / h$} & \textbf{ $\omega / 2 \pi $} & \textbf{Temperature} \\ \hline
					
					 (a) & JJ qubits & 1~$\mu$m to 1~mm & charge, current & voltage, flux & \rule{0pt}{3ex} 10~MHz to 10~GHz & 1~GHz & 50~mK \\[1ex] \hline
					
					(b) & quantum dots & 10~nm to 1~$\mu$m & charge, spin & \rule{0pt}{3ex} voltage, magnetic field & 0.1 to 10~GHz & 1 to 10~GHz & 50~mK \\[1ex] \hline
					
					(c) & atomic qubits & \begin{tabular}{@{}l@{}}  1~\AA \\ \\ \hline \\ 1~$\mu$m\end{tabular} & \rule{0pt}{2ex} \begin{tabular}{@{}l@{}} electron charge or spin \\ \\ \hline\\ nuclear spin \end{tabular}  & optical and microwave fields  & 0.1~GHz & 1~MHz to 10~GHz & \begin{tabular}{@{}l@{}} 50~mK to 10~K \\ \\ \hline\\ 1K to room \end{tabular}\\[2ex] \hline
					
					 (d) & graphene & 1 $\mu$m & \rule{0pt}{3ex} \begin{tabular}{@{}l@{}} conduction bands\\ \\\hline\\ valence bands  \end{tabular} & electric field & 100~THz to 1~PHz & 100~THz & room \\[2ex] \hline
					
					(e) & \rule{0pt}{3ex} ultracold molecules & \begin{tabular}{@{}l@{}}  1~\AA \\ \\ \hline\\ 40~$\mu$m\end{tabular} & \begin{tabular}{@{}l@{}} molecular states\\ \\\hline\\ lattice bands \end{tabular} & \begin{tabular}{@{}l@{}} magnetic field\\ \\\hline \\ lattice tilt \end{tabular} & 10~kHz & 10~kHz & 0.01 to 100~$\mu$K \\[2ex] \hline
					
					(f) & \rule{0pt}{3ex} classical resonators & 50~$\mu$m & oscillation modes & bias voltage  & 10~kHz & 10~kHz & room \\[2ex] \hline
					
					(g) & \rule{0pt}{3ex} quantum phase transitions & 300~$\mu$m & defect orientation & confining voltage & 100~kHz & 100~kHz & $10~\mu$K \\[2ex] \hline
					
				\end{tabularx}
			\end{center}
			\caption{\textbf{Characteristic two-level systems }(TLSs) and their
				parameters, including minimal energy-level splitting $\Delta$ and characteristic driving frequency $\omega$. The respective systems are described in the main text, while
				details can be found in the references in the main text. The numbers listed above are characteristic
				values or ranges. The table lists both the size of the core quantum system
				and the size of the host. For example, for ultracold molecules, the
				characteristic size of the atoms is of the order of several Angstroms, while
				the size of the localized Bose–Einstein condensate (BEC) is typically a few
				dozens of micrometers. }
			\label{Table:TLSs}
		\end{table}
		\item[*] Atomic impurities, such as nitrogen-vacancy (NV) color centers in
		diamond and phosphorous impurities in silicon, allow for the manipulation of single
		electron spins and/or nuclear spins. These can be conveniently coupled with
		each other, nicely isolated from the environment, can be controlled by optical
		and microwave fields, and can be integrated in solid-state devices. We
		illustrate this with the device from Ref.~\cite{Dupont-Ferrier2013}, which is based
		on a silicon nanowire. The source-drain current was then defined by the
		electron transport through two tunnel-coupled donor atoms, of which the
		electronic-state populations created the charge qubit.
		
		\item[*] Energy bands with avoided-level crossings, which are relevant for our
		consideration, also take place in graphene. When driven by an external
		electromagnetic field, the Dirac Hamiltonian for graphene results in
		LZSM phenomena near the Dirac points \cite{Higuchi2017, Heide2018}. It has been
		shown that thin films of a Weyl semimetal subjected to a strong AC
		electromagnetic field should behave similarly to graphene \cite{Rodionov2016}. 
		It has also been discussed that there is a profound similarity between the
		effects of spatial and temporal periodicity, which is one more argument why
		the avoided-level-crossing structures appear in many different contexts. For
		a review of other related materials, the so-called \textquotedblleft artificial
		graphenes\textquotedblright, see \cite{Montambaux2018}.
		
		\item[*] The theory of LZSM transitions is closely related to the 
		Kibble-Zurek mechanism (KZM) \cite{Damski2005} which will be considered in
		section~\ref{subsubsection:KZM}. This describes second-order phase
		transitions, which occurs when one of the system parameters passes through a critical
		point. The universality of second-order phase transitions makes their
		dynamics independent of their microscopic nature. This results in a long
		list of related realizations, from cosmology to condensed matter. Leaving
		this intriguing issue for later, here we illustrate the realization of the
		KZM with chains of ions confined in harmonic traps~\cite{Pyka2013}. In this situation,
		weakening the triaxial confining potential in the transverse direction makes
		the chain buckle and form a zig-zag shape. This second-order phase transition can
		lead to the formation of topological defects, which is illustrated by a “zig”,
		followed by another “zig”, rather than by a “zag”. LZSM theory
		quantitatively describes the formation of such topological defects. Note
		that the characteristic parameters for defect formation in Table~\ref%
		{Table:TLSs} are used for this very realization; parameters for other phase
		transitions may be completely different.
		
		\item[*] Two Majorana works meet when Majorana qubits are described by the
		LZSM Hamiltonian. These are formed by the Majorana bound states that reside
		in topological superconducting systems. A realization of this could be an rf (radio frequency)
		superconducting quantum interference device (SQUID) with a topological
		JJ that is formed by a one-dimensional nanowire with spin-orbit
		coupling, quantum spin–Hall edge states, or ferromagnetic atomic chains \cite%
		{You2014,Huang2015, Wang2018, Feng2018, Zhang2020, Zhang2021}. The energy level
		structure is controlled by an external magnetic flux. Alternatively, a
		topological superconductor can be a weak link between quantum dots \cite%
		{Zazunov2020}. Besides being fundamentally interesting, Majorana qubits
		provide the basis of topological quantum computation. For more on engineering gauge
		fields and triggering topological order in periodically driven systems, see 
		\cite{Goldman2014}.

		\item[*] Somewhat unexpectedly in this context, some classical systems can
		also be described as TLSs. This arises because what is needed for LZSM
		physics (superposition and transition between discrete states) appears not
		only in the quantum world, but also in classical physics. To this issue, we
		devote Sec.~\ref{Sec:Classics}; but here, we illustrate this with a
		nanomechanical resonator in the classical regime \cite{Faust2012}. This
		system is based on the coherent energy exchange between two strongly coupled
		high-quality modes of a nanomechanical resonator placed in a vacuum at
		ambient temperature. 		
		
		\item[*] To emphasize the variety of TLSs, for which LZSM physics
		matters, we kaleidoscopically mention a few other realizations: electron
		spin-polarized $^{4}$He$^{+}$ ion scattering \cite{Suzuki2010, Suzuki2016},
		low-dimensional conductors \cite{Montambaux2016, Benito2016}, and
		charge-density-wave insulators \cite{Shen2014a}. Overall, nonadiabatic
		transitions are relevant in physics, chemistry, biology, economics, and some
		other—sometimes unexpected—research fields \cite{Nakamura2012}. As an
		exotic example, the LZSM model can be useful in describing decision making in which there are
		a few possible outcomes; in Ref.~\cite{Levi2013}, the author applies the
		model to describe free will with afterthoughts.
	\end{itemize}

	
	\section{Linear drive: Landau-Zener-St\"{u}ckelberg-Majorana (LZSM)
		transition}
	
	\label{Sec:Linear}
	
	\subsection{Hamiltonian and bases}
	\begin{figure}[b]
		\centering{\includegraphics[width=1.0%
			\columnwidth]{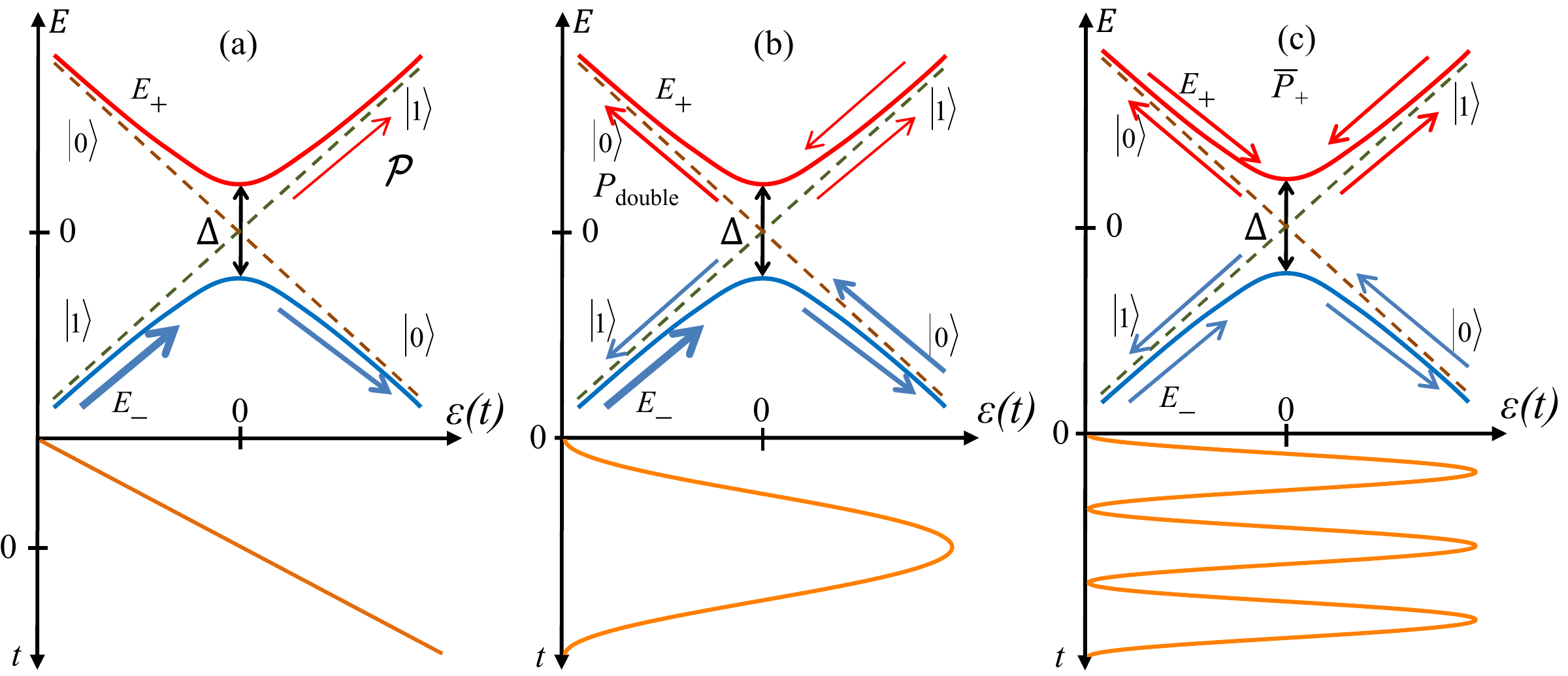}}
		\caption{\textbf{Schematic of a driven two-level system (TLS)}. Here, $E_{\pm
			}$ are the upper and lower energy levels, $\Delta $ is the minimal distance
			between energy levels, and $\protect\varepsilon $~describes the energy bias.
			The avoided-crossing region corresponds to the vicinity of $\protect%
			\varepsilon =0$. Three problems are considered in this work: (a) single
			passage of the avoided-crossing region, which is described by the probability $%
			\mathcal{P}$; (b) double-passage problem for $P_{\mathrm{double}}$; and (c)
			multiple-passage problem with the solution for the time-averaged probability 
			$\overline{P}_{+} $.}
		\label{Fig:Energy_level_Scheme}
	\end{figure}
	
	We now present various approaches to derive the formula for the excitation
	probability of a TLS. To this end, we briefly introduce the main steps,
	while details are presented in Appendix~\ref{Sec:AppendixA}. Interestingly,
	this can be done in several ways within different theoretical formalisms~%
	\cite{DiGiacomo2005}. We aim to study and compare different techniques by
	applying these to the classical LZSM problem with a linear drive to the TLS.
	
	Consider a TLS described by the Hamiltonian 
	\begin{equation}
	\color{red}\boxed{\color{black}	H(t)=-\frac{\Delta }{2}\sigma _{x}-\frac{\varepsilon (t)}{2}\sigma _{z}=-%
		\frac{1}{2}\left( 
		\begin{array}{cc}
			\varepsilon & \Delta \\ 
			\Delta & -\varepsilon%
		\end{array}%
		\right), \color{red}}\color{black}  \label{H(t)}
	\end{equation}%
	with the linear bias 
	\begin{equation}
		\varepsilon (t)=vt,  \label{eps(t)}
	\end{equation}
	$\Delta $ being a time-independent quantity and time $t\in (-\infty, \infty
	)$. Let us now define the \textit{diabatic} states, which are the
	Hamiltonian eigenfunctions at $\Delta =0$: $\left\vert 0\right\rangle =%
	\binom{1}{0}$ and $\left\vert 1\right\rangle =\binom{0}{1}$. The respective
	(diabatic) energy levels are $E_{0,1}=\mp \varepsilon /2$. These are plotted
	by the dashed lines in Fig.~\ref{Fig:Energy_level_Scheme}. In general, the
	wave function is a superposition state 
	\begin{equation}
		\left\vert \psi (t)\right\rangle =\alpha \left\vert 0\right\rangle +\beta
		\left\vert 1\right\rangle =\binom{\alpha }{\beta }  \label{psi(t)}
	\end{equation}%
	with $\alpha $ and $\beta $ being the time-dependent coefficients.

	The \textit{adiabatic} eigenvalues $E_{\pm }(t)$ and eigenstates $\left\vert
	E_{\pm }(t)\right\rangle $ are given by the Schr\"{o}dinger equation, where
	time is a parameter, $H(t)\left\vert E_{\pm }(t)\right\rangle =E_{\pm
	}(t)\left\vert E_{\pm }(t)\right\rangle $. We obtain the \textit{adiabatic}
	energy levels%
	\begin{equation}
		\color{red}\boxed{\color{black}E_{\pm }(t)=\pm \frac{1}{2}\sqrt{\Delta ^{2}+\varepsilon (t)^{2}}=\pm \frac{1%
		}{2}\Delta E(t).\color{red}}\color{black}  \label{Epm(t)}
	\end{equation}%
	Here, $\Delta E=E_{+}-E_{-}$ is the distance between the energy levels, which
	are presented by the solid curves in Fig.~\ref{Fig:Energy_level_Scheme}. Now
	we can see the meaning of the parameter $\Delta $ (the minimal energy
	spacing or gap), while the parameter $\varepsilon $ is the energy bias.
	The energy gap is 
	 smallest at $\varepsilon =0$; 
	accordingly, we say that at this point, we have an
	avoided-level crossing.
	
	The \textit{adiabatic} energy eigenstates are%
	\begin{eqnarray}
		\left\vert E_{\pm }(t)\right\rangle &=&\gamma _{\mp }\left\vert
		0\right\rangle \mp \gamma _{\pm }\left\vert 1\right\rangle, 
		\label{eigenstates} \\
		\gamma _{\pm } &=&\frac{1}{\sqrt{2}}\sqrt{1\pm \frac{\varepsilon (t)}{\Delta
				E(t)}}.  \label{gammas}
	\end{eqnarray}%
	In particular, at the point of the avoided-level crossing, $\varepsilon =0$, 
	$\left\vert E_{\pm }(0)\right\rangle =\frac{1}{\sqrt{2}}\left( \left\vert
	0\right\rangle \mp \left\vert 1\right\rangle \right) $. For $\varepsilon \gg
	\Delta $, the adiabatic energy levels approach the diabatic ones.
	
	Now the problem is in finding the probability of a TLS to be in the upper
	state after passing the avoided-level crossing. Let us assume that we
	start from the ground state $\left\vert E_{-}\right\rangle $ in the
	left-hand side of Fig.~\ref{Fig:Energy_level_Scheme}(a), that is, at $%
	t\rightarrow -\infty $. We are interested in the probability $\mathcal{P}$
	of finding the system in its excited state $\left\vert E_{+}\right\rangle $
	after passing the avoided-crossing region, that is, at $t\rightarrow +\infty $.
	Alternatively, the problem can be formulated in terms of diabatic states:
	what is the probability $\mathcal{P}$ to stay in the same diabatic state $%
	\left\vert 0\right\rangle $, or what is the probability $\left(1-\mathcal{P}%
	\right)$ of changing the state from $\left\vert 0\right\rangle $ to $%
	\left\vert 1\right\rangle $?
	
	The time-dependent Schr\"{o}dinger equation gives us the solution, which is known as
	the LZSM formula:%
	\begin{equation}
		\color{red}\boxed{\color{black}\mathcal{P}=\exp{\left\{-2\pi \delta \right\}},\color{red}}\color{black}  \label{P}
	\end{equation}%
	where%
	\begin{equation}
		\color{red}\boxed{\color{black}\delta =\frac{\Delta ^{2}}{4\hbar v}\color{red}}\color{black}  \label{delta}
	\end{equation}%
	is the adiabaticity parameter. For slow changes, $\delta \gg 1\,$\ (i.e., $%
	v\ll \Delta ^{2}/\hbar $), we have an adiabatic evolution, where the two-level system (TLS)
	mostly stays in the ground state, $\mathcal{P}\approx 0$.
	
	For fast changes, $\delta \ll 1\,$, we have the diabatic evolution, where the
	system dominantly follows the diabatic state, $(1-\mathcal{P})\approx 0$;
	this means that by starting from the $\left\vert 0\right\rangle $ state at $%
	t=-\infty $ in Fig.~\ref{Fig:Energy_level_Scheme}(a), we end up with an
	almost unit probability in the same $\left\vert 0\right\rangle $ state at $%
	t=\infty $. We emphasize that the LZSM formula, Eq.~(\ref{P}), describes the
	transition probability if starting from an eigenstate; the case when the
	system starts in a superposition state will be considered later.
	
	Besides the absolute value of the wave function, the phase obtained during the LZSM transition becomes crucial for interferometry and quantum control. This phase is known as the Stokes phase,
	\begin{equation}
		\color{red}\boxed{\color{black}\phi _{\text{S}}(\delta )=\frac{\pi }{4}+\delta (\ln {\delta }-1)+\mathrm{Arg} \left[\Gamma (1-i\delta )\right],\color{red}}\color{black} \label{StocesPhase}
	\end{equation}
	where $\Gamma$ here refers to the Gamma function.
	
	In what follows, we present the derivation of the formula~\eqref{P} as used in four
	different methods.
	
	
	\subsection{Brief overview of the original works of Landau, Zener, St\"{u}ckelberg, and Majorana}
	
	Let us briefly consider the approaches of LZSM, which are summarized in
	Table~\ref{Table:LZSM} and of which the details are presented in Appendix~\ref%
	{Sec:AppendixA}. Importantly, all four of them published the very same year
	papers where one of the key results was exactly Eq.~\eqref{P}.From the dates in 
	\ref{Table:LZSM} we can see that if
	one follows the dates of publication, \textbf{the correct ordering would be
		MLZS}. Concerning Majorana's contribution, see also~\cite{Wilczek2014,Kofman2022}. 	
	\begin{table}[h]
		\begin{center}
			\begin{tabularx}{500pt}{|C{3cm}|C{1.9cm}|C{1.9cm}|C{5cm}|c|C{0.97cm}|}
				\hline
				\rule{0pt}{2ex} \textbf{ Article by} & \textbf{Submission} & \textbf{Publication} & \textbf{System} & \textbf{Method} & \textbf{Phase} \\ \hline
				 \rule{0pt}{3ex} E. Majorana \cite{Majorana1932} & ?-1931 & 02-1932 & Spin 1/2 in a magnetic field &
				Laplace transform  & Yes\footnote{Note that Majorana in his work obtained only the probability and did not pay attention to the phase change. Also note that he published Eq.~\eqref{P} before others, which is not known to many. For detailed derivations of the full wave function, including the phase change, within Majorana's approach see Appendix \ref{App_Majorana} as well as \cite{Rodionov2016} and \cite{Kofman2022}.}
				\\[2ex] \hline
				\rule{0pt}{3ex} L.D. Landau \cite{Landau1932b} & 12-1931 & 06-1932 & Inelastic adiabatic atomic collisions & Quasiclassical approach & No \\[2ex] \hline
				\rule{0pt}{3ex} C. Zener\:\:\:\:\:\:\:\:\: \cite{Zener1932} & 07-1932 & 09-1932 & Crossing polar and homopolar states in molecule & Parabolic cylinder function & Yes\footnote{Zener obtained the full wave function in terms of the parabolic-cylinder functions. However, in his work, the author discussed only the absolute value, that is the probability, Eq.~\eqref{P}. For detailed discussion of the solution, including the phase, see Appendix \ref{LZSM_by_Zener} and \cite{Child1974a,Kayanuma1997}.}
				\\[1ex] \hline \rule{0pt}{3ex}
				 E.C.G. St\"{u}ckelberg \cite{Stueckelberg1932} & ?-1932 & 11-1932 & Inelastic adiabatic collision & WKB
				approximation & No \\[1ex] \hline
			\end{tabularx}
		\end{center}
		\caption{ The works of Landau, Zener, St\"{u}ckelberg, and Majorana at a
			glance.}
		\label{Table:LZSM}
	\end{table}  
	\subsubsection{Near-adiabatic limit (Landau)}
	
	In his first work concerning the nonadiabatic transitions \cite{Landau1932a}, L.D.~Landau studied
	adiabatic nonelastic atomic collisions. He derived a general expression for
	the probability of nonadiabatic transitions within perturbation theory, which 
	was applied for the near-sudden limit, with $\delta \ll 1$. The
	resulting excitation probability for the double-passage process was
	\begin{equation}
		P_{\mathrm{double}}^{(\mathrm{L})}=8\pi \delta \sin ^{2}\Phi _{\mathrm{L}},
	\end{equation}
	which is a rapidly oscillating function. Being averaged over a large
	dynamical phase $\Phi _{\mathrm{L}}$, accumulated during double passage
	evolution [see Fig.~\ref{Fig:Energy_level_Scheme}(b)], this would give $%
	\overline{P}_{\mathrm{double}}^{(\mathrm{L})}=4\pi \delta $. Indeed, from
	Eq.~\eqref{P} at $\delta \ll 1$, we have $\mathcal{P}\approx 1–2\pi \delta $.
	These are consistent results, if we note that the latter gives the
	probability $P_{0\rightarrow 1}\approx 2\pi \delta $ of staying in the
	ground state after the first passage; then, there are two possibilities to be
	excited during the second passage: $P_{0\rightarrow 1\rightarrow 1}\approx
	2\pi \delta $ and $P_{0\rightarrow 0\rightarrow 1}\approx 2\pi \delta $,
	which would add up to Landau's value of $4\pi \delta $.
	
	In his second related paper \cite{Landau1932b}, the author applied the
	general formula of the transition to a generic case of almost-crossing
	potential curves in the near-adiabatic limit, that is, for $\delta \gg 1$ and
	the obtained excitation probability in the form of Eq.~\eqref{P}, but with
	the prefactor $C$ being presumably of the order of unity. If analyzed by the
	other (more precise) methods, which are presented below, this constant becomes exactly
	equal to 1.
	
	\subsubsection{Using parabolic cylinder functions (Zener)}
	
	The second relevant approach is by Clarence~Zener~\cite{Zener1932}. The
	author studied the crossing of the polar and homopolar states of a molecule.
	The energy bias for the electronic states was the slow variable of the
	nuclei position. The task was reduced to the very same problem formulated
	in Fig.~\ref{Fig:Energy_level_Scheme}(a): What is the probability of
	excitation if starting from the ground state to the left and linearly
	driving to the right when passing the avoided crossing? The respective Schr\"{o}dinger equation was transformed into a second-order differential equation, of
	which the solution was the parabolic cylinder Weber functions. This exact
	solution, after taking the asymptotes, resulted in Eq.~\eqref{P}.
	
	Here, a note about Zener tunneling/effect/diode is in order. Two years later,
	Zener published another paper \cite{Zener1934}, in which he studied the
	dielectric breakdown, which is the electrical breakdown in solid insulators
	when applying a strong constant electric field. The breakdown occurs because of the
	tunneling between the conduction bands through a forbidden band. Later, such
	sort of electric breakdown was studied for semiconductors \cite{Kane1960}
	and is the basis of the \textit{Zener diode} (stabilitron). 
	
	Some authors have analyzed the analogy between Zener tunneling and LZSM transitions, for example,~\cite{Romanova2011}; however, we differentiate Zener tunneling from LZSM
	transitions because the former does not involve an avoided-level
	crossing but instead needs strong fields, while the avoided-level crossing is
	the origin for LZSM interferometry. Based on this, the two cases can also be called
	nonresonant and resonant (Zener) tunneling, respectively \cite{Glutsch2004}.
	As a special case, one can mention here the so-called Bloch(-Landau)-Zener
	dynamics \cite{Rotvig1995, Holthaus2000, Wu2003, Ke2015, Khomeriki2016,
		Xia2021}, which involves LZSM transitions between energy bands when
	these display Bloch oscillations with avoided-level crossing.
	
	\subsubsection{Using the WKB approximation (St\"{u}ckelberg); double-passage
		solution}
	
	Much like the above, E.C.G. St\"{u}ckelberg also considered atomic collisions, for which he used the Wentzel-Kramers-Brillouin (WKB) approximation and the phase integral method \cite{Stueckelberg1932}. As a result, St\"{u}ckelberg
	obtained the formula for the double-passage problem%
	\begin{equation}
		\color{green}\boxed{\color{black}P_{\mathrm{double}}=4\mathcal{P}(1-\mathcal{P})\sin ^{2}\Phi _{\mathrm{St}},\color{green}}\color{black}
		\label{P_double}
	\end{equation}%
	where the single-passage probability $\mathcal{P}$ is again given by Eq.~\eqref{P}, and $\Phi _{\mathrm{St}}$ is the phase, that is, the so-called St\"{u}ckelberg phase, which is accumulated by the wave function during evolution. We
	will see that this consists of two parts: the one accumulated during the
	adiabatic motion and the other (called dynamical or Stokes phase) $\phi_\text{S}$
	acquired during the single passage of the avoided-crossing region \cite%
	{Nikitin1999}. Interestingly, St\"{u}ckelberg pointed out that, particularly
	for $\delta \ll 1$, his result gives what Landau obtained in the work \cite%
	{Landau1932a} with 
		$\Phi _{\mathrm{St}}=\Phi _{\mathrm{L}} \:\: \text{ and } \:\:2\mathcal{P}(1-\mathcal{P})=4\pi \delta.$
    In Appendix \ref{App_Stuckelberg}, we present
	some details about the St\"{u}ckelberg approach. In particular, we
	see that even with all the complications and generalities of this approach,
	the expression for the dynamical part of the St\"{u}ckelberg phase cannot be
	obtained within this formalism \cite{Child1974a}.
	
	\subsubsection{Using contour integrals (Majorana)}
	
	In the fourth approach, Ettore~Majorana considered an oriented atomic beam
	passing a point of a vanishing magnetic field \cite{Majorana1932}. The problem
	was reduced by the author to a spin-1/2 particle in a linearly
	time-dependent magnetic field, exactly as described by the Hamiltonian~\eqref{H(t)} with the bias~\eqref{eps(t)}. Much like the approach by
	Zener, Majorana reduced the problem to a mathematical treatment of
	a second-order differential equation. This time, the author solved the
	equation using the direct and inverse Laplace transform by calculating the
	respective contour integrals in the limits of $t\rightarrow \pm \infty $,
	resulting again in Eq.~\eqref{P}. Expectedly, that integral is similar to
	the integral representation of the parabolic cylinder function.
	
	We note that, previously, most of the papers on the subject of nonadiabatic
	transitions called these either LZ or LZS transitions. Paradoxically enough, to some extent, the
	paper by Majorana is even more relevant and better
	suited for the problem: \begin{itemize}
		\item Majorana's derivation does not contain undefined exponential prefactors
		or limitations for the value of the adiabaticity parameter $\delta $, as in the derivations by Landau.
		\item It does not refer to special functions that require using
		asymptotics from books or numerics, as in Zener's approach.
		\item The Majorana's derivation is less complicated than the one by St\"{u}ckelberg.
	\end{itemize}    
	
	The work of Majorana was both stimulated and verified by experimental
	observation \cite{Frisch1933}. For the history of this, see \cite%
	{Esposito2014, Esposito2017, Inguscio2020}. With similar arguments,
	F.~Di~Giacomo and E.E.~Nikitin \cite{DiGiacomo2005} proposed, first, to
	make Majorana's approach a central problem for textbooks on quantum
	mechanics and, second, to denote the problem and formula, Eq.~\eqref{P}%
	, using all four names: LZSM problem and LZSM formula,
	respectively. From the dates in Table~\ref{Table:LZSM} we can see that \textbf{the correct ordering would be
		MLZS}. However, to avoid introducing confusion, we will call this
	LZSM, as almost all other authors who acknowledge Majorana's role.
	Concerning Majorana's contribution, see also~\cite{Wilczek2014,Kofman2022}. 
	
	As an additional advantage of Majorana's formulation, we note that he (in
	contrast from LZS) formulated the problem in terms of the spin-1/2
	Hamiltonian, exactly in the form employed in quantum information nowadays.
	
	Finally, Majorana's approach allows for explicitly obtaining the
	phase acquired during the transition, like in Zener's approach, while this
	cannot be done in the semiclassical calculations by Landau and St\"{u}ckelberg.
	
	See Appendix~\ref{Sec:AppendixA} for further details, where we present the
	approaches developed by LZSM. Among other approaches, we can mention the one
	by \cite{Wittig2005}, which was also presented in \S 1.5.2 of the textbook 
	\cite{Zagoskin2011}, and the Zhu--Nakamura theory \cite{Nakamura2012,Nakamura2019}. See also \cite{Hagedorn1991, Chichinin2013, Ho2014,Liu2019,
		Rodriguez-Vega2020,Wang2022}. 
	
	Hence, there are different ways to find
	the LZSM transition probability, including shortcuts to finding the
	solutions without solving the differential Schr\"{o}dinger equation.
	However, being interested in the complete wave function—not only in the
	transition probability—we emphasize that this can be done only by one of the
	differential equation methods \cite{Child1974a, Nikitin1999}. We illustrate
	this in the last column of Table~\ref{Table:LZSM}, which responds to the following
	question: Can the method be directly applied to derive the phase factor
	acquired after the transition? Only two answers are positive, and we address
	these in Appendices~\ref{LZSM_by_Zener}~and~\ref{App_Majorana}. Namely, we examine the
	approaches by Zener and Majorana, where the former is quite known and the
	latter much less so. For these reasons, we present Majorana's approach briefly
	in Appendix~\ref{Sec:AppendixA}, for the readers' convenience, with details
	given elsewhere \cite{Kofman2022}.
	
	\subsection{Different properties of the transition}
	
	\subsubsection{Adiabatic theorem}
	
	The adiabatic theorem is one of the oldest and most important theorems in
	quantum mechanics. It provides the foundation for various techniques
	(such as the adiabatic-impulse method described below) and for emergent devices (such as
	adiabatic quantum computers, also discussed below). The adiabatic theorem is
	limited by nonadiabatic transitions, making this natural to be discussed
	here.
	
	The adiabatic theorem states that in a system with a discrete energy
	spectrum under certain conditions, an infinitely slow—or adiabatic—change
	of the Hamiltonian does not change the level populations; for example, see Chapter
	1.5 in \cite{Zagoskin2011}. Let us now discuss this formulation and clarify
	those conditions.
	
	First, we note that it is not enough to formulate the adiabatic theorem as
	it is often formulated: a physical system remains in its instantaneous
	eigenstate if a given perturbation is acting on it slowly enough and if
	there is a gap between the eigenvalues \cite{Albash2018}. Even under slow
	perturbation, resonant and interference effects may result in significant
	changes in the energy-level populations. 
	  Below, in Sec.~\ref{Sec:AIM_Main}, we show that even with a small LZSM probability of excitation during a single passage  ($\mathcal{P}\ll1$), under a condition of constructive interference, the upper-level occupation probability would increase in a step-like manner. Then, during many driving periods, the occupation probability could reach significant values, up to unity, displaying as a result slow Rabi-like oscillations.  
	This could be
	termed the \textquotedblleft inconsistency\textquotedblright\ of 
	the adiabatic theorem~\cite{Marzlin2004}. Hence, we should add \textquotedblleft
	certain conditions\textquotedblright\ \cite{Amin2009, Tagliaferri2018,
		Hatomura2020} that can be formulated as either absence of resonance or as a limitation on the time duration of the process, which, in our example, means that the time span should be much less than the period of the Rabi-like oscillations.
	
	In the general case of a multilevel system, the eigenstates are defined by 
	\begin{equation}
		H(t)\left\vert E_{n}(t)\right\rangle =E_{n}(t)\left\vert
		E_{n}(t)\right\rangle .  \label{schrod}
	\end{equation}%
	Then, the adiabatic condition is usually quantified in either one of the two
	equivalent forms (e.g., \citet{Silveri2017}):
	
	\begin{equation}
		\left\vert \left\langle E_{m}(t)\right\vert \left. \dot{E}%
		_{n}(t)\right\rangle \right\vert \ll \left\vert \omega _{nm}(t)\right\vert
		\label{adiabaticity1}
	\end{equation}%
	or
	
	\begin{equation}
		\max_{t\in \left[ t_{0},t_{0}+\Delta t\right] }\frac{\left\langle
			E_{m}(t)\right\vert \dot{H}(t)\left\vert E_{n}(t)\right\rangle }{\omega
			_{nm}(t)^{2}}\ll 1,  \label{adiabaticity2}
	\end{equation}%
	where $\hbar \omega _{nm}(t)=E_{n}(t)-E_{m}(t)$ and $m\neq n$; the evolution
	is considered from $t=t_{0}$ until $t_{0}+\Delta t$. One can derive
	that Eq.~(\ref{adiabaticity2}) follows from Eq.~(\ref{adiabaticity1}) by
	differentiating Eq.~(\ref{schrod}). The interpretation of the adiabaticity
	condition~(\ref{adiabaticity1}) is that for all pairs of energy levels, the
	expectation value of the time rate of change of the Hamiltonian must be
	small compared with the gap \cite{Sarandy2004}. To be more precise, we could
	add the max and min, with respect to time, to the two sides of this inequality,
	respectively. The value standing in the left-hand-side of Eq.~(\ref{adiabaticity2})
	can be considered as the quantitative measure of adiabaticity \cite%
	{Skelt2018}.
	
	In particular, for a TLS, from Eq.~(\ref{adiabaticity2}) with Eqs.~(\ref%
	{Epm(t)},~\ref{eigenstates}), we obtain $\hbar v/\Delta ^{2}\ll 1$. This
	means $\delta \gg 1$, and explains why $\delta $ is called the \textit{adiabaticity
	parameter}. Indeed, in this \textit{adiabatic limit}, the nonadiabatic transitions
	are suppressed: $\mathcal{P}=\exp{\{-2\pi \delta \}}\rightarrow 0$ when $\delta
	\rightarrow \infty$. In general, the adiabaticity parameter $\delta $
	changes from zero with the diabatic transition, $\mathcal{P}=1$, and to
	infinity when the evolution is adiabatic and without nonadiabatic
	transitions, $\mathcal{P}=0 $. However, recall that for the
	adiabatic theorem to be fulfilled, the time step $\Delta t$ should be
	shorter than any possible resonance time, such as the Rabi period.
	
	\subsubsection{Dynamics and times of a transition}
	
	\label{Dynamics and times of a transition} We stated above that for linear
	driving, $\varepsilon =vt$, a TLS starting from the ground state with the
	LZSM\ probability $\mathcal{P}$ can then be found in the excited state. For
	a graphical representation of the problem with linear drive, see Fig.~\ref%
	{Fig:Energy_level_Scheme}(a). We consider this now in more detail by addressing
	the following questions: What is the system's dynamics $P_{+}(t)$? How does
	it change when not starting from the ground state? What are the
	characteristic times describing how $P_{+}(t)$ tends to $\mathcal{P}$? What
	changes if the driving is nonlinear?
	
	To start with, the dynamics depends on the representation.
	Both theoretically and experimentally, we can study evolution in various
	bases \cite{Tayebirad2010}. The most important bases are the diabatic and
	adiabatic ones, which we have introduced above. 
	Given the relevance of these two bases, both for theory and 
	measurements, we consider the dynamics and its characteristic features
	for the two cases. For further studies, see Ref.~\cite{Sun2015} on the
	experimental visualization of the single-passage dynamics; Refs.~\cite%
	{Barra2016, Thingna2017, Thingna2018} on many-level crossing in open
	quantum systems; and Refs.~\cite{Vitanov1996, Ribeiro2013a} on the finite
	coupling solution where $\Delta $ is a step function.
	
\begin{figure}[t]
		\centering{\includegraphics[width=1.0\columnwidth]{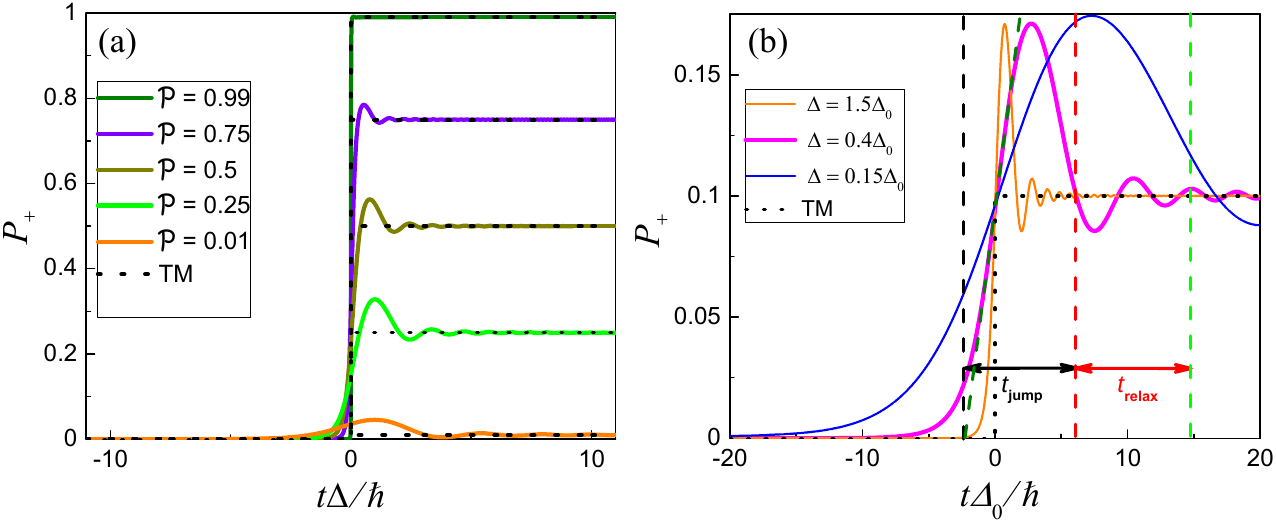}}
		\caption{\textbf{Single-passage Landau-Zener-St\"{u}ckelberg-Majorana (LZSM)
				transition} with different probabilities (a) and with different transition
			times for a fixed value of the transition probability $\mathcal{P}$ (b). In
			(a), we plot the time dependence of the upper-level occupation probability $%
			P_{+}(t)$, hence demonstrating dynamics for a given value of the final LZSM
			probability $\mathcal{P}$. Black dashed curves are given by the transfer
			matrix (TM) method, while the solid colored curves show the numerical
			solution of the Schr\"{o}dinger equation. In (b), all the curves are for a
			fixed LZSM probability $\mathcal{P}=0.1$, while both $\Delta $ and $v$ are
			varied; this demonstrates the impact of these parameters on the
			transient dynamics. Here, it is convenient to normalize time with a fixed
			value $\Delta _{0}$, to which also the $\Delta $ in the legend are related.
			The evolution is characterized by the two transition times: the jump time $%
			t_{\mathrm{jump}}$ and the relaxation time $t_{\mathrm{relax}}$, which are
			shown in the figure for the thick magenta curve.}
		\label{Fig:LZ transition}
	\end{figure}

	In the simplest approach, the dynamics is described by what we call the
	adiabatic-impulse model (AIM). Given its importance, we consider this in
	much more detail in the next section when we examine periodic driving. For the
	single-passage problem, the model consists of the adiabatic state following
	the ground state, then resonant (impulse-type) excitation to the upper level
	at the quasicrossing point, and then again the adiabatic state, now with a
	certain probability at occupying the excited state. These dynamics are shown
	in Fig.~\ref{Fig:LZ transition}(a) with dashed lines. Mathematically, such
	step-function behavior is conveniently described by the transfer matrix
	(TM) method, where each type of evolution is attributed to the respective
	matrices.
	
	Let us now clarify how accurate the TM approach is and what are the
	limitations on the application of the AIM \cite{Mullen1989}. Accordingly, we
	will solve the Schr\"{o}dinger equation exactly and describe the transient
	behavior by introducing relaxation times.
	
	There are two different ways to obtain the exact solution. The first one
	consists of the numerical solution for the Schr\"{o}dinger equation
	
	\begin{equation}
		\color{orange}\boxed{\color{black}i\hbar \frac{d}{dt}\left\vert \psi (t)\right\rangle =H(t)\left\vert \psi
		(t)\right\rangle, \color{orange}}\color{black}  \label{TDSE}
	\end{equation}%
	which can be used for all cases, including different nonlinear excitation
	signals and different initial conditions.
	
	The second approach involves solving the Schr\"{o}dinger equation with
	a linear excitation in terms of the parabolic cylinder functions $D_{\nu }(x)$ (see Appendix~\ref{LZSM_by_Zener}). This approach gives a simple expression for the
	probability in the diabatic basis $\left\{ \left\vert 0\right\rangle
	,\left\vert 1\right\rangle \right\} $. When solving the Schr\"{o}dinger
	equation following Zener, it is natural to first introduce the
	dimensionless time, $\tau $, and then the related complex value, $z$, which
	can be called the \textquotedblleft Zener\textquotedblright\ variable: 
	\begin{equation}
		\tau =t\sqrt{\frac{v}{2\hbar }},\quad z=\tau e^{i\pi /4} \sqrt{2}e^{i\pi /4}=t\sqrt{\frac{v}{\hbar}}.
		\label{tau}
	\end{equation}%
	Then, starting the evolution from the ground level, $\left\vert
	E_{-}\right\rangle $ to the left in Fig.~\ref{Fig:Energy_level_Scheme}(a),
	we can obtain the time-dependent solution for the upper level occupation probability in diabatic basis (see Appendix~\ref{LZSM_by_Zener})%
	\begin{equation}
		P_{\text{d}}(z)=\delta \exp\left(-\pi \delta/2\right)\left\vert D_{-1-i\delta
		}(-z)\right\vert ^{2}.  \label{P_d}
	\end{equation}%
	 To obtain transition dynamics for the upper level occupation probability in the adiabatic basis $\left\{ \left\vert E_{-}(t)\right\rangle, \left\vert E_{+}(t)\right\rangle \right\} $, we use formulas Eqs.~(\ref{eigenstates},\ref{gammas}),
	\begin{equation}
		P_{\text{a}}(z)=\exp{\left(-\pi \delta/2\right)}\left\vert D_{-i\delta }(-z)\gamma
		_{+}-\sqrt{\delta }e^{-\frac{i\pi }{4}}D_{-1-i\delta }(-z)\gamma
		_{-}\right\vert ^{2}. \label{P_a}
	\end{equation}%
	An analytical solution like this has the advantage that one does not need to
	find all the values of the wave function from the initial time to the desired
	moment of time; this is in contrast to the numerical solution, where we need to
	calculate all the previous values of the wave function between the current and
	initial times.
	
	In Fig.~\ref{Fig:LZ transition}, we illustrate the evolution of the
	upper-level occupation probability, emphasizing several different aspects.
	In Fig.~\ref{Fig:LZ transition}(a), we first fix several values of
	the final LZSM transition probability $\mathcal{P}$. These are defined by
	the adiabaticity parameter $\delta $, Eq.~\eqref{delta}. Inverting the
	relation for $\mathcal{P}$, Eq.~(\ref{P}), we obtain the expression 
	\begin{equation}
		\frac{\hbar v}{\Delta ^{2}}=-\frac{\pi }{2\ln{(\mathcal{P})}},
	\end{equation}
	which defines the ratio between $v$ and $\Delta $ for a given value of $%
	\mathcal{P}$. With the defined values of $v$ and $\Delta $, we plot in Fig.~%
	\ref{Fig:LZ transition}(a) both the analytical solution (dashed lines),
	which is the step function from $0$ to $\mathcal{P}$, and the numerical
	solution (shown with the solid lines). We emphasize that, for the numerical
	approach, we can equally use either the direct solution for the Schr\"{o}%
	dinger equation or the formulas above, that is, Eqs.~(\ref{P_d},~\ref{P_a}). Note
	that with an increasing $\mathcal{P}$, the evolution becomes more similar to
	the analytical solution: the step function. Importantly, Fig.~\ref{Fig:LZ
		transition}(a) vividly shows that the LZSM\ formula is robust and is valid
	in the whole range of TLS parameters, which was theoretically grounded
	in Refs.~\cite{Hagedorn1991, Joye1994, Vitanov1999,Nakamura2019}.
	
	In Fig.~\ref{Fig:LZ transition}(b), we take the fixed value of the LZSM\
	transition probability, $\mathcal{P}=0.1$. Then, for different curves, we
	simultaneously vary both $\Delta $ and $v$ to keep this $\mathcal{P}$
	constant; the values of $\Delta $ are displayed. Figure \ref{Fig:LZ
		transition}(b) demonstrates that for a given $\mathcal{P}$, the other
	parameters drastically influence the dynamics. We characterize this using two
	transition times: $t_{\mathrm{jump}}$ and $t_{\mathrm{relax}}$. Consider now
	the definition and calculation of these important values; the
	details are presented in Appendix~\ref{Time of the LZSM transition}.
	Importantly, the time scales of the transition processes are very different
	in the adiabatic and diabatic bases; thus, we describe the duration of
	the LZSM transition in both bases, following \cite{Vitanov1999b}.
	
	\begin{figure}[t]
		\centering{\includegraphics[width=1.0		%
			\columnwidth]{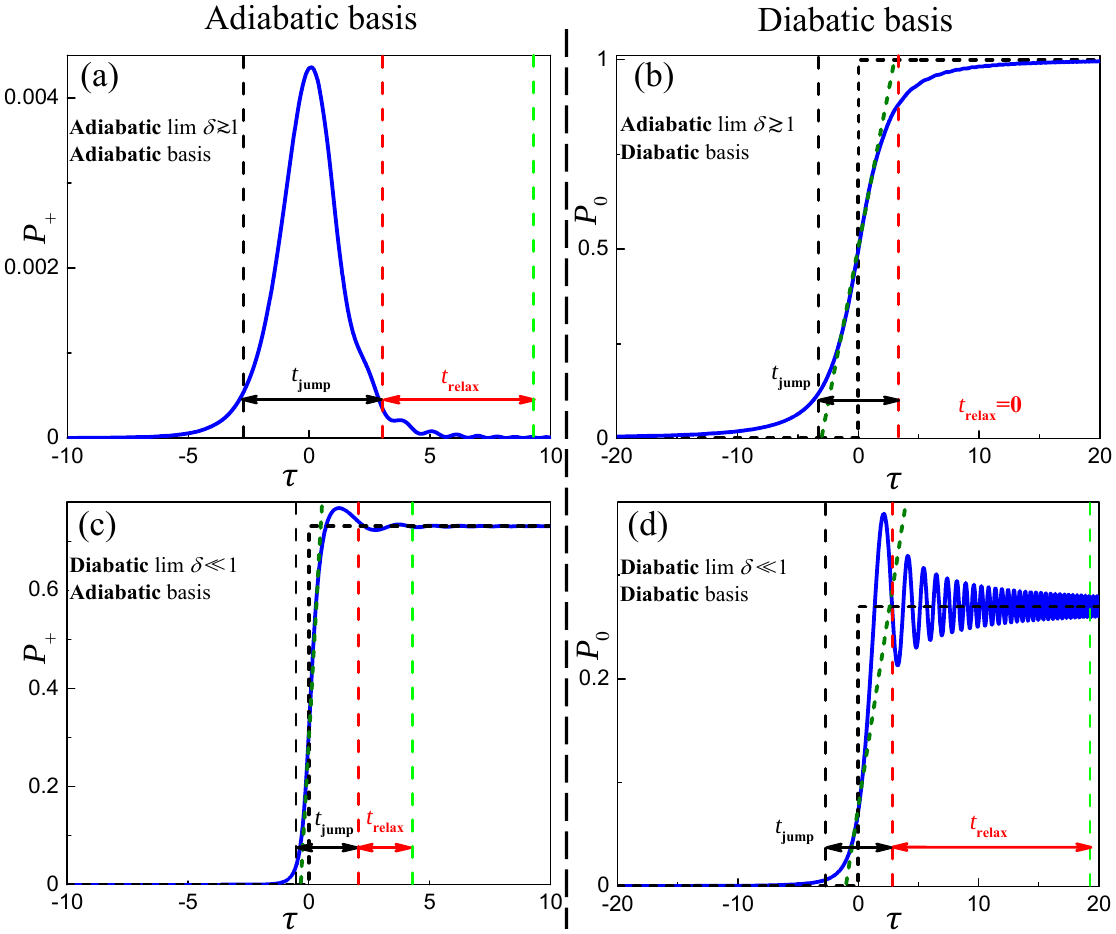}}
		\caption{\textbf{Dynamics of a single LZSM transition in the adiabatic basis
				(a,c) and in the diabatic basis (b,d).} We illustrate two limit cases of the
			adiabaticity parameter: $\protect\delta =2\gtrsim 1$ in $P_+(\protect\tau)$
			(a,b) and $\protect\delta =0.05\ll 1$ in $P_0(\protect\tau)$ (c,d). Horizontal black
			double arrows show $t_{\text{jump}} $ and red double arrows $t_{%
				\text{relax}}$. Inclined dashed green lines show the derivative at the
			avoided crossing point, $\protect\tau =0$. Recall that the dimensionless time is 	$\tau =t\sqrt{\frac{v}{2\hbar }}$. }
		\label{Fig:Singletrans in diff basises}
	\end{figure}
	
	Therefore, the transition process has two subsequent phases. The first one
	is when the probability jumps from the initial value $P(-\infty )$ to the
	vicinity of the final value $P(\infty )$. For the adiabatic basis, we have $P(\infty
	)=\mathcal{P}$, while for the diabatic basis, we have $P(\infty )=1-\mathcal{P%
	}$. To quantify this time span $t_{\text{jump}}$, we note that the slope
	at zero is approximated by $P^{\prime }(0)\sim \Delta P/\Delta t$; replacing $%
	\Delta P$\ with $P(\infty )$\ and $\Delta t$\ with $t_{\text{jump}}$, we
	come to the definition 
	\begin{equation}
		t_{\text{jump}}=\frac{P(\infty )}{P^{\prime }(0)}.
	\end{equation}
	However, this definition is not always appropriate, as we discuss in
	Appendix~\ref{Time of the LZSM transition}.
	
	The second phase of evolution is the time when the probability exhibits
	damping oscillations around its final value $P(\infty )$; the duration of
	this process is denoted as $t_{\text{relax}}$. This relaxation time can be
	quantified by introducing the small parameter $\eta \ll 1$, which describes
	that, after $t_{\text{relax}}$, the amplitude of the oscillations becomes less
	than $\eta P(\infty )$.
	
	We must distinguish the upper-level occupation probability in the
	adiabatic basis $P_{+}$ from the one in the diabatic basis $P_{0}$. The
	dynamics of the probabilities in different bases was theoretically
	investigated in \cite{Wubs2005, Danga2016} and experimentally in \cite%
	{Zenesini2009,Tayebirad2010}. We demonstrate the dynamics in these two
	bases in Fig.~\ref{Fig:Singletrans in diff basises}. For the adiabaticity
	parameter $\delta $, which describes the dynamics, we take two
	opposite limits: $\delta \ll 1$ (\textit{diabatic limit}) and $\delta \gg 1$
	(\textit{adiabatic limit}). To be more precise, in the adiabatic limit, we 
	take $\delta \gtrsim 1$ because with $\delta \gg 1$, the LZSM probability $%
	\mathcal{P}=\exp{(-2\pi \delta )}$ becomes too small. With these four
	possibilities, in Fig.~\ref{Fig:Singletrans in diff basises}(a-d), we can
	observe quantitatively different types of dynamics. 
	
	Note that the relaxation
	times are also very different in the adiabatic and diabatic bases. Namely,
	the total transition time $\left( t_{\mathrm{LZSM}}=t_{\mathrm{jump}}+t_{\mathrm{
			relax}}\right)$ in the adiabatic limit ($\delta \gtrsim 1$) is much longer in the
	adiabatic basis than in the diabatic one. The opposite is true, as well, where the transition
	time in the \textit{diabatic limit} ($\delta \ll 1$) is much longer in the diabatic
	limit than in the adiabatic one. In particular, for $\delta \gtrsim 1$, in
	the diabatic basis, there are no oscillations, which means there is zero relaxation
	time, $t_{\mathrm{relax}}^{\mathrm{d}}=0$.
	
	\begin{figure}[t]
		\centering{\includegraphics[width=1.0		%
			\columnwidth]{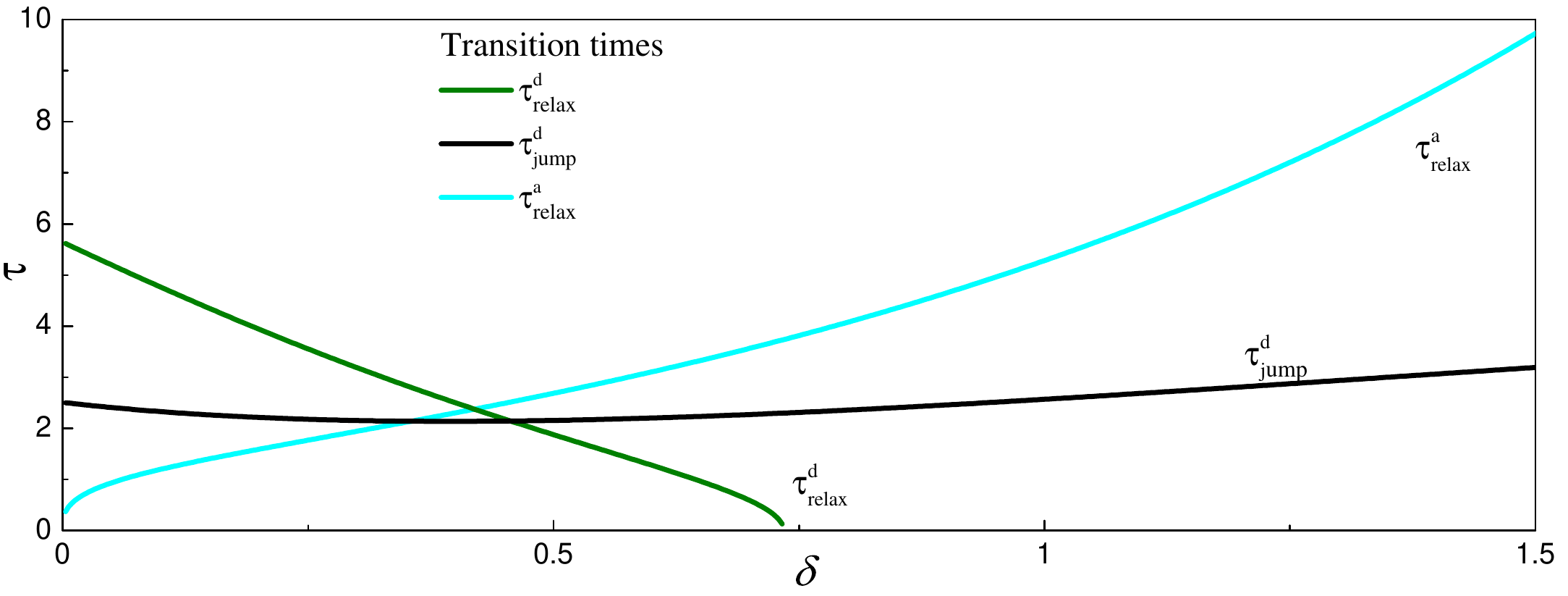}}
		\caption{\textbf{Transition times\ versus the adiabaticity parameter }$%
			\protect\delta $\textbf{. }We plot both $\tau_{\text{relax}}^{\text{d}}$ and $%
			\tau_{\text{jump}}^{\text{d}}$ for the transition in the diabatic basis and
			the relaxation time $\tau_{\text{relax}}^{\text{a}}$ in the adiabatic basis. We
			do not plot the jump time in the adiabatic basis because this is very close to
			the relaxation time, $\tau_{\text{jump}}^{\text{a}}\approx \tau_{\text{relax}}^{%
				\text{a}}$. Here, the dimensionless time is $\protect\tau =t\protect
				(v/2\hbar)^{1/2}$. See Appendix~\protect\ref{Time of the LZSM transition}
			for more details.}
		\label{Fig:Transition times in diff bases 1}
	\end{figure}
	
	The jump and relaxation times can be obtained analytically from the exact
	solution that describes a single LZSM transition with linear excitation; the
	details are presented in Appendix~\ref{Time of the LZSM transition}. Here, we
	define the simplified transition times $t_{\mathrm{LZSM}}$, which allow us
	to check the validity of the TM method. The characteristic time for a
	single-passage process is the sweeping time from the initial to the final
	state; if the driving is periodic, then the sweeping time equals half the
	period, $T_{\mathrm{d}}/2$. Hence, this characteristic time should be much
	larger than the LZSM\ transition time, which can be compactly written as
	follows (see Appendix~\ref{Time of the LZSM transition}):
	\begin{subequations}\begin{eqnarray}
		t_{\mathrm{LZSM}}^{\text{d}}\sim 4\sqrt{\frac{\hbar }{v}}\max \left\{ 1,%
		\sqrt{\delta }\right\},   \label{t_d} \\
		t_{\mathrm{LZSM}}^{\text{a}}\sim \pi \sqrt{\frac{\hbar }{v}}\max \left\{
		\left( \frac{\delta }{\eta }\right) ^{1/3},\left( \frac{\delta }{\eta }%
		\mathcal{P}\right) ^{1/6}\right\},  \label{t_a}
	\end{eqnarray}\end{subequations}
	in the diabatic and adiabatic bases, respectively. In Eq.\eqref{t_a}, $\eta \ll 1$ is the small
	parameter that describes the magnitude of the vicinity near the initial and
	final probabilities. These formulas are
	illustrated in Fig.~\ref{Fig:Transition times in diff bases 1}, using $\protect\tau =t\protect
	(v/2\hbar)^{1/2}$.
	
	\subsubsection{Problems with nonlinearities}
	
	In general, the bias $\varepsilon (t)$ is not a linear function of time. To
	obtain the linear model, Eq.~(\ref{eps(t)}), which we considered before, we
	need to linearize the otherwise nonlinear bias around the point $t_{0}$,
	where $\varepsilon (t_{0})=0$: 
	\begin{equation}
		\varepsilon (t)=\left(
		t-t_{0}\right)\left. \left(\frac{d\varepsilon }{dt}\right)\right\vert _{t=t_{0}} +o\left( t-t_{0}\right) .  \label{linearizing}
	\end{equation}%
	The linearization is appropriate if both the first derivative is nonzero
	and other terms are negligible. Then, the probability of the nonadiabatic
	transition is given by Eq.~(\ref{P}), with 
	\begin{equation}
		\delta =\frac{\Delta ^{2}}{4\hbar \left. \left(\frac{d\varepsilon }{dt}\right)\right\vert _{t=t_{0}}}.
	\end{equation}%
	Alternatively, instead of $\varepsilon (t)$, we can write the
	distance between the diabatic states: $\varepsilon (t)\rightarrow
	(E_{0}-E_{1}) $. Even more generally, instead of $\Delta /2$, we can write the
	off-diagonal part of the Hamiltonian as the corresponding matrix element $%
	V_{10} $; then, the adiabaticity parameter is written, as, for example, in Ref.~\cite%
	{Bendersky2013}, 
	\begin{equation}
		\delta =\left\vert \frac{V_{10}^{2}}{\hbar \left. \frac{d}{dt}\left(
			E_{0}-E_{1}\right) \right\vert _{t=t_{0}}}\right\vert .
	\end{equation}
	
	Let us now consider the cases where the bias cannot be linearized or
	nonlinear corrections are relevant. Different nonlinear-level-crossing
	models were analyzed in Refs.~\cite{Suominen1991,Vitanov1999,Ashhab2022}. In that study, the authors used the
	Dykhne–Davis–Pechukas formula [this appears in the Appendix~\ref{App:Landau}
	as Eq.~(\ref{fromLL})] to calculate the nonadiabatic transition probability
	when driven by different nonlinear biases $\varepsilon (t)$. 
	
	The biases can be grouped in two types: perturbative and essential nonlinearities.
	The former relates to the case when the linear term in Eq.~(\ref{linearizing}%
	) is dominant and the nonlinear corrections result in perturbative changes
	of the LZSM formula, while the latter relates to the case where there is no
	linear term and the essentially nonlinear bias is analyzed for the cases $%
	\varepsilon (t)\propto t^{N}$ with $N=2,3,...~$\cite{Shimshoni1991,
		Vitanov1999, Lehto2012, Lehto2015, Kapralova2022}.
	
	A special case of nonlinear LZSM\ tunneling relates to a TLS where the
	energy levels depend on the occupation of these levels. This may arise in a
	mean-field treatment of many-body systems where the particles predominantly
	occupy two energy levels \cite{Liu2002}. Then, the system is described by the
	Hamiltonian
	
	\begin{equation}
		H_{C}(t)=-\frac{\Delta }{2}\sigma _{x}-\frac{\varepsilon (t)}{2}\sigma _{z}-%
		\frac{C}{2}\left( P_{+}-P_{-}\right) \sigma _{z},
	\end{equation}%
	where $C$ is the parameter of nonlinearity, which describes the dependence
	of the energy levels on the state populations $P_{\pm }$. This parameter is
	not necessarily small and can be tunable. This form of the Hamiltonian is
	characteristic for the systems described by the Gross–Pitaevskii equation 
	\cite{Ishkanyan2010b, Liu2016c}. This Hamiltonian —both with the linear
	driving, $\varepsilon (t)\propto t$, and periodic driving, $\varepsilon
	(t)\propto \sin \omega t$ —may be experimentally realized in several ways
	with BECs~\cite{Zhang2008}. If taking the nonlinearity into
	account for such systems, all the features of the nonadiabatic transitions
	should be revisited, including the quantum adiabatic theorem \cite{Wu2003}.
	When $\varepsilon (t)$ is a sinusoidal driving field, we should focus
	on nonlinear LZSM interferometry \cite{Li2018b}.
	
	One more development of the linear LZSM problem can be obtained using an asymmetric linear
	bias, where the slopes are different on the left and right of the
	avoided-level crossing \cite{Damski2006}. When the bias is a
	nonanalytic function, similar cases were studied in Ref.~\cite{Garanin2002}. Various
	aspects of the nonlinearity of nonadiabatic transitions have been studied
	recently for realizations in such systems as a periodic lattice system \cite%
	{Takahashi2018}, spin qubit in a quantum wire \cite{Tchouobiap2018},
	superconducting qubits \cite{Wu2019b}, and topological systems \cite{Kam2020}.
	
	Therefore, one can consider versatile nonlinear biases in the contexts of 
	LZSM-like problems. At the end of this subsection, we 
	illustrate this also with the idea of reverse engineering \cite{Kang2022}. This is
	formulated as finding a Hamiltonian $\widetilde{H}(t)$ generating a given
	dynamics, here evolving in states that are the instantaneous eigenstates of a given
	Hamiltonian $H(t)$ \cite{Berry2009}. One formulation of this is the inverse
	LZSM problem, which is formulated as finding the bias $\varepsilon (t)$ resulting in
	any required time dependence of the level populations. This problem was
	formulated and solved in Ref.~\cite{Garanin2002a}. Then, in Ref.~\cite%
	{Shevchenko2012a}, a similar problem was studied for a qubit-resonator system
	as restoring a driven qubit Hamiltonian, provided its stationary behavior is
	known; see also \cite{Barnes2013}. Another problem related to the reverse
	engineering approach is transitionless quantum driving, which
	is analogous to the explanation of reflectionless potentials. This was
	studied both theoretically \cite{Berry2009, Villazon2021} and experimentally 
	\cite{Bason2012, Xu2018}. We address this in more detail in Sec.~\ref%
	{Sec:STA}, where we demonstrate that a Hamiltonian with appropriate
	nonlinear driving results in a transitionlesness  system dynamics.
	
	\subsection{Some experimental observations}
	
	\label{Sec:exprt_single}
	
	Driven TLSs are ubiquitous, which is true for the observations of the LZSM transition.
	In this section, we present illustrative examples of observations of the 
	\textit{single passage} LZSM transition.
	
	\begin{itemize}
		\item[*] Historically, the first works by Landau, Zener, and St\"{u}ckelberg
		related to inelastic atomic and molecular collisions. These described
		energy and charge exchange, as well as predissociation and associative
		recombination. The patterns observed in the scattering form the subject of
		collision spectroscopy. This was demonstrated for the inelastic
		scattering of He$^{+}$ by Ne \cite{Coffey1969}. In that work of more than half-century
		ago, the authors demonstrated both LZSM transitions and St\"{u}ckelberg
		oscillations.
		
		\item[*] As one realization of Majorana's problem, a spin moves
		through the point of a vanishing magnetic field, also considered 
		experimentally in Ref.~\cite{Betthausen2012}. In this case, a spin moved in a spin transistor over a
		distance of 50 micrometers while experiencing an adiabatically variable magnetic
		field. Alternation of adiabatic evolution and nonadiabatic transitions
		allows for accurate transistor control, making the spin transistor
		tolerant against disorder.
		
		\item[*] Nonadiabatic transitions were experimentally shown in a strong
		electric field between the Stark states of highly excited (Rydberg) states
		in lithium \cite{Rubbmark1981}. In contrast to molecular collisions, all the parameters there can be controlled accurately, which allowed
		obtaining quantitative agreement with theory in the two-level approximation,
		regarding their multilevel energy diagram. This allowed to better understand the
		dynamics of Rydberg atoms in rapidly rising electric fields.
		
		\item[*] The tunneling dynamics of a BEC of
		ultracold atoms in a tilted periodic potential is realized by accelerating
		the lattice \cite{Zenesini2009, Tayebirad2010}. Researchers studied LZSM
		tunneling both in the diabatic and adiabatic bases. Another study of an
		ultracold Fermi gas in a tunable honeycomb lattice was presented in Ref.~%
		\cite{Uehlinger2013}. The authors realized two successive LZSM transitions
		without interference after sequentially passing through two Dirac points.
		The authors of Ref.~\cite{Thalhammer2006} experimentally demonstrated
		the association and dissociation of the so-called Feshbach molecules, which are
		the molecules of a BEC formed by means of a Feshbach resonance.
		
		\item[*] In quantum systems, it is often the case that many levels are
		relevant, and this will be a subject of one of the following sections. However,
		sometimes, only the dynamics of two close levels is relevant. We find this in the
		experiment of \cite{Zhao2017} on a large-spin system, $S=7/2$. The
		authors observed the nonadiabatic dynamics around one of the avoided-level
		crossings controlled with an external magnetic field in a Gd$^{3+}$ impurity
		ion ensemble, which makes a qubit system with \textquotedblleft a virtually
		unlimited relaxation time\textquotedblright .
		
		\item[*] Qubit states in the NV center in diamond also have
		the advantage of good isolation. A single NV center electronic spin was
		coupled with a single nitrogen nuclear spin, creating a hybridized
		electronic-nuclear state \cite{Fuchs2011}. In this case, the LZSM transition
		can transfer the excitation between the two subsystems, which was proposed
		as a basis for a room temperature quantum memory. To create the avoided-level crossing 
		with the NV centers, the authors of Ref.~\cite{Xu2017} first
		applied a resonant microwave and considered the RWA, and then they explored the
		adiabatic evolution and nonadiabatic transitions.
		
		\item[*] Micrometer-size superconducting qubits allow the realization of
		macroscopically distinct quantum states. The nonadiabatic transitions
		between them were demonstrated for a variety of JJ-based
		qubits, including the flux \cite{Izmalkov2004}, quantronium \cite{Ithier2005},
		charge (Cooper pair sluice) \cite{Gasparinetti2012}, and phase \cite{Tan2015}
		qubits. In these works, it was demonstrated that such measurements are useful
		for probing and controlling both the qubits themselves and the coupled
		microscopic systems, hence providing a fast and sensitive tool to study and control qubits.
		
		\item[*] Another mesoscopic-size platform is provided by (double-)quantum
		dots. LZSM transitions were studied in singlet-triplet qubits in silicon
		in \cite{Harvey-Collard2019,Khomitsky2022}. In this case, tunneling was demonstrated
		to be useful to extract the spin-orbit coupling. Besides the electronic degrees
		of freedom, quantum dots can be used to manipulate the nuclear spin ensemble
		with chirped magnetic resonance pulses \cite{Munsch2014}.
		
		\item[*] Somewhat unexpectedly, LZSM physics is related to second-order
		phase transitions, of which the dynamics is described by the KZM. We discuss this in the
		section~\ref{subsubsection:KZM}, but here, to better 
		describe the second-order phase transitions dynamics, we
		present a variety of systems in which the KZM was experimentally studied in laboratories. The
		KZM correctly predicts the creation of topological defects during a
		single passage through a symmetry-breaking transition. Following the
		original proposal by \cite{Zurek1985}, defects (vortices) in
		superfluid $^{4}$He were created during the phase transition induced by fast
		expansion through the critical density, crossing the $\lambda $--line on the
		pressure--temperature\ phase diagram\ \cite{Hendry1994}.\ The vortices in
		superfluid $^{3}$He were created using neutron-induced nuclear reaction ($n+$
		$^{3}$He$\rightarrow $ $^{3}$H$+p$) to heat small regions of superfluid $%
		^{3} $He above the superfluid transition temperature\ \cite{Baeuerle1996,
			Ruutu1996}.\ The probability to trap a single flux line in annular JJs was 
		shown to work because of a causal KZM rather than because of thermal
		activation \cite{Monaco2009}. Other examples include zig-zag defects in
		buckled chains of ions (introduced above, in Fig.~\ref{Fig:Fig1}), defect
		textures in liquid crystals, flux lines in superconductors quenched through
		the critical temperature, and spin domains in Bose condensates realized in a
		shaken optical lattice. For a review, see \cite{Pyka2013, Hedvall2017,Dziarmaga2010}.
		
		\item[*] The analogy between nonadiabatic transitions in classical and
		quantum mechanics has been known for quite a long time \cite{Maris1988}. However,
		this idea was realized only recently on mechanical resonators \cite%
		{Faust2012}, in which the authors demonstrated that the energy transfer between
		the two modes of a nanomechanical resonator obeys LZSM behavior. We
		explore analogies and studies in a separate section~\ref{Sec:Classics}.
	\end{itemize}
	
	For a description of additional experimental observations, see Sec.~\ref%
	{Sec:Interferometry}. We would like to note that in many cases it is
	possible to observe either single-passage LZSM transitions or
	multiple-passage LZSM interference, which are described here and in Sec.~\ref%
	{Sec:Interferometry}, respectively. For this reason, we describe the related
	experiments in these two places. Before moving to the latter observations, let
	us consider the underlying physics first.
	
	\subsection{Transfer matrix (TM) method}
	
	\label{TM}
	
	Consider now the dynamics during a single-passage transition as a sequence of
	three stages:\begin{itemize} 
	\item (i) adiabatic evolution starting from the time $t_{\mathrm{i}}<0$
	until an avoided-level crossing is reached, 
	\item (ii) a nonadiabatic transition very near $t=0$, and 
	\item (iii) adiabatic evolution starting from an avoided-level crossing passed until the time $t_{\mathrm{f}}=-t_{\mathrm{i}}>0$. \end{itemize}
	
	The wave function can be expanded in the basis of the \textit{adiabatic} eigenstates $%
	\left\vert E_{\pm }(t)\right\rangle $:%
	\begin{equation}
		\left\vert \psi (t)\right\rangle =\alpha (t)\left\vert E_{-}(t)\right\rangle
		+\beta (t)\left\vert E_{+}(t)\right\rangle =\binom{\alpha (t)}{\beta (t)}.
	\end{equation}%
	The normalization condition results in all the transfer matrices being
	unitary ones. Consider this for both (i,iii) adiabatic and (ii) nonadiabatic
	evolutions. For the former, from a nonstationary Schr\"{o}dinger
	equation, Eq.~(\ref{TDSE}), we obtain the adiabatic time-evolution operator 
	\begin{equation}
		\color{orange}\boxed{\color{black}U(\zeta(t,t_{\mathrm{i}}))=%
		\begin{pmatrix}
			\exp{\left[-i\zeta (t,t_{\mathrm{i}})\right]} & \ 0 \\ 
			0 & \exp{\left[i\zeta (t,t_{\mathrm{i}})\right]}%
		\end{pmatrix}%
		=\exp{\left[-i\zeta(t,t_\text{i}) \sigma _{z}\right]}, \color{orange}}\color{black} \label{AdiabaticEvolution}
	\end{equation}%
	where $\zeta (t,t_{\mathrm{i}})$ is the phase accumulated during the adiabatic evolution from the time $t_\text{i}$ until the time moment $t$
	\begin{equation}
		\zeta (t,t_{\mathrm{i}})=\frac{1}{2\hbar }\int_{t_{\mathrm{i}}}^{t}\Delta
		E(t)\text{ }dt. \label{zeta}
	\end{equation}%
	Hence, the adiabatic evolution is described by the relation%
	\begin{equation}
		\left\vert \psi (t)\right\rangle =U(\zeta(t,t_{\mathrm{i}}))\left\vert \psi (t_{%
			\mathrm{i}})\right\rangle, 
	\end{equation}%
	which corresponds to no transitions between the adiabatic states, with only
	the phase difference accumulated.
	
	Next, consider an impulse-type transition at $t=0$. In Appendix~\ref{Sec:AppendixA}, we obtain the transition matrix in the diabatic basis; after
	transferring from the diabatic basis to the adiabatic one, using Eq.~\eqref{gammas} and assuming $\varepsilon (t)\gg \Delta $, we obtain the
	transition matrix%
	\begin{equation}
		\color{orange}\boxed{\color{black}N=%
		\begin{pmatrix}
			Re^{-i\phi _{\text{S}}} & -T \\ 
			T & Re^{i\phi_{\text{S}}}%
		\end{pmatrix}%
		,\color{orange}}\color{black}  \label{TMDiabatic}
	\end{equation}%
\begin{subequations}
	where \begin{eqnarray}
		T=\sqrt{\mathcal{P}}=\text{  Transition coefficient,}\\
		R=\sqrt{1-\mathcal{P}}=\text{  Reflection coefficient;} \label{Transition_and_Reflection}
	\end{eqnarray}\end{subequations}  
	$\phi _{\text{S}}$  is the Stokes phase \eqref{StocesPhase}, which appears in the theory of second-order
	differential equations. For more details, see \cite{Nikitin1972, Child1974,
		Kayanuma1997, Gasparinetti2011}, particularly for how this phase
	appears in terms of the Bloch vector evolution and Berry phase
	accumulation. Hence, the impulse-type nonadiabatic transition at around $t=0$
	is described by 
	\begin{equation}
		\left\vert \psi (+0)\right\rangle =N\text{ }\left\vert \psi
		(-0)\right\rangle .
	\end{equation}%
	Note that given the asymptotics of the gamma function, the monotonous
	function ${\phi }_{\text{S}}(\delta )$ changes from $0
	$ in
	the \textit{adiabatic limit} ($\delta \gg 1$) to $\pi /4$ in the \textit{diabatic limit} ($%
	\delta \ll 1$).
	
	Now, we can define the total single-transition evolution matrix in the general case $\zeta_\text{i}\equiv\zeta(0,t_\text{i})\neq\zeta(t_\text{f},0)\equiv\zeta_\text{f}$
		\begin{equation}
		U(\zeta_\text{f})NU(\zeta_\text{i})=%
		\begin{pmatrix}
			R\exp{[-i(\phi _{\text{S}}+\zeta_\text{i}+\zeta_\text{f} )]} & -T\exp{[-i(\zeta_\text{i}-\zeta_\text{f})]} \\ 
			T\exp{[i(\zeta_\text{i}-\zeta_\text{f})]} & 
			R\exp{[i(\phi _{\text{S}}+\zeta_\text{i}+\zeta_\text{f} )]}
		\end{pmatrix}.
		\label{SingleTRansMatrixDiffZeta}
	\end{equation}
	 We also can find the total single-transition evolution matrix in the case of symmetric adiabatic evolution $\zeta(0,t_\text{i})=\zeta(t_\text{f},0)\equiv\zeta$, 

	\begin{equation}
		U(\zeta)NU(\zeta)=%
		\begin{pmatrix}
			R\exp{[-i(\phi _{\text{S}}+2\zeta )]} & -T \\ 
			T & R\exp{[i(\phi _{\text{S}}+2\zeta )]}
		\end{pmatrix}. 
	\label{SingleTRansMatrix}
	\end{equation}
	Note that when we consider the inverse transition we should replace the direct-transition matrix $N$ with the inverse transition matrix \begin{equation}
		N_\text{inverse}\equiv N^{\top}, \label{N_Inverse}
	\end{equation}  which is the transposed matrix, see Eq.~\eqref{LZSM_back_transition_matrix}, 
	and Ref.~\cite{Teranishi1998}.
	
	As a generic initial condition at $t=t_{\mathrm{i}}$, we take a superposition
	state 
	\begin{equation}
		\left\vert \psi (t_{\mathrm{i}})\right\rangle =\alpha (t_{\mathrm{i}%
		})\left\vert E_{\_}(t_{\mathrm{i}})\right\rangle +\beta (t_{\mathrm{i}%
		})\left\vert E_{+}(t_{\mathrm{i}})\right\rangle,
	\end{equation}
	where $\left\vert E_{\pm }(t)\right\rangle $ are the instantaneous
	eigenstates of the time-dependent Hamiltonian $H(t)$ and 
	\begin{equation}
		\begin{cases}
			\alpha (t_{\mathrm{i}})=\sqrt{P_{-}(t_{\mathrm{i}})}, \\ 
			\beta (t_{\mathrm{i}})=\sqrt{P_{+}(t_{\mathrm{i}})}e^{i\phi _{\text{i}}},%
		\end{cases}%
	\end{equation}%
	where $P_{\pm }(t_{\mathrm{i}})$ are the occupation probabilities of the
	respective states and $\phi _{\mathrm{i}}$ describes the initial phase
	difference. Importantly, for a superposition state, the phase difference
	significantly influences the dynamics \cite{Emmanouilidou2000, Wubs2005}.
	Now, with the evolution matrix $UNU$, we can obtain the final upper-level
	occupation probability 
	\begin{equation}
		P_{+}(t_{\mathrm{f}})=T^{2}P_{-}(t_{\mathrm{i}})+R^{2}P_{+}(t_{\mathrm{i}%
		})-2RT\sqrt{P_{-}(t_{\mathrm{i}})P_{+}(t_{\mathrm{i}})}\cos (-{\phi 
		}_{\text{S}}-2\zeta +\phi _{\text{i}}).  \label{P_+Final}
	\end{equation}%
	This formula describes several important aspects. First, when the
	cosine equals $+1$ or $-1$, we have maximal and minimal excitation
	probabilities $P_{+}^{\max /\min }(t_{\mathrm{f}})$, respectively. These
	correspond to the constructive and destructive interference of the incoming
	states. The respective conditions are 
	\begin{equation}
		\begin{cases}
			-{\phi }_{\text{S}}-2\zeta +\phi _{\text{i}}=2\pi n\:\:\:\:\:\:\:\:\:\:\:\:\:\:\:\text{ for }
			P_{+}^{\max }, \\ 
			-{\phi }_{\text{S}}-2\zeta +\phi _{\text{i}}=2\pi (n+\frac{1}{2})\:\:\:
			\text{ for }P_{+}^{\min },%
		\end{cases}
		\label{Cond_for_Constr_Destr_Interf}
	\end{equation}%
	where $n$ is an integer.
		\begin{figure}[t]
		\centering{\includegraphics[width=1.0				%
			\columnwidth]{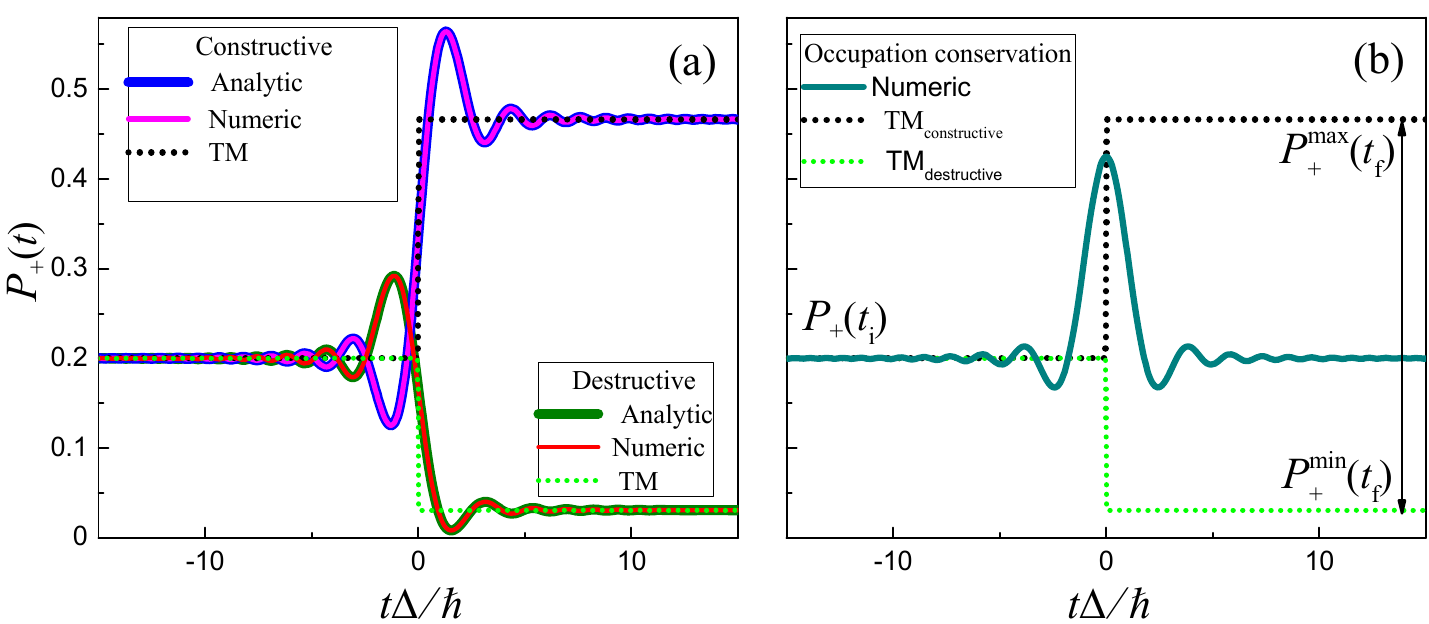}}
		\caption{\textbf{Nonadiabatic transition with a superposition state.}
			Comparison of the numerical, analytical, and transfer matrix (TM) solutions for
			a single LZSM transition starting from a superposition state. For this, we took $\protect\alpha (t_{\text{i}})=\protect\sqrt{0.2}$
			and $\protect\beta (t_{\text{i}})=\protect\sqrt{0.8}e^{i\protect\phi _{\text{%
						i}}}$, here with two cases for the phase difference: one for a destructive
			interference case, $\protect\phi _{\text{i}}\ =5\protect\pi /3$, and another
			case for constructive interference, $\protect\phi _{\text{i}}\ =2%
			\protect\pi /3$ in (a), which can realize the maximum and minimum possible
			values of the final probability, Eqs.~(\protect\ref{P_+Final},~\protect\ref{Cond_for_Constr_Destr_Interf}), and the probability-conserving case in
			(b), here with the initial
			phase difference defined in Eq.~\protect\eqref{Probability_conservation_phase_difference} and a transition without changes in the probability. The black double-arrow
			shows the range of the possible final values of the probability. This
			demonstrates the dramatic dependence of the evolution on the initial phase
			difference and the role of interference.}
		\label{Fig:Exact_LZ_Transition}
	\end{figure}
	Second, note that the range between the extremal values $P_{+}^{\min }$
	and $P_{+}^{\max }$ includes the initial probability $P_{+}(t_{\mathrm{i}})$. 
	This means that we can select the value of the initial phase difference $%
	\phi _{\text{i}}$, which gives us the transition without any change of the
	probability so that 
	\begin{equation}
		P_{+}(t_{\mathrm{f}})=P_{+}(t_{\mathrm{i}}).
	\end{equation}%
	This process can be named \textit{occupation-conserving transition}
	(OCT). This takes place for the phase difference $\phi _{\text{i}}=\phi _{%
		\text{i}}^{\mathrm{OCT}}$ 
	\begin{equation}
		\phi _{\text{i}}^{\mathrm{OCT}}=-{\phi }_{\text{S}}-2\zeta +\arccos %
		\left[ \frac{T\left( P_{+}(t_{\mathrm{i}})-1/2\right) }{R\sqrt{P_{+}(t_{%
					\mathrm{i}})P_{-}(t_{\mathrm{i}})})}\right] .
		\label{Probability_conservation_phase_difference}
	\end{equation}%
	Note that this is possible only for a superposition initial state; when
	starting from a ground state, there is no effect on the phase difference.

	In the case of starting from a superposition initial state,
	all these dynamical features are illustrated in Fig.~\ref%
	{Fig:Exact_LZ_Transition}. In this figure, we compare the numerical solution with the
	analytical one, which is given by Eq.~\eqref{Exact_Zener_Solution}, with a perfect
	agreement between the two. In addition, we show the asymptotic solution
	described by the transfer matrix method with a step-like transition, as
	described by the equations above. For the calculations, we take the
	adiabaticity parameter $\delta =0.4$, which corresponds to the LZSM
	probability $\mathcal{P}=0.08$. In Fig.~\ref{Fig:Exact_LZ_Transition}, we
	present the solutions for three different cases: for constructive
	interference with initial phase difference $\phi _{\text{i}}\ =\phi
	_{1}=2\pi /3$, for destructive interference with $\phi _{\text{i}}\ =\phi
	_{2}=5\pi /3$, and for the probability-conserving case with $\phi _{\text{i}%
	}=\phi _{\text{i}}^{\mathrm{OCT}}$, Eq.~\eqref{Probability_conservation_phase_difference}. Note that $\mathcal{P}$
	is the probability of excitation if starting from the ground state, while
	now, we can demonstrate a more general case of starting from the superposition
	state. This demonstrates that the upper-level occupation probability 
	$P_{+}(\infty )$ is essentially different from $\mathcal{P}$ and 
	that this is defined by not only $\delta $, but also by the initial
	condition.
	
	\section{Repetitive passage: interference}
	
	\label{Sec:Repetitive}
	
	In the previous section, we considered a TLS when driven by a linear
	drive $\varepsilon (t)=vt$. From now on, we consider the evolution for a generic
	periodic bias with an offset $\varepsilon_{0}$,%
	\begin{equation}
		\color{red}\boxed{\color{black}\varepsilon (t)=\varepsilon _{0}-A\cos \omega t. \color{red}}\color{black} \label{eps_with_cos}
	\end{equation}%
	We now consider several approaches.
	
	\subsection{Adiabatic-impulse model (AIM)}
	
	\label{Sec:AIM_Main}
	
	\subsubsection{Double-passage and multiple-passage cases}
	
	The adiabatic-impulse model is possibly the most intuitive model for
	describing the repetitive LZSM passage \cite{Damski2006,Tomka2018}. In this model,
	we consider the evolution of the TLS beyond the avoided-crossing region as
	adiabatic evolution, and in the avoided-crossing region, we consider the diabatic
	evolution of the TLS. Also, we approximate the avoided-crossing region by the
	point of minimum distance between energy levels. Therefore, we have the
	adiabatic evolution everywhere save for the points of the minimal distance
	between energy levels, where nonadiabatic LZSM transitions occur. Thus, the AIM consists in that the evolution is modeled (approximated) as adiabatic one besides the non-adiabatic transitions in the avoided-level-crossing points (where the driving is approximated as a linear one); given these approximations, instead of AIM this technique can alternatively be called \textit{adiabatic-impulse approximation}. 
	Essentially, the AIM is described by using the TM method. The
	latter was developed by \cite{Bychkov1970,Averbukh1985, Kayanuma1993, Vitanov1996, Garraway1997, Teranishi1998,Delone2012}.
	
	In the section \textquotedblleft Transfer Matrix Method\textquotedblright\,
	we introduced the matrices $U$ and $N$ for the adiabatic evolution and
	nonadiabatic impulse-type transition, respectively. This describes
	the evolution during half the period. When
	speaking about driven systems, it is illustrative to sequentially consider 
	three cases: single-passage transition probability, double-passage case, and
	the multiple-passage transition probabilities \cite{Nikitin2006}. We now
	consider the evolution during one full period and then during many periods,
	to which we refer to as double-passage and multiple-passage cases,
	respectively.
	
\begin{figure}[tbp]
		\centering{\includegraphics[width=1.0\columnwidth]{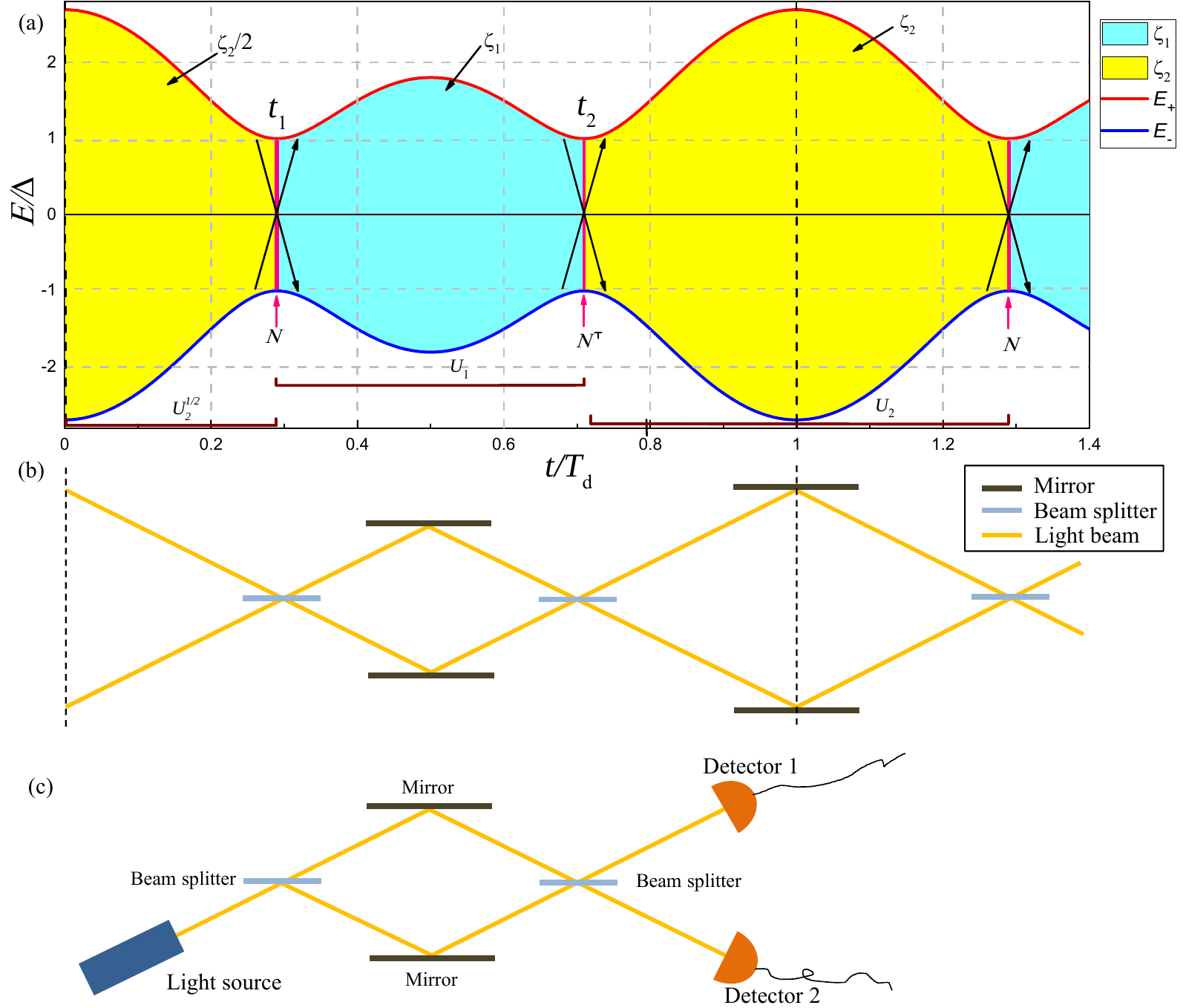}}
		\caption{\textbf{Evolution of energy levels in the  adiabatic-impulse model (AIM).} (a) The time-dependent
			adiabatic energy levels $E_{\pm }(t)$ define two stages of evolution: first,
			the adiabatic evolution with the transfer matrices $U_{1}$ and $U_{2}$ and,
			second, the transition in the region of avoided level crossing as determined by
			the matrix $N$.  Here we consider a single-period evolution from $t=0$ to $t=T_\text{d}$, which is the time between the two vertical dashed black lines in panel (a). This evolution is described by $\sqrt{U_{2}}N_{\text{inverse}}U_{1}N\sqrt{U_{2}}$, in terms of time-evolution matrices, where $\sqrt{U_2}=U(\zeta_2/2)$. 
		(b) Equivalent optical scheme based on Mach-Zehnder interferometer, where the optical beam splitters are analogous to the transition points in (a). The trajectories can have different lengths (e.g. due to the moving upper mirror) resulting in the relative phases of the two partial rays, in analogy with the wave-function phase difference accumulated during the LZSM transitions in (a). (c) Typical scheme of the optical Mach-Zehnder interferometer, analogous to the double-passage LZSM problem.}
		\label{Fig:AIMscheme}
	\end{figure}

	The adiabatic energy levels 
	\begin{equation}
		E_{\pm }(t)=\pm \frac{\Delta E(t)}{2}=\pm \frac{1}{2}\sqrt{\Delta
			^{2}+\varepsilon (t)^{2}}
	\end{equation}%
	have a minimum distance (equal to $\Delta $) at times $t_{1,2}+nT_{\mathrm{d}%
	}$, where $\omega t_{1}=\arccos (\varepsilon _{0}/A)$ and $\omega t_{2}=\pi
	+\omega t_{1}$, see Fig.~\ref{Fig:AIMscheme}(a). 
	
	The adiabatic evolution is described by%
	\begin{equation}
		U_{1,2}=\exp (-i\zeta _{1,2}\sigma _{z})
	\end{equation}%
	with the phase differences%
	\begin{equation}
		\zeta _{1}=\frac{1}{2\hbar }\int\limits_{t_{1}}^{t_{2}}\Delta E(t)dt,\ \ \ \
		\ \zeta _{2}=\frac{1}{2\hbar }\int\limits_{t_{2}}^{t_{1}+T_{\mathrm{d}%
		}}\Delta E(t)dt.  \label{zeta_1_2}
	\end{equation}%
	The nonadiabatic transitions between the states $\left\vert E_{\pm
	}(t)\right\rangle $ are described by the transfer matrix~$N$, which is true for both
	transitions and corresponds to the sweeping occurring both to the right and left
	in Fig.~\ref{Fig:Energy_level_Scheme}, respectively. Note that the evolution
	in the diabatic basis should be described differently \cite{Ashhab2007,
		Shevchenko2010}.  In some papers \cite{Shevchenko2010} double passage evolution is described starting at the quasicrossing point and finishing also at the quasicrossing point, see \cite{Vitanov1996,Cucchietti2007}. And that way gives a correct result for averaged level occupations under the periodic driving $\Xi = N_{\text{inverse}}U_{2}NU_{1}=\left( 
	\begin{array}{cc}\Xi _{11} & \Xi _{12} \\ -\Xi _{12}^{\ast } & \Xi_{11}^{\ast }\end{array}\right)$. 
	
	It is more convenient to describe the evolution starting far from the quasicrossing point and finishing also far from it. 
	Then the double-passage evolution takes place after the full period and is
	described by the double-passage transfer matrix 
	\begin{eqnarray}
		\color{orange}\boxed{\color{black}\Xi \equiv \sqrt{U_{2}}N_{\text{inverse}}U_{1}N\sqrt{U_{2}}=\left( 
		\begin{array}{cc}
			\Xi _{11} & \Xi _{12} \\ 
			\Xi _{12} & \Xi _{11}^{\ast }%
		\end{array}%
		\right), \color{orange}}\color{black} \label{SingleTransitionMatrix}
	\end{eqnarray}%
 where
	\begin{subequations}\begin{eqnarray}
		\Xi _{11} &=&-R^2e^{-i\zeta _{+}}-T^2e^{-i\zeta _{-}}\\
		\Xi _{12} &=&
		-2iRT\sin(\Phi_\mathrm{St})=-\Xi_{12}^{\ast}, \label{Xi_12St}\\
		\zeta _{+}&=&\zeta _{1}+\zeta _{2}+2\phi_{\mathrm{S}},\text{ \ }\zeta _{-}=\zeta _{1}-\zeta _{2},\text{ \ 
	}\Phi _{\mathrm{St}}=\phi_{\mathrm{S}}+\zeta_1, 
	\end{eqnarray}\end{subequations} 
for the inverse transition matrix $N_\text{inverse}$ see Eq.~\eqref{N_Inverse}.
 We obtain the same $\Xi_{11}$ as in Ref.~\cite{Shevchenko2010} for the double transition, but $\Xi_{12}$ is different due to using a shifted driving signal. 
	
From Eq.~\eqref{Xi_12St}, one can see that the upper-level occupation probability, if
	starting from the ground state, becomes
	
	\begin{eqnarray}
		\color{green}\boxed{\color{black}P_{+}^{\mathrm{double}} =\left\vert \Xi _{12}\right\vert ^{2}=4\mathcal{P}%
		(1-\mathcal{P})\sin ^{2}\Phi _{\mathrm{St}}. \color{green}}\color{black} \label{with_Fi_St}
	\end{eqnarray}
	In this way, following Zener's approach we confirmed the St\"{u}ckelberg formula, Eq.~\eqref{P_double}.
	
	Here, an instructive \textit{analogy with the Mach–Zehnder interferometer} is
	appropriate \cite{Oliver2005, Petta2010, Burkard2010,Ma2011}. We illustrate this analogy graphically: Fig.~\ref{Fig:AIMscheme}(b) shows that our dynamics in Fig.~\ref{Fig:AIMscheme}(a) is analogous to multiple periodic Mach-Zehnder interferometers \cite{Oliver2005}, and Fig.~\ref{Fig:AIMscheme}(c) shows that our double-passage LZSM problem is analogous to an optical Mach-Zehnder interferometer \cite{Burkard2010}. Namely, passing the avoided-level crossing is analogous to light passing a partly transparent mirror functioning as a beam splitter with the coefficients $R$ and $T$. After the two beams meet, the outcome is the result of the interference, which depends on the phase difference $\Phi _{\mathrm{St}}$. For more about double-passage regime theory, see, e.g., \cite{Saxon1975,Garraway1992,Garraway1995, Nagaya2007,Gasparinetti2011}.
	
\begin{figure}[tbp]
		\centering{\includegraphics[width=1.0\columnwidth]{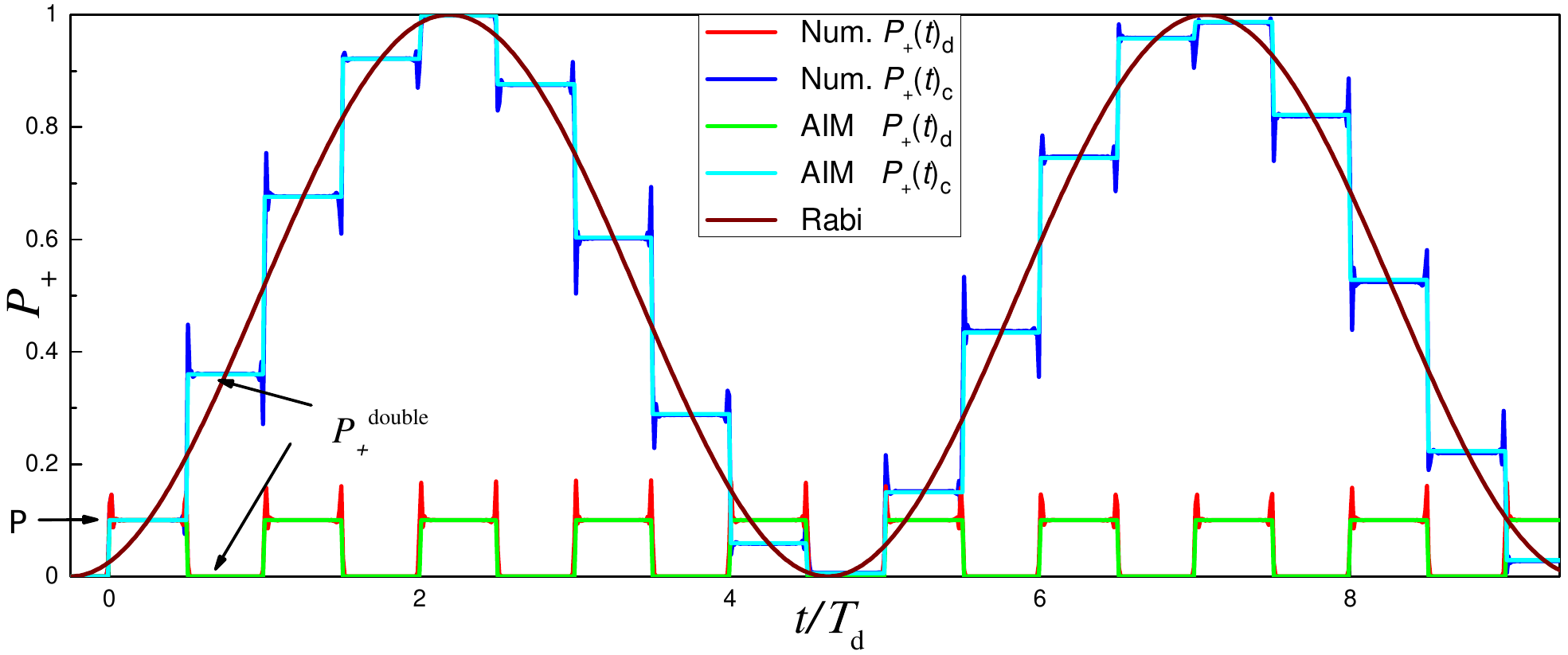}}
		\caption{ \textbf{Constructive versus destructive interference.} Comparison
			of the upper-level occupation probability obtained numerically and from the AIM
			in the regime of constructive interference (blue and light blue
			curves: $P_{+}(t)_{\mathrm{c}}$ calculated numerically and in the AIM,
			respectively) and destructive interference (red and green curves: $P_{+}(t)_{%
				\mathrm{d}}$ calculated numerically and in the AIM, respectively). The light
			blue and green curves present the analytical solutions, while the blue and red
			curves show numerical solutions. We take the following parameters: $%
			\protect\varepsilon _{0}=0$ (then $\protect\zeta _{1}=\protect\zeta _{2}$)
			and $\mathcal{P}=0.1$; and the conditions for the constructive and
			destructive interference, $\protect\zeta _{1}+{\protect\phi }_{%
				\mathrm{S}}=\protect\pi k$ and $=\protect\pi k+\frac{\protect\pi }{2}$,
			respectively; these conditions define $\protect\omega $ and $A$.  The brown line is plotted with Rabi frequency obtained in  Eq.~\eqref{AIM_Rabi_Frequency}. }
		\label{Fig:AIMRez}
	\end{figure}

	For the multiple-passage case, after $n$ full periods, the time evolution is
	described by the following evolution matrices:
	\begin{equation}
		U\left( t,t_{1}+nT_{\mathrm{d}}\right) \Xi ^{n}\:\:\ \ \text{ for }\:\:t-nT_{\mathrm{d}%
		}\in \left( t_{1},t_{2}\right),   \label{(I)}
	\end{equation}%
	\begin{equation}
		U\left( t,t_{2}+nT_{\mathrm{d}}\right) NU_{1}\Xi ^{n}\:\: \ \text{ for }\:\:t-nT_{%
			\mathrm{d}}\in \left( t_{2},t_{1}+T_{\mathrm{d}}\right) .  \label{(II)}
	\end{equation}%
	Hence, the system state after $n$ full periods of evolution is given by $\Xi
	^{n}$, which reads as \cite{Bychkov1970} 
	\begin{subequations}
		\label{zeta_pm}
		\begin{eqnarray}
			\Xi ^{n} &=&\left( 
			\begin{array}{cc}
				\Xi _{n11} & \Xi _{n12} \\ 
			-\Xi _{n12}^{\ast } & \Xi _{n11}^{\ast }%
			\end{array}%
			\right),   \label{Ksi_n} \\
			\Xi _{n11} &=&\cos n\phi +i\text{Im}\Xi _{11}\frac{\sin n\phi }{\sin \phi },%
			\text{ \ \ }\:\: \Xi _{n21}=\Xi _{21}\frac{\sin n\phi }{\sin \phi },\text{ \ }%
			\phi =\arccos \text{Re}\Xi _{11}.\text{\ }
		\end{eqnarray}%
		Then, for the respective upper-level occupation probability, if starting
		from the ground state, we obtain \newline $P_{+}(n)=\left\vert \Xi _{n12}\right\vert
		^{2}$, which gives 
	\end{subequations}
	\begin{equation}
		\color{green}\boxed{\color{black}P_{+}(n)=\left\vert \Xi _{12}\right\vert ^{2}\frac{\sin ^{2}n\phi }{\sin
			^{2}\phi }=\underset{P_{+}^{\mathrm{double}}}{\underbrace{4\mathcal{P}(1-%
				\mathcal{P})\sin ^{2}\Phi _{\mathrm{St}}}}\frac{\sin ^{2}n\phi }{\sin
			^{2}\phi }. \color{green}}\color{black} \label{P+AIM}
	\end{equation}%
	This describes the time evolution, with $n$ denoting the integer
	number of periods passed, as shown in Fig.~\ref{Fig:AIMRez}. Note the
	impressive agreement between the results of the AIM and numerics; see also 
	\cite{Mukherjee2018, Kuno2019}. For a similar description of the multilevel
	systems, see \cite{Qin2016, Neilinger2016, Niranjan2020, Suzuki2022}.
	
	\paragraph*{\textbf{Rabi oscillations from LZSM transitions}} \mbox{} \\
	
	Adiabatic dynamics is
	characterized by small steps which, under resonance condition, result in 
	Rabi-like oscillations \cite{Garraway1992a,Pu2000, Shevchenko2005}. This can be seen in
	Fig.~\ref{Fig:AIMRez}. The frequency of these Rabi-like oscillations can be
	found from Eq.~(\ref{P+AIM}) if we identify $\sin ^{2}(n\phi) $ with $\sin ^{2}%
	\left(\frac{\Omega _{\mathrm{R}}}{2}t\right)$ \cite{Ashhab2007, Neilinger2016, Liu2021}.
	Then, we note that during one driving period, the integer $n$ changes by unity,
	and this corresponds to changing the time $t$ by $2\pi /\omega $. With this,
	we obtain the relation for the coarse-grained oscillations%
	\begin{equation}
		\color{orange}\boxed{\color{black}\Omega _{\mathrm{R}}=\frac{\omega }{\pi }\left\vert \phi \right\vert =\frac{%
			\omega }{\pi }\arccos \left[ (1-\mathcal{P})\cos \zeta _{+}-\mathcal{P}\cos
		\zeta _{-}\right] .\color{orange}}\color{black}
		\label{AIM_Rabi_Frequency}
	\end{equation}%
	This formula correctly describes the \textit{Rabi oscillations induced by strong driving}, as studied in Refs.~\cite{Zhou2014, Neilinger2016}. Indeed, we
	note that for a small offset $\varepsilon _{0}/A\ll 1$, the expressions for $%
	\zeta _{\pm }$ can be simplified: 
	\begin{equation}
		\zeta _{1}+\zeta _{2}\approx 2A/\hbar \omega \:\:\text{ \ and \ }\:\:\zeta
		_{1}-\zeta _{2}\approx \pi \varepsilon _{0}/\hbar \omega.  \label{zetapm}
	\end{equation}%
	With this, in the \textit{adiabatic limit} ($\mathcal{P}\ll 1$), we obtain $\Omega
	_{\mathrm{R}}\sim A/\hbar $. This correctly describes the resonant Rabi
	frequency and justifies the term \textquotedblleft Rabi
	oscillations\textquotedblright, which we used above.
	
	The long-time averaged occupation probability,
	is given by averaging over large $n$, 
	\begin{equation}
		\color{green}\boxed{\color{black}\overline{P_{+}}=\frac{\left\vert \Xi _{21}\right\vert ^{2}}{2\sin ^{2}\phi }%
		=\frac{1}{2}\frac{\left\vert \Xi _{21}\right\vert ^{2}}{\left\vert \Xi
			_{21}\right\vert ^{2}+(\text{Im}\Xi _{11})^{2}}. \color{green}}\color{black} \label{P_I_avrgd}
	\end{equation}%
	It follows that the upper-level occupation probability $\overline{P_{+}}$ is maximal at Im$\Xi
	_{11}=0$. This results in the resonance condition:%
	\begin{equation}
		(1-\mathcal{P})\sin \zeta _{+}-\mathcal{P}\sin \zeta _{-}=0.
		\label{res_cond_gen}
	\end{equation}%
	In particular, in the adiabatic (slow) and diabatic (fast) limits, the
	resonance condition takes the following forms: 
	\begin{subequations}
		\begin{eqnarray}
			\zeta _{1}+\zeta _{2} &=&k\pi\:\:\:\:\: \text{ for }\delta \gg 1\:\:\:\:\:\text{(adiabatic)},  \label{B12a} \\
			\zeta _{1}-\zeta _{2} &=&k\pi\:\:\:\:\: \text{ for }\delta \ll 1\:\:\:\:\:\text{(diabatic)}.
		\end{eqnarray}%
	
			\begin{figure}[t]
		\centering{\includegraphics[width=1.0		%
			\columnwidth]{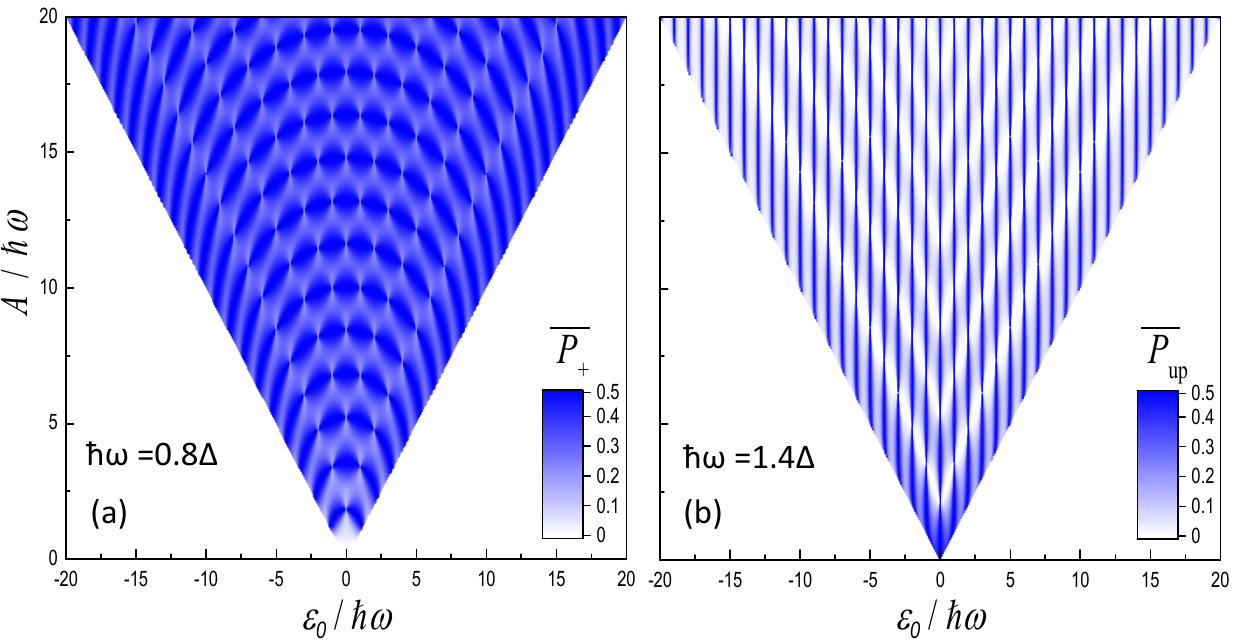}}
		\caption{\textbf{LZSM interferograms calculated within the adiabatic-impulse model (AIM)}. (a) The
			low-frequency driving corresponding to the slow-passage regime, Eq.~(\protect
			\ref{ShIF}), with $\hbar \protect\omega /\Delta =0.8<1$. (b) The
			high-frequency driving corresponds to the fast-passage limit, Eq.~(\protect
			\ref{Pupk}), with $\hbar \protect\omega /\Delta =1.4>1$.}
		\label{Fig:interferogramsAIM}
	\end{figure}
		With the limiting expressions (\ref{zeta_pm}), the resonance condition for
		the \textit{adiabatic limit} reads $2A\approx k\pi \hbar \omega $, and for the
		\textit{diabatic limit}, this gives $\varepsilon _{0}\approx k\hbar \omega $. Because in
		the \textit{diabatic limit} $\Delta $ is relatively small $(\Delta E\approx
		\left\vert \varepsilon _{0}\right\vert), $ the latter condition can be
		interpreted as an exchange of $k$ photons between the driving field and our
		two-level system.
		
		In the \textit{slow passage limit}, where $\delta \gg 1$ and $\mathcal{P}\ll
		1$, we directly obtain the time-averaged occupation probability of the upper
		state:
		
	\end{subequations}
	\begin{equation}
		\color{green}\boxed{\color{black}
		\overline{P_{+}}=\frac{\mathcal{P}(1+\cos \zeta _{+}\cos \zeta _{-})}{\sin
			^{2}\zeta _{+}+2\mathcal{P}(1+\cos \zeta _{+}\cos \zeta _{-})},\color{green}}\color{black}  \label{ShIF}
	\end{equation}%
	which describes the dependence on the variable and controllable parameters $%
	\varepsilon _{0}$, $A$, and $\omega $.
	
	In the case of \textit{fast passage}, where $\left( 1-\mathcal{P}\right)
	\approx 2\pi \delta \ll 1$, there is a large
	probability ($\mathcal{P}\sim 1$) for transitions between the adiabatic
	states in a single passage, while the transition probability between
	diabatic states is small, $\left( 1-\mathcal{P}\right) \ll 1$. Hence, we
	consider the time-averaged probability of the upper diabatic
	state $P_{\mathrm{up}}$. Then, one can obtain: 
	\begin{equation}
		\overline{P}_{\mathrm{up}}=\frac{1}{2}\frac{4\pi\delta\cos ^{2}(\zeta
			_{2}-\pi /4)}{\sin ^{2}\zeta _{-}+4\pi\delta\cos ^{2}(\zeta _{2}-\pi /4)}.
	\end{equation}%
	On resonance, we have $\zeta _{-}=k\pi $ and $\overline{P}_{\mathrm{up}}=1/2$. Then,
	in particular, for a small offset, with $\zeta _{-}\approx \pi \varepsilon
	_{0}/\hbar \omega $, we obtain the resonance frequency $\hbar \omega
	^{(k)}=\varepsilon _{0}/k$, meaning that the resonance transitions are
	described by their multi-photon relation. In the vicinity of the $k$-th
	resonance, for $\omega \sim \omega ^{(k)}$, to first approximation in $%
	\varepsilon _{0}/A$, we obtain 
	\begin{eqnarray}
		\color{green}\boxed{\color{black}\overline{P}_{\mathrm{up}}^{(k)} =\frac{1}{2}\frac{\Delta _{k}^{2}}{\Delta
			_{k}^{2}+\left( k\hbar \omega -\varepsilon _{0}\right) ^{2}}, \color{green}}\color{black} \label{Pupk}
		\\
		\Delta _{k} =\Delta \sqrt{\frac{2\hbar \omega }{\pi A}}\cos{\left( \frac{A%
		}{\hbar \omega }-\frac{\pi }{4}(2k+1)\right) }.  \notag
	\end{eqnarray}%
	The total probability $\overline{P}_{\mathrm{up}}$ is obtained as the sum of
	the partial contributions $\overline{P}_{\mathrm{up}}^{(k)}$. Note that the
	above derivation within the AIM assumes that the excitation probability may
	become nonzero when the energy quasicrossing is reached, that is, at $%
	\left\vert \varepsilon _{0}\right\vert <A$; otherwise, at $\left\vert
	\varepsilon _{0}\right\vert >A$, this model gives a zero transition
	probability.
	\paragraph*{\textbf{Coherent destruction of tunneling (CDT)}} \mbox{} \\
	
	From the formulas above, we
	can see that there are conditions under which the driven system can stay
	unexcited, with no tunneling between the states, even under the impact of a
	strong resonant drive. This phenomenon is known as coherent destruction of tunnelling or CDT \cite{Grossmann1991}
	and can be applied for controlling tunneling in TLSs \cite{Llorente1992,Hu2022}. It can
	be understood and described as either a degeneracy of the Floquet
	quasienergies \cite{Hijii2010} or, equivalently, as a result of
	destructive LZSM interference \cite{Kayanuma1994, Kayanuma2008}. Indeed,
	from Eq.~(\ref{P+AIM}), looking at a general case, there are no
	transitions between the states under the antiresonant condition, $\Xi
	_{21}=0$. Another case is Eq.~(\ref{Pupk}), which gives zero
	excited-state populations if driven with the amplitude $A=A_{k}\equiv (\pi
	/4)(2k-1)\hbar \omega $. 
	
	Complete CDT is only possible for isolated systems,
	such as the ones we consider here; taking into account dissipation spoils
	the effect \cite{Miao2016}. CDT happens to be important for the
	description of various systems: transitions in graphene \cite{Gagnon2016,
		Gagnon2017} or in biomolecular protein—solvent reservoirs in photosynthetic
	light harvesting complexes \cite{Eckel2009}; for a review of CDT in different
	systems, see \cite{Wubs2010}.

	\paragraph*{\textbf{Latching modulation}}\mbox{}\\
	 
	Our derivations in this section are mainly
	developed for sinusoidal driving. These can be adapted to a
	description of any other periodic driving, $\varepsilon (t+T_{\mathrm{d}%
	})=\varepsilon (t)$.\ Particularly, consider now the situation when a TLS is
	driven so that the level separation is switched abruptly between two values
	and is kept constant otherwise \cite{Silveri2015}. In this case, we have the
	bias 
	\begin{equation}
		\varepsilon (t)=\varepsilon _{0}+A\text{ sgn}\left[\cos (\omega t)\right],
	\end{equation}
	which
	results in the periodic \textit{latching modulation} of the qubit energy levels.
	Direct application of the LZSM approach would give infinite speed $v$,
	resulting in exact unit probability $\mathcal{P}=1$ with no interference.
	In this case, with $v=\infty $, the AIM is not directly applicable because
	then the width of the transition region becomes infinite: $vt^\text{d}_{\mathrm{LZSM}%
	}\sim \sqrt{\hbar v}$, see Eq.~(\ref{t_d}). This has been analyzed in
	Ref.~\cite{Silveri2015}, where the generalization of the AIM is presented.
	Interestingly, most of the formulas above, which describe the upper-level
	occupation probability, remain valid with only one important substitution:
	now, instead of the LZSM probability~$\mathcal{P}$, we write the
	sudden-switch transition probability $p_{s}=\langle \psi _{+}^{(l)}|\psi
	_{-}^{(r)}\rangle $, which occurs between the lower state in one latch
	position $\left\vert \psi _{-}^{(r)}\right\rangle $ and the upper state in
	the other latch position $\left\vert \psi _{+}^{(l)}\right\rangle $. The
	successful application of the AIM with this substitution $\mathcal{P}%
	\rightarrow p_{s}$ in Ref.~\cite{Silveri2015} was not only compared with the
	numerical solution and experiment, but also with the rotating-wave
	approximation (RWA). We return to this later when discussing the RWA.
	\paragraph*{\textbf{Interferograms}}\mbox{}\\
	
	 With the formulas above, Eq.~(\ref{P_I_avrgd}), as well as its limiting expressions, Eqs.~(\ref{ShIF}) and~(\ref{Pupk}), we can
	graphically visualize the interference. These dependencies,
	say, on $\varepsilon _{0}$ and $\omega$, or on $\varepsilon _{0}$ and $A$,
	can be called interferograms. These are shown in Fig.~%
	\ref{Fig:interferogramsAIM}(a) and (b) for $\hbar \omega /\Delta =0.8$ and $%
	1.4$, respectively.
	
	\subsubsection{Kibble–Zurek mechanism (KZM)}
	
	\label{subsubsection:KZM}
	
	Making analogies can bring us very far from where we started. 
	The KZM is much like this, bringing us from qubits to the Big Bang. The
	KZM started from the proposition by T.W.B.~Kibble to model the physics of
	the early universe as cosmological phase transitions that result in the formation
	of topological defects in the form of monopoles or cosmic strings \cite%
	{Kibble1976}. This was shown by W.H.~Zurek to be a universal feature for
	second-order phase transitions and related to the adiabatic-impulse
	approximation \cite{Zurek1985}. The latter can be equally applicable to
	two-level systems, hence relating the KZM and LZSM
	transitions \cite{Damski2005}.
	
	It was shown that for what we call a single-passage
	process, LZSM theory can be used to accurately describe the KZM \cite{Damski2005,
		Zurek2005, Dziarmaga2005}. Namely, the AIM for an
	avoided-level crossing is a general model that describes both qubit
	dynamics and symmetry-breaking second-order phase transitions. This allows
	us to pass from the LZSM evolution to phase transition dynamics and back
	again~\cite{Damski2006}.
	
	Following \cite{Damski2005}, consider this for a pressure quench that drives
	liquid $^{4}$He from a normal phase to a superfluid one. The process is
	described by the distance $\lambda $ from the critical point. The quench
	comes with the linear increase 
	\begin{equation}
		\lambda =\frac{t}{\tau _{\mathrm{Q}}},  \label{quench}
	\end{equation}%
	with the rate $\tau _{\mathrm{Q}}^{-1}$ and time taken such that the
	transition point is crossed at $t=0$. In the case of quenching liquid
	helium, $\lambda $ is the relative temperature, such that $\lambda (t=0)=0$.
	Changes in pressure translate into changes of the dimensionless parameter $%
	\lambda $. So, we start at $t\rightarrow-\infty$, here with helium in the liquid phase.
	The dynamics is described by the relaxation time  $\tau _{\mathrm{r}}$, which is
	the time needed for the system to adjust to new thermodynamic conditions.
	Far from the critical point, $\tau _{\mathrm{r}}$ is small and the evolution
	is adiabatic. 
	
\begin{figure}[tbp]
		\centering{\includegraphics[width=0.8\columnwidth]{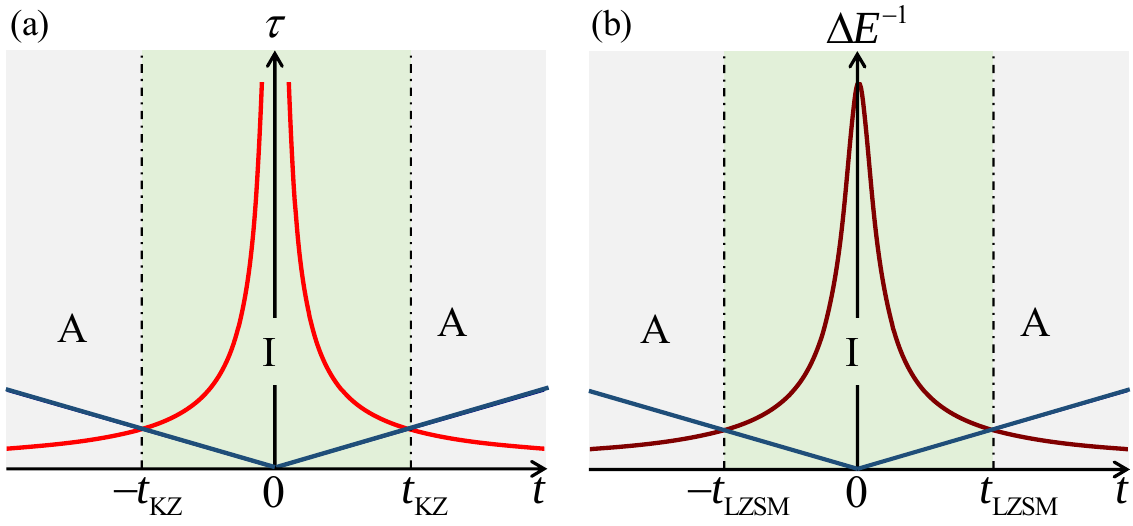}}
		\caption{\textbf{ Adiabatic-impulse model (AIM) which relates the Landau-Zener-St\"{u}ckelberg-Majorana (LZSM) with the Kibble-Zurek mechanism (KZM)}. In~(a) we show the time
			dependence of the relaxation time $\protect\tau _{\mathrm{r}}(t)$ of
			second-order phase transitions. The inclined blue lines at $\protect\tau _{\mathrm{r}%
			}=\left\vert t\right\vert $ give the moments of time $t_{\mathrm{KZ}}$,
			which separate the adiabatic~(A) and impulse~(I) stages of evolution.
			In~(b), we plot the inverse energy-level gap $\Delta E^{-1}$ for a two-level system.}
		\label{Fig:KZM}
	\end{figure}

	Moving closer to the transition point, critical
	slowing down (longer relaxation) occurs, which is the divergence of the
	relaxation time, $\tau _{\mathrm{r}}=\tau _{0}/|\lambda| $, with the
	characteristic time value $\tau _{0}$. This dynamics is shown in Fig.~\ref%
	{Fig:KZM}(a); the adiabatic and impulse regions are separated by $t_{\mathrm{%
			KZ}}$, which is called the freeze-out time. This characteristic time is defined by
	the condition 
	\begin{equation}
		\tau _{\mathrm{r}}(t_{\mathrm{KZ}})=\alpha\, t_{\mathrm{KZ}},  \label{Zurek}
	\end{equation}%
	as graphically shown by the inclined lines in Fig.~\ref{Fig:KZM}(a). Here, $%
	\alpha =O(1)$ is the system-specific parameter, which is taken as unity in the
	figure. From Eq.~(\ref{Zurek}), we obtain $t_{\mathrm{KZ}}=\sqrt{\tau
		_{0}\tau _{\mathrm{Q}}/\alpha }$. Knowledge of $t_{\mathrm{KZ}}$ allows us to
	find the density of topological defects, which appear as a result of the
	nonequilibrium phase transition, interestingly without solving those equations
	describing the dynamics of the system. For this, the analogy with the LZSM model
	is beneficial.
	
	Considering the analogy with a TLS, in Fig.~\ref{Fig:KZM}(b), we plot the
	inverse distance between the energy levels of a TLS, which is $\Delta E=\sqrt{%
		\Delta ^{2}+\left( vt\right) ^{2}}$. This analogy is based on the adiabatic
	theorem, which states that a system stays in the ground state as long as the
	inverse gap $\Delta E^{-1}$ is small enough. Hence, the inverse of the gap
	can be considered a quantum-mechanical equivalent of the relaxation time, $%
	\tau _{\mathrm{r}}=\hbar\Delta E^{-1}$. Then, the equivalent of the quench time $%
	\tau _{\mathrm{Q}}$ is $\Delta /v$, while $\Delta $ is identified with $\tau
	_{0}^{-1}$. There, the LZSM transition is analogous to a phase transition.
	Indeed, by solving Eq.~\eqref{quench} with $\tau_\text{r}=\hbar\Delta E^{-1}$, $\tau_\text{Q}=\Delta/v$, and $\hbar\Delta=\tau_0^{-1}$, one exactly reproduces the above-mentioned $^4\text{He}$ result for $t_\text{KZ}$ in the fast quench/transition limit.
	
	Analogously to the above-considered quenched $^{4}$He, a quantum Ising model
	can be used to describe the paramagnet-ferromagnet quantum phase transition 
	\cite{Zurek2005, Dziarmaga2005, Polkovnikov2005, Dutta2010}. In a more
	general context, the characteristic transition time $t_{\mathrm{KZ}}$ and
	length $\xi _{\mathrm{KZ}}$ (size of regions in which the order
	parameter is smooth) are defined by the universal critical exponents $z$ and 
	$\nu $: $t_{\mathrm{KZ}}\sim \tau _{\mathrm{Q}}^{1/\left( 1+\nu z\right) }$
	and $\xi _{\mathrm{KZ}}\sim \tau _{\mathrm{Q}}^{\nu /\left( 1+\nu z\right) }$
	\cite{Zurek1985, Dziarmaga2010}. Hence, this shows that equilibrium
	critical exponents can be used to predict the nonequilibrium dynamics and that
	the KZM correctly describes the results of this dynamics by giving the density
	of residual topological defects. We emphasize that the deep analogy between
	the LZSM and KZM is in the \textit{adiabatic-impulse approximation}, which has been shown to
	quantitatively well describe both the dynamics of quantum TLSs (which is the
	subject of the present review article) and those of quantum phase
	transitions.
	
	There are some difficulties in observing the time evolution of
	second-order phase transitions, and these are related to their rapid speed (sufficient
	range of quench time scales) or controlling and 
	counting the defects. Then the
	quantum simulation can effectively be used by means of a convenient
	controllable quantum system. 
	
	Making use of the interrelation between the LZSM
	and KZM, this simulation was done recently with such diverse systems as an
	optical interferometer \cite{Xu2014a}, a semiconductor charge qubit in a
	double quantum dot \cite{Wang2014}, superconducting phase and transmon
	qubits \cite{Gong2016a}, 
	a single trapped $^{171}$Yb$^{+}$ ion  \cite%
	{Cui2016,Cui2020}, spin-1 Bose-Einstein condensate \cite{Damski2009,Anquez2016}, and NMR based studies \cite{Zhang2017a}.  These simulations correctly reproduced the main KZM results: the
	boundary between the adiabatic and impulse regions, the freeze-out
	phenomenon, and the dependence of the topological defect density on the
	quench rate.
	
	Nowadays, the KZM is a general model that provides a description of the
	nonequilibrium dynamics and creation of topological defects such as strings,
	vortices, and domain walls. Here, using an analogy with the LZSM
	theory, these appear during symmetry-breaking phase transitions in the
	following systems: Ising chains \cite{Quan2010, Das2010, Henriet2016} and spin-1/2 XX
	and XY chains when the transverse field or anisotropic interaction is
	quenched \cite{Mukherjee2007, Divakaran2009, Roosz2014}, graphene in a
	time-dependent electric field \cite{Fillion-Gourdeau2016}, and biaxial
	paramagnet in an external magnetic field \cite{Zvyagin2018}. For deviations
	of realistic systems from the paradigmatic KZM, see for example ~\cite{Gao2017}.
	
	Even though there are some studies on periodic driving, for example, \cite{Mukherjee2009, Setiawan2015, Dutta2015, Kar2016,HigueraQuintero2022}, we note
	that in the context of phase transitions, the vast majority of research is
	devoted to the single-passage transition/quench. In quantum simulations
	of the KZM, as mentioned above, it is straightforward to realize double 
	or even multiple passages. This introduces interference, in addition to the
	possibility of nonadiabatic transitions. This may become one of the new
	twists in the interrelation between LZSM physics and 
	symmetry-breaking second-order phase transitions. Much like how the KZM for a
	single passage was used to describe the Big Bang, the respective development
	may be useful for speculating about the Big Bounce theory.
	
	\subsection{Rotating-wave approximation (RWA)}
	
	\subsubsection{Multi-photon Rabi oscillations}
	
	Consider now the situation of resonant driving, with those parameters close to
	where the energy distance $\Delta E$ equals the photon energy $\hbar
	\omega $ or, more generally, close to the energy of $k$ photons, $\Delta
	E\sim k\hbar \omega $. The former (the single-photon resonant excitation) is
	critical for microscopic systems where there are electron
	paramagnetic (spin) resonance and nuclear magnetic resonance \cite{Rabi1937}. In this case, the amplitude $A$ is usually small, and the $k$-photon
	resonances appear within perturbation theory \cite
	{Shirley1965, Krainov1976, Krainov1980}. 
	
	In contrast, for mesoscopic
	systems, the strong-driving regime is both accessible and important. Hence, we
	will first consider the solution of the Schr\"{o}dinger equation for a weak
	driving, afterwards for a strong driving, and then finally the solution of the Bloch
	equation. These solutions are based on the RWA
	(also called \textit{secular approximation}), where the terms that are quickly varying
	in time are neglected.
	
	For weak resonant driving, with $A\ll \Delta $ and $\delta \omega \equiv
	\omega -\Delta E/\hbar \ll \omega $, the upper-level occupation probability
	exhibits Rabi oscillations [see e.g.~\citet{Neilinger2016, Shevchenko2019}]:%
	\begin{subequations}\begin{eqnarray}
		P_{\mathrm{up}}(t) &=&\overline{P_{\mathrm{up}}}(1-\cos \Omega _{\mathrm{R}%
		}t),  \label{Rabi} \\
		\Omega _{\mathrm{R}} &=&\sqrt{\Omega _{\mathrm{R0}}^{2}+\delta \omega ^{2}}%
		\text{, \ \ \ }\Omega _{\mathrm{R0}}=\frac{A\Delta }{2\hbar \Delta E}\text{, \
			\ \ }\overline{P_{\mathrm{up}}}=\frac{1}{2}\frac{\Omega _{\mathrm{R0}}^{2}}{%
			\Omega _{\mathrm{R0}}^{2}+\delta \omega ^{2}}.  
	\end{eqnarray}\end{subequations}
	Here, $\overline{P_{\mathrm{up}}}$ describes the time-averaged occupation
	probability; it is maximal in resonance at $\delta \omega =0$.
	
	For stronger driving, one usually assumes $A\gg \Delta ^{2}/\hbar \omega $,
	which corresponds to what we call the \textit{diabatic limit}, with $\delta \ll 1$. This
	condition also means that the minimal energy distance $\Delta $  is small; we
	then, have $\Delta E\approx \left\vert \varepsilon _{0}\right\vert $.
	The solution of the Schr\"{o}dinger equation can be obtained close to the $k$-th
	resonance, with 
	\begin{equation}
		\delta \omega ^{(k)}\equiv k\omega -\frac{\left\vert \varepsilon
		_{0}\right\vert}{\hbar} \ll \omega,
	\end{equation} as in Refs.~\cite{Henry1977,Kmetic1985, Lopez-Castillo1992},%
	\begin{subequations}\begin{eqnarray}
		P_{\mathrm{up}}^{(k)}(t) &=&\overline{P_{\mathrm{up}}^{(k)}}(1-\cos \Omega _{%
			\mathrm{R}}^{(k)}t),  \label{kRabi} \\
		\Omega _{\mathrm{R}}^{(k)} &=&\sqrt{\Omega _{\mathrm{R0}}^{(k)2}+\delta
			\omega ^{(k)2}}\text{, \ \ }  \label{OmegaRk} \\
		\Omega _{\mathrm{R0}}^{(k)} &=&\frac{\Delta }{\hbar }J_{k}\!\!\left( \frac{A}{%
			\hbar \omega }\right) \text{, \ \ }  \label{OmegaR0k} \\
		\overline{P_{\mathrm{up}}^{(k)}} &=&\frac{1}{2}\frac{\Omega _{\mathrm{R0}%
			}^{(k)2}}{\Omega _{\mathrm{R0}}^{(k)2}+\delta \omega ^{(k)2}}.
		\label{Pupk_ave}
	\end{eqnarray}\end{subequations}
	These describe \textit{multi-photon Rabi oscillations}. Here, the Rabi frequency
	is modulated by the Bessel function of the first kind $J_{k}$. If the
	system's parameters vary, we must take into account all the resonances:
	\begin{equation}
		\overline{P_{\mathrm{up}}}=\sum_{k}\overline{P_{\mathrm{up}}^{(k)}}.
	\end{equation}
	We could study the moment of time $t_{\max }$, when the
	upper-level occupation reaches its maximum for the first time and the
	probability $P_{\mathrm{up}}^{(k)}(T_{\mathrm{d}})$ after the double passage
	of the avoided crossing (i.e., after a full period) in resonance~\cite%
	{Lopez-Castillo1992}. From Eq.~(\ref{kRabi}), these are given by $t_{\max
	}=\pi /\Omega _{\mathrm{R}}^{(k)}$ and%
	\begin{equation}
		P_{\mathrm{up}}^{(k)}(T_{\mathrm{d}})=1-\cos \left( 2\pi \frac{\Delta }{%
			\hbar \omega }J_{k}\left( \frac{A}{\hbar \omega }\right) \right) .
		\label{after_period}
	\end{equation}
	
	In the next approximation, in the small parameter $\Delta $ (to be more
	precise $\Delta ^{2}/A\hbar \omega \ll 1$ here), we can obtain the shift of
	the resonance frequency, which is known as the Bloch–Siegert shift \cite%
	{Lopez-Castillo1992}. Although for small $\Delta $
	the resonance frequency for the first resonance ($k=1$) from Eq.~(\ref%
	{Pupk_ave}) is $\omega =\left\vert \varepsilon _{0}\right\vert /\hbar $, the
	resonance for larger $\Delta $ is given by Eq.~(\ref{Rabi}): $\omega =\Delta
	E/\hbar =\sqrt{\varepsilon _{0}^{2}+\Delta ^{2}}/\hbar $. Hence, when either $%
	\Delta $ is large or the frequency $\omega $ is small, the results of the RWA
	can be improved by adding the Bloch–Siegert shift and higher corrections
	(generalized Bloch–Siegert shift) \cite{Tuorila2010}. See Refs.~\cite%
	{Abovyan2016, Sun2016, Wang2017, Huang2017, Saiko2018a, Kohler2018} for those
	cases beyond the RWA.
	
	More generally, taking into account relaxation, the RWA for a
	periodically driven system is presented in Appendix~\ref{Sec:RWA}. This
	is considered for a \textit{generic periodic drive} $\varepsilon
	(t)=\varepsilon _{0}+\widetilde{\varepsilon }(t)$, with $\widetilde{%
		\varepsilon }(t)=A\cos \omega t$ as a particular case. The TLS is considered
	as being coupled to a dissipative environment, the impact of which is taken into
	account by introducing the relaxation and decoherence rates $\Gamma
	_{1,2}=T_{1,2}^{-1}$ \cite{Silveri2012, Silveri2015}. Then, the solution of
	the Bloch equations gives the upper-level occupation probability $P_{\mathrm{%
			up}}$, which for the stationary case reads 
	\begin{equation}
		\color{green}\boxed{\color{black}\overline{P_{\mathrm{up}}}=\frac{1}{2}\sum_{k=0}^{\infty }\frac{\Omega _{%
				\mathrm{R0}}^{(k)2}}{\Omega _{\mathrm{R0}}^{(k)2}+\frac{\Gamma _{1}}{\Gamma
				_{2}}\left( k\omega -\left\vert \varepsilon _{0}\right\vert /\hbar \right)
			^{2}+\Gamma _{1}\Gamma _{2}},\color{green}}\color{black}  \label{PupRWA}
	\end{equation}%
	where%
	\begin{equation}
		\Omega _{\mathrm{R0}}^{(k)}=\frac{\Delta }{\hbar }\frac{\omega }{2\pi }%
		\int\limits_{0}^{2\pi /\omega }dt\exp \left[ \frac{i}{\hbar }%
		\int_{0}^{t}dt^{\prime }\widetilde{\varepsilon }\left( t^{\prime }\right)
		-ik\omega t\right] .  \label{Deltak}
	\end{equation}%
	These are quite general results that can be applicable to any driving $%
	\widetilde{\varepsilon }\left( t\right) $. In particular, for $\widetilde{%
		\varepsilon }\left( t\right) =A\cos \omega t$, we obtain the Rabi frequency (%
	\ref{OmegaR0k}).

	\paragraph*{\textbf{Role of the driving shape}} \mbox{}\\ 
	
	Although most of this review is devoted to
	sinusoidal driving, we consider here the role of driving shape on
	the LZSM interference \cite{Blattmann2015}. In general, equations like (\ref%
	{PupRWA}, \ref{Deltak}) can be useful for describing any periodic
	perturbation $\widetilde{\varepsilon }\left( t\right) $, including
	multiharmonic drivings. Although the qualitative picture is similar for any
	periodic signal, the overall interference fringes can differ. This was
	studied for such pulses as triangular \cite{Xu2010}, hyperbolic tangent and
	Gaussian \cite{Cao2010}, rectangular \cite{Silveri2015, Shi2020}, secant 
	\cite{Zhao2018b}, and other ones \cite{Mukherjee2016, Xie2018}. 
	
	As a special
	and illustrative case, we now consider biharmonic driving \cite%
	{Blattmann2015}. This allows to study the effect of commensurate versus
	incommensurate driving frequencies \cite{Forster2015}.  It has been demonstrated
	that depending on the phase difference between its two components, 
	bichromatic driving with commensurable frequencies may break time-reversal
	symmetry, which is visible in the Fourier transform of the LZSM interference
	pattern and which is useful for quantum simulation. Depending on the desired
	properties of the transitions, we can tailor the requested driving signal. We
	will return to this in the section on quantum control. Here, we 
	illustrate this by referring to the development of the Lyapunov-based
	control method. In Ref.~\cite{Cong2015}, it was demonstrated that an 
	ultrafast excitation can be gained by using a signal based on the Lyapunov
	control method, which is a design control method based on the 
	Lyapunov indirect stability theorem.
	
	Now, with the equations above, Eqs.~(\ref{PupRWA}) and (\ref{OmegaR0k}%
	), we can plot the interferograms for sinusoidal driving, as shown in Fig.~\ref%
	{Fig:Different_w}. Regarding the dependence of the time-averaged occupation of the
	upper diabatic level $\overline{P_{\mathrm{up}}}$ on the bias $\varepsilon
	_{0}$ and driving amplitude $A$, we can see that the resonances are along
	the lines $\left\vert \varepsilon _{0}\right\vert =k\hbar \omega $, and
	these are interrupted by the zeros of the Bessel functions, resulting in the
	CDT. The width of the resonances, as defined by Eq.~(\ref{PupRWA}), depends on
	the interrelation of the parameters; this is illustrated in the figure by
	varying the driving frequency $\omega $.
	
	\begin{figure}[tbp]
		\centering{\includegraphics[width=1.0\columnwidth]{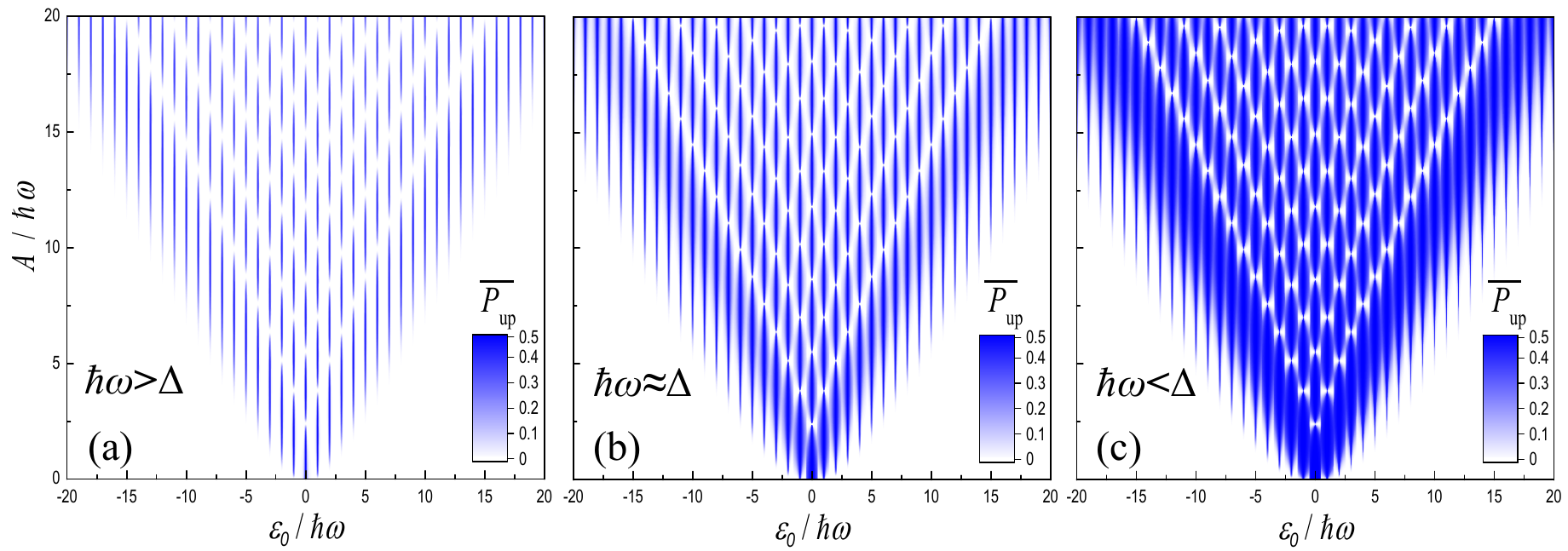}}
		\caption{\textbf{LZSM interferogram for different frequencies}. For this, we
			took $\hbar \protect\omega /\Delta =10$, $1.14$, $0.32$, for (a-c) and $%
			\Gamma _{1}=0.1T_{\mathrm{d}}^{-1}$, $\Gamma _{2}=\Gamma _{1}/2$.}
		\label{Fig:Different_w}
	\end{figure}
	
	For strong driving, $A>\hbar \omega $, from Eqs.~(\ref{PupRWA}, \ref{Deltak}%
	), we arrive at Eq.~(\ref{Pupk}). For this, we use the asymptotics of the
	Bessel functions and neglect the dissipation by considering the case when $%
	T_{1}=T_{2}=\infty $. Importantly, we now see that two different theoretical
	approaches, the AIM and RWA, which we redeveloped in different parameter regimes, lead to the
	very same result: Eq.~\eqref{Pupk}. Moreover, in the following
	subsections, we arrive at at the same result while presenting two other
	approaches.
	
	\subsubsection{Multi-photon transitions analyzed by the LZSM theory}
	
	It is well known that \textquotedblleft an atom submitted to sufficiently
	intense radiation can absorb several incident photons and go from a discrete
	level $a$ to another discrete level $b$ located at a higher energy, at a
	distance equal to the sum of the energies of the absorbed photons. Such a
	process is called multi-photon absorption, the reverse process being
	multi-photon stimulated emission\textquotedblright\ \cite{Cohen-Tannoudji1998}. 
	The observation in the previous
	subsection demonstrates that multi-photon transitions are essential and
	appear consistently in different pictures, both within resonant RWA and
	within LZSM theory. In particular, from Eq.~(\ref{PupRWA}), we can see
	that the resonances take place at $k\hbar \omega =\left\vert \varepsilon
	_{0}\right\vert \approx \Delta E$. 
	
	Multi-photon processes have been
	studied in various systems: superconducting qubits \cite{Wallraff2003,
		Saito2004, Tornes2008, Shevchenko2012}, two-dimensional electron systems 
	\cite{Zudov2006}, atomic and molecular structures \cite{Arbo2010}, Rydberg
	atoms \cite{Forre2004, Ditzhuijzen2009}, TLSs in superconducting circuits 
	\cite{Burin2014}, quantum dots \cite{Giavaras2019, Giavaras2019b}, and
	electromechanical systems \cite{Heinrich2010}.
	
	A TLS can be subjected to resonant driving with the energy of $k$
	photons matching the distance between the energy levels $k\hbar \omega
	\approx \Delta E$; this induces a transition from one level to another.
	These can be described as the exchange of energy quanta (here, photons) with
	the oscillating driving field, which is known as \textit{photon-assisted
		tunneling} (PAT); see, for instance, Ref.~\cite{Li2013}, where this was
	studied for a superconducting qubit. PAT was studied extensively for other
	various quantum systems with tunneling, especially for semiconductor
	nanostructures \cite{Platero2004}, including driven quantum dots \cite%
	{Mavalankar2016, Osika2017} and qubit-resonator dimers \cite{Zheng2021}. Through 
	a double quantum dot, PAT is described by the tunneling current
	defined by a formula analogous to Eq.~(\ref{PupRWA}) \cite{Stoof1996,
		Gallego-Marcos2015}. 
	
	Historically, the first experiments with PAT were performed
	on AC-driven Josephson contacts \cite{Tien1963}. In this context, 
	we can note the similarity and difference of the PAT (or, equivalently,
	multi-photon Rabi oscillations) and AC Josephson effect with the Shapiro
	steps, both of which appear for an irradiated superconducting tunnel
	junction. The former relates to the tunneling of electrons (quasiparticles)
	and results in steps in the current-voltage curve displaced in voltage
	by $k\hbar \omega /e$; the latter assumes the supercurrent response, which
	occurs as the Shapiro steps at voltages $k\hbar \omega /2e$ \cite%
	{Nakamura2001}. Both sets of resonances were observed experimentally and
	described theoretically in \cite{Boris2015,Shaikhaidarov2022}. In that study, it was demonstrated that
	the Shapiro steps and PAT states originate from Cooper-pair and
	quasiparticle currents; and if a JJ is subjected to an intense
	microwave signal, both sets of resonances can be observed in the current, 
	changing both the voltage and microwave power \cite{Boris2015,
		Snyder2018}.
	
	\subsection{Floquet theory}
	
	It is natural to use the Floquet theorem for solving the Schr\"{o}dinger
	equation for a periodically driven system. The theory based on this—
	Floquet theory—has been covered in many reviews \cite{Chu1989,Grifoni1998,
		Chu2004, Son2009, Shevchenko2010, Eckardt2017, Rodriguez-Vega2020,
		Sen2021}. We now consider this for our problem of describing the dynamics and
	stationary state of a driven TLS; the details are given in Appendix \ref%
	{Sec:Floquet}.
	
	For a time-dependent Schr\"{o}dinger equation, Eq.~(\ref{TDSE}), for the
	wave function $\left\vert \psi (t)\right\rangle $ with a periodic
	Hamiltonian $H(t)=H(t+T_{\mathrm{d}})$, we can use the Floquet theorem.
	Accordingly to this, the solution (so-called Floquet-state solution) is 
	\begin{equation}
	\left\vert \psi (t)\right\rangle =\Sigma _{j=1,2}C_{j}\exp \left(
	-i\epsilon_{j}t/\hbar \right) \left\vert \Phi_{j}(t)\right\rangle, 	
	\end{equation} where $\left\vert \Phi_{j}(t)\right\rangle$, is a periodic state, called the Floquet mode. The
	index $j=1,2$ appears because we are dealing with a two-dimensional Hilbert
	space. The real-valued $\epsilon{j}$ are called quasienergies, to reflect the
	formal analogy with a quasimomentum $\mathbf{k}$, which characterizes electron
	Bloch eigenstates in a periodic solid state \cite{Zeldovich1967,
		Zeldovich1973, Averbukh1985}. The quasienergies are unique
	up to multiples of $2\pi n/T_{\mathrm{d}}$. It follows that these are the
	eigenstates in the problem%
	\begin{equation}
		\color{orange}\boxed{\color{black}\left( H(t)-i\hbar \frac{d}{dt}\right) \left\vert \Phi_{j}(t)\right\rangle
		\equiv \mathcal{H}(t)\left\vert \Phi_{j}(t)\right\rangle =\epsilon_{j}\left\vert
		\Phi_{j}(t)\right\rangle .\color{orange}}\color{black}
	\end{equation}%
	The quasienergy states for time-dependent problems play a role analogous
	to the stationary states for time-independent problems.
	
	Using the Floquet formalism allows us to reduce the problem of periodic
	perturbation to the stationary problem \cite{Shirley1965, Barone1977,Ikeda2022}, which
	is known as the Floquet Hamiltonian method \cite{Chu1989, Chu2004}. For
	this, one needs to expand the quasienergy function into a Fourier series, $%
	\left\vert \Phi_{j}(t)\right\rangle =\Sigma _{n}e^{in\omega t}\left\vert
	\Phi_{j,n}\right\rangle $. Hence, the quasienergy states become expressed as a
	superposition of stationary states $\left\vert \Phi_{j,n}\right\rangle $, with
	energies equal to $\epsilon{j}+n\hbar \omega $. Using the Fourier series
	expansion, from the Schr\"{o}dinger equation, we obtain the relation%
	\begin{equation}
		\epsilon{j}\left\vert \Phi_{j,n}\right\rangle =\left( -\frac{\Delta }{2}\sigma _{x}-%
		\frac{\varepsilon _{0}}{2}\sigma _{z}+n\omega \right) \left\vert
		\Phi_{j,n}\right\rangle -\frac{A}{2}\sigma _{z}\left( \left\vert
		\Phi_{j,n-1}\right\rangle +\left\vert \Phi_{j,n+1}\right\rangle \right) .
	\end{equation}%
	Multiplying this to the left by $\left\langle \Phi_{i,m}\right\vert $, we obtain
	the matrix equation for the eigenvalues $\epsilon{j}$ and eigenfunctions $%
	\left\vert \Phi_{j,n}\right\rangle $. Then, we truncate the matrix to
	solve the equation numerically, say, with $n$ ranging from $-50$ to $50$ \cite%
	{Deng2016}.
	
	Close to the resonance, $\delta \omega ^{(k)}\equiv k\omega -\left\vert
	\varepsilon _{0}\right\vert /\hbar \ll \omega $, the contribution of the
	nonresonant states can be neglected. This means using the RWA, for example, as
	in \cite{Autler1955, Aravind1984, Silveri2013}. Then, the Floquet Hamiltonian
	consists just of the copies of 
	\begin{equation}
		H_{\mathrm{RWA}}^{(k)}=\left( 
		\begin{array}{cc}
			-\varepsilon _{0}/2 & -\hbar \Omega _{\mathrm{R0}}^{(k)}/2 \\ 
			-\hbar \Omega _{\mathrm{R0}}^{(k)}/2 & \varepsilon _{0}/2+k\hbar \omega%
		\end{array}%
		\right), 
	\end{equation}%
	where the Rabi frequency $\Omega _{\mathrm{R0}}^{(k)}$ is the one defined in
	Eq.~(\ref{OmegaR0k}). Diagonalization of this Hamiltonian produces the
	quasienergy difference, which we denote as $\hbar \Omega _{\mathrm{R}%
	}^{(k)} $: $H_{\mathrm{RWA}}^{(k)\prime }=\sigma _{z}\hbar \Omega _{\mathrm{R%
	}}^{(k)}/2$. This brings us to the same result as RWA; see Eq.~(\ref%
	{OmegaRk}). Hence, Floquet theory in the secular approximation (i.e., in
	the RWA) gives us again the upper-level occupation probability in Eq.~(\ref%
	{Pupk_ave}) \cite{Son2009}.
	
	These RWA results are accurate, provided $\Delta /\hbar \omega \ll 1$. The
	higher-order terms can be obtained within the generalized van Vleck
	perturbation theory \cite{Hausinger2010}, resulting in a shift of the
	resonance frequency by 
	\begin{equation}
		\delta _{k}=\frac{1}{2}\sum_{l\neq -k}\frac{\Omega _{\mathrm{R0}}^{(l)2}}{%
			\varepsilon _{0}/\hbar +l\omega }.  \label{delta_AC}
	\end{equation}%
	This means that the resonances are situated at $\varepsilon _{0}=k\hbar
	\omega -\delta _{k}$, and that the Rabi frequency becomes $\delta _{k}$-shifted, $%
	\Omega _{\mathrm{R}}^{(k)}=\sqrt{\Omega _{\mathrm{R0}}^{(k)2}+\left( \delta
		\omega ^{(k)}-\delta _{k}\right) ^{2}}$. Note that these formulas are
	calculated in the diabatic basis, while for the adiabatic basis, some
	modifications must be made \cite{Silveri2013}.
	
	In particular, consider the first-order correction from Eq.~(\ref{delta_AC}%
	), which is $\delta _{1}=\Delta ^{2}/2\varepsilon _{0}$. Then, we expect the
	first resonance at $\varepsilon _{0}=\hbar \omega -\Delta ^{2}/2\varepsilon
	_{0}$ (where we can apply $\varepsilon _{0}=\hbar \omega $ in the right-hand-side).
	This matches the original Rabi approach. Indeed, from Eq.~(\ref{Rabi}),
	we have the resonances at $\hbar \omega =\sqrt{\Delta ^{2}+\varepsilon
		_{0}^{2}} $, where the expansion gives $\varepsilon _{0}=\hbar \omega
	-\Delta ^{2}/2\varepsilon _{0}$. Hence, generalized van Vleck perturbation
	theory gives the correction exactly consistent with the Rabi RWA, which is valid for
	small driving amplitudes \cite{Sambe1973}. For more about the resonance
	shift, termed AC Stark shift or Bloch-Siegert shift, see \cite%
	{Autler1955, Aravind1984, Yan2017}.
	
	Because they are related to the Rabi frequency, the Floquet quasienergies can be
	visualized in experiments, known as Floquet spectroscopy.
	This was realized recently with qubits in cavities. In this case, the cavity is a
	superconducting microwave resonator and qubits are based either on
	superconducting circuits \cite{Silveri2013, Deng2015}\ or on double quantum
	dots \cite{Koski2018, Chen2020}. Other possible realizations include such
	systems as a strongly driven Anderson insulator \cite{Agarwal2017} or pumping
	in a Cooper pair sluice \cite{Russomanno2011}. As a further development of
	the theory, this approach can be used to study the low-frequency limit \cite%
	{Rodriguez-Vega2018}, include the dissipation \cite{Henriet2014,
		Restrepo2016, Kohler2017,Mori2022} (Floquet-Markov theory), and considering
	multilevel systems \cite{Denisenko2010, Ganeshan2013, Satanin2014,Han2019,
		Han2020, Munyaev2021, Zhou2021}.
	
	\subsection{Driving fields in the quantum regime}
	
	Our approach above, in which the driving field is treated as a classical field and
	enters as $\varepsilon (t)$, is essentially the semiclassical approach.
	This takes place when the system of interest is described by the Schr\"{o}dinger equation and a field satisfies classical Maxwell equations. In the
	other, fully quantum-mechanical approach, both the system and 
	electromagnetic field are treated quantum mechanically. Importantly, 
	semiclassical theory gives rise to results that are equivalent to those
	obtained from fully quantized theory in intense fields \cite%
	{Aravind1984, Grifoni1998, Chu1989}. We discuss this below.
	For experimental realizations, see \cite{Nakamura2001, Saito2004,
		Wilson2007, Kervinen2019}.
	
	A TLS coupled to a quantum field is described by the Janes–Cummings
	Hamiltonian with driving:%
	\begin{equation}
		\color{orange}\boxed{\color{black}H_{\mathrm{Q}}(t)=-\frac{\Delta }{2}\sigma _{x}-\frac{\varepsilon _{0}}{2}%
		\sigma _{z}+\hbar \omega _{\mathrm{r}}a^{\dagger }a-g\sigma _{z}\left(
		a+a^{\dagger }\right) +\xi \left( a^{\dag }e^{-i\omega t}+ae^{i\omega
			t}\right), \color{orange}}\color{black}  \label{J-C_Ham}
	\end{equation}%
	where $a$ and $a^{\dagger }$ are the annihilation and creation operators for
	photons in the electromagnetic field, $g$ is the coupling constant, $\xi $
	describes the amplitude of the driving field, and the driving frequency $\omega $
	is close to the resonator frequency $\omega _{\mathrm{r}}$. For example, for
	a flux qubit coupled to a transmission line resonator, the coupling constant
	is proportional to the persistent current in a flux qubit $I_{\mathrm{p}}$,
	which is the current constant in the resonator $I_{\mathrm{r0}}$, and the mutual
	inductance $M$: $g=MI_{\mathrm{p}}I_{\mathrm{r0}}$; the driving amplitude is
	defined by the amplitude of the applied microwave voltage $V_{A}$, the
	voltage constant in the resonator $V_{\mathrm{r0}}$, and a coupling
	capacitance $C_{0}$: $\xi =C_{0}V_{A}V_{\mathrm{r0}}$ \cite{Greenberg2007,
		Shevchenko2019}. Note that the last term in Eq.~(\ref{J-C_Ham}) describes
	driving through the resonator; if the driving is through the qubit, this term
	would be proportional to $\sigma _{z}\cos \omega t$ \cite{Zhao2015}.
	
	Next, we need to get rid of the temporal dependence in the driving term
	with the transformation $U=\exp \left( i\omega ta^{\dagger }a\right) $,
	resulting in $H_{\mathrm{Q}}\rightarrow \widetilde{H}_{\mathrm{Q}}$, and to
	average this over the coherent states $\left\vert \alpha \right\rangle $ 
	\cite{Aravind1984, Sun2012}. One way to define the coherent states is to
	assign them as the eigenstates of the annihilation operator, $a\left\vert
	\alpha \right\rangle =\alpha \left\vert \alpha \right\rangle $; it follows
	that the value $\alpha $ is given by the mean number of photons in the
	resonator: $\left\langle n\right\rangle =\left\vert \alpha \right\vert ^{2}$. 
	The resulting Hamiltonian $H=\left\langle \alpha \right\vert \widetilde{H}%
	_{\mathrm{Q}}\left\vert \alpha \right\rangle $ becomes our
	quasiclassical qubit Hamiltonian $H(t)$ in Eq.~(\ref{H(t)}), with the cosine
	bias with the amplitude $A=4\alpha g$ in Eq.~(\ref{eps_with_cos}), \textit{%
		quod erat demonstrandum}, which is one of the cornerstones of
	circuit quantum electrodynamics (QED).
	
	Alternatively, one can show that $H_{\mathrm{Q}}(t)$ in matrix form is
	exactly reduced to the Floquet Hamiltonian $H_{\mathrm{F}}$ for an intense
	driving field \cite{Aravind1984}. For this, the annihilation
	and creation operators satisfy the following relations: $a\left\vert n\right\rangle =%
	\sqrt{n}\left\vert n-1\right\rangle $ and $a^{\dagger }\left\vert
	n-1\right\rangle =\sqrt{n}\left\vert n\right\rangle $. Then, with $\Delta =0$, 
	the Hamiltonian can be diagonalized exactly \cite{Cohen-Tannoudji1998,
		Nakamura2001, Saito2004, Wilson2007, Wilson2010} with the eigenstates%
	\begin{equation}
		\left\vert 0,n\right\rangle =\left\vert 0\right\rangle \otimes D\left( 
		\widetilde{g}\right) \left\vert n\right\rangle \text{, \ \ }\left\vert
		1,n\right\rangle =\left\vert 1\right\rangle \otimes D\left( -\widetilde{g}%
		\right) \left\vert n\right\rangle \text{, where\ }D(\widetilde{g})=\exp
		\left( \widetilde{g}\left( a^{\dagger }-a\right) \right) \:\:\text{ and\ }\:\:\:%
		\widetilde{g}=\frac{g}{\hbar \omega }.  \label{dressed_basis}
	\end{equation}
	These states are called the dressed-state basis; their eigenenergies are 
	\begin{equation}
		E_{j,n}=n\hbar \omega -\frac{g^{2}}{\hbar \omega} +(-1)^{j}\varepsilon _{0}.
	\end{equation}
	Here, $%
	D $ is the displacement operator, the parameter $\widetilde{g}$ defines the
	strength of the coupling, while $j=0,1$ is attributed to the TLS, and $n$
	to the number of photons in the cavity. Then, in this dressed-state
	representation, the matrix elements of the full Hamiltonian (with nonzero $%
	\Delta $) coincide with the ones of the Floquet Hamiltonian $H_{\mathrm{F}}$%
	, Eq.~(\ref{FloquetHamiltonian}). When $E_{0,n+k}\approx E_{1,n}$, that is,
	when the $k$-photon resonance condition is met, $k\hbar \omega =\varepsilon
	_{0}$, the two dressed-states experience avoided-level crossing with the
	coupling (off-diagonal element) $\frac{1}{2}\Delta J_{k}(\alpha )$,
	resulting in $k$-photon Rabi oscillations with frequency $\Omega _{%
		\mathrm{R}}^{(k)}=\Delta J_{k}(\alpha )/\hbar $, as above in the RWA.
	\paragraph*{\textbf{Ultra-strong coupling regime}} \mbox{}\\
	
	The picture above is in agreement with the semiclassical approach. However, there is a particular regime with
	very strong coupling and small photon numbers where some differences may
	appear \cite{Saiko2016, Ashhab2017}. To understand this, note
	that with the driving field in the quantum regime, when we consider 
	strong driving, we mean large $A=4\alpha g$; but this can be reached with either a
	large coupling $g$ or a large number of photons $\left\langle n\right\rangle $. 
	Accordingly, we can consider several limiting cases of the strong driving 
	\cite{Li2013, Ashhab2017, Bonifacio2020}:%
	\begin{equation}
		A=4\sqrt{\left\langle n\right\rangle }g \ \gg \  \hbar \omega \text{ \ }%
		\leftrightarrow \text{ \ }
		\begin{cases}
			\widetilde{g}\ll 1,\:\:\:\text{ weak coupling with }\left\langle n\right\rangle
			\ggg 1\text{,}\:\:\:\:\:\:\:\:\:\:\:\:\:\: \\ 
			\widetilde{g}\sim 1,\:\:\:\text{ (deep-)strong coupling with }\left\langle
			n\right\rangle \gg 1\text{,} \\ 
			\widetilde{g}\gg 1,\:\:\:\text{ ultra-strong coupling with }\left\langle
			n\right\rangle \sim 1\text{.}\:\:\:\:\:%
		\end{cases}
	\end{equation}%
	For all these regimes, the Rabi frequency in the quantum case, around the $k$%
	-photon resonance, $k\hbar \omega =\varepsilon _{0}$, reads \cite{Ashhab2017}
	\begin{equation}
		\Omega _{\mathrm{R,Q}}^{(k)}=\frac{\Delta }{\hbar }\left( 2\widetilde{g}\right)^{k}\exp\left[-2\widetilde{g}%
			^{2}\right] \sqrt{\frac{n!}{\left( n+k\right) !}}%
		L_{n}^{k}\left( 2\widetilde{g}^{2}\right),   \label{with_Laguerre}
	\end{equation}%
	where $L_{n}^{k}$ are the associated Laguerre polynomials. The numerical
	calculations in Ref.~\cite{Ashhab2017} demonstrate that the semiclassical
	dynamics is fully consistent with the quantum dynamics with the Rabi frequency
	in Eq.~(\ref{with_Laguerre}), but only for small coupling $\widetilde{g}$ and
	large photon number $n$. In the ultra-strong coupling regime, from Eq.~(\ref{with_Laguerre}), where the fully quantum approach provides decaying oscillations because of quantum fluctuations and where the Rabi frequency deviates from the result
	of the semiclassical approach.
	\paragraph*{\textbf{Dressed-states and interferometry for systems which are either doubly driven or driven by a chirped microwave}}\mbox{}\\
	
	The notion of dressed-states, which we just
	considered for the quantum driving fields, appears when describing driving
	systems in a different context. Here, we mean the case when both the
	Hamiltonian of a TLS and driving can be rewritten as an effective TLS;
	hence describing entangled light-matter states, which are also known as dressed-states 
	\cite{Magazzu2018, Chang2020}. This notion appears because the
	qubit is now dressed by the driving field. This approach is useful for
	systems that do not have avoided energy-level crossing, or where their
	energy-level spacing does not depend on the external bias parameters \cite%
	{Gong2016}. Here, a typical case with these two properties is a transmon
	qubit \cite{Fink2009}; to make its effective energy levels controllable and,
	in particular, for realizing the LZSM regime, this can be achieved by using either a
	chirping field or a second signal \cite{Garraway1992a,Gauthey1997,Vitanov2001, Sun2011, Silveri2015,Garraway2016,Gong2016}.
	
	To illustrate dressed-states and interferometry, consider now a doubly driven TLS, described by the
	\textquotedblleft pump-probe\textquotedblright\ Hamiltonian, for example \cite%
	{Wen2020},%
	\begin{equation}
		H_{\mathrm{pp}}=-\sigma _{x}\Delta \cos \omega _{\mathrm{probe}}t+\frac{%
			\sigma _{z}}{2}\left( \hbar \omega _{10}+\delta \cos \omega _{\mathrm{pump}%
		}t\right) .  \label{pump-probe}
	\end{equation}%
	This corresponds to a TLS with distance between the energy levels equal
	to $\hbar \omega _{10}$ and driven by two signals, with amplitudes
	proportional to $\Delta $ and $\delta $, assuming $\omega _{\mathrm{pump}%
	}\ll \omega _{\mathrm{probe}}$. Using the transformation $U=\exp \left(
	-i\omega _{\mathrm{probe}}\sigma _{z}t/2\right) $ and the RWA lead to
	the Hamiltonian~(\ref{H(t)}) with 
	\begin{equation}
		\varepsilon (t)=\hbar \left( \omega _{10}-\omega _{\mathrm{probe}}\right)
		+\delta \cos \omega _{\mathrm{pump}}t.  \label{effective_epsilon}
	\end{equation}%
	Now, after arriving at the pseudo-spin Hamiltonian for our driven system
	(which we can now call the dressed Hamiltonian), we can understand the two
	options above. 
	
	First, we apply two signals, the pump (dressing) signal
	and probe signal, obtaining the bias (\ref{eps_with_cos}) with $%
	\varepsilon _{0}=\hbar \left( \omega _{10}-\omega _{\mathrm{probe}}\right) $
	(for a transmon, this is controlled by the DC magnetic flux, changing $\omega
	_{10}$), $A=\delta $, and $\omega =\omega _{\mathrm{pump}}$ \cite%
	{Silveri2015}. 
	
	Second, we do not have a second signal, $\delta =0$, while the
	\textquotedblleft probe\textquotedblright\ frequency is time dependent, $%
	\hbar \omega _{\mathrm{probe}}(t)\rightarrow \varepsilon _{0}+A\cos \omega t$; see Ref.~\cite{Gong2016} for chirping with triangular pulses, Refs.~\cite%
	{Childress2010, Blattmann2014, Ono2019} for sinusoidal frequency modulation,
	and Refs.~\cite{Saiko2007, Saiko2014, Saiko2019, Wang2021} for the impact of
	bichromatical driving.
	\paragraph*{\textbf{LZSM spectroscopy of multilevel systems}}\mbox{}\\
	
	For a multilevel system,
	a similar dressing can also be  applied to create an avoided-level structure
	of energy levels \cite{Mi2018, Shevchenko2018}. For this, we apply
	the driving signal with frequency $\omega _{\mathrm{d}}$ close to the
	energy-level separation $\Delta _{0}$ at $\varepsilon =0$, to obtain
	the dressed avoided-level gap $\Delta \approx (\Delta _{0}-\hbar \omega _{%
		\mathrm{d}})\ll \Delta _{0}$. Then, the dressed energy levels are related to
	the bare ones as $\widetilde{E}_{i}\approx E_{i}\pm \hbar \omega _{\mathrm{d}%
	}$. As a result, in addition, this dressing can significantly increase the
	distance between the lower two energy levels and upper ones, providing
	an instrument for creating controllable TLSs out of multilevel ones. In
	such case, note that features of the TLS's spectrum, such as the curvature
	of its energy levels, essentially depend on the entire energy spectrum of the
	bare multilevel system \cite{Mi2018, Shevchenko2018}. Hence, the LZSM
	interferogram of the effective dressed TLS is a convenient tool for the
	spectroscopy of multilevel systems.
	
	\subsection{Impact of dissipation and temperature}
	
	Any real quantum system is coupled with the environment, and for mesoscopic
	systems, it is particularly important to include dissipation.
	Many authors study the impact of temperature, relaxation, and decoherence on the
	dynamics of TLSs \cite{Leggett1987, Grifoni1998}. The general approach is to
	start from the Liouville--von Neumann equation for the density matrix $\rho
	_{\mathrm{tot}}$ of the system, here comprising of our quantum system and the
	dissipative environment 
	\begin{equation}
		\frac{d}{dt}\rho _{\mathrm{tot}}(t)=-\frac{i}{\hbar }\left[ H_{\mathrm{tot}%
		}(t),\rho _{\mathrm{tot}}(t)\right] .  \label{vonNeumann}
	\end{equation}%
	The total Hamiltonian, $H_{\mathrm{tot}}=H+H_{\mathrm{env}}+H_{\mathrm{int}}$,
	consists of our system's part $H$, the environment Hamiltonian $H_{\mathrm{%
			env}}$, and the interaction between them $H_{\mathrm{int}}$, for example, \cite%
	{Nalbach2009}. Provided the coupling is weak, the environment can be
	represented as a set of harmonic oscillators with a coupling linear in the
	oscillator coordinates. Within this \textquotedblleft
	spin-boson\textquotedblright\ model, the environment is characterized by the
	Hamiltonian $H_{\mathrm{env}}=\sum \hbar \omega _{i}b_{i}^{\dag }b_{i}$ with
	the frequencies $\omega _{i}$ and annihilation operators $b_{i}$; a bosonic
	reservoir can represent phonons if our system is coupled with a crystal
	lattice. The general form of the interaction is given by $H_{\mathrm{int}}=-\frac{1%
	}{2}\mathcal{S}\sum \hbar \lambda _{i}(b_{i}+b_{i}^{\dag })$, with $\mathcal{S%
	}$ representing a spin operator; usually, $\mathcal{S}=\sigma _{z}$ \cite%
	{Kayanuma1998, Saito2002,Chen2015,Wertnik2018,Lambert2019,Lambert2020,Funo2021}. 
	
	The environment is described by the spectral
	density function $J(\omega )=\pi \sum \lambda _{i}^{2}\delta (\omega -\omega
	_{i})$ \cite{Leggett1987}. At small energies, this is characterized by a
	power-law dependence and can be further approximated by introducing the
	cutoff frequency $\omega_{\mathrm{c}}$: $J(\omega )=\alpha \omega (\omega
	/\omega _{\mathrm{c}})^{s-1}\exp \left( -\omega /\omega _{\mathrm{c}}\right) 
	$, where the dimensionless parameter $\alpha $ defines the strength of the
	coupling and the power law is described by $s$, with $s=1$ for the ohmic
	environment \cite{Ao1989, Ao1991}. For nonohmic cases, see \cite{Leggett1987}.
	As a simple approach, we can consider the environment being modeled as
	only one (finite-temperature) harmonic oscillator \cite{Saito2007,
		Ashhab2014, Malla2018}.
	
	Tracing out the environment’s degrees of freedom, we can obtain a master
	equation for the reduced density matrix $\rho (t)$ of our system, which can
	be written as the Lindblad equation (also known as the Lidblad-Gorini–Kossakowski–Sudarshan equation, \cite{Chruscinski2017}) 
	\begin{equation}
		\color{orange}\boxed{\color{black}\dot{\rho}=-\frac{i}{\hbar }\left[ H(t),\rho \right] +\sum_{k}\left(
		L_{k}\rho L_{k}^{\dag }-\frac{1}{2}L_{k}^{\dag }L_{k}\rho -\frac{1}{2}\rho
		L_{k}^{\dag }L_{k}\right).\color{orange}}\color{black}  \label{Lindblad}
	\end{equation}%
	Different channels of relaxation are described by the Lindblad operators $L_{k}$. 
	Note that Lindblad equation is correct under certain conditions such as:  
	\begin{itemize}
		\item separability (no correlations between the system and its environment at $t=0$),
		\item Born approximation (environment is static despite interaction with quantum system, and weak coupling between quantum system and environment),
		\item Markov approximation (time scale of decay is much shorter than the smallest characteristic time of the system),
		\item secular approximation (all fast rotating terms in the interaction picture can be neglected).
	\end{itemize}
	For the limitations of the Lindblad equation, see \cite{Teixeira2021}.

	In particular, for a TLS, the impact of the environment results in
	relaxation and dephasing, as described by $L_{\mathrm{relax}}=\sqrt{\Gamma _{1}}%
	(\sigma_x - i\sigma_y)/2$ and $L_{\phi }=\sqrt{\Gamma _{\phi }/2}\sigma _{z}$.
	Instead of the relaxation rates, we can use the respective times: $%
	T_{1}=\Gamma _{1}^{-1}$ and $T_{\phi }=\Gamma _{\phi }^{-1}$. Then, the
	Lindblad equation is reduced to the Bloch equations for the components of
	the Bloch vector $\mathbf{s}=\text{Tr}(\mathbf{\sigma }\rho )$; writing the
	Hamiltonian in the form $H=-\mathbf{B\cdot \sigma }/2$, with the effective
	\textquotedblleft magnetic\textquotedblright\ field $\mathbf{B}$, these read as 
	\begin{eqnarray}
		&&\frac{d}{dt}\mathbf{s=-B\times s-}\frac{1}{T_{1}}\left( \mathbf{s}_{\Vert }%
		\mathbf{-s}_{\Vert }^{\mathrm{eq}}\right) -\frac{1}{T_{2}}\mathbf{s}_{\bot },
		\label{Bloch} \\
		T_{2}^{-1} &=&\frac{1}{2}T_{1}^{-1}+T_{\phi }^{-1}\text{, \ \ }\:\:\: \: s_{z}^{%
			\mathrm{eq}}=\tanh (\Delta E/2k_{B}T).
	\end{eqnarray}%
	Importantly, the Bloch equations are obtained based on the qubit
	eigenstates \cite{Krzywda2021} (see also \cite{Xu2014}). In particular, this
	assumes a relaxation to the ground state at low temperatures. Thus, the
	Hamiltonian $H(t)$ in the Lindblad and Bloch equations should be written 
	in the representation of the adiabatic basis. The diagonalization of the
	Hamiltonian is done by a rotation on the angle $\eta $. The
	relaxation rates are related to the spectral density \cite{Makhlin2001} and
	depend on the energy offset $\varepsilon_0$ and temperature $T$ 
	\begin{equation}
		\Gamma _{1}=\pi \alpha \sin ^{2}\eta \frac{\Delta E}{\hbar }\coth \frac{%
			\Delta E}{2k_{B}T}\text{, \ \ }\:\:\:\: \Gamma _{\phi }=\pi \alpha \cos ^{2}\eta 
		\frac{2k_{B}T}{\hbar }.  \label{Gammas}
	\end{equation}%
	The Bloch equations can be extended to a multilevel system
	\cite{Shimshoni1991, Gefen1987}%
	\begin{subequations}\begin{eqnarray}
		\dot{\rho}_{ii}(t) &=&-\frac{i}{\hbar }\left[ H(t),\rho \right] _{ii}-\frac{%
			\rho _{ii}(t)-\rho _{ii}^{\mathrm{eq}}(t)}{T_{1,i}}, \\
		\dot{\rho}_{ij}(t) &=&-\frac{i}{\hbar }\left[ H(t),\rho \right] _{ij}-\frac{%
			\rho _{ij}(t)}{T_{2,ij}},\text{ \ \ }i\neq j\text{.}  
	\end{eqnarray}\end{subequations}
	
	\begin{figure}[tbp]
		\centering{\includegraphics[width=1.0%
			\columnwidth]{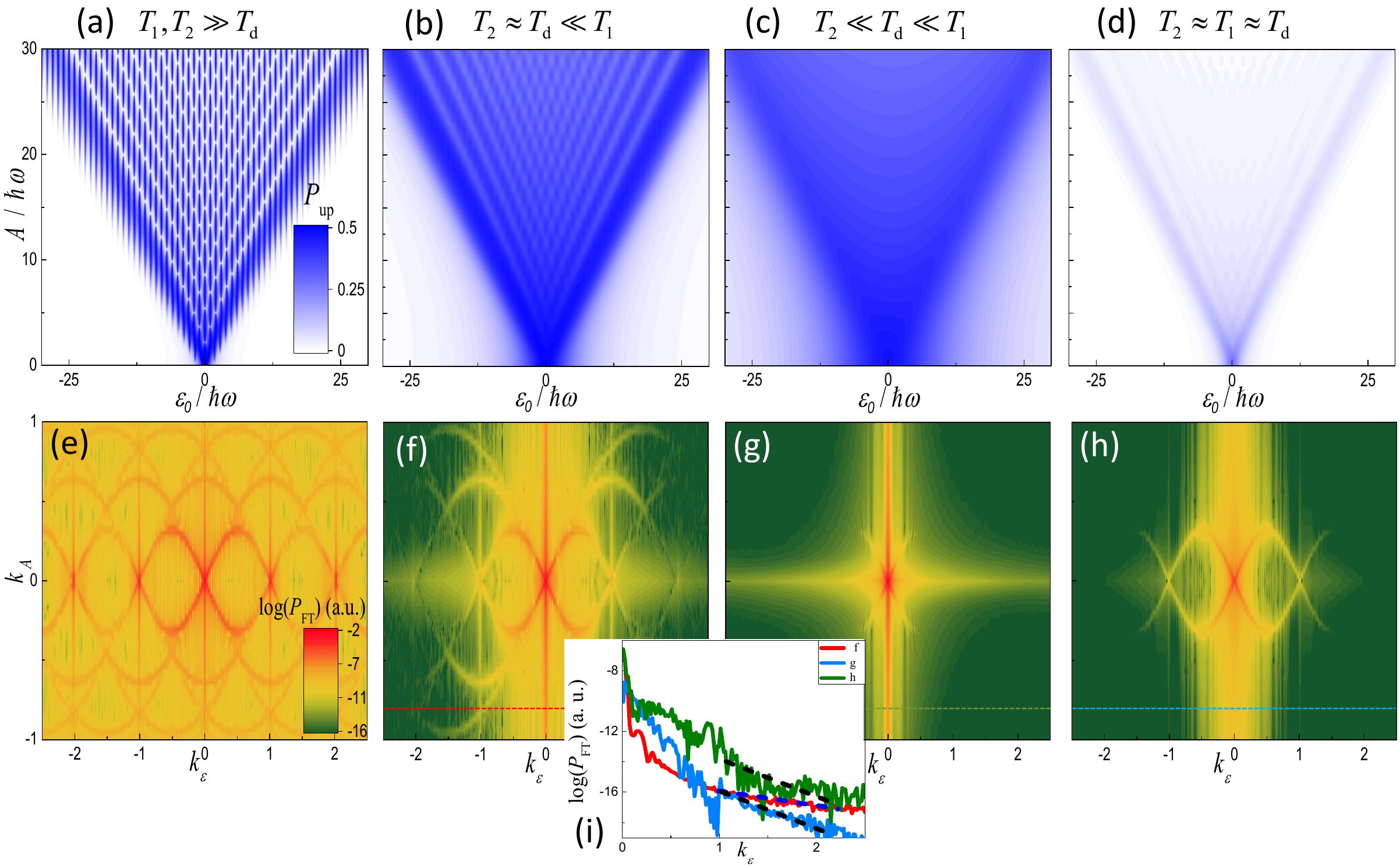}}
		\caption{\textbf{Impact of dissipation and dephasing on interferograms and
				tomograms.} The upper panels present the LZSM interferograms and the lower
			panels their respective 2D Fourier transforms also known as tomograms. These interferograms
			and their tomograms are demonstrated in several cases: (a) and (e) low
			relaxation and dephasing, $T_{1}=10T_{\mathrm{d}}$, $T_{2}=20T_{\mathrm{d}}$%
			; (b) and (f) low relaxation and the dephasing time comparable with the
			driving period, $T_{1}=10T_{\mathrm{d}}$, $T_{2}=0.5T_{\mathrm{d}}$; (c) and
			(g) weak relaxation and strong dephasing, $T_{1}=10T_{\mathrm{d}}$, $T_{2}=0.1T_{\mathrm{d}}$; and (d) and (h) both the relaxation time and the
			dephasing time comparable with the driving period, $T_{1}=0.25T_{\mathrm{d}}$, $T_{2}=0.5T_{\mathrm{d%
			}}$. For all panels, the driving frequency is $%
			\protect\omega =1.14\Delta /\hbar $. The inset (i) shows that fitting the
			linear part of the dependence can be used for defining the decoherence time $%
			T_{2}$. }
		\label{Fig:Dissipation}
	\end{figure}
	
	Therefore, the most straightforward approach to demonstrate the impact of both
	dissipation and temperature is to solve the Bloch equations; see this also in
	Refs.~\cite{Kayanuma1993, Zueco2008, Nalbach2009, Scala2011, Whitney2011,
		Orth2013, Xu2014,Dai2022}. 
		In Fig.~\ref{Fig:Dissipation}, we present the
	interferograms for different relaxation and dephasing rates and their
	Fourier images. In panel (a), we show the case of low dissipation and
	relaxation rates, in which all the resonances are distinguishable. In panel (b),
	we show the case of the average dephasing regime, which we also refer to as the
	double-passage regime, when several passages occur before dephasing
	dominates, which leads to merging resonances into lines. 
	
	The next regime is with large dephasing, which is shown in panel (c), where dephasing dominates
	over excitation and resonances merge into one region $A>|\varepsilon |$.
	Panel (d) presents the strong relaxation case when relaxation dominates
	over excitations. The characteristic pattern for this case is the peak of
	probability near $(\varepsilon_{0}=0, A=0)$, where an increasing in the driving amplitude results in the resonances
	disappearing.
	
	Dissipative LZSM transitions were studied for quantum dots \cite%
	{Fouokeng2014, Ota2017, Huang2021b}, superconducting qubits \cite%
	{Gramajo2019}, and molecular nanomagnets that include the high-spin case \cite%
	{Calero2005, Vogelsberger2006, Foeldi2008}. The transition in a system
	subject to continuous measurement was considered in Ref.~\cite{Haikka2014}.
	When the environment can be treated as a continuum of states, LZSM
	transitions can be formulated within the open-multistate model \cite%
	{Dodin2014}. The impact of dissipation on CDT is the suppression of 
	those processes \cite{Hausinger2010}.
	
	\subsection{Rate equation}
	
	The master equation which describes a TLS can be written in the form of a rate
	equation. (This was not considered in \cite{Shevchenko2010}, but this
	approach is also useful and important.) Consider here the microscopic analysis of
	the dynamics of a TLS based on the rate equation and adding classical noise to
	model decoherence \cite{Berns2006}. Details are presented in the Appendix~%
	\ref{Sec:Rate}. 
	
	First, we calculate the transition rate $W$ for a bias $%
	\varepsilon (t)=\varepsilon _{0}+A\cos \omega t+\delta \varepsilon (t)$, in
	the presence of classical noise $\delta \varepsilon (t)$. For concreteness,
	we present the derivations and results for harmonic driving. This approach can
	be used for any other periodic driving, such as sawtooth-like and
	biharmonic ones \cite{Rudner2008,Berns2008,Cao2013}.
	
	Then, the averaging over $\delta \varepsilon (t)$ using the white-noise
	model gives the rate of transitions between the TLS states
	
	\begin{equation}
		W=\frac{1}{2}\sum_{n=-\infty }^{\infty }\frac{\Gamma _{2}\Delta _{n}^{2}}{%
			\left( \varepsilon _{0}/\hbar -n\omega \right) ^{2}+\Gamma _{2}^{2}},
		\label{W_main}
	\end{equation}%
	where the splitting is modulated with the Bessel function, $\Delta
	_{n}=\Delta J_{n}(A/\hbar \omega )$ and $\Gamma _{2}=T_{2}^{-1}$ is the
	decoherence rate \cite{Du2010, Du2013}.
	
	Next, we construct the rate equation, which includes possible transitions
	between the states, with rate $W$, and relaxation, with rate $\Gamma
	_{1}=T_{1}^{-1}$. From the stationary solution of the rate equation, we
	obtain%
	\begin{equation}
		P_{+}=\frac{W}{2W+\Gamma _{1}}\text{.}  \label{P+_with_W}
	\end{equation}%
	The summation of all possible resonances gives us exactly the same solution
	as the one above: Eq.~(\ref{PupRWA}). This derivation of the formula~(\ref%
	{PupRWA}) reveals its robustness.
	
	The formalism presented here has the advantage that it is straightforward to
	be generalized for multilevel systems \cite{Wen2009}. For a system with $N$
	energy levels, we must solve a system of $N$ rate equations (like we do
	for a TLS in Appendix~\ref{Sec:Rate}) to describe the transitions between
	levels $i$ and $j$ with the rates $W_{ij}$, given by Eq.~(\ref{W_main}),
	replacing the splitting $\Delta \rightarrow \Delta _{ij}$ and offsets from
	them $\varepsilon _{0}\rightarrow \varepsilon _{ij}$. This approach was
	developed and applied to multilevel systems, including the ones based on
	superconducting qubits~\cite{Wen2010, Du2010a, Wang2010, Chen2011} and double quantum dots (DQD). 
	\cite{Chatterjee2018, Liul2022}.
	
	\subsection{Quantum phase tomography}
	
		The Fourier transform is a useful tool for analyzing different periodic
	patterns. In the case of LZSM interferograms, the Fourier transform results
	in a highly ordered structure of one-dimensional arcs \cite{Berns2008}.
	Because the Fourier image gives the structured mapping of the interferograms,
	it presents the tomographic imaging and the respective graphs give rise to a technique called quantum phase tomography \cite{Rudner2008}. This
	approach gives additional information about the physical processes in a
	driven quantum system and is useful for defining system parameters,
	including for probing the dephasing mechanisms \cite{Rudner2008, Cao2013}.
	
	Having an analytical expression for $\overline{P_{\mathrm{up}}}$, we can
	use the continuous 2D Fourier transformation 
	\begin{equation}
		P_{\mathrm{FT}}(k_{\varepsilon },k_{A})=\int \!\!\!\!\int_{-\infty }^{\infty
		}\overline{P_{\mathrm{up}}}(\varepsilon _{0},A)\exp \left[ {-ik}%
		_{\varepsilon }\varepsilon _{0}-ik_{A}A{)}\right] d\varepsilon _{0}dA,
		\label{FT}
	\end{equation}%
	where $k_{\varepsilon }$ and $k_{A}$ are the reciprocal-space variables
	corresponding to the variables $\varepsilon _{0}$ and $A$, respectively. In
	the case when the matrix of the data is given instead by a formula for $%
	\overline{P_{\mathrm{up}}}$, we can use the discrete 2D Fourier transform.
	In the lower panels of Fig.~\ref{Fig:Dissipation}, we present the tomograms
	related to the interferograms in the upper panels; these were calculated by
	the discrete Fourier transform of the data for the interferograms. We can
	see that the relaxation suppresses the low-frequency resonance curves; when
	increasing the relaxation rate, only the main curve (with $l=1$) survives, as
	can be seen in Fig.~\ref{Fig:Dissipation}(h). Note that the vertical and
	horizontal lines of increased $P_{\mathrm{FT}}$ along the axes, especially at 
	$k_{\varepsilon }=0$ and $k_{A}=0$, are artifacts because of the finite size of the
	interferograms.
	
	The overall tomograms in Fig.~\ref{Fig:Dissipation}(e-h) are
	lemon-shaped structures formed by the 1D sinusoids of the form 
	\begin{equation}
		k_{A}(k_{\varepsilon })=\pm \frac{2l}{\omega }\sin \left( \frac{\omega
			k_{\varepsilon }+2\pi l^{\prime }}{2l}\right),   \label{sinusoids}
	\end{equation}%
	with $l=1,2,3,...$, $l^{\prime }=0,1,2,...$, and $l^{\prime }<l$. This can
	be seen by either analysis of the phase acquired during the periodic
	evolution \cite{Rudner2008} or by directly transforming Eq.~(\ref{PupRWA}) 
	\cite{Forster2014,Blattmann2015}. From the lower panels in Fig.~\ref{Fig:Dissipation}, we
	can see how relaxation and dephasing impact the Fourier images. With low
	relaxation and decoherence rates, there are many resonances in panel (e). From
	panels (f) and (g), we can see that increasing the dephasing rate leads
	to increasing the slope rates along the $|k_{\varepsilon }|$ axis. Then, when we
	compare Fourier images on panels (f) and (h), we can see that for the same
	dephasing rate, increasing the relaxation rate leads to the disappearance of the
	higher resonances. In panel (i), we show the cross-sections of panels
	(f,g,h) along $k_{A}=0.85$; the slopes are given by the dephasing rates, as
	shown by the dashed lines. Note that in this panel, this is very noisy because of
	resonances and artifacts; to better see the slopes, one can consider the
	response along the resonance lines, as in \cite{Forster2014}.
	
	From Eqs.~(\ref{W_main}-\ref{P+_with_W}), we can find that the slope of
	the Fourier image is proportional to the decoherence rate:\newline $\ln P_{\mathrm{FT%
	}}\propto -\Gamma _{2}|k_{\varepsilon }|$ \cite{Rudner2008}. This is
	demonstrated as the inset, Fig.~\ref{Fig:Dissipation}(i). This
	observation provides a convenient tool for obtaining the decoherence rate
	without additional measurement or fitting \cite{Cao2013, Gonzalez-Zalba2016}.
	If this theory accounts for inhomogeneous broadening $\Gamma _{2}^{\ast
	}$, this results in 
\begin{equation}
	\ln P_{\mathrm{FT}}(k_{\varepsilon },k_A({\varepsilon }%
	))\propto -\Gamma _{2}|k_{\varepsilon }|-\Gamma _{2}^{\ast 2}k_{\varepsilon
	}^{2}/2, 
\end{equation}
where $k_A(k_{\varepsilon })$
 is defined by the relation Eq.~(\ref%
	{sinusoids}) \cite{Forster2014}. In addition to $\Gamma _{2}$, this provides a
	convenient tool for defining $\Gamma _{2}^{\ast }$, most importantly without any
	additional measurements, such as spin-echo experiments \cite{Cao2013}.
	
	Hence, the Fourier transform provides a useful tool, quantum-phase tomography, providing an additional visualization of the interference and the
	system parameters, such as the decoherence rate. This methodology can be
	further developed and applied to other systems, specifically for bichromatic driving 
	\cite{Forster2015} and multilevel systems \cite{Berns2006, Liul2022}.
	
	\subsection{Comparison of different methods}
	
		\begin{table}[b]
	\begin{tabularx}{500pt}{|X|c|c|c|c|c|c|}
		\hline
		\textit{}&$\:\:\:\:$ \rule{0pt}{3ex}  Numerical ODE  $\:\:\:\:$ &$\:\:\:$  Numerical QuTiP  $\:\:\:\:$ & $\:\:\:\:$  Floquet method  $\:\:\:\:$ & $\:\:\:$  RWA  $\:\:\:$ & $\:\:\:\:$  Optimized RWA $\:\:\:\:$  &$\:\:\:$ AIM $\:\:\:$ \\[1ex] \hline
		
		$(\varepsilon_0,A)$ & \rule{0pt}{3ex} 3144 & 2050 & 1300 &16.4 & 4.7 & 9.4 \\[1ex] \hline
		
		$(\varepsilon_0,\omega)$ & \rule{0pt}{3ex} 1400 & 1241 & 774 & 9.3 & 4 & 8.6  \\[1ex] \hline
		
	\end{tabularx}
	\caption{\textbf{Relative efficiency of different methods}. This is quantified 
		with the time (in seconds) needed to calculate the respective interferograms
		shown in Fig.~\protect\ref{Fig:Different_Methods_Interferograms}. The first line
		relates to the interferograms in the coordinates $(\protect\varepsilon %
		_{0},A)$ and the second line to the coordinates $(\protect%
		\varepsilon _{0},\protect\omega )$. The numbers in the table were obtained using an eight-core CPU Intel 10875H using eight threads.}
	\label{Table:comparison}
\end{table}
	
	Above we have considered in detail different methods for describing
	driven quantum systems. Now, we would like to compare these different
	approaches; see also \cite{Ashhab2007}. We summarize their key aspects and
	then compare them by computing the interferograms.
		\begin{figure}[h]
		\centering{%
			\includegraphics[width=1.0\columnwidth]{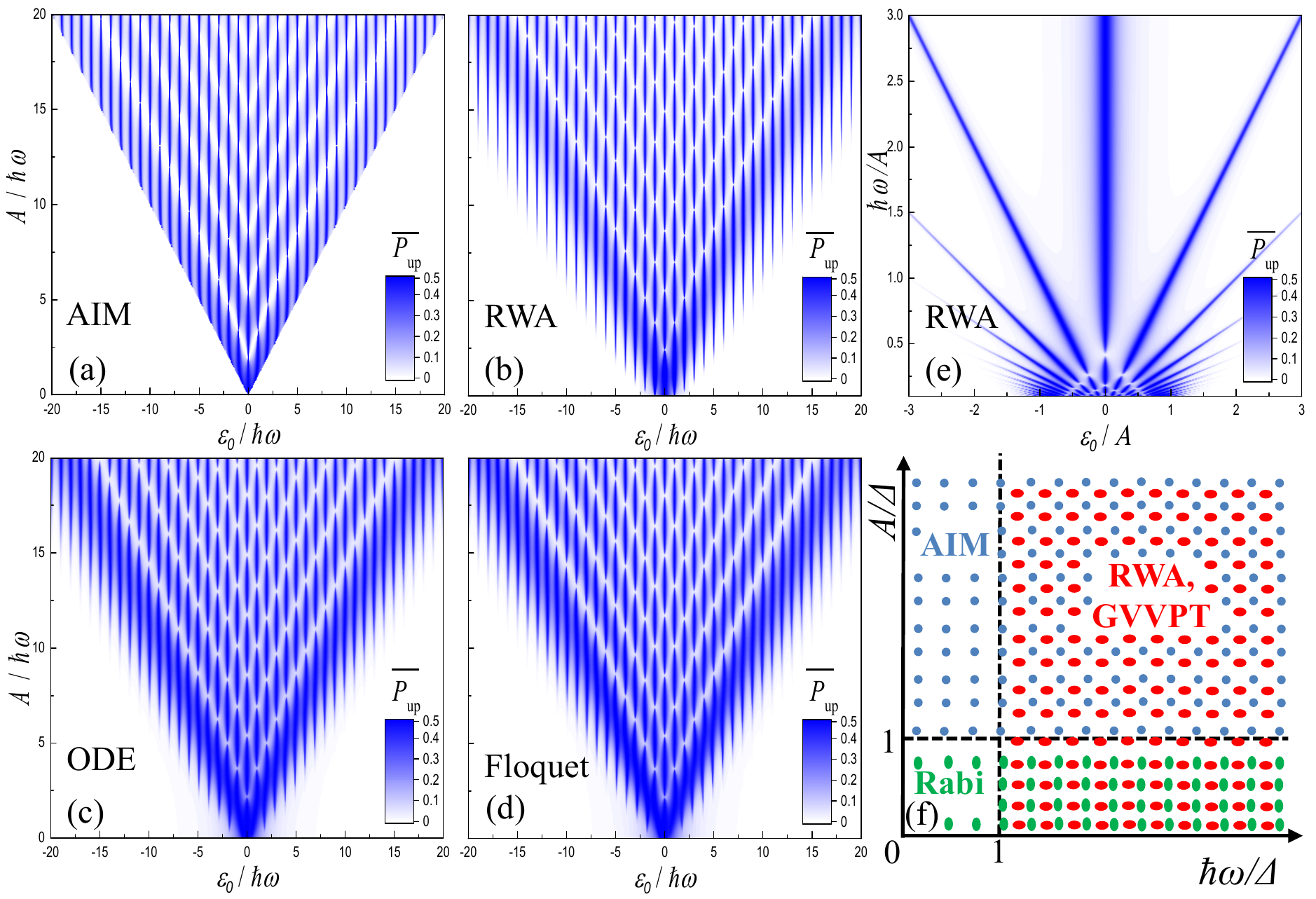}}
		\caption{\textbf{LZSM interferograms calculated by different methods}. (a-d)
			and (e) present the time-averaged upper-level occupation probability as a
			function of the energy offset $\protect\varepsilon _{0}$, the driving
			amplitude $A$ for (a-d) and the driving frequency $\protect\omega $ for (e).
			Comparison between the regions of validity for different methods with small
			relaxation and dephasing rates (f). The methods used are the adiabatic-impulse
			model (AIM) (blue circles in panel (f)), rotating-wave approximation (RWA), and the second approximation by the Floquet method given by the generalized Van-Vleck perturbation theory (GVVPT), which has the same region of validity (red ovals in panel (f)). Numerical solution of the ordinary differential
			equations (ODE), and the Floquet approach can be applicable everywhere.  Also, panel (f) demonstrates the
			region of validity for the Rabi oscillations method (green ovals).}
		\label{Fig:Different_Methods_Interferograms}
	\end{figure}

	\paragraph{\textbf{Numerically solving the Schr\"{o}dinger or Lindblad equation}}\mbox{}\\
	
	This method is the most commonly used because it is based on the integration of a
	system of differential equations. This gives the best-quality
	results and can be used in all range of parameters. We compare all other
	methods with this one, because the numerical integration can be performed accurately. However, the numerical integration has one significant
	disadvantage: it needs considerable computing power, so it requires considerable CPU time
	to compute interferograms. We use the numerical solution based on the modified
	Runge-Kutta method, which is a standard approach for solving systems of
	ordinary differential equations (ODE) in Python.
	
	\paragraph{\textbf{Numerically solving using QuTiP}}\mbox{}\\
	
	The QuTiP (Quantum Toolbox in Python) framework \cite{Johansson2012,Johansson2013,Shammah2018} is a very useful tool to calculate the dynamics of
	quantum systems. We need only to introduce a Hamiltonian and set of
	parameters; as a result, we have the full evolution of the system (all terms
	of the density matrix, in all the defined moments of time) or, alternatively,
	the averaged value of population of a certain level. This library also has a
	detailed manual and many different solvers.
	\paragraph{\textbf{Floquet method}}\mbox{}\\
	
	This approach can be used only for periodic Hamiltonians and is based
	on the Floquet theorem. The Floquet method is faster than the direct numerical
	solution, but slower than other analytical methods. Importantly, relaxation and dephasing can be included. This method is analogous to a
	decomposition on eigenfunctions in methods of mathematical physics. 

	\paragraph{\textbf{Rotating-wave approximation (RWA)}} \mbox{}\\
	
	The RWA corresponds to the first term in a
	Floquet formalism by the small distance between levels $\Delta\ll\omega,%
	\varepsilon$. It is the most commonly used analytical method because it combines
	adequate accuracy with a moderate need in computing power. Here, for
	calculating interferograms, the range of application is limited by the
	overlap of resonances, which occurs at increasing relaxation and dephasing
	or decreasing driving frequency. In the simplest approach, we can use
	the formula Eq.~\eqref{PupRWA}, which assumes calculating a number of terms
	in each point. However, if we take a few close resonances only, that is, with $%
	k$ close to the resonance condition $\varepsilon_0=k\hbar\omega$, this
	significantly speed-up the calculation; this approach can be  called \textit{optimized RWA}.
	
	\paragraph{\textbf{Adiabatic-impulse model (AIM)}} \mbox{}\\
	
	The Adiabatic-impulse model can provide interferograms in different
	bases. This is generally done neglecting relaxation and decoherence;  thus the time-averaged result obtained by this method depends on the initial condition.  This correctly gives
	the location of the resonances; but between them this method may
give different values from those obtained using more precise numerical calculations, which
	is a result of not including relaxation and dephasing. This means that the time-averaged result depends on the initial condition; but when we use the ground-state initial condition in this AIM and small dissipation in other methods, then the AIM gives the correct result.  The range of usability of
	this method is defined by a high amplitude of excitation $A>\Delta$, which is
	needed for the adiabatic evolution between transitions. The second limitation is
	the times of the transition process: $T_\text{d}/2>(\tau_\text{relax}+\tau_\text{jump})$. See the transition times described in paragraph \ref{Dynamics
		and times of a transition}.
	\hspace{1cm}
	\paragraph{\textbf{Interferograms and time of calculation}} \mbox{}\\

	We illustrate our conclusions in Fig.~\ref%
	{Fig:Different_Methods_Interferograms} and table \ref{Table:comparison}.
	First, in Fig.~\ref{Fig:Different_Methods_Interferograms}: (a,b,c,d), we
	present the interferograms using the coordinates $(\varepsilon _{0},A)$; and in (e)
	using $(\varepsilon _{0},\omega )$. Even the former shows not so many
	differences, and these become even less for the latter in (e). For the calculations, we
	chose the driving frequency $\hbar \omega /\Delta =1.14$ and the relaxation
	rates for (b-d), $\Gamma _{1}=0.1T_{\mathrm{d}}^{-1}$ and $\Gamma
	_{2}=\Gamma _{1}/2$. Note that the adiabatic-impulse model in (a) is valid under the condition $%
	A>\varepsilon _{0}$, with no relaxation and decoherence included. The RWA in (b)
	provides a result more similar to the numerical solution in (c), except
	the shifting of the resonances when increasing the driving amplitude, here in the
	region $A<\varepsilon _{0}$.
	
	In table \ref{Table:comparison}, as a quantitative measure of the
	efficiency, we have chosen the time needed to compute the interferograms. For this,
	we used an eight-core CPU Intel 10875H using eight threads. Calculations
	were performed as in Fig.~\ref{Fig:Different_Methods_Interferograms}, with
	401 points in each axis.  Note that the most significant impact on calculation time is given by the number of points, while the dependence on other parameters vary for different methods. 
	  As we see from table \ref{Table:comparison},
	the analytical methods are much faster than the numerical calculations, with a
	difference of about two orders of magnitude in calculation time. However, the
	numerical calculations is more flexible and can be easily used for more
	complicated systems, including many levels, adding noise, and so forth. For describing
	the time-averaged occupation probability, the most convenient method is the RWA. For
	describing the dynamics, one of the most illustrative methods is the adiabatic-impulse model, but it is
	limited by the time of the transition process; for more related information, see Sec.~\ref{Dynamics and times
		of a transition}.
	
	\section{Interferometry}
	
	\label{Sec:Interferometry}
	
	The interference in a driven two- or multi-level system is essentially
	defined by the problem parameters. It is important for applications to
	demonstrate this through calculations and to formulate recipes of how to extract not only 
	the quantum system parameters from the interferograms, but also the
	information about systems coupled with them, as well as about the dissipative
	environment \cite{Shytov2003, Yang2017}.
	
\begin{figure}[tbp]
		\centering{\includegraphics[width=0.8\columnwidth]{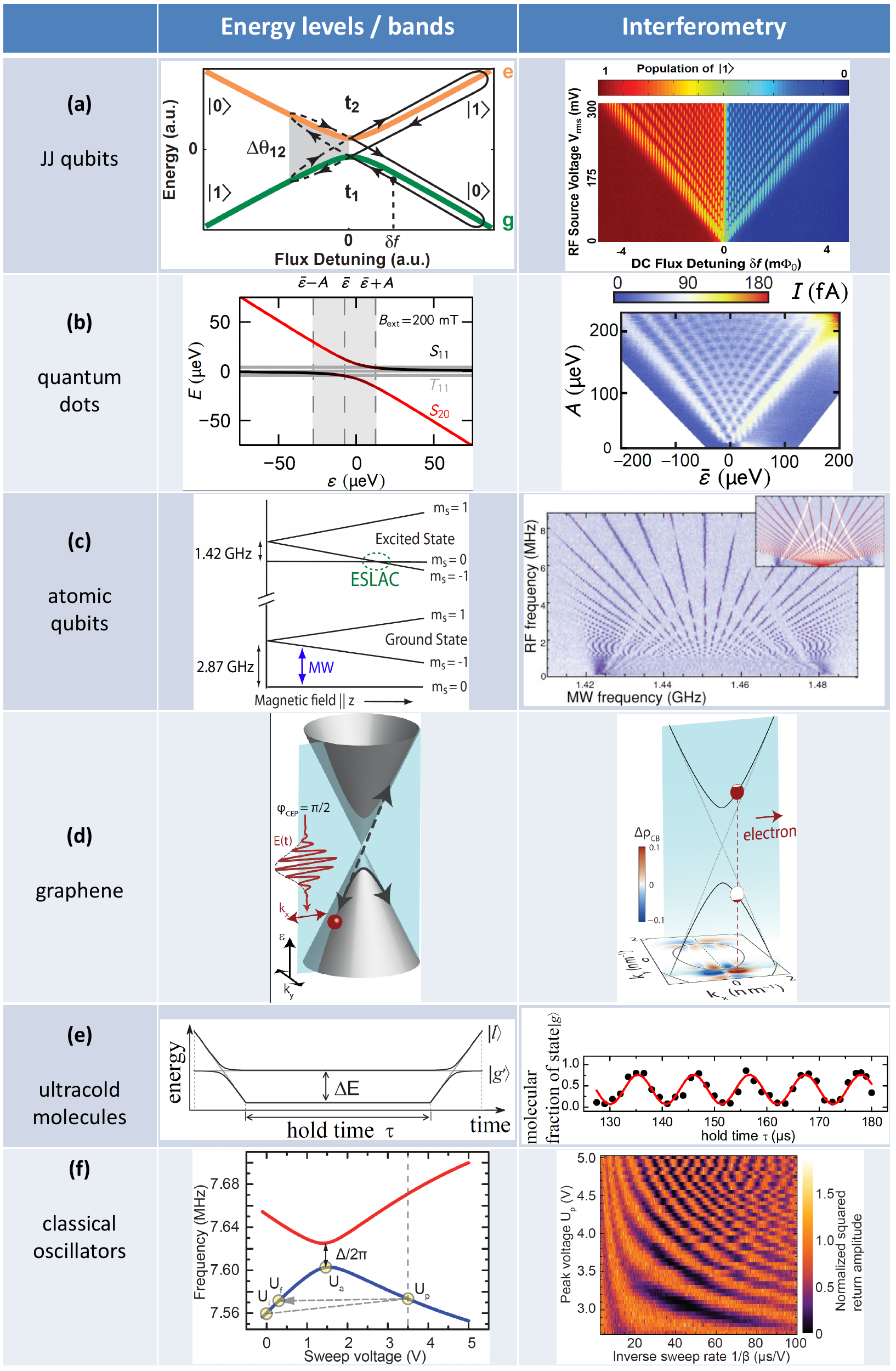}}
		\caption{\textbf{LZSM interferograms in different systems. }See the extended explanation in the  text for
			a detailed description. Figure (a) is reprinted from \protect\cite{Berns2006}
			with permission; copyright (2006) by APS. Figure (b) is reprinted from 
			\protect\cite{Forster2014} with permission; copyright (2014) by APS. Figure
			(c) is reprinted from \protect\cite{Childress2010} with permission;
			copyright (2010) by APS. Figure (d) is reprinted from \protect\cite%
			{Heide2018} with permission; copyright (2018) by APS. Figure (e) is
			reprinted from \protect\cite{Mark2007} with permission; copyright (2007) by
			APS. Figure (f) is reprinted from \protect\cite{Seitner2016} with
			permission; copyright (2016) by APS. }
		\label{Fig:interferograms}
	\end{figure}

	Although we have given a detailed description of various systems with LZSM interference,
	we display the diversity of these, presenting
	several systems in Fig.~\ref{Fig:interferograms}. (Please note that the
	choice of works and figures for Fig.~\ref{Fig:interferograms} are used only for 
	illustrative purposes, while many other works are analyzed below and
	throughout this review paper.)  A detailed description of Fig.~\ref%
	{Fig:interferograms} follows.
\paragraph*{\textbf{Extended explanation of Fig.~\eqref{Fig:interferograms}:}}
	\paragraph*{ \textbf{(a) Superconducting circuits}} \mbox{}\\
	
	Although JJ-based
	quantum superconducting circuits are diverse, we
	would like to illustrate the work on these systems with Ref.~\cite{Berns2006}. A superconducting flux
	qubit, also called a persistent-current qubit, was realized as a niobium (but it could
	also be aluminum) superconducting loop with three JJs. The
	basis states are created by a persistent current flowing either clockwise or
	in the counterwise direction. Strong driving by an AC magnetic flux induces
	transitions between the two states. In Ref.~\cite{Oliver2005}, these qubit
	states were read out with an outer DC-SQUID, a sensitive magnetometer that
	distinguishes the flux generated by the persistent currents. Up to 20-photon
	transitions were observed, with an impressive increase up to 45-photon
	transitions in Ref.~\cite{Berns2006}. The energy-level distance was
	controlled by a DC magnetic flux; the tunneling amplitude was rather small, $%
	\Delta \simeq 13$~MHz$\cdot h$, and the driving frequency $\omega /2\pi
	=0.27~\mathrm{GHz}\gg \Delta /h$.
	\paragraph*{\textbf{(b) Quantum dots}} \mbox{}\\
	
	These are based on electrons localized in
	semiconductor low-dimensional systems. Specifically, a two-electron charge
	qubit, which is defined in a lateral double quantum dot, is considered in Ref.~\cite%
	{Forster2014}. The source and drain leads, tunnel-coupled with dots, allow current
	flow by single-electron tunneling. This current is used to detect the
	properties of the driven double quantum dot (DQD) device. The charge qubit
	is formed by the singlet states $S_{11}$ (one electron in each dot) and $%
	S_{20}$ (two electrons in the left dot). The source-drain current is
	proportional to the occupation probability of $S_{20}$. The three triplets $%
	T_{11}$ are hindered by a Pauli-spin blockade; however, blockade is
	lifted using an on-chip nanomagnet to quickly initialize the qubit. In the
	energy diagram, the singlets (the qubit states) are represented as black and
	red lines; the triplets, which are Zeeman split, are shown as gray
	lines. The interdot coupling is $\Delta \simeq 3.1$~GHz$\cdot h$, and the
	energy detuning $\varepsilon $ is defined by a DC+AC gate voltages $%
	\varepsilon (t)=\overline{\varepsilon }+A\cos \omega t$. The driving
	frequency for the interferogram shown is $\omega /2\pi =2.5~$GHz, which is smaller than $\Delta /h$.
		\paragraph*{\textbf{(c) Atomic qubits (impurity or donor based)}} \mbox{}\\
	
	These may involve
	electrons, holes, and/or nuclear spins in impurities within silicon, diamond,
	or other materials. For illustration purposes, we now consider an NV
	center in diamond, which consists of a substitutional nitrogen atom
	replacing a carbon atom and neighboring one vacancy. Its electronic ground
	state has a spin $S=1$, which, in the absence of a magnetic field, has the
	spin singlet sublevel $m_{s}=0$ situated $2.87$~GHz below the $m_{s}=\pm 1$
	two-fold-degenerate triplet sublevels. The optically excited state has the
	zero-field splitting $1.42$~GHz. These can be further changed by the
	magnetic field. We illustrate this with an energy diagram displaying the
	excited-state level anticrossing (ESLAC) and the interferogram from Ref.~%
	\cite{Childress2010}. The system was doubly driven by a microwave and rf
	signals, resulting in characteristic dressed-state interference fringes.
	This was displayed by measuring a spin-dependent fluorescence intensity,
	which is proportional to the excitation probability in the TLS
	formed by the $m_{s}=0$\ and\ $m_{s}=-1$ states; on the color scale, darker
	indicates lower fluorescence. Versatile analysis, which included
	impressively elaborated experimental, numerical, and analytical aspects,
	allowed the authors \cite{Childress2010} to describe the dressed-state interferometric picture
	with aspects such as multi-photon transitions and  coherent destruction of tunnelling (CDT).
	\paragraph*{ \textbf{(d) Graphene structures}} \mbox{}\\
	
	Electrons in graphene have a
	Dirac-cone dispersion, with a slope $v_{\mathrm{F}}$ around the K and K$%
	^{\prime }$ points. If electrons are driven by an electric field $\mathbf{E}%
	=E_{0}\mathbf{e}_{x}$ of linearly $x$-polarized light with an amplitude of $%
	E_{0}$, only the $k_{x}(t)$ component of the wave vector is affected, meaning that
	the relevant electron dispersion is a hyperbola \cite{Higuchi2017,
		Heide2018}. This dispersion can be described with the TLS Hamiltonian: $H=-%
	\left[\Delta \sigma _{x}+\varepsilon (t)\sigma _{z}\right]/2$, with $%
	\varepsilon (t)=2\hbar v_{\mathrm{F}}k_{x}(t)=2v_{\mathrm{F}}e\omega
	^{-1}E_{0}(t)$ \cite{Heide2021a}. The avoided-level spacing $\Delta =2\hbar v_{\mathrm{F}%
	}k_{y}$ is continuous and can be swept with $E_0$ for optimum splitting of the occupation probability. The central driving photon energy was $\hbar \omega
	\approx 1.55~$eV, which is $\omega /2\pi \approx 375~$THz; $v_{\mathrm{F}%
	}\approx 1$~nm$\cdot $fs$^{-1}$ and $E_{0}$ from $ 0 $ to $ 3$~V$\cdot $nm$^{-1}$ \cite{Heide2020,Heide2021}. Hence, 
	in a strong-field regime, the electron dynamics is governed
	by sub-optical-cycle LZSM interference, here on a femtosecond timescale \cite{Ishikawa2013}. For
	this, a graphene strip on a SiC substrate was coupled with Au/Ti leads and
	illuminated by two-cycle laser pulses with a controlled carrier-envelope
	phase, in a vacuum chamber. Since the LZSM interference takes place on a time scale where the coherence of the electron wave function is preserved, this experiment can even be performed at room temperature. Adiabatic evolution and
	nonadiabatic transitions correspond to intraband and interband processes
	for the electron wavefunction. The excitation probability of the upper
	(conduction) band results in the generation of a residual persistent
	current after the laser pulse ended. The right figure shows the conduction band population imbalance $\Delta\rho$ with respect to $k_x = 0$ (dashed line) generated with $E_0 = 2.5\,V\cdot $nm$^{-1}$, resulting in a net electron flow to $k_x>0$. An increase of $E_0$ leads to an inversion of the imbalance and hence, the current direction reverses, which is a direct experimental indication for LZSM interference. The curved lines in the $k_x$--$k_y$ plot indicate the multi-photon resonances where the energy difference between the two bands is $n\cdot\hbar\omega$, with $n = 1$ on the innermost circle and the subsequent ring with $n = 2$. 
	
	\paragraph*{ \textbf{(e) Ultracold atoms and molecules}} \mbox{}\\
	
	When a bound molecular dimer
	state is magnetically tuned near a two-atom scattering state, the Feshbach
	resonance leads to a resonant atom–molecule transition. The internal state
	structure of ultracold Feshbach molecules has several avoided crossings,
	which is considered in the example of weakly bound Cs$_{2}$ dimers \cite%
	{Mark2007}. The crossing used for the interferometer is the one between $%
	\left\vert g^{\prime }\right\rangle $ and $\left\vert l\right\rangle $
	states, which can be determined upon molecular dissociation. Two
	subsequent passages through a weak avoided crossing between these two
	different orbital angular momentum states result in LZSM interference,
	which is defined by the variable hold time $\tau $. The driving is realized
	by ramping the magnetic field $B$ around the point of the avoided crossing, here
	at $B=B_{\mathrm{c}}$. The acquired phase difference is defined by the
	product of $\Delta E$ and $\tau $. The interference is visualized in the $g$-wave molecular fraction of the state $\left\vert g\right\rangle $ as a
	function of the hold time $\tau $. The oscillation frequency of the
	sinusoidal fit is about $100$~kHz; and up to $100$ oscillations were observed.
	
	\paragraph*{ \textbf{(f) Mechanical oscillators}} \mbox{}\\
	
	One classical system that
	behaves as a two-level system is a two-mode mechanical oscillator. This system
	could be either two coupled mechanical oscillators or two modes of one
	mechanical oscillator. The latter case is realized by \cite{Seitner2016} as
	two flexural modes of a string resonator. The modes represent in-plane and
	out-of-plane oscillations with an eigenfrequency of about 7~MHz. The resonator is
	made of a high-stress silicon nitride and has the cross-section of
	100~x~270~nm, here with the length being 50~$\mu $m. The sweep voltage controls
	the distance between the eigenfrequencies, with the minimal splitting $\Delta
	/2\pi $ of the order of 10~kHz. Triangular changes of the sweep voltage
	result in transgressing the avoided crossing twice. Then, the interference
	results in the fringes shown as amplitude of the transitions between the
	modes versus the sweep rate $\beta ^{-1}$ and amplitude $U_{\mathrm{p}}$.
	
	\subsection{Superconducting circuits}
	
	\paragraph*{\textbf{Flux qubits}} \mbox{}\\
	
	This is presented above in relation to Fig.~\ref%
	{Fig:interferograms}, with references to~\cite{Oliver2005, Berns2006}. Based
	on such interferometry, these authors also developed several techniques:
	amplitude spectroscopy \cite{Berns2008}, quantum phase tomography \cite%
	{Rudner2008}, and pulse imaging \cite{Bylander2009}. These works were summarized
	in Ref.~\cite{Oliver2009} and scrutinized theoretically in Refs.~\cite%
	{Ferron2010, Ferron2012, Ferron2016}. Other close observations were realized:
	an rf-SQUID-based flux qubit with the persistent-current states probed
	by the nearby dc-SQUID magnetometer \cite{Sun2009} and three-junction aluminum
	flux qubit probed by a niobium $LC$ (tank) circuit \cite{Ilichev2004,Izmalkov2008}. In
	this latter case, the sign-changing response was explained in terms of the
	impedance measurement technique \cite{Ilichev2007,Shevchenko2008, Shevchenko2008b}, where the quantum inductance is related to the qubit
	occupation probability and its derivative. In a flux qubit, the role of a
	weak link can be played not only by a conventional JJ, but
	also by a phase slip based on nanowires made from thin films of niobium
	nitride, for which the interferometry was observed in Ref.~\cite%
	{Neilinger2016}.

	\paragraph*{\textbf{Charge qubits}} \mbox{}\\
	
	A superconducting charge qubit is based on a
	Cooper pair box that is created using JJs. In Ref.~\cite%
	{Sillanpaeae2006}, this was controlled by the magnetic flux piercing the loop
	and biased by the gate voltage; the state was read out by an $LC$ circuit. The
	latter feels changes in the (quantum) capacitance, which probes the
	charge qubit state \cite{Sillanpaeae2005, Sillanpaeae2007}. 
	
	Alternatively,
	the qubit was probed via an inductive coupling to the resonant circuit \cite%
	{Tuorila2013}. Two types of multi-photon transitions in this system were
	further studied by \cite{Paila2009}. A similar aluminum Cooper pair box was
	probed via the reflection coefficient in a coupled electric resonator in
	Refs.~\cite{Wilson2007, Wilson2010}, where an emphasis was made on the
	dressed-state description, and in Refs.~\cite{Graaf2013, Leppaekangas2013},
	where the emphasis was made on quasiparticle tunneling. 
	
	An analogous qubit was
	also probed by a nanomechanical resonator \cite{LaHaye2009}, where a
	charge qubit state influenced the displacement of the resonator, of which
	the resonant frequency was probed. The sign-changing behavior of the
	response in that case was explained in Ref.~\cite{Shevchenko2012a} in terms
	of the quantum capacitance.
	\paragraph*{\textbf{Transmon qubits}} \mbox{}\\
	
	This refers to a charge qubit 
	shunted by a capacitance \cite{Xiang2013,Krantz2019}. This results in a small sensitivity to the charge
	noise because of the alignment of the energy levels as a function of a gate
	voltage, meaning there is \textit{no intrinsic avoided-level crossing}. Because
	this is needed for LZSM interferometry, the solution is to create
	dressed-states by applying a microwave signal; then, the slow driving would
	result in traversing these dressed-states around the avoided-level crossing.
	LZSM interference was realized by several groups with the following
	features: driven by sinusoidal and noisy signals \cite{Li2013}, performing a
	time-resolved state tomography measurement \cite{Gong2016}, resolving
	multi-photon processes \cite{Chang2020} with variable coupling between the
	qubit and the transmission line \cite{Wen2020}, and coupling to a mechanical
	resonator \cite{Bera2020}.
	
	\subsection{Semiconductor quantum dots}
	
	Quantum dots are based on an electron charge, spin, or even valley degree of
	freedom (for brevity, we refer to electrons; however, sometimes, these could be
	holes). LZSM interferometry was realized and studied on different types
	of quantum dots, as we describe below. Mainly, these are 
	DQDs exploiting either an electron spin (namely, a singlet-triplet
	transition) or an electron charge (being placed in one of the quantum dots).
	\paragraph*{\textbf{Spin qubits}} \mbox{}\\
	 
	 A system of two electrons in a DQD can be in a singlet
	or triplet state, which form a singlet-triplet or spin qubit \cite{Jirovec2021,Burkard2021}. The states are coupled via the hyperfine interaction between
	the trapped electron spins and surrounding nuclear spins. The singlet
	and triplet states form an avoided-level crossing controlled by a detuning
	gate voltage and external magnetic field (which splits a triplet state
	into three states). This was realized in Refs.~\cite{Petta2010, Stehlik2012}
	in a triple quantum dot geometry, where one of the dots was used as a
	high-sensitivity quantum-point-contact charge sensor. 
	
	Related studies were
	performed in a linear triple-dot device in Refs.~\cite{Gaudreau2011,
		Studenikin2012, Granger2015}, realizing a spin qubit on either two
	three-spin states or three two-spin states; in a three-qubit chain, the
	transitions in the avoided-crossing regions were used to shuttle the
	entanglement \cite{Nakajima2018}; in a DQD fabricated in a Si/SiGe
	heterostructure, here with an integrated micromagnet \cite{Wu2014}; and in a silicon
	metal-oxide-semiconductor DQD with both single-spin addressability and
	single-shot readout \cite{Fogarty2018}. For a theoretical analysis of the dynamics in singlet-triplet transitions in DQDs, see \cite{Brataas2011, Saerkkae2011,
		Ribeiro2013a, Mehl2013, Zhao2018}. For extreme harmonic generation in a multielectron double quantum dot, displaying multi-level interference resonances, see \cite{Stehlik2014,Danon2014}.

	\paragraph*{\textbf{Charge qubits}} \mbox{}\\
	 
	 These can be based on a DQD that is laterally defined from a
	GaAs/AlGaAs-heterostructure double quantum dot with the voltage pulses
	applied to depletion gates~\cite{Petersson2010, Nalbach2013}. There, the two
	states of a charge qubit correspond to one electron in the left and right
	dots; the charge occupation of the dots can be continuously detected via the
	electric current through a capacitively coupled quantum point contact. With
	a similar layout, the authors of Ref.~\cite{Cao2013} used LZSM
	interference for ultrafast universal quantum control at the picosecond
	scale, which is important given that the decoherence times in semiconductor
	quantum dots are typically less than a few nanoseconds. 
	
	A similarly driven DQD
	was realized in Ref.~\cite{Braakman2013} in an array of three dots 
	to demonstrate coherent coupling via long distance, which involved
	two three-dot states. The charge qubit in Ref.~\cite{Forster2014} was probed
	through the DC source-drain current, as discussed above. The effect of the pulse
	shape and distortion was studied in Ref.~\cite{Ota2018}. Further development
	is a coupling of charge qubits: two DQDs were strongly coupled capacitively
	by the authors of Ref.~\cite{Ward2016}, where the LZSM interference in one
	DQD (right one) was shown to be defined by the charge state of the other DQD
	(left one), which was called charge-state-conditional coherent quantum
	interference.
	
	Charge qubits can be defined not only on a two-dimensional gas, but also
	based on a nanowire. In Ref.~\cite{Stehlik2014}, this was an InAs nanowire that was
	placed on top of prepatterned depletion gates; the authors studied 
	strong harmonic generation involving up to eight photons for qubit
	excitation. In the circuit quantum electrodynamics (cQED) architecture, such
	DQD can be placed in a microwave cavity (with a resonant frequency in a GHz
	domain), where driving the qubit results in interferometric cavity power
	gain \cite{Stehlik2016}. 
	
	Using silicon allows for exploiting the advances of
	complementary metal-oxide-semiconductor (CMOS) technology, as in Ref.~\cite%
	{Gonzalez-Zalba2016}, in which the DQD was realized on the corner states in a
	silicon nanowire transistor. There, an electrical resonator response probed
	the differential capacitance of a DQD, which contained two competing terms: one,
	called the quantum capacitance, is proportional to the occupation
	probability, and the other one, which is the tunneling capacitance, is proportional to
	the derivative of the occupation probability; in resonance, the former gives
	the Lorentzian peak and the latter the alternation of a peak and a
	dip. Because of the quantum capacitance, similar peak-and-dip features were
	observed on a DQD charge qubit based on a carbon nanotube \cite%
	{Penfold-Fitch2017}. The authors of Ref.~\cite{Khivrich2020} constructed a
	TLS in a carbon nanotube single quantum dot from electron wavefunctions with
	different magnetic moments and spatial charge distributions.
	
	\subsection{Donors and impurities (atomic qubits)}
	
	The electrically controlled electronic states in some donors and impurities have
	much in common with quantum dots, both conceptually and in terms of qubit
	specifications and hardware requirements. Similar to quantum dots, a donor
	atom has one excess electron, here compared with the atoms in the surrounding
	lattice (less frequently in this context, these could be acceptors with
	single excess holes). It is the electron spin states that form the spin
	qubit basis. These spin qubits have many key assets that are well separated from
	the noisy environment and that use nuclear spins in addition to electronic ones 
	\cite{Vandersypen2017}. Good isolation from the environment is characterized
	by long coherence times ($\sim 1$~ms for electron spins and $\sim 1$~s for
	nuclear spins). Similar to the gate-defined quantum dots above, quantum
	information can be encoded in the charge states of an electron in a
	double-well potential of two nearby impurities (a charge qubit) or in a spin
	state of an electron or nuclear spin in a single impurity (spin qubit).
	We consider several works, where LZSM interference was realized.
	
	Our first example is phosphorous (P) or arsenium (As) donor atoms in silicon
	(Si). The authors of Refs.~\cite{Dupont-Ferrier2013, Jehl2015} studied a
	system of two tunnel-coupled donor atoms in silicon. A TLS was formed by a
	single electron located on either of the two donors. The donors were
	implanted in a silicon nanowire, and the coherent interference pattern was
	observed by measuring the source-drain current through the double-donor
	system. In Ref.~\cite{Ono2019}, a combination of a shallow impurity and a deep
	impurity in a silicon field-effect transistor created a DQD-like structure
	controlled by a magnetic field and rf and mw signals, of which the layout allows
	quantum coherent dynamics at temperatures above one Kelvin.
	
	In Ref.~\cite%
	{Miao2019}, the authors demonstrated electrically driven coherent quantum
	interference in the optical transition of a divacancy in commercially
	available 4H silicon carbide (SiC). The divacancy consisted of a carbon
	vacancy adjacent to a silicon vacancy; the TLS was formed by the ground and
	excited states of electronic orbital levels in the divacancy; the monochromatic
	and bichromatic LZSM interference patterns involved up to 15-photon
	excitations.
	
	The NV centers in diamond have a nearly spinless
	environment, leading to extremely long electronic and nuclear spin coherence
	times. In Ref.~\cite{Childress2010}, the authors simultaneously excited both
	the electronic transition by weak microwaves and nuclear spin resonance by
	strong rfs, while the experimentally observable fluorescence was dominated
	by the response of the electronic spin qubit. In Ref.~\cite{Scheuer2017}, the
	electron spin polarization was transferred from a single NV center to the
	surrounding carbon nuclear spins; the polarization dynamics of the
	nuclear spin ensemble displayed the LZSM oscillations. 
	
	In Ref.~\cite{Huang2011}, the authors
	considered the doubly driven system of the two states, $m_{s}=0$ and $%
	m_{s}=+1$, coupled with a nearby $^{14}$N nuclear spin. First, a microwave
	field matched those two bare energy levels and created the dressed-states
	with the avoided-level distance defined by the microwave amplitude.
	Second, the repeated passage around the avoided-level crossing was made by
	applying rectangular rf pulses. Hence, these demonstrated the quantum coherent
	control of electronic single spins at room temperature.
	
	\subsection{Graphene}
	
	Nonadiabatic transitions between Dirac cones in the vicinity of the Dirac
	points take place if these display avoided-level crossing and are driven appropriately. This was realized in graphene by illuminating it with
	a strong linearly polarized light field \cite{Higuchi2017, Heide2018,
		Li2020a}. This was described above in relation to Fig.~\ref%
	{Fig:interferograms}, here in terms of the one-dimensional trajectories of electrons
	in reciprocal space; see also Refs.~\cite{Syzranov2013, Lefebvre2018}. If
	the light becomes circularly polarized, the intraoptical-cycle LZSM
	interference cannot occur and there is no respective change in the current.
	The intermediate situation takes place for laser pulses with various degrees
	of ellipticity \cite{Heide2018, Heide2019, Heide2020, Boolakee2020}.
	
	The theory of nonadiabatic transitions in the vicinity of a conical
	intersection and related interference were addressed in several works for
	graphene \cite{Rozhkov2016} and other related materials \cite{Montambaux2018, Li2020a,
		Wu2020, Huang2021, Heide2021}. Here, we note that the Dirac cones and Dirac
	points are important to many objects, including massless electrons in
	graphene and in conducting edge states in topological insulators. In
	particular, transitions that happen between two massive Dirac cones were
	shown to create an interferometer, revealing information on the band
	eigenstates, such as the chirality and mass sign \cite{Fuchs2012, Lim2014}.
	Similar descriptions of the driving near the Dirac points are applicable to
	various systems, such as Bloch oscillations of ultracold atoms in a honeycomb
	optical potential \cite{Xu2014b}, surface states of 3D topological
	insulators \cite{Lim2012}, ultrafast transfer of excitation energy in
	photoactive molecules \cite{Nalbach2016}, and electron-hole pair production in
	an electric field \cite{Fillion-Gourdeau2016, Taya2021}.
	
	\subsection{Microscopic systems}
	
	In microscopic systems, the energy levels often experience avoided crossing,
	which is relevant here. In particular, ultracold atomic and
	molecular quantum gases have an extraordinary degree of control \cite{Chin2010}.
	Ultracold gases, which are produced by laser cooling and trapping techniques, have
	characteristic temperatures in the (sub-)$\mu $K range and typical densities
	below $10^{15}$~cm$^{-3}$. Under these conditions, the atomic de Broglie
	wavelength exceeds the average interatomic distance, resulting in the
	quantum-degenerate state of a BEC. In such
	systems, when the bound molecular state energetically approaches the
	scattering state, two colliding atoms couple with a molecular bound state,
	a phenomenon known as Feshbach resonance.
	
	Several phenomena associated with the Feshbach resonance, such as molecular
	association and dissociation, are conveniently described in terms of the
	LZSM model \cite{Koehler2006}. The energy levels of the molecular states
	(produced by atoms in the BEC state) around the Feshbach resonance are often
	manipulated using time-dependent magnetic fields. Tunneling at the
	avoided-level crossing leads to the formation of stable dimer molecules \cite%
	{Mies2000, Goral2004}. Using the repetitive passage of the avoided-level
	crossing allows for a high degree of control over the interferometric
	dynamics, which was demonstrated on caesium ultracold Feshbach molecules
	confined in a CO$_{2}$-laser trap \cite{Mark2007, Mark2007a} and presented
	above, here in relation to Fig.~\ref{Fig:interferograms}. Such a BEC-based
	tunable \textquotedblleft beam splitter\textquotedblright\ allows for
	controllably transferring the molecular states. The driving of ultracold atoms
	was studied theoretically \cite{Li2020, He2020}.
	
	The dynamics of ultracold atoms in an optical lattice can be described in terms
	of LZSM transitions between Bloch bands of the optical lattice \cite%
	{Kling2010}. The authors loaded an atomic rubidium BEC into a regular
	optical lattice so that the atoms were transferred into the lowest energy band;
	then, the lattice was accelerated, which was equivalent to the application of
	an external force, resulting in interferometric tunneling between the
	two bands. Analogous St\"{u}ckelberg oscillations were studied in Ref.~\cite%
	{Zenesini2010}, where the respective double-passage model was used for a
	description of the experiment. These systems can also be used for
	multipassage interferometry \cite{Ploetz2011, Liang2020, Beguin2021}.
	
	Among other microscopic systems, Rydberg atoms play a special role because they
	allow for early observation of the Bessel-function modulated multi-photon
	transitions and St\"{u}ckelberg oscillations. This was studied by \cite%
	{Baruch1992}, where a thermal beam of K atoms passed through a microwave
	cavity, to which a static electric field was applied for tuning and a pulse
	for field ionization by two laser beams. In the same year, \cite{Yoakum1992}
	studied a monoenergetic beam of helium atoms in the Rydberg state and drove
	them using a microwave electric field, with nice agreement between experimental
	St\"{u}ckelberg oscillations and a numerical solution of the Schr\"{o}dinger
	equation. The St\"{u}ckelberg oscillations were observed in the
	dipole–dipole interactions between Rydberg atoms and described theoretically
	within the Floquet approach in Ref. \cite{Ditzhuijzen2009}.
	
	LZSM interferometry was also realized with single ions and free radicals.
	A single $^{171}$Yb$^{+}$ ion was trapped in a four-rod radio-frequency trap
	and driven by microwaves by \cite{Zhang2014}. The molecular free radical $%
	^{138}$BaF, with one unpaired electron in the ground state, was Zeeman tuned
	and electrically driven by \cite{Cahn2014}.
	
	\subsection{Other systems}
	
	For classical systems, mechanical resonators included, we devote a separate
	section. This is because to the best of our knowledge, no review on this
	important subject exists. Here, we would rather mention several other
	possible realizations of LZSM interferograms.
	
	In principle, any TLS subjected to a periodic driving can display LZSM
	interference. However, there are limitations; in particular, the system needs
	to preserve coherence long enough. This may take place in
	superconducting weak links, where the Andreev levels as a function of time
	display avoided-level crossing~\cite{Gorelik1998, Melin2019, Melin2020,
		Doucot2020, Oriekhov2021}; there, it was pointed out that the
	microwave-induced LZSM interference should be revealed as a significant source of
	energy absorption; for a mini review, see \cite{Roy2013}. Similar studies of
	coherent interference effects were done for Schwinger vacuum pair production~%
	\cite{Akkermans2012}, many body localized systems~\cite{Gopalakrishnan2016}, and
	photoisomerization reaction \cite{Duan2016}.
	
	If in place of energy levels we have energy bands, that is, Bloch bands, we can
	call these \textit{Bloch–LZSM transitions} or interference. As in the other
	cases, in the literature we can meet shorter names such as
	Bloch-(Landau-)Zener transitions~/ oscillations~/ interference. In the case
	of the Bloch bands, these effects also include Bloch oscillations, which
	are the periodic motion of particles in a superlattice subjected to a
	constant external field. These take place in Bose–Einstein condensates in
	tilted and driven optical superlattices~\cite{Witthaut2011, Reid2016}, in
	graphene and topological insulators~\cite{Krueckl2012, Sun2018}, in photonic
	Lieb lattices~\cite{Long2017}, for trapped ions~\cite{Gagge2018}, and for the
	high-order harmonic generation in solids~\cite{Jin2018}, and in moir\'{e} systems under a uniform magnetic field \cite{Paul2022}.
	
	\subsection{Multilevel systems}
	
	Having considered two-level systems, it is important to extend the study to
	possible multilevel systems, with multiple avoided-level crossings. These
	may relate to diverse systems such as a chain of interacting spins-1/2 \cite%
	{Ostrovsky2006, Larson2014,Grimaudo2022}, a spin or multiple spins coupled to a resonator 
	\cite{OKeeffe2013, Singh2020}, molecular nanomagnets \cite{Foeldi2007,
		Garanin2008, Pavlyukh2020}, interacting bosons in optical lattices \cite%
	{Tschischik2012}, broad periodic photonic waveguide \cite{Benisty2011}, or
	NV center coupled with a substitutional nitrogen center in
	diamond \cite{Zangara2019, Band2020}, to name a few.
	
	\begin{figure}[tbp]
		\centering{\includegraphics[width=1.0\columnwidth]{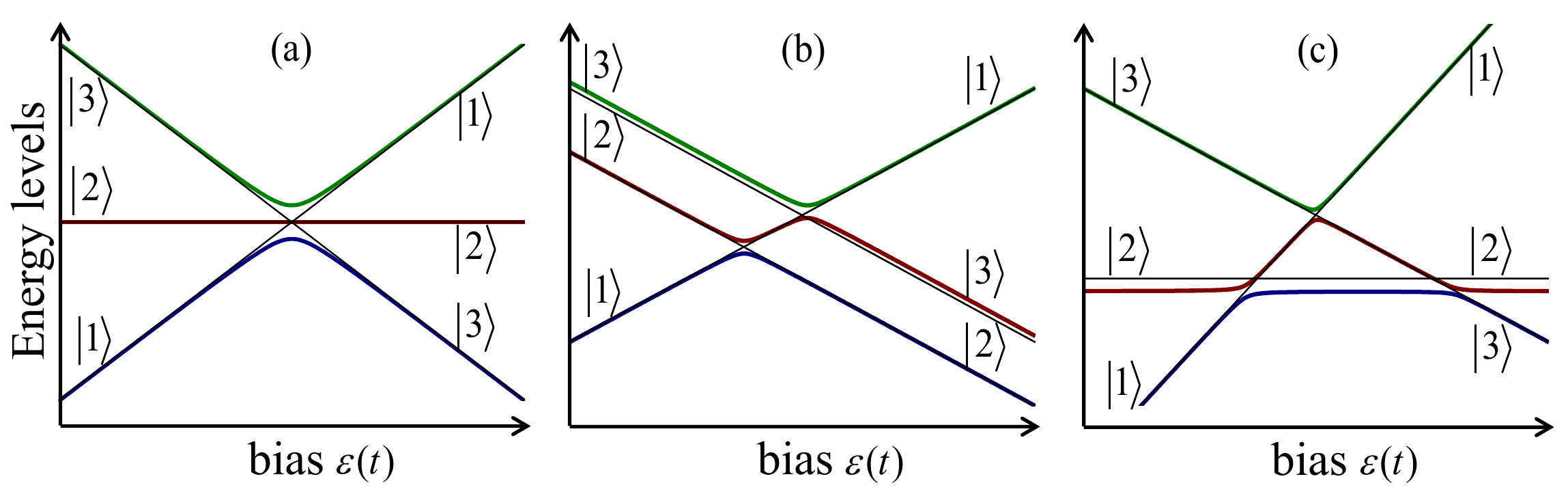}}
		\caption{\textbf{Three-level systems. }\textit{Diabatic} energy levels of states $%
			\left\vert i\right\rangle $ (thin black lines) intersect in one (a), two
			(b), or three (c) points. Eigenenergies of the full Hamiltonian $H(t)$, as
			shown with thick color lines, exhibit avoided-level crossing. Note that in
			(c) there are two possible trajectories to get from $\left\vert
			1\right\rangle $ to the left into $\left\vert 3\right\rangle $ to the right,
			which implies interference during the single-passage process.}
		\label{Fig:3LS}
	\end{figure}
	
	Let us start here from the simplest realization of a multilevel system: the
	three-level one. If the system diabatic levels $\left\vert i\right\rangle $
	depend linearly on the energy bias $\varepsilon (t)$ with possible offset
	between levels $h$, then the respective diabatic energy levels can be drawn
	as straight lines, as in Fig.~\ref{Fig:3LS}, with one, two, or three
	intersections; here, $h=0$ is used for the case with a single intersection and $%
	h\neq 0$ for the other cases. Taking into account possible tunneling,
	proportional to $\Delta $, lifts the degeneracies and results in
	avoided-level crossings. We can describe this with the Hamiltonian 
	\begin{equation}
		H(t)=-\frac{\varepsilon (t)}{2}\widehat{A}-\frac{h}{2}\widehat{B}-\frac{%
			\Delta }{2}\widehat{C},
	\end{equation}%
	with $\widehat{A}$, $\widehat{B}$, and $\widehat{C}$ being $3\times 3$%
	~matrices; these can be expanded with Gell–Mann matrices \cite%
	{Kiselev2013}. Then, taking these matrices with the parameters, say, close to
	the ones of Ref.~\cite{Ashhab2016}, we display in Fig.~\ref{Fig:3LS} several
	situations: (a) the \textquotedblleft bow-tie\textquotedblright\ model, when
	all three levels (quasi-)intersect at one point, (b) the \textquotedblleft
	equal-slope\textquotedblright\ model with the (avoided) crossing of the three
	levels in two points, and (c) the \textquotedblleft triangle\textquotedblright\
	model with three pairwise (quasi-) intersections. The most important aspect
	that distinguishes (c) from (a) and (b) is the possibility of
	interference during a single-passage dynamics. For example, there are two
	possible trajectories that lead from $\left\vert 1\right\rangle $ to the
	left into $\left\vert 3\right\rangle $ (or alternatively to $\left\vert
	2\right\rangle $) to the right\ in Fig.~\ref{Fig:3LS}(c). This means that
	the final occupation probability is essentially defined by the phase
	difference between the two trajectories. Note that a three-level system can
	be described with a spin-1 Hamiltonian, so all theory for both
	single- and multiple-passage problems can be extended from two-level systems
	to three-level ones \cite{Zhang2011}.
	
	Let us now consider previous work done for driven multilevel systems, here separating the
	subject into three topics.
	
	\paragraph{ \textbf{Theory}} \mbox{}\\
	
	 From the above example of a three-level system, we can
	formulate a generalized LZSM problem for a multilevel system, for example,
	as follows: what is the occupation probability of a state $\left\vert
	j\right\rangle $ if starting from a state $\left\vert i\right\rangle $? More
	generally, the evolution can be described by the matrix $S$ linking the
	final vector-state to an initial one \cite{Shytov2004, Suzuki2022}. Here,
	as for the evolution of a TLS, it is instructive and most common to
	consider either a single-passage problem or periodic evolution with
	multiple passages. The \textquotedblleft bow-tie\textquotedblright\ model
	describes a quasi-intersection of $N$ energy levels \cite{Ostrovsky1997},
	while the \textquotedblleft equal-slope\textquotedblright\ model is the
	(quasi-) intersection of the energy band, here with $N-1$ levels, by one more
	level \cite{Demkov1968}. The single passage for a \textquotedblleft
	bow-tie\textquotedblright\ and \textquotedblleft
	equal-slope\textquotedblright\ models is described by sequential pairwise-level crossings \cite{Brundobler1993}. Magnetic sub-levels of two atomic levels with nonzero total angular momenta can be described by degenerate LZSM model, which involves two crossing sets of degenerate energy levels with quasi-crossing between them \cite{Vasilev2007}. 
	
	As for a three-level case, for an 
	$N$-level system, we can also have an extended \textquotedblleft
	triangular\textquotedblright\ model, with several sequential crossings
	making trajectories meet \cite{Demkov1995, Fai2015}. In this case, even for
	a single-passage evolution, we can take possible interference into
	account \cite{Ostrovsky2007, Sinitsyn2015, Malla2021}. 
	
	The description of the $S$%
	-matrix approach for single-passage problems is reviewed in \cite%
	{Sinitsyn2017}. As an extension of this, we can interpret the method of
	solving rate equations, which we consider in a separate subsection. This
	method is applicable to periodically driven systems, as shown by \cite%
	{Wen2010}, for the example of multilevel devices based on
	superconducting circuits \cite{Berns2008, Oliver2009}. Importantly, this
	latter approach takes a dissipative environment into account. An intuitive
	picture of open multilevel systems can be obtained by modeling the
	external environment by a harmonic oscillator \cite{Ashhab2016}. In the case
	of a weak interaction between $N$ spins-$%
	{\frac12}%
	$, the LZSM transitions can be described by a mean-field approach \cite%
	{Garanin2003}.
	
	\paragraph{\textbf{Single passage in multilevel systems}} \mbox{}\\
	
	 Although a three-level system
	is useful as a prototypical model for introducing the features of multilevel
	systems, more realistic and widely used are those models with four levels.
	These describe coupled two-level systems or genuine four-level systems,
	such as double quantum dots with two electrons. The single passage problem with
	the four-level model was addressed, for instance, for two
	antiferromagnetically coupled tunneling systems, such as Mn$_{4}$ dimers 
	\cite{Garanin2004}, Rydberg atoms \cite{Mallavarapu2020, Niranjan2020}, and
	quantum dots \cite{Sinitsyn2002, Reynoso2012, Krzywda2020}. Not only qubits,
	but also qutrits, can be coupled, as in Refs.~\cite{Grimaudo2019,
		Grimaudo2020}, which demonstrates that such coupling is useful for 
	entanglement control.
	\paragraph{\textbf{Multiple-passage transitions in multilevel systems}}\mbox{}\\
	
	 Periodically
	driven three-level systems may describe a qubit coupled with two microscopic
	systems \cite{Sun2010} or a superconducting charge qubit that is taking a
	higher energy level into consideration \cite{Parafilo2018a}. A three-level
	model for a driven system is convenient for describing the impact of the
	external noise \cite{Kenmoe2013, Kenmoe2015, Kenmoe2016, Li2018, Band2019}.
	Periodically driven four-level systems are useful for describing the charge
	states of two particles in DQDs, as realized in \cite{Shi2014,
		Chatterjee2018, Mi2018} and described in \cite{Pasek2018, Shevchenko2018,
		Zhao2022, Zhou2021}. A five-level model describes singlet-triplet states in
	spin-based DQDs \cite{Stehlik2016a, Chen2017, Qi2017}, which was also
	addressed by \cite{Zhao2018a, Karami2019b, Ginzel2020}. 
	
	As a further
	development, triple quantum dots were also studied \cite{Aers2012,
		Poulin-Lamarre2015, Gallego-Marcos2016, Luczak2016}. These systems present
	more flexibility and controllability than DQDs while representing a
	minimal model for a chain of quantum dots, which can be considered a
	highly controllable quantum metamaterial \cite{Smirnov2007,Rakhmanov2008}. The interference effects, here
	appearing in periodically driven coupled qubits, were extensively studied by 
	\cite{Denisenko2010, Satanin2012, Gramajo2017, Gramajo2018, Gramajo2021,
		Munyaev2021}.
	
	\section{Quantum control}
	\label{Sec:Quantum}
	
	To summarize what we discussed above, LZSM transitions act as a tool to
	control a TLS's occupation; that is, to control single-qubit states and
	the states of multi-level systems. As our system is
	analogous to a two-slit layout, a single-passage process presents a beam
	splitter for the energy-level occupation probabilities where the ratio of
	splitting can be controlled. Besides creating the coherent superposition, repeated passage of the avoided-level crossing gives the option to
	control the quantum system energy-level occupation probabilities by
	varying the accumulated dynamical phase.
	
	Having obtained information about the quantum system, we can go further by
	studying opportunities for its control. In addition to other trends such as quantum simulations \cite{Georgescu2014,Buluta2009} and quantum sensing \cite{Degen2017}, coherent quantum control can be viewed as another direction in quantum engineering. Since it is important for applications to have individual
	control of single quantum systems, we will present here studies of LZSM
	non-adiabatic transitions and interference for the quantum control of
	few-level natural and artificial quantum systems.
	
	\subsection{Coherent control of microscopic and mesoscopic structures}
	
	The 2012 Nobel Prize in Physics was awarded for \textquotedblleft
	experimental methods that enable measuring and manipulation of individual
	quantum systems\textquotedblright . To clarify, these systems were
	microscopic and consisted of atoms and photons. More recently, the variety of related quantum systems increased
	and also includes mesoscopic systems behaving as very large atoms \cite{Buluta2011}. To demonstrate this, we present the following
	variety of systems and approaches where the underlying effective Hamiltonian
	can be readily modified or engineered and their coherent quantum control was
	studied in the frame of LZSM physics.
	
	* In superconducting structures, the properties of non-adiabatic LZSM
	transitions can be used to control \textit{superposition states} in single
	qubits \cite{Salmilehto2011} and the \textit{entanglement} in coupled-qubit
	systems \cite{Quintana2013}. Several techniques were developed:
	\begin{itemize}
		\item\textit{amplitude spectroscopy} \cite{Berns2008}, which is applicable to multi-level
	systems, where increasing the driving amplitude results in reaching more
	avoided-level crossings for a single driving frequency that may be orders
	of magnitude smaller than the energy scales being probed; \item\textit{quantum
		phase tomography} \cite{Rudner2008}, which is based on the Fourier
	transforms of the interferograms that provide additional visibility and
	information; \item\textit{pulse imaging} \cite{Bylander2009}, where the measured
	interferograms are used to image the actual wave form of the driving signal
	at the device, which in turn can be used to engineer the desired time evolution
	of a quantum system. These and other similar methods provide means to
	characterize and manipulate states of superconducting and other quantum
	systems.\end{itemize}
	
	* Quantum dots, which can be charge or spin qubits, can be controlled. The
	hyperfine interactions can be harnessed for quantum control in GaAs
	semiconductor quantum dots \cite{Ribeiro2010}. The \textit{visibility of
		quantum oscillations} can be increased by enhancing the adiabatic passage
	probability. For this purpose the researchers designed tailored pulses that combined both fast
	and slow rise-time ramps to minimize dissipation and enhance adiabaticity \cite{Ribeiro2013}.
	
	* A qubit can be coupled to a bath of TLSs describing noise in quantum
	circuits. Properly driving such TLSs can help to dynamically decouple such
	systems, thus \textit{reducing dielectric losses}. This was demonstrated with
	TLSs in deposited aluminum oxide by using it as the dielectric in a
	lumped-element $LC$-resonator \cite{Matityahu2019}. Using LZSM\
	interference in quantum dynamics can help to isolate a single particle
	(in a quantum dot) from a Fermi sea by closing a tunnel barrier \cite%
	{Kashcheyevs2012}. The sensitivity of LZSM transitions to the parameters of the
	dissipative environment makes them useful for \textit{gauging a quantum
		heat bath }\cite{Wubs2006}. Using repetitive transitions with underlying
	interference allows accurately obtaining all relevant information about
	complex influences. This was observed in Ref.~\cite{Forster2014} on
	the example of a two-electron charge qubit defined in a lateral double
	quantum dot.
	
	* In composite or hybrid systems, non-adiabatic transitions can bring a
	system from one state to a desired state that might be entangled
	\cite{Li2011}. For a system of qubits, \textit{the transfer state between
		qubits} was studied with a photonic qubit (localized exciton in an optically
	active quantum dot) coupled to a spin qubit hosted in gate-defined quantum
	dots \cite{Joecker2019}, with a composite system of Majorana-hosted semiconductor nanowire and superconducting flux qubits \cite{Zhang2013b}, and with three superconducting qubits \cite{Li2019}. For a hybrid system consisting of a qubit coupled
	to a photonic cavity, the transition results in the entanglement of the
	system and, the \textit{generation of Schr\"{o}dinger cat
		states} in the photonic cavity \cite{Lidal2020}.
	
	* Adiabatic passages and nonadiabatic transitions in a macroscopic,
	cylindrically shaped water sample were studied using standard \textit{%
		nuclear magnetic resonance} (NMR) techniques with controlled linear ramping of the external magnetic field \cite{Hurlimann2020}%
	. The pulse sequence used in this experiment was the standard in NMR
	applications: the Carr-Purcell-Meiboom-Gill sequence.
	
	* The similarity between driven TLS dynamics and other processes can provide a basis for simulating these phenomena with a highly controllable TLS. \textit{The emulation of mesoscopic phenomena}, such as weak
	localization and universal conductance fluctuations, was demonstrated on a
	system of superconducting transmon qubits by \cite{Gramajo2020}.
	
	* Using the acquired phase of the wave function during single and repeated
	non-adiabatic transitions provides the basis for \textit{optimal control
		theory} \cite{Zhdanov2015}. Depending on the purpose (e.g., to obtain
	complete population transfer, which we will also address later),  the driving signal can be tailored \cite{Larocca2018}.
	The efficiency can be quantified with the methods of \textit{quantum
		metrology}, and calculations of quantum Fisher information \cite{Yang2017}.
	
	* A driven TLS provides a testbed quantum thermodynamics \cite{Ono2020},
	where the non-adiabatic transitions serve to change states between
	different regimes. This was studied in Otto heat engines \cite{Halpern2019,
		Son2021}, and in calorimetric methods and measurements of work \cite{Solinas2013,
		Pekola2013}.
	
	The following three subsections present specific chapters in the quantum
	control with LZSM transitions.
	
	\subsection{Universal single- and two-qubit control}
	
	Most often, quantum control (logic gates including) is associated with
	resonant driving that then results in Rabi oscillations \cite{Kwon2021,Krantz2019}. However, under resonant
	driving, certain limitations may arise, including achievable gate speed and
	nonidealities, such as counter-rotating terms \cite{Vitanov2003,Saito2006,
		Longhi2019, Campbell2020}. Coherent LZSM interference allows for
	a complementary approach to quantum control based on nonresonant driving with
	the alternation of adiabatic evolution and nonadiabatic transitions. This
	enables ultrafast qubit gates that are controlled solely using baseband pulses, hence
	alleviating the need for pulsed-microwave control signals \cite{Cao2013,
		Ota2018, Campbell2020, Zhang2021,Li2022}.
	
	Qubit evolution can be conveniently described by trajectories on the Bloch
	sphere. For this, the evolution matrices in the transfer matrix (TM) approach should be
	rewritten using the Euler-angle decomposition \cite{Sillanpaeae2006,
		Sillanpaeae2007}. First, to describe the nonadiabatic transition, we
	rewrite the $N$ matrix from Eq.~(\ref{TMDiabatic}) in the form%
	\begin{equation}
		N=\left( 
		\begin{array}{cc}
			\cos (\theta /2)e^{-i\phi _{\mathrm{S}}} & -\sin (\theta /2) \\ 
			\sin (\theta /2) & \cos (\theta /2)e^{i\phi_{\mathrm{S}}}%
		\end{array}%
		\right) =U_{z}\left( \phi _{\mathrm{S}}\right)
		U_{x}\left( \theta \right) U_{z}\left( \phi _{\mathrm{S}}\right), 
		\label{U1}
	\end{equation}%
	where $\sin ^{2}(\theta /2)=\mathcal{P}$ and the matrices $U_{x,y,z}(\alpha
	)=\exp \left( -i\alpha \sigma _{x,y,z}/2\right) $ describe their respective
	rotations. The adiabatic evolution is characterized with the $z$-rotation $%
	U(t,t_{\mathrm{i}})=\exp{(-i\zeta \sigma _{z})}$ in Eq.~(\ref{AdiabaticEvolution}%
	). Then, the single-passage evolution is described by $UN$, Eq. (\ref%
	{SingleTRansMatrix}). The repetition of these can correspond to a one-period
	evolution (with doubly passing the avoided crossing), $\Xi =N^\top U_{2}NU_{1}$,
	Eq.~(\ref{SingleTransitionMatrix}), or to an $n$-cycle evolution with $\Xi
	^{n}$ in Eq.~(\ref{Ksi_n}). After writing this, we can equate the evolution
	matrix with the matrix for the Euler-angle decomposition $U(\alpha, \beta
	,\gamma )=U_{z}(\gamma )U_{y}(\beta )U_{x}(\alpha )$ \cite{LandauLishitz_QM}%
	; as a result, we obtain the formulas for the designated rotation angles $%
	\alpha $, $\beta $, $\gamma $. With this, we can obtain a desired protocol
	and the parameters for single-qubit operations \cite{Campbell2020}. Here, the
	important claim is that with the pure LZSM interference technique, we can
	obtain any rotation and, correspondingly, the efficient single-qubit and
	two-qubit control \cite{Ribeiro2010}. This approach allows for defining
	the parameters for a desired evolution or a unitary logic gate and for
	analyzing the resonance conditions \cite{Ashhab2007}.
	
	\begin{figure}[tbp]
		\centering{\includegraphics[width=1.0\columnwidth]{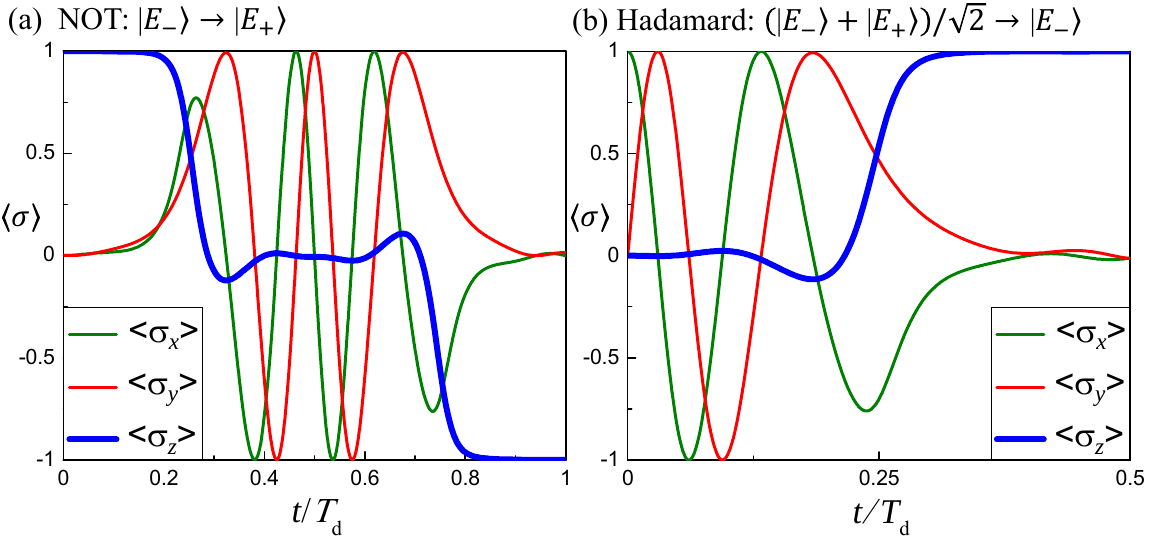}}
		\caption{\textbf{LZSM gates.} Dynamics of $\protect\sigma $-matrix
			coefficients $\left\langle \protect\sigma _{x,y,z}\right\rangle $, which
			define the position on the Bloch sphere, under $\protect\varepsilon %
			(t)=A\cos \protect\omega t$ driving. (a) NOT gate, which is the transfer $%
			\left\vert E_{-}\right\rangle \rightarrow \left\vert E_{+}\right\rangle $,
			here as performed by two consecutive LZSM transitions. For this, we assume the
			constructive-resonance condition $\protect\zeta +{\protect\phi }%
			_{S}=2\protect\pi n$ (this defines the amplitude $A$) and take the
			transition probability $\mathcal{P}=0.5$. Specifically, we found $%
			A=3.8872\Delta $ and the frequency was defined from $\protect\omega =-\pi \Delta ^{2}\left[2 \hbar A\ln \mathcal{P}\right]^{-1}$. (b) Hadamard gate
			performed with the following parameters: $\mathcal{P}=0.5$, $A=4.309\Delta $,
			which is defined from the condition for destructive interference $\protect%
			\zeta +{\protect\phi }_{S}=\frac{3\protect\pi }{2}+2\protect\pi n$%
			. }
		\label{Fig:LZSM_Gates}
	\end{figure}
	
	As an illustrative example, consider first driving a TLS with pulses
	that result in a double passage of the avoided crossing \cite{Cao2013}. This
	is described by the matrix $\Xi =N^\top U_{2}NU_{1}$, here with the alternation of
	adiabatic evolution ($z$-rotation) and nonadiabatic transition with the $N$
	rotation matrix. We note that the rotation angles, both $\theta $ and $%
	{\phi }_{\mathrm{S}}$, are defined by the adiabaticity
	parameter $\delta $ and, consequently, by the passage velocity $v$. Hence,
	adjusting the pulse shape (which defines this velocity $v$), one can obtain
	any rotation and respective logic gate \cite{Cao2013}. For a fixed velocity,
	the evolution is limited between the two trajectories defined by destructive
	and constructive interference \cite{Ota2018}. 
	
	In Fig.~\ref{Fig:LZSM_Gates}
	(a) we demonstrate the NOT operation, which is the rotation around the $x$ axis
	by the angle $\pi $. For this, we consider driving with a cosine pulse and
	choose the parameters to perform the NOT operation. Namely, we use $%
	\mathcal{P}=0.5$ and the total phase changing during one LZSM transition $%
	{\phi }_{\mathrm{S}}+\zeta =2\pi n$ to have the
	constructive interference, see Eqs.~(\ref{zeta},~\ref{StocesPhase}). As a
	result, one can perform a fast and robust NOT operation, which in Fig.~\ref%
	{Fig:LZSM_Gates}(a) illustrates the transition from the ground state
	(logical~$0$) to the excited state (logical~$1$): $\left\vert
	E_{-}\right\rangle \rightarrow \left\vert E_{+}\right\rangle $.
	
	Next, consider the Hadamard gate. As above, this can be implemented with the
	double-passage process. However, with the chosen value $\mathcal{P}=0.5$, we
	can do this faster, here with a single-passage transition, which occurs during half
	the period. We illustrate this in Fig.~\ref{Fig:LZSM_Gates}(b). For
	performing the Hadamard gate, we use a single LZSM transition, with the
	total phase changing ${\phi }_{\mathrm{S}}+\zeta =3\pi /2+2\pi n$,
	resulting in $\left\{ \left\vert E_{-}\right\rangle +\left\vert
	E_{+}\right\rangle \right\} /\sqrt{2}\rightarrow \left\vert
	E_{-}\right\rangle $. Hence, we have demonstrated two gates, and this can be
	supplemented by, for example, the phase gate to obtain the universal set of single-qubit operations. The phase gate can be realized by $U_{z}(\zeta
	)=\exp{(-i\zeta \sigma _{z})}$, which is idling (in the absence of driving).
	
	In addition to single-qubit operations, two-qubit operations can also be
	done with LZSM transitions and interference. Two-qubit operations are
	conveniently demonstrated by the conditional phase gate or controlled-Z
	(CZ) gate between two qubits; four-level dynamics was studied by 
	\cite{Rol2019, Campbell2020} for superconducting qubits and by \cite%
	{Huang2018} for Rydberg atoms. Similarly, another two-qubit operation—the
	controlled-NOT (CNOT) operation—can be realized \cite{Ribeiro2010, Wu2021}.
	Importantly, this approach allows for fast execution of the two-qubit operations
	with high fidelity. Provided one has single- and two-qubit operations, the
	next step could be the realization of algorithms.
	
	Destructive interference can be harnessed to perform unitary
	transformations based on cyclic evolutions, which refer to the
	process where the state of the system returns to its original state after
	gate operations \cite{Wang2016a}. As a result of this evolution with
	destructive interference (i.e. coherent destruction of tunneling [CDT]), the state vector accumulates a
	phase only. This phase acquired in the evolution contains both dynamic and
	geometric components; see the detailed and pedagogical explanations in the
	supplementary material of \cite{Wang2016a}, and for relations to the Berry phase,
	see \cite{Lim2015, Lim2015a, Bleu2018}. The geometric phase can be related
	to the solid angle subtended by the evolution curve on the Bloch sphere.
	This analysis opens up new ways of interpretation and control; this is
	called \textit{geometric LZSM interferometry} \cite{Gasparinetti2011}.
	Because we can perform any logic operations using this, it provides the
	instruments for \textit{geometric quantum computation.} This is a
	promising approach to achieve robust control of a quantum system \cite%
	{Tan2014, Zhuang2022, Barnes2021}.
	
	\subsection{Shortcuts to adiabaticity}
	
	\label{Sec:STA}
	
	Two major aims of quantum control can be defined as either reaching a given
	target state or tracking the instantaneous ground state of a system during
	its evolution \cite{Bason2012}. For each task, an optimum strategy can be
	designed based on requirements such as the highest possible fidelity, fastest
	possible operation, and so forth.
	
	For illustration, the authors of Ref.~\cite{Bason2012} considered both a
	shortcut protocol that reaches the maximum speed compatible with the
	Heisenberg uncertainty principle and the opposite limit of nearly perfectly
	following the instantaneous adiabatic ground state. For the LZSM-form
	Hamiltonian $H(t)=\left[-\Delta (t)\sigma _{x}-\varepsilon (t)%
	\sigma _{z}\right]/2$, the authors designed a protocol that drives the system
	through avoided crossing in such a way that at the end of evolution, the
	final state $\left\vert \psi _{\mathrm{fin}}\right\rangle $ is as close as
	possible to the adiabatic ground state $\left\vert E_{-}(t_{\mathrm{fin}%
	})\right\rangle $, which here aims to reach a unit’s final fidelity $\left\vert \left\langle
	\psi _{\mathrm{fin}}\right. \left\vert E_{-}(t_{\mathrm{fin}})\right\rangle
	\right\vert ^{2}=1$. In general, there are infinitely numerous paths in Hilbert
	space connecting the initial state, for example, $\left\vert \psi _{\mathrm{ini}%
	}\right\rangle =\left\vert E_{-}(t_{\mathrm{ini}})\right\rangle $ with the
	final one $\left\vert E_{-}(t_{\mathrm{fin}})\right\rangle $. First,
	fixing $\Delta (t)=\text{const}$, minimizing the evolution time results in a
	problem known as the \textit{shortcut to adiabaticity} (STA) \cite{Guery2019,Chen2021b}
	or also by analogy with the equivalent classical case, “quantum
	brachistochrone” \cite{Oh2016}. Here, the minimum time for reaching the target state is
	known as the \textit{quantum speed limit} time:
	\begin{equation}
	T_{\mathrm{qsl}}=2\arccos \left\vert
	\left\langle \psi _{\mathrm{fin}}\right. \left\vert \psi _{\mathrm{ini}%
	}\right\rangle \right\vert /\Delta.
	\end{equation} 
    In the opposite formulation, aiming at perfect following the ground state $\left\vert E_{-}(t)\right\rangle $ is
	known as the \textit{counterdiabatic} protocol \cite{Demirplak2003} or \textit{%
		superadiabatic} protocol \cite{Berry2009}. Details of this transitionless
	adiabatic protocol  are considered below.
	
	As mentioned, there are two major routes for
	manipulating the state of a quantum system with interacting fields: either
	using resonant pulses or adiabatic methods \cite{Chen2010}. The most popular
	methods for efficient population transfer are the adiabatic rapid passage
	(ARP; sometimes RAP), chirped-pulse excitation or Stark-chirped RAP (SCRAP),
	and stimulated Raman adiabatic passage (STIRAP) \cite{Vitanov2001,
		Goswami2003,Wei2008}. The latter technique permits the precise control of population
	transfer using partially overlapping pulses, from pump and Stokes lasers 
	\cite{Bergmann1998}. In its simplest form, for a three-level system, the
	pump pulse links the initial state $\left\vert 1\right\rangle $ with an
	intermediate state $\left\vert 2\right\rangle $, which, in turn, interacts via
	a Stokes (or dump) pulse with a target state $\left\vert 3\right\rangle $;
	the aim of this pump–dump technique is to achieve complete transfer of a
	population between states $\left\vert 1\right\rangle $ and $\left\vert
	3\right\rangle $. 
	
	The STIRAP technique can be interpreted as pairs of LZSM
	transitions in the areas of avoided crossings of quasienergies \cite%
	{Yatsenko1998, Malinovsky2001}. If the avoided crossing is passed
	repeatedly, the interference has to be taken into account \cite{Vitanov2018}.
	This was also analyzed by \cite{Zhang2019}, in which the nonadiabatic
	transitions were quantified by the upper-bound function, hence providing the
	optimal criteria for adiabatic control. In relation to our previous
	discussion on the interrelation between LZSM interference and multi-photon
	processes, coherent population transfer can also be described
	as multi-photon ARP \cite{Maeda2006}. The ARP strategy can be used for a high-quality source of single photons that are emitted because of spontaneous emission from
	the excited state, which is robust against control errors and environmental
	fluctuations \cite{Miao2016a}.
	
	To understand the idea of counter-diabatic driving, consider this, following \cite{Demirplak2003} and \cite{Theisen2017}: 
	let our arbitrary $n$-level system be described by the Hamiltonian $H(t)$, and let $U(t)$ be a
	time-dependent unitary transformation that makes the matrix $%
	U(t)H(t)U^{\dag }(t)$ diagonal; it consists of the eigenvectors of $%
	H(t) $. Consider now the full Hamiltonian $\widetilde{H}(t)=H(t)+H_{\mathrm{%
			CD}}(t)$. Then, retaining the definition of $U(t)$, consider the new
	Hamiltonian after such transformation 
	\begin{equation}
		\widetilde{H}^{\prime }=U\widetilde{H}U^{\dag }-i\hbar U\dot{U}^{\dag
		}=UHU^{\dag }+\left( UH_{\mathrm{CD}}U^{\dag }-i\hbar U\dot{U}^{\dag
		}\right) .
	\end{equation}%
	Because a priori, $UHU^{\dag }$ is diagonal, then the condition 
	\begin{equation}
		H_{\mathrm{CD}}=i\hbar \dot{U}^{\dag }U  \label{HCD1}
	\end{equation}%
	guarantees that at all times, the population, which was placed initially in
	one of the adiabatic states, remains exactly in that particular state.
	
	Alternatively, the counter-diabatic Hamiltonian $H_{\mathrm{CD}}(t)$\ can be
	written following \cite{Berry2009} in terms of the instantaneous eigenvalues 
	$E_{n}(t)$ and eigenstates $\left\vert \psi _{n}(t)\right\rangle $ of the
	system Hamiltonian $H(t)$
	\begin{equation}
		H_{\mathrm{CD}}=i\hbar \sum_{m\neq n}\sum_{n}\frac{\left\vert \psi
			_{m}\right\rangle \left\langle \psi _{m}\right\vert \dot{H}\left\vert \psi
			_{n}\right\rangle \left\langle \psi _{n}\right\vert }{E_{n}-E_{m}}.
		\label{HCD2}
	\end{equation}%
	Consider this for a TLS with a generic Hamiltonian $H(t)=-\left[\Delta (t)
	\sigma _{x}+\varepsilon (t)\sigma _{z}\right]/2$. This is written in the
	diabatic (time-independent) basis $\left\{ \left\vert 0\right\rangle
	,\left\vert 1\right\rangle \right\} $, and the transformation to the
	instantaneous basis $\left\{ \left\vert E_{-}(t)\right\rangle, \left\vert
	E_{+}(t)\right\rangle \right\} $ is defined by the angle $\theta (t)=\,$%
	$\arctan\frac{\Delta (t)}{\varepsilon (t)}$. Then, Eq.~(\ref{HCD2})
	gives 
	\begin{equation}
		H_{\mathrm{CD}}(t)=\frac{\hbar }{2}\dot{\theta}(t)\sigma _{y}.
		\label{HCD4TLS}
	\end{equation}%
	We emphasize that this formula is valid for any $\Delta (t)$ and $%
	\varepsilon (t)$, leaving the system in an eigenstate of $H(t)$.
	In particular, for the linear LZSM\ problem, with $\dot{\Delta}=0$ and $\dot{%
		\varepsilon}=v$, we obtain 
	\begin{equation}
		\widetilde{H}(t)=-\frac{\Delta }{2}\sigma _{x}-\frac{vt}{2}\sigma _{z}+\frac{%
			v\Delta }{\Delta ^{2}+v^{2}t^{2}}\sigma _{y}.  \label{H_tilde_TLS}
	\end{equation}%
	Note that the time integration of $\dot{\theta}(t),$ from $-\infty $ to $%
	\infty, $ gives $\pi $ so that $\dot{\theta}(t)$ represents a $\pi $-pulse 
	\cite{Theisen2017}. More generally, for pulse areas equal to odd multiples
	of $\pi $, a complete population transfer to the excited state takes place,
	whereas for pulse areas equal to even multiples of $\pi $, the system
	returns to the initial state \cite{Vitanov2001,Teranishi1999}.
	
	Therefore, for a perfect adiabatic behavior, the nonadiabatic transitions should be
	suppressed, for which the LZSM model is central for testing
	control protocol. This was discussed for TLSs \cite{Chen2010,
		Ahmadinouri2019, Funo2021}, for three-level systems \cite{Theisen2017},
	and for $n$-level ones \cite{Poggi2015a, Xu2018}. In particular, it is
	possible to approximately decompose the transitionless quantum driving into
	the sum of separate single-crossing corrections \cite{Theisen2017}. Analysis
	of the adiabatic condition can help further improve the coherent
	population transfer, resulting in tangent-shaped pulses \cite{Xu2019}.
	
	In some cases, it is possible to reach diabaticity by excluding
	the $\sigma _{y}$ component by certain system-specific transformation \cite%
	{Bason2012} or by applying additional fast and strong oscillating terms to a
	linear one, as in $\varepsilon (t)$ \cite{Chasseur2015}. Indeed, as it was
	scrutinized by \cite{Petiziol2018}, adding a fast oscillation in the control
	parameters can approximately (to arbitrary precision) cancel nonadiabatic
	effects. Hence, with this, transitionless dynamics can be realized for
	finite-dimensional quantum systems without requiring additional Hamiltonian
	components that are not included in the initial control setup; this was further
	developed for single and coupled qubits (creating entanglement between them),
	as well as for three-level systems \cite{Petiziol2018, Petiziol2019a,
		Petiziol2019}.
	
	The techniques of population transfer that we considered here were tested
	and developed for diverse physical realizations: atoms and molecules
	controlled by lasers \cite{Vitanov2001}, Bose–Einstein condensate \cite%
	{Bason2012}, Rydberg states \cite{Shi2016,Bengs2022}, photonic structures \cite%
	{Longhi2009} (see therein for classical analogies), NMR \cite{Herrera2014},
	electron spin of a single NV center in diamond \cite{Zhang2013}%
	, electron spin in a QD \cite{Shafiei2013} (see for chirping back and forth
	resulting in a double-passage dynamics; for theory see \cite{Teranishi1999, Nagaya2002}), ensemble of QDs to demonstrate an
	ideal molecular switch with the LZSM formalism \cite{Kaldewey2018},
	superconducting qubits \cite{Wei2008}, and topologically protected edge states
	in a dimeric spin chain \cite{Longhi2019}.
	
	\subsection{Adiabatic quantum computation}
	
	Adiabatic quantum computation (AQC) is an approach to quantum computing
	that is an important alternative to the standard circuit (gate)
	model. In principle AQC is expected to be as powerful as the circuit model of quantum computation 
	\cite{Albash2018,Ashhab2006}. The former is not very sensitive to dephasing, yet the huge
	cost of this approach is the lack of universality and problems with 
	accurate calculations \cite{Zagoskin2011}. AQC is best suited for
	optimization problems, where the requirement is to find the global minimum
	of a cost function. In AQC, the computation proceeds from a simple initial
	Hamiltonian $H_{\mathrm{i}}$, whose ground state is easy to prepare, to a
	final Hamiltonian $H_{\mathrm{f}}$, whose ground state encodes the solution
	to the computational problem. Hence, the task is to find the ground state of $%
	H_{\mathrm{f}}$, whose unknown eigenvalues determine the cost function. The
	evolution is described by the Hamiltonian 
	\begin{equation}
		H(s)=(1-s)H_{\mathrm{i}}+sH_{\mathrm{f}},  \label{H(s)}
	\end{equation}%
	so that at $t=0$ with $s=0$, we start from $H_{\mathrm{i}}$, and at the end of
	computation at $t=t_{\mathrm{f}}$ with $s=1$, we have the desired final
	Hamiltonian $H_{\mathrm{f}}$ \cite{Cullimore2012}. The eigenvalues and
	eigenstates of the Hamiltonian $H(s)$ describing a system of $M$ qubits are
	given by $H(s)\left\vert \varphi _{m}(s)\right\rangle =E_{m}(s)\left\vert
	\varphi _{m}(s)\right\rangle $, with $m$ ranging from $0$ to $2^{M}-1$. The
	suitable figure of merit is the closeness of the state vector $\left\vert
	\psi (s)\right\rangle $ at the end of the evolution to the desired state $%
	\left\vert \varphi _{0}(s)\right\rangle $. This is quantified by the success
	probability $\left\vert \left\langle \varphi _{0}(1)\!\right. \left\vert
	\psi (1)\right\rangle \right\vert ^{2}$.
	
	For adiabatic evolution, the cornerstone of this is the adiabatic
	theorem, which assumes no LZSM transitions \cite{Johansson2009}.\ The
	avoided crossings are defined by the minima of $\Delta
	E(s)=E_{1}(s)-E_{0}(s) $. Hence, in the vicinity of these minima, the
	low-energy Hamiltonian is equivalent to the one describing two wells
	that are connected by a tunneling rate and detuned from each other by a bias \cite%
	{Wild2016}. In this case, LZSM transitions are usually considered a nuisance for
	AQC \cite{Grajcar2005}. Also, LZSM theory tells us that,
	akin to a double-slit experiment, two consecutive transitions
	(at the minima of $\Delta E(s)$) can generate interference effects \cite%
	{Lubin1990,Munoz-Bauza2019}.
	
	Therefore, studying LZSM physics in an open system with the Hamiltonian $H(s)$
	can help minimize undesired nonadiabatic transitions \cite{Wild2016}.
	Because, on the one hand, the system has to evolve adiabatically slow to avoid
	nonadiabatic transitions and, on the other hand, we want to make the
	calculations faster, there are several ways to satisfy these
	two conflicting requirements. For this, we must make the control of the
	feedback in calculations, monitoring the curvature of the ground state $%
	E_{0}(s)$. Then, the feedback-control scheme speeds up the computation
	process by allowing for faster evolution of $s(t)$ when the ground state is well
	separated from the excited one \cite{Wilson2012}. Alternatively, the success
	probability can be increased by tailoring the driving pulses \cite%
	{Karanikolas2019}. These quantum annealing protocols, called pulsed
	quantum annealing, are reached by optimizing the pulse parameters. Also, in
	the context of AQC, an additional periodic drive can be applied to suppress
	control errors by using a destructive LZSM interference, which is known as coherent destruction of tunneling (CDT) \cite%
	{Atia2019}.
	
	Because the minimal gap between the ground and first excited state can
	decrease with an increasing number of qubits $M$, in general, the
	probabilities of nonadiabatic transitions are nonzero and should be taken
	into account \cite{Santoro2002}. Then, the question arises regarding how far the
	evolution can deviate from the ground state and still obtain useful results.
	Taking these into account results in the \textit{approximate AQC}.
	Particularly, estimating the efficiency in this case, here for a large number
	of possible interlevel transitions, can be done by using the analogy of the
	evolution with the standard \textit{random walk} \cite{Zagoskin2011}.
	Indeed, let us denote by $N\gg 1$ the average number of avoided-level
	crossings per one energy level, $\kappa =sN$ an estimate number of
	avoided-level crossings, $q=q(n,\kappa )$ the probability of finding the
	system in the eigenstate $\left\vert \varphi _{n}(s)\right\rangle $, and $p$
	the average LZSM transition probability. Then, assuming $n\gg 1$, after
	passing an avoided-level crossing, the system can be found in the $n$-th
	state if either it stays in this state or if it underwent a nonadiabatic
	transition from a neighboring energy level:%
	\begin{equation}
		q(n,\kappa +1)=(1-p)q(n,\kappa )+\frac{p}{2}\left[ q(n-1,\kappa
		)+q(n+1,\kappa )\right] \text{.}
	\end{equation}%
	This is almost the equation for the standard random walk \cite{Zagoskin2011},
	which allows us to use standard methods to describe the \textit{statistics of
		multiple LZSM transitions in a multilevel system}. For instance, the final
	dispersion reads
	\begin{equation}
		\left\langle \left( n-n_{0}\right) ^{2}\right\rangle
		_{s=1}=pN,
	\end{equation} which is proportional to the average transition probability $p$
	and the average number of quasicrossings $N$. Furthermore, we note that
	for a linear bias, $s=t/t_{\mathrm{f}}$, the driving velocity $v$ is
	inversely proportional to the evolution duration $t_{\mathrm{f}}$. Then, the
	LZSM probability gives an estimation for the accuracy of the approximate
	AQC:
	\begin{equation}
		\sqrt{\left\langle \Delta n^{2}\right\rangle }\propto \exp \left(
		-\alpha t_{\mathrm{f}}\right),
	\end{equation} which is defined by the exponential
	dependence on $t_{\mathrm{f}}$ with some nonuniversal rate $\alpha $.
	
	\section{Related classical coherent phenomena}
	
	\label{Sec:Classics}
	
	On many occasions, the similarities between classical wave optics and quantum wave mechanics have
	been highlighted \cite{Dragoman2004, Longhi2009,Bondar2013,Bliokh2013,Dressel2014}. In
	particular, in our context, we can find numerous examples in the literature
	where nonadiabatic transitions between classical states have been interrelated
	with LZSM physics. Some examples that can be mapped to the precession
	of a spin-1/2 particle exhibiting LZSM-type nonadiabatic transitions
	between two states, modes, and so forth, include the following: transformation of electromagnetic
	waves \cite{Zheleznyakov1983, Bliokh2001}, topological boundary states in acoustic waveguides \cite{Chen2020, Chen2021}, two coupled photonic
	crystal nanocavities (forming a photonic molecule) that are tuned by RF surface
	acoustic waves \cite{Kapfinger2015,Bliokh2019}, magnetoresistance in
	quasi-one-dimensional organic conductors \cite{Cooper2006}, the mixing of two
	tones in a ferrimagnetic sphere resonator \cite{Mathai2020}, optical modes
	coupled by a moveable membrane in optomechanical systems~\cite{Wu2013,
		Jiang2020}, two modes of a mechanical resonator or two directly-coupled resonators \cite{Zhang2020a}, and a rolling sphere on a Cornu spiral \cite{Rojo2010}. In these
	systems, different realizations of classical analogues of Rabi oscillations and
	LZSM transitions and interference were demonstrated.
	
\begin{figure}[tbp]
		\centering{\includegraphics[width=1.0		%
			\columnwidth]{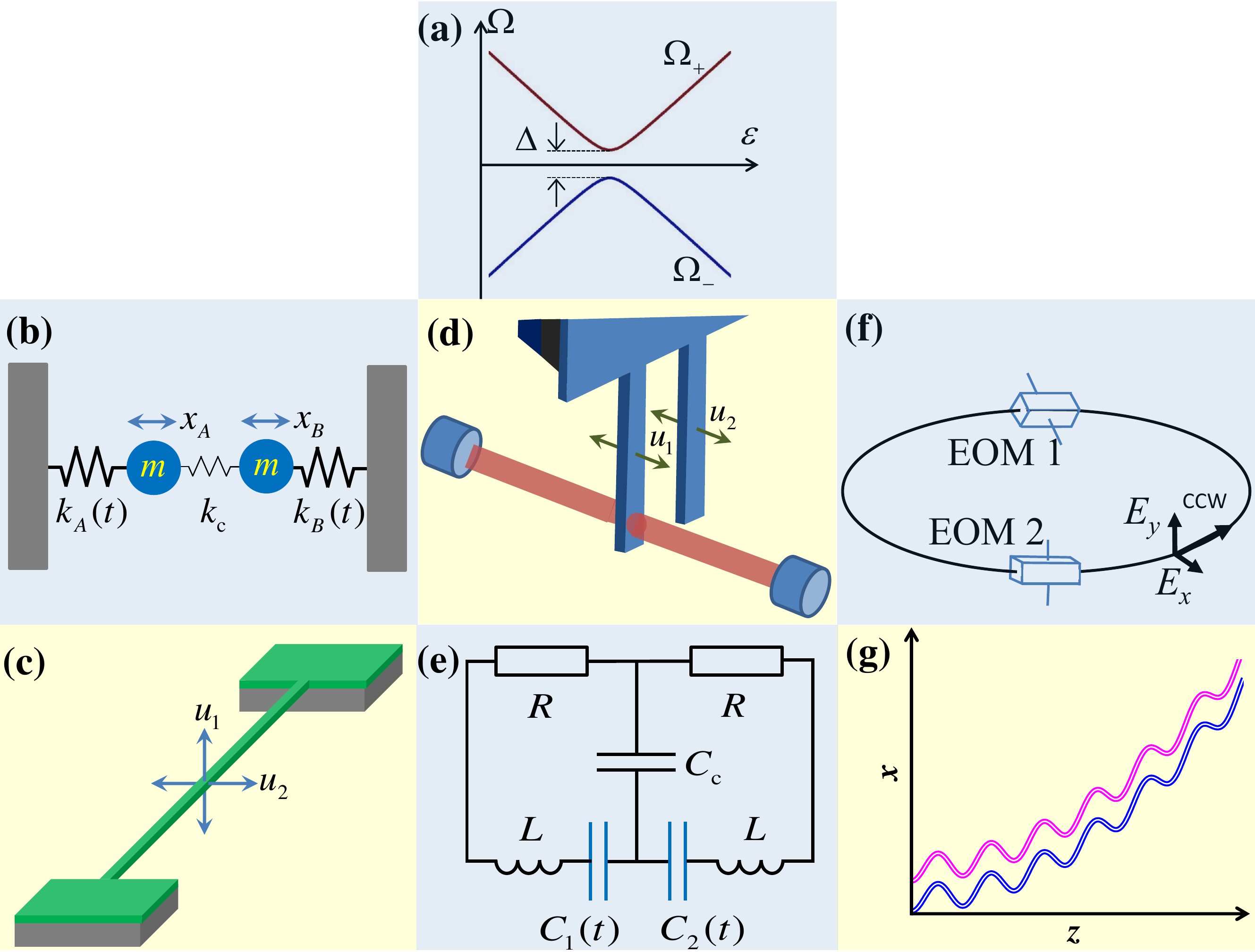}}
		\caption{ \textbf{Different classical systems that behave analogously to
				quantum TLSs}. In (a) the two eigenfrequencies, $\Omega _{\pm }$ are shown to
			depend on the bias $\protect\varepsilon $ and display avoided-level
			crossing at $\protect\varepsilon =0$ with a minimal distance $\Delta $.
			(b-g)~Several possible classical systems that have eigenfrequencies as
			in (a). Namely, (b)~two weakly coupled spring oscillators~\protect\cite%
			{Frimmer2014}, (c)~two-mode nano-beam~\protect\cite{Faust2012, Seitner2016},
			(d)~optomechanical system with two cantilevers, one of which is coupled with
			an optical cavity~\protect\cite{Fu2016, Fu2017}, (e)~two coupled electrical
			resonators~\protect\cite{Garrido2002, Muirhead2016}, (f)~two polarization
			modes of light propagating in the counter-clockwise (ccw) direction 
			tuned by the electro-optic modulators, EOM1 and EOM2, with the
			tuning parameter being the electric field inside EOM1 \protect\cite%
			{Spreeuw1990, Beijersbergen1992}, and (g)~two coupled curved waveguides in which
			an electromagnetic wave is spread between them \protect\cite{Longhi2009,
				Liu2019}.}
		\label{Fig:DiffClasSyst}
	\end{figure}

	To be specific, in this section, we demonstrate that a system of two weakly
	coupled resonators can be described analogously to a quantum TLS. Namely, we
	demonstrate that under appropriate conditions, the Newton equations for
	this classical system are reduced to an equation formally analogous to the
    Schr\"{o}dinger equation for a TLS. As a consequence, the classical
	system can display a classical analogue of LZSM transitions and
	interference.  In addition, any physical system that can be mapped on a
	two-resonator system with weak coupling could behave similarly.  In Fig.~\ref{Fig:DiffClasSyst},
	we demonstrate a variety of such systems~\cite{Ivakhnenko2018}.
	
	\paragraph{\textbf{From Newton to Schr\"{o}dinger }}\mbox{}\\
	
	We now consider how the classical Newton
	equations for two coupled resonators can be mapped to the Schr\"{o}dinger
	equation for a quantum TLS. Here, we follow Refs.~\cite{Novotny2010,
		Frimmer2014}. Consider the system of classical Newton equations for two
	weakly coupled oscillators $A$ and $B$, Fig.~\ref{Fig:DiffClasSyst}(b), 
	\begin{equation}
		\begin{cases}
			m_{A}\ddot{x}_{A}+\gamma _{A}\dot{x}_{A}+k_{A}x_{A}+k_{c}\left(
			x_{A}-x_{B}\right) =0, \\ 
			m_{B}\ddot{x}_{B}+\gamma _{B}\dot{x}_{B}+k_{B}x_{B}+k_{c}\left(
			x_{A}-x_{B}\right) =0,%
		\end{cases}%
	\end{equation}%
	where $x_{A,B}$ are the coordinates, $m_{A}=m_{B}=m$ are the masses, $\gamma
	_{A}=\gamma _{B}=\gamma $ are the damping coefficients, $k_{A,B}$ are
	the spring coefficients, and $k_{c}$ is a~coupling coefficient that is assumed
	to be weak, $k_{c}\ll k_{A,B}$. To obtain an analogy with a quantum TLS, we
	choose the excitation in the form $k_{A,B}=k_{0}\pm \Delta k(t)$ with $%
	\Delta k\ll k_{0}$. We denote the interaction-shifted eigenfrequency $\Omega
	_{0}^{2}\equiv (k_{0}+k_{c})/m$, and rewrite this system of
	classical Newton equations in matrix form, here using Pauli matrices: 
	\begin{equation}
		\left( \frac{d^{2}}{dt^{2}}+\frac{\gamma }{m}\frac{d}{dt}+\Omega
		_{0}^{2}\right) 
		\begin{bmatrix}
			x_{A} \\ 
			x_{B}%
		\end{bmatrix}%
		-\left( \frac{k_{c}}{m}\sigma _{x}+\frac{\Delta k(t)}{m}\sigma _{z}\right) 
		\begin{bmatrix}
			x_{A} \\ 
			x_{B}%
		\end{bmatrix}%
		=0.  \label{ClaNewtSystEq}
	\end{equation}%
	We use the ansatz 
	\begin{equation}
		\frac{\tilde{x}_{A,B}(t)}{x_{0}}=\psi _{A,B}\exp {\left( i\Omega
			_{0}t\right) },~~~~~~x_{A,B}=x_{0}\text{Re}~\tilde{x}_{A,B}.
		\label{ClasAnsatz}
	\end{equation}%
	Here, $x_{0}$ is the initial deviation of the springs, which is used to
	normalize $\tilde{x}_{A,B}$. We obtain the equation 
	\begin{equation}
		\left( \frac{d^{2}}{dt^{2}}+\left( \gamma +2i\Omega _{0}\right) \frac{d}{dt}%
		+i\Omega _{0}\gamma \right) 
		\begin{pmatrix}
			\psi _{A} \\ 
			\psi _{B}%
		\end{pmatrix}%
		-\left( \frac{k_{c}}{m}\sigma _{x}+\frac{\Delta k(t)}{m}\sigma _{z}\right) 
		\begin{pmatrix}
			\psi _{A} \\ 
			\psi _{B}%
		\end{pmatrix}%
		=0.
	\end{equation}%
	We can simplify this equation by using two assumptions
	\begin{equation}
		\gamma \ll \Omega _{0}\text{ \ and \ }k_{c},\text{ }\Delta k\ll k_{0}.
		\label{conditions}
	\end{equation}%
	The former assumption of small dissipation allows for neglecting $\gamma $ in
	comparison with $\Omega _{0}$. The latter assumption, together with the
	ansatz~\eqref{ClasAnsatz}, allows to use the slowly varying
	envelope approximation, which consists of neglecting the second derivative.
	This means that $\psi _{A,B}$ have small changes during the time span $2\pi
	/\Omega _{0}$; in other words, the characteristic evolution rate for $\psi
	_{A,B}$ is much smaller than $\Omega _{0}$. Also, the assumption $\Delta k\ll
	k_{0}$ means that we cannot reach a fast transition, where $v\rightarrow
	\infty $ and $\left\vert \psi _{A}^{2}(t)\right\vert \rightarrow \theta (t)%
	\mathcal{P}$. After using these assumptions, we obtain a Schr\"{o}%
	dinger-like equation 
	\begin{equation}
		\color{orange}\boxed{\color{black}i\frac{d}{dt}\left\vert \psi \right\rangle =H(t)\left\vert \psi
		\right\rangle -i\frac{\gamma }{2}\left\vert \psi \right\rangle,\color{orange}}\color{black}
		\label{ClasSchrEq}
	\end{equation}%
	where $\left\vert \psi \right\rangle =\left( \psi _{A},\psi _{B}\right) ^{\top}$%
	 and $H(t)$ is the Hamiltonian of Eq.~(\ref{H(t)}) with 
	\begin{equation}
		\Delta =\frac{k_{c}}{m\Omega _{0}}\approx \frac{k_{c}}{\sqrt{mk_{0}}},\text{
			\ \ }\:\:\:\:\:\varepsilon (t)=\frac{\Delta k(t)}{m\Omega _{0}}\approx \frac{\Delta
			k(t)}{\sqrt{mk_{0}}}.
	\end{equation}%
	Much like qubits, the mechanical resonators can be driven via $\Delta
	k(t) $, to have both an offset and a periodic excitation, we have $%
	\varepsilon (t)=\varepsilon _{0}+A\cos \omega t$. Note that the problem of
	Eq.~(\ref{ClasSchrEq}) can be described in terms of a non-Hermitian
	Hamiltonian \cite{Shen2019}.
	
	In the absence of dissipation, $\gamma =0$, Eq.~\eqref{ClasSchrEq}
	formally coincides with the Schr\"{o}dinger equation for a TLS, Eq.~\eqref{TDSE}, in the natural measuring system, that is, assuming $\hbar =1$.
	Dissipation can be eliminated by the substitution $\left\vert \psi
	\right\rangle =\left\vert \overline{\psi }\right\rangle \exp {\left( -
			\gamma t/2\right) }$; then, the classical Schr\"{o}dinger-like equation~\eqref{ClasSchrEq} becomes $i\frac{d}{dt}\left\vert \overline{\psi }%
	\right\rangle =H(t)\left\vert \overline{\psi }\right\rangle $. In addition, the
	\textquotedblleft density matrix\textquotedblright\ can be introduced as $%
	\rho =\left\vert \psi \right\rangle \left\langle \psi \right\vert $, where $%
	\left\langle \psi \right\vert :=\left( \psi _{A}^{\ast },\psi _{B}^{\ast
	}\right) $. Then, for the derivative, we obtain%
	\begin{equation}
		\dot{\rho}=-i\left[ H,\rho \right] -\gamma \rho .
	\end{equation}%
\begin{figure}[tbp]
	\centering{\includegraphics[width=1		%
		\columnwidth]{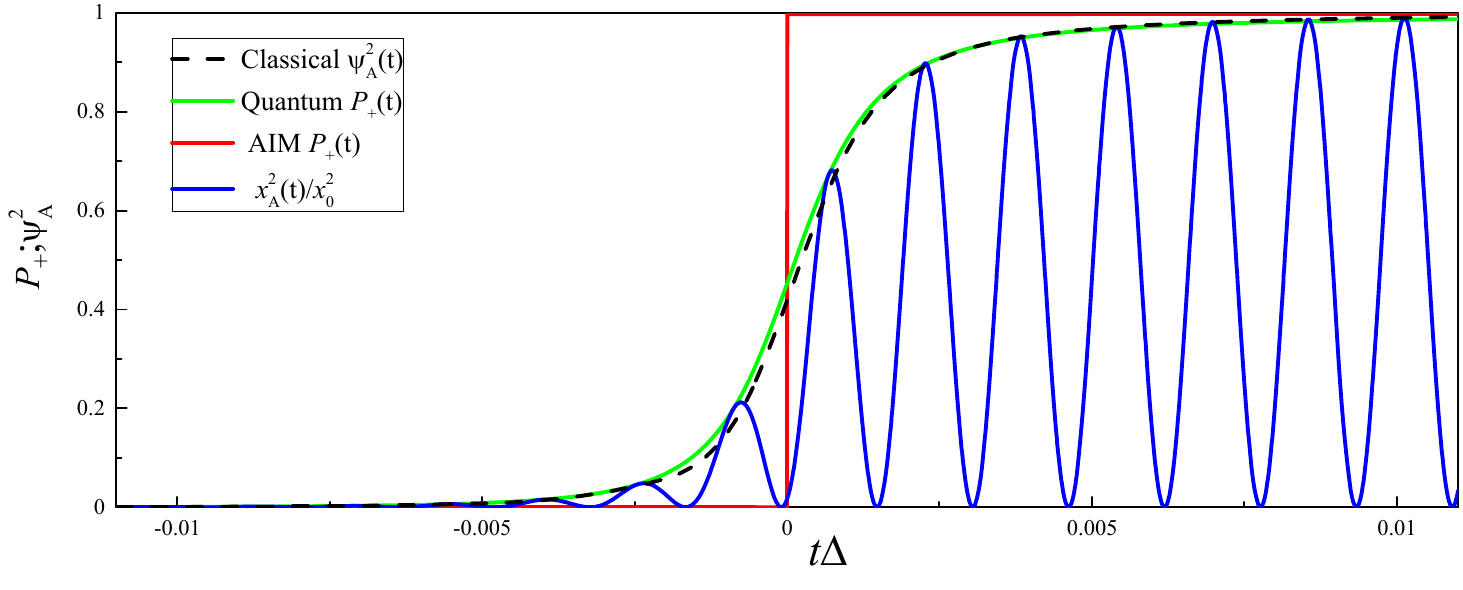}}
	\caption{\textbf{Single-passage transition in classical and quantum cases}.
		We here compare the solutions of the classical Newton equations, Eq.~\eqref{ClaNewtSystEq}, the Schr\"{o}dinger equation, and Eq.~\eqref{TDSE} for a
		LZSM transition. For illustration purposes, we take the parameters so that the final
		excitation probability is close to unity, $\mathcal{P}\approx 1$. The
		classical transfer of energy from one harmonics to another is given by the
		black line $\left\vert \protect\psi _{A}^{2}(t)\right\vert $, which virtually
		coincides with the quantum upper-level occupation probability $P_{+}(t)$, which is the
		green line. Furthermore, the red curve shows the AIM, which is considered
		above and is given here by $\protect\theta (t)\mathcal{P}$; the blue curve shows
		the classical oscillations $x_{A}^{2}(t)/x_{0}^{2}$.}
	\label{Fig:CompClasQuant1}
\end{figure}
	This coincides with the Bloch equation for a TLS, Eq.~%
	\eqref{Bloch}, for the Hamiltonian $H(t)$, and assuming that $\hbar =1$ and with equal relaxation rates, $T_{1}=T_{2}=1/\gamma $.
	
	This means that we can use the same methods for solving the
	classical and quantum systems. Interestingly, historically, the system of nonadiabatic transitions between two states was first studied and
	developed for a quantum TLS, and only later it was recognized that it has a classical doppelg\"{a}nger. Particularly, the theory by Zener helped 
	describe the coupled pendulums in Ref.~\cite{Maris1988}.
	
	In a system of two coupled classical oscillators, two eigenfrequencies are
	analogous to the two energy levels for the quantum TLS; these are the
	eigenvalues of the Hamiltonian above%
	\begin{equation}
		\Omega _{\pm }= \pm \sqrt{\Delta ^{2}+\varepsilon _{0}^{2}}.
	\end{equation}%
	The probability of the occupation of the energy levels in the quantum
	TLS is analogous to the squared amplitude of the oscillations $\left\vert
	\psi _{A,B}\right\vert ^{2}$; in other words, it is proportional to the
	amount of energy in that oscillation mode. To further demonstrate the
	similarity of the classical problem to its quantum analogue, in Fig.~\ref%
	{Fig:CompClasQuant1}, we consider a single-passage LZSM-like transition with $%
	\varepsilon (t)=vt,$ and choose the following parameters: $\Delta
	_{c}/k_{0}=5\times 10^{-4}$ for the classical case and $\delta =\delta
	_{cl}\equiv k_{c}^{2}/4v=3.57\times 10^{-4}$ for both cases. This
	demonstrates that under the conditions (\ref{conditions}), the classical
	envelope $\left\vert \psi _{A}(t)\right\vert ^{2}$ corresponds to the
	quantum occupation probability $P_{+}(t)$. This principle can be applied for any classical system of two coupled resonators with weak coupling. 

	Coherent phenomena in classical mechanical systems were reliably
	demonstrated experimentally; see the caption of Fig.~\ref{Fig:DiffClasSyst}. In
	particular, LZSM interferometry was realized with mechanical resonators
	in Refs.~\cite{Seitner2017} and \cite{Zhou2019,Lorenz2022}. For further theoretical
	study of such processes, see Refs. \cite{Chotorlishvili2011a, Parafilo2012,
		Parafilo2018, Villazon2019}.
	\paragraph{\textbf{Engineered photonic waveguides }}\mbox{}\\
	
	Curved waveguides have recently provided
	a rich laboratory tool to visualize quantum coherent phenomena with
	classical optical waves in the spatial domain \cite{Longhi2009,Menchon2016}. To understand
	how this becomes possible, consider the propagation of monochromatic light
	waves of wavelength $\lambda $ along a curved waveguide structure. This
	could be a single waveguide, coupled ones, or multiple waveguides \cite%
	{Longhi2009, Morales-Molina2011, Reyes2012, Liu2019}. Let the waveguide
	structure be planar, in the plane $\left( x,z\right) $; see Fig.~\ref%
	{Fig:DiffClasSyst}(g). There, $z$ is the paraxial propagation distance, 
	the axis bending profile 
	$n_{s}$, and the refractive index profile of the guiding structure $n(x)$, $%
	\omega =kc_{0}$. Using the scalar and paraxial approximations, we can write
	the electric field amplitude as $\psi (x,z)\exp \left[ i\left(
	kn_{s}z-\omega t\right) +c.c.\right] $, where the slow-varying field envelope 
	$\psi $ is described by the paraxial wave equation 
	\begin{equation}
		i\lambdabar \frac{\partial \psi }{\partial z}=-\frac{\lambdabar ^{2}}{2n_{s}}%
		\frac{\partial ^{2}\psi }{\partial x^{2}}+U(x)\psi,   \label{waveguide}
	\end{equation}%
	with $\lambdabar =\lambda /2\pi $ and $U(x)=\left[ n_{s}^{2}-n^{2}(x)\right]
	/2n_{s}\simeq n_{s}-n(x)$. This equation, Eq.~(\ref{waveguide}), can be
	formally written in the form of a 1D Schr\"{o}dinger equation with the
	following substitutions: $\lambdabar \rightarrow \hbar $, $n_{s}\rightarrow
	m $, and $z\rightarrow t$. Furthermore, the eigenmodes of two coupled
	waveguides can be described by a Hamiltonian formally analogous to the one
	of a driven TLS \cite{Liu2019}. 
	
	To conclude, electromagnetic waves
	propagating along waveguides have different advantages, such as long
	coherent lengths and high control in preparation and measurement in a
	room-temperature environment. Given that the Maxwell equation for these takes the
	form of the Schr\"{o}dinger equation and the interwaveguide interactions are
	analogous to the interaction between the quantum states, the light
	propagating along curved waveguides (namely, its intensity distributions) in
	the spatial domain can simulate quantum driven dynamics in the time domain \cite%
	{Longhi2009, Liu2019, Shen2019}.
	
	\section{Conclusion}
	
	\label{Sec:Conclusion}
	
	An avoided-level crossing is described by the LZSM model, which includes
	adiabatic evolution, nonadiabatic transitions, and quantum
	interference. We considered this by starting from the original LZSM works.
	It was demonstrated that this evergreen problem provides important tools for
	characterizing and controlling quantum systems.
	
	Our detailed consideration of the works by Landau, Zener, St\"{u}ckelberg, and
	Majorana demonstrates that the nonadiabatic transitions should be
	attributed to all four names. When presenting these four approaches here, we
	wanted not only to attract researchers' attention to this, but to
	make these diverse approaches easily accessible and clear for modern
	readers. We focused on Zener's approach because this
	describes both the final wave function, including its phase, and the
	dynamics. Both 
	Zener's and Majorana's 
	approaches can be developed to describe a few or many times
	passing of avoided-level crossing, providing the basis for the
	AIM in terms of transfer matrices.
	
	In describing single- and multiple-passage problems, we supplemented
	previous works and the review \cite{Shevchenko2010}. In that review, besides theory, mainly experiments with superconducting quantum
	systems were described. Here, we presented many more works, particularly the
	ones that have appeared since then, including on semiconducting devices,
	graphene, ultracold gases, and many other systems.
	
	Single and multiple LZSM transitions display rich physics of quantum
	systems, both on microscopic and mesoscopic scales. Related physics can be
	harnessed to characterize their quantum dynamics (interferometry)\ and to
	control their states. Classical systems can also display analogous phenomena.
	
	\begin{acknowledgments}
		We acknowledge useful discussions with S.~Ashhab, T.~Boolakee, B.~Damski, S.~Esposito, M.F.~Gonzalez-Zalba, P. Hommelhoff, P.~Kofman, M.~Liul, H.~Nakamura, A.~N.~Omelyanchouk, K.~Ono, A.~Ryzhov, and A.~Sotnikov. The research
		of O.V.I. and S.N.S. was sponsored by the Army Research Office and was
		accomplished under Grant Number W911NF-20-1-0261. O.V.I. was supported by
		the RIKEN International Program Associates (IPA). S.N.S. was
		partially supported by Grant No.~2/21-H from the NAS of Ukraine.
		F.N. is supported in part by:
		Nippon Telegraph and Telephone Corporation (NTT) Research,
		the Japan Science and Technology Agency (JST) [via
		the Quantum Leap Flagship Program (Q-LEAP), and
		the Moonshot R\&D Grant Number JPMJMS2061],
		the Japan Society for the Promotion of Science (JSPS)
		[via the Grants-in-Aid for Scientific Research (KAKENHI) Grant No. JP20H00134],
		the Army Research Office (ARO) (Grant No. W911NF-18-1-0358),
		the Asian Office of Aerospace Research and Development (AOARD) (via Grant No. FA2386-20-1-4069), and
		the Foundational Questions Institute Fund (FQXi) via Grant No. FQXi-IAF19-06. 
	\end{acknowledgments}
	
	\newpage
	
	\appendix
	\addcontentsline{toc}{section}{Appendix} 
	
	\section{LZSM transition probability}
	
	\label{Sec:AppendixA}
	
	Here we present four ways to derive the LZSM formula, Eq.~(\ref%
	{P}), following the four classic papers from 1932. In our presentation we  follow the original works and make those derivations
	clear for a broad audience. For this sake, some derivations are simplified and
	some notations changed. Readers interested in the original works can
	find them in Refs.~\cite{Landau1932a, Landau1932b, Zener1932,
		Stueckelberg1932, Majorana1932} and translated into English in Refs.~%
	\cite{Landau_EN, Cifarelli2020, Stueckelberg_EN}. Presenting these together, allows to see the interrelation between these approaches. For example,
	C.~Zener and E.~Majorana solved the very same second-order differential equation, both by referring to the same book \textquotedblleft Modern
	Analysis\textquotedblright\ \cite{Whittaker1920}, but Zener made use of special functions, while Majorana used the Laplace transform. Also,
	while following the LZSM papers, we will make several remarks; for example, we
	will not only be interested in the upper-level excitation probability, but
	also in the changes of the wave function phase, which is important for
	quantum-mechanical interference.
	
	\subsection{Near-adiabatic limit (Landau)}
	
	\label{App:Landau}
	
	First, we consider the LZSM transition following L.D.~Landau's
	two original papers \cite{Landau1932a, Landau1932b}, which
	have been translated into English in~\cite{Landau_EN}, and the textbook \cite%
	{LandauLishitz_QM}. This derivation of the LZSM formula is arguably the
	simplest one.
	
	Consider a slowly varying (with $v\ll \Delta ^{2}/\hbar $) time-dependent
	Hamiltonian (\ref{H(t)}). To describe the adiabatically slow variations,
	one can consider time $t$ as a parameter, and the system dynamics
	described by the Schr\"{o}dinger equation 
	\begin{equation}
		H(t)\left\vert \psi (t)\right\rangle =E\left\vert \psi (t)\right\rangle .
		\label{HamEq}
	\end{equation}%
	We substitute the wave function~\eqref{psi(t)} in the Schr\"{o}dinger equation~
	\eqref{HamEq} and obtain 
	\begin{equation}
		(H-E)%
		\begin{pmatrix}
			\alpha \\ 
			\beta%
		\end{pmatrix}%
		=-\frac{1}{2}%
		\begin{pmatrix}
			\ vt+2E & \ \Delta \\ 
			\Delta & \ -vt+2E%
		\end{pmatrix}%
		\begin{pmatrix}
			\alpha \\ 
			\beta%
		\end{pmatrix}%
		=0.  \label{Systek}
	\end{equation}%
	For a nontrivial solution of this matrix system of equations, we need to
	equate the matrix determinant to zero. As a result, we obtain the two energy
	levels of this TLS 
	\begin{equation}
		E_{\pm }(t)=\pm \frac{1}{2}\sqrt{\Delta ^{2}+\varepsilon (t)^{2}}=\pm \frac{1%
		}{2}\sqrt{\Delta ^{2}+(vt)^{2}}\equiv \pm \frac{\Delta E(t)}{2}.  \label{Epm}
	\end{equation}
	
	For slow excitations ($v\ll \Delta ^{2}/\hbar $), it is assumed that the
	evolution is adiabatic everywhere except for the avoided-crossing region. Then,
	we determine the probability of the upper-energy-level occupation $%
	P_{+}(t\rightarrow +\infty )$, provided that the system was initially in its
	ground state $\left\vert 0\right\rangle $, which means $P_{-}(t\rightarrow
	-\infty )=1$. 
	
	Slow perturbation implies that at large times of the
	\textquotedblleft transition process\textquotedblright\, the change of the
	action $S=\int E(t)dt$ is large. This means that the dynamics has a
	quasi-classical character and it is fully analogous to the over-barrier reflection. Note that we have the
	quasiclassical wave function as a function of time, not versus coordinates. Therefore, the
	avoided-crossing region determines the transition probability between the
	energy levels. In this region, the energy levels do not have an intersection
	in \textit{real} space, but they have an intersection in \textit{complex} space, so
	the equation $E_{+}(t_{0})=E_{-}(t_{0})$ is fulfilled at $t_{0}=\pm
	i\Delta /v$. Therefore, we can obtain the probability of the transition by
	following section 53 in Ref.~\cite{LandauLishitz_QM}: 
	\begin{equation}
		P_{+}=\exp {\left( -\frac{2}{\hbar }\mathrm{Im}%
			\int_{0}^{t_{0}}[E_{+}(t)-E_{-}(t)]dt\right) .}  \label{fromLL}
	\end{equation}%
	Using the substitution $u=vt/\Delta $, we obtain the integral 
	\begin{equation}
		\int_{0}^{t_{0}}\Delta E(t)\;dt=\frac{\Delta ^{2}}{v}\int_{0}^{i}\sqrt{1+u^{2}}%
		\;du=i\frac{\pi \Delta ^{2}}{4v}.
	\end{equation}%
	Therefore, we obtain the excitation probability of the TLS for the linear
	perturbation, $P_{+}=\mathcal{P}=\exp{(-2\pi \delta )}$, which is Eq.~(\ref{P}).
	
	In fact, in the original paper~\cite{Landau1932b}, in the right-hand side of the
	formula (\ref{fromLL}), there should be a prefactor $C\sim 1$. This happened
	to be exactly unity for the LZSM formula, as we can see from each of the
	three following subsections.
	
	The formula~(\ref{fromLL}) is very convenient for calculating the
	single-transition probability for any bias signal, and we consider this in
	the main text for a nonlinear drive. Because the approach by Landau was
	generalized in Refs.~\cite{Dykhne1962, Davis1976} for calculating
	nonadiabatic transitions regarding the form of the Hamiltonian, this
	formula is sometimes referred to as a \textit{Landau-Dykhne or
		Dykhne-Davis-Pechukas formula} \cite{Vitanov1999, Lehto2015}. Particularly,
	see the latter references for a generalization to the case when there are
	multiple zero points.
	
	\subsection{Using parabolic cylinder functions (Zener)}
	
	\label{LZSM_by_Zener}
	
	In this section, we consider how to derive the probability of the LZSM
	transition by following the work by Clarence~Zener, Ref.~\cite{Zener1932}. In fact we develop Zener's approach by finding analytically the full wave function with the time dependence \cite{Vitanov1996,Shevchenko2010}. Here we pay more attention to find the wave function's phase change after the transition. In this approach,
	we obtain a straightforward solution of the Schr\"{o}dinger equation with
	time dependence in terms of a special function, known as the parabolic
	cylinder function. This method is not so simple as the previous
	one, but it provides a more general and precise solution of this problem. In
	particular, there will be no unknown prefactor $C$, as in the approach by
	Landau. For a theoretical description involving a generalization
	for non-Hermitian Hamiltonians, see \cite{Shen2019,Kam2021}.
	
	Consider the time-dependent Schr\"{o}dinger equation, Eq.~(\ref{TDSE}).
	Using the wave function Eq.~\eqref{psi(t)} and Hamiltonian in the matrix
	form Eq.~(\ref{H(t)}), we obtain the system of equations 
	\begin{equation}
		\begin{cases}
			i\hbar \dot{\alpha}=-\frac{v}{2}t\alpha -\frac{\Delta }{2}\beta,  \\ 
			i\hbar \dot{\beta}=-\frac{\Delta }{2}\alpha +\frac{v}{2}t\beta .%
		\end{cases}
		\label{ZenSystEq}
	\end{equation}%
	This system of equations can be rewritten in the form of two second-order
	Weber equations, for either $\alpha $ or $\beta $:
\begin{subequations} 
 
	\begin{eqnarray}
		\frac{d^{2}\alpha }{d\tau ^{2}}+\left( 2\delta -i+\tau ^{2}\right) \alpha
		&=&0,  \label{ZenSystEqSolve} \\
		\frac{d^{2}\beta }{d\tau ^{2}}+\left( 2\delta +i+\tau ^{2}\right) \beta &=&0,
	\end{eqnarray}%
\end{subequations}
	where 
	\begin{equation}
		\tau =t\sqrt{\frac{v}{2\hbar }},  \label{z}
	\end{equation}%
	and $\delta $ is the adiabaticity parameter, Eq.~(\ref{delta}). The
	following replacement%
	\begin{equation}
		z=\tau \sqrt{2}\exp{(i\pi /4)}  \label{ZenerZprime}
	\end{equation}%
	allows us to obtain the so-called Weber equation in the canonical form 
	\begin{subequations} 
	 
	\begin{eqnarray}
		\frac{d^{2}\alpha }{dz^{2}}+\left( -i\delta -\frac{1}{2}-\frac{z^{2}}{4}%
		\right) \alpha &=&0, \\
		\frac{d^{2}\beta }{dz^{2}}+\left( -i\delta +\frac{1}{2}-\frac{z^{2}}{4}%
		\right) \beta &=&0.  
	\end{eqnarray}%
	\end{subequations}
	The solutions of the Weber equation are the combinations of the parabolic
	cylinder functions \cite{GradshteynRyzhik},%
	\begin{subequations} 
	
	\begin{eqnarray}
		\alpha &=&A_{+}D_{-1-i\delta }\left( z\right) +A_{-}D_{-1-i\delta }\left(
		-z\right),   \label{ZenerAnsatz} \\
		\beta &=&B_{+}D_{-i\delta }\left( z\right) +B_{-}D_{-i\delta }\left(
		-z\right) .    \label{ZenExactSolving}
	\end{eqnarray}
	\end{subequations} 
	We find relations between the coefficients $A_{\pm }$ and $B_{\pm }$ by
	inserting the solutions in the first-order equations~\eqref{ZenSystEq}: 
	\begin{equation}
		B_{\pm }=\mp \frac{\exp {\left( -i\pi /4\right) }}{\sqrt{\delta }}A_{\pm }.
		\label{ARelatB}
	\end{equation}%
	Now, using the relation~\eqref{ARelatB} and the exact solution~\eqref{ZenerAnsatz}, we can find $A_{\pm }$ from the initial conditions at $%
	z=z_{\text{i}}$ with certain $\alpha (z_{\text{i}})$ and $\beta (z_{\text{i}%
	})$ \cite{Vitanov1996}:
	\begin{subequations} 
	\begin{eqnarray}
		A_{+} &=&\frac{\Gamma (1+i\delta )}{\sqrt{2\pi }}\left[ \alpha (z_{\text{i}%
		} )D_{-i\delta }(-z_{\text{i}})-\beta (z_{\text{i}})e^{i\frac{\pi }{4}}%
		\sqrt{\delta }D_{-1-i\delta }(-z_{\text{i}})\right],  \\
		A_{-} &=&\frac{\Gamma (1+i\delta )}{\sqrt{2\pi }}\left[ \alpha (z_{\text{i}%
		})D_{-i\delta }(z_{\text{i}})+\beta (z_{\text{i}})e^{i\frac{\pi }{4}}\sqrt{%
			\delta }D_{-1-i\delta }(z_{\text{i}})\right] .
	\end{eqnarray}%
	\end{subequations}
	This gives the \textit{exact analytical solution at any moment of time} $%
	\tau >\tau _{\text{i}}$ for a single-passage LZSM transition \textit{for any
		initial condition} in terms of the evolution matrix $\Xi ^{\text{e}}$ 
	\begin{equation}
		\color{green}\boxed{\color{black}\begin{pmatrix}
			\alpha (z) \\ 
			\beta (z)%
		\end{pmatrix}%
		=%
		\begin{pmatrix}
			\Xi _{11}^{\text{e}} & \Xi _{12}^{\text{e}} \\ 
			\Xi _{21}^{\text{e}} & \Xi _{22}^{\text{e}}%
		\end{pmatrix}%
		\begin{pmatrix}
			\alpha (z_{\text{i}}) \\ 
			\beta (z_{\text{i}})%
		\end{pmatrix}%
		\equiv \Xi ^{\text{e}}%
		\begin{pmatrix}
			\alpha (z_{\text{i}}) \\ 
			\beta (z_{\text{i}})%
		\end{pmatrix}%
		, \color{green}}\color{black} \label{LZSM_ExactSol}
	\end{equation}%
	where the time-dependent matrix elements of $\Xi ^{\text{e}}$ are as follows 
	\begin{subequations}
		\begin{eqnarray}
			\Xi _{11}^{\text{e}} &=&\frac{\Gamma (1+i\delta )}{\sqrt{2\pi }}\left[
			D_{-i\delta }(-z_{\text{i}})D_{-1-i\delta }(z)+D_{-i\delta }(z_{\text{i}%
			})D_{-1-i\delta }(-z)\right],   \label{Exact_Zener_Solution} \\
			\Xi _{12}^{\text{e}} &=&\frac{\Gamma (1+i\delta )}{\sqrt{2\pi }}e^{\frac{%
					i\pi }{4}}\sqrt{\delta }\left[ -D_{-1-i\delta }(-z_{\text{i}})D_{-1-i\delta
			}(z)+D_{-1-i\delta }(z_{\text{i}})D_{-1-i\delta }(-z)\right],  \\
			\Xi _{21}^{\text{e}} &=&\frac{\Gamma (1+i\delta )}{\sqrt{2\pi }}\frac{e^{- 
					\frac{i\pi }{4}}}{\sqrt{\delta }}\left[ -D_{-i\delta }(-z_{\text{i}%
			})D_{-i\delta }(z)+D_{-i\delta }(z_{\text{i} })D_{-i\delta }(-z)\right],  \\
			\Xi _{22}^{\text{e}} &=&\frac{\Gamma (1+i\delta )}{\sqrt{2\pi }}\left[
			D_{-1-i\delta }(-z_{\text{i}})D_{-i\delta }(z)+D_{-1-i\delta }(z_{\text{i}%
			})D_{-i\delta }(-z)\right] .
		\end{eqnarray}%
	\end{subequations}
		From this transfer~matrix~\eqref{LZSM_ExactSol}, we can obtain the \textit{%
			exact} solution Eq.~\eqref{ZenExactSolving} for the LZSM problem in terms of the wave function~\eqref{psi(t)} at any moment of time for linear drives. Of course, this exact solution coincides with the numerical solution of Eq.~\eqref{ZenSystEq}.  However if we want
		to know the wave function \eqref{psi(t)} at some moment of time, we do not need to
		calculate it for all previous moments as we do for the numerical solution of Eq.~\eqref{ZenSystEq}. The exact
		solution transfer matrix \eqref{LZSM_ExactSol} provides the basis for the transfer matrix method describing
		the dynamics of the LZSM transition. Hence, now we explain how the exact transfer
		matrix $\Xi ^{\text{e}}$ results in the transition matrix $N$, Eq.~%
		\eqref{TMDiabatic}. For this, let us take the initial and final moments of
		the time symmetrically far from the quasicrossing region 
	\begin{eqnarray}
		\tau _{\text{i}} &=&-\tau _{\text{a}},\text{ }\tau _{\text{f}}=\tau _{\text{a%
		}},\ \ \ \text{ \ where }\:\: \tau _{\text{a}}\gg 1; \\
		z_{\text{i}} &=&\tau _{\text{a}}\sqrt{2}e^{-3i\pi /4},\text{ }\:\:\:\:\:\  z_{\text{f}}=\tau _{\text{a}}\sqrt{2}e^{i\pi /4}.  \label{InfAsympt}
	\end{eqnarray}%
	We now use approximations for the parabolic cylinder functions, with the
	argument tending to infinity \cite{GradshteynRyzhik} (see the chapter on parabolic
	cylinder functions, pp. 1092–1094), 
	\begin{subequations} 
	 
	\begin{eqnarray}
		\lim\limits_{\left\vert z\right\vert \rightarrow \infty }D_{-i\delta
			-1}\left( z\right) &\approx &e^{-z^{2}/4}z^{-i\delta -1}-\frac{\sqrt{2\pi }}{%
			\Gamma (i\delta +1)}e^{i\pi (i\delta +1)}e^{z^{2}/4}z^{i\delta }, \ \ \ \text{ Arg}%
		(z)=-\frac{3\pi}{4},  \label{asympt1} \\
		\lim\limits_{\left\vert z\right\vert \rightarrow \infty }D_{-i\delta
			-1}\left( -z\right) &\approx &e^{-z^{2}/4}z^{-i\delta -1},\ \ \ \text{ Arg}(z)=%
		\frac{\pi}{4},  \label{asympt2}
	\end{eqnarray}%
	\end{subequations}
	and the following properties of the Gamma function 
	\begin{subequations} 
	\begin{eqnarray}
		\Gamma (i\delta )=\sqrt{\frac{\pi }{\delta \sinh \pi \delta }}\exp{\left(i\text{Arg}[
			\Gamma (i\delta )]\right)},  \label{about_Gamma} \\
		\text{Arg}[\Gamma (i\delta)]=-\text{Arg}[\Gamma (-i\delta)]=-\frac{\pi}{2}-\text{%
			Arg}[\Gamma (1-i\delta)].  
	\end{eqnarray}%
	\end{subequations}
	As a result, we find the asymptotic values of the single-passage evolution
	matrix in the diabatic basis
	\begin{subequations} 
	  
	\begin{eqnarray}
		\Xi _{11}^{\text{e}} &\approx &\exp \left[-\pi \delta \right]=T, \\
		\Xi _{12}^{\text{e}} &\approx &\frac{\sqrt{2\pi }}{\Gamma (1+i\delta )}\sqrt{%
			\delta }e^{\frac{i\pi }{4}}e^{i\tau _{\text{a}}^{2}}\left( \sqrt{2}\tau _{%
			\text{a}}\right) ^{2i\delta }e^{-\frac{\pi \delta }{2}}=R\exp i{\left\{ 
			\frac{\pi }{4}+\text{Arg}\left[ \Gamma (1-i\delta )\right] +\tau _{\text{a}%
			}^{2}+2\delta \ln {\left( \sqrt{2}\tau _{\text{a}}\right) }\right\} }.
		\label{Xi_12}
	\end{eqnarray}%
	\end{subequations}
	Other elements, $\Xi _{21}^{\text{e}}$ and $\Xi _{22}^{\text{e}}$, can be
	found from these two. Here,
	\begin{equation}
		T\;=\;e^{-\pi \delta },\;\;\;
		R\;=\;\sqrt{1-e^{-2\pi\delta }}
	\end{equation} stand for the transition and reflection coefficients.
	
	Let us now consider how to relate this to the adiabatic-impulse model;
	namely, let us find the nonadiabatic transition matrix $N$. According to
	the TM method, Sec.~\ref{TM}, the full single-passage evolution
	for any initial condition consists of three stages: 
	
	(a)~adiabatic evolution before the transition, 
	
	(b)~the transition itself,
	and 
	
	(c)~adiabatic evolution after the transition. 
	
	Hence, for this, the natural
	description appears in the adiabatic basis. The three-stage evolution in the
	adiabatic basis is described by the product of the three respective matrices 
	$\Xi ^{\mathrm{a}}=U_{2}NU_{1}$. Here, the adiabatic-stage matrices are $%
	U_{1,2}=\exp \left[ i\zeta (\pm \tau _{\text{a}})\sigma _{z}/2\right] $ with 
	\begin{subequations}\begin{equation}
		\zeta (\pm \tau _{\text{a}})=\frac{1}{2\hbar }\int_{0}^{\pm \tau _{\text{a}}}%
		\sqrt{\Delta ^{2}+2\hbar v\tau ^{2}}\;d\tau \approx \Phi (\tau )-\Phi _{\delta
		},
	\end{equation}%
	where 
	\begin{eqnarray}
		\Phi (\tau ) =\frac{\tau ^{2}}{2}+\delta \ln {\left( \sqrt{2}\tau \right) }%
		,\:\:\:\:\:\:\:\:
		\Phi _{\delta } =\frac{1}{2}\delta \left( \ln \delta -1\right) .  
		\label{Linear_Adiabatic_Evolution}
	\end{eqnarray}\end{subequations}
	Therefore, we obtain the single-passage evolution matrix in the adiabatic
	basis%
	\begin{equation}
		\Xi ^{\mathrm{a}}=U_{2}NU_{1}=%
		\begin{pmatrix}
			R\exp i{\left\{ -\frac{\pi }{4}-\text{Arg}\left[ \Gamma (1-i\delta )\right]
				-2\Phi (\tau _{\mathrm{a}})\right\} } & T \\ 
			-T & R\exp i{\left\{ \frac{\pi }{4}+\text{Arg}\left[ \Gamma (1-i\delta
				)\right] +2\Phi (\tau _{\mathrm{a}})\right\} }
		\end{pmatrix}.
	\label{Zener_Tot_single_trans_ev_matrix}
	\end{equation}
	
	The evolution matrices in the adiabatic and diabatic bases are linked by the
	transition matrix
	\begin{equation}
		S(\tau )=%
		\begin{pmatrix}
			\gamma _{+} & \gamma _{-} \\ 
			\gamma _{-} & -\gamma _{+}%
		\end{pmatrix},%
	\end{equation} where the coefficients $\gamma _{\pm }$ were defined in Eq.~\eqref{gammas}.
The asymptotes of the matrix $S(\tau)$ have the form 
	\begin{equation}
		S(-\tau _{\text{a}})=%
		\begin{pmatrix}
			0 & 1 \\ 
			1 & 0%
		\end{pmatrix}%
		,\text{ }\ \ \ S(\tau _{\text{a}})=%
		\begin{pmatrix}
			1 & 0 \\ 
			0 & -1%
		\end{pmatrix}%
		.
	\end{equation}%
    Then,
	we match the two representations%
	\begin{equation}
		S(\tau _{\text{a}})\Xi ^{\mathrm{e}}S(-\tau _{\text{a}})=\Xi ^{\mathrm{a}%
		}=U_{2}NU_{1},
	\end{equation}%
	from which we obtain the desired matrix for the \textit{nonadiabatic} transition%
	\begin{equation}
		N=U_{2}^{-1}S(\tau _{\text{a}})\Xi ^{\mathrm{e}}S(-\tau _{\text{a}%
		})U_{1}^{-1}=%
		\begin{pmatrix}
			Re^{-i{\phi }_{\text{S}}} & -T \\ 
			T & Re^{i{\phi }_{\text{S}}}%
		\end{pmatrix}%
		,\label{Zener_transfer_matrix}
	\end{equation}%
	where the Stokes phase $\phi_\text{S}(\delta)$ was defined in Eq.~\eqref{StocesPhase}.
	For more on this phase shift, see also Chapter
	4 in \cite{Child1974}.
	
	Note that the transition matrix $N$ is given here in the adiabatic
	representation. In the diabatic one, we obtain%
	\begin{equation}
		N_\mathrm{d}=S(\tau _{\text{a}})NS(-\tau _{\text{a}})=%
		\begin{pmatrix}
			T & Re^{i{\phi }_{\text{S}}} \\ 
			-Re^{-i{\phi }_{\text{S}}} & T%
		\end{pmatrix}.%
	\end{equation}
	
	Consider the inverse transition, from $\tau =+\tau _{\text{a}}$ to $\tau
	=-\tau _{\text{a}}$. Then, we have $z_{\text{i}}=\tau _{\text{a}}\sqrt{2}%
	e^{i\pi /4}$ and$\text{ }z_{\text{f}}=\tau _{\text{a}}\sqrt{2}e^{-3i\pi /4}$.
	We obtain that the transition matrices for this inverse transition are transposed to direct transitions matrices in the adiabatic and diabatic bases,%
	\begin{subequations}\begin{eqnarray}
		N_{\text{inverse}} &=&N^\top=%
		\begin{pmatrix}
			Re^{-i{\phi }_{\text{S}}} & T \\ 
			-T & Re^{i{\phi }_{\text{S}}}%
		\end{pmatrix}, \label{LZSM_back_transition_matrix}\\
		N_{\mathrm{d}{\text{,inverse}}} &=&N_\mathrm{d}^{\top}=%
		\begin{pmatrix}
			T & -Re^{-i{\phi }_{\text{S}}} \\ 
			Re^{i{\phi }_{\text{S}}} & T%
		\end{pmatrix},
	\end{eqnarray}\end{subequations}
	where $\top$ denotes transposition. We emphasize that the inverse transition is described by the transposed matrices, which can be obtained both from inverting time $(t\rightarrow-t)$ in the Schr\"{o}dinger equation and from direct solution, as we checked in Eq.~\eqref{LZSM_back_transition_matrix}.
	Importantly, the above consideration bears a general character and can be given
	for any initial condition. Particularly, in the case in which initially our
	TLS was in the lower (ground) energy level, with
	
	\begin{equation}
		\begin{cases}
			\alpha (z_{\text{i}})=0,\\%
			\beta (z_{\text{i}})=1, 
		\end{cases}
		\label{ZInitCond}
	\end{equation}%
	we have 
	\begin{equation}
		\begin{cases}
			\alpha (z)=\Xi _{12}^{\text{e}}, \\ 
			\beta (z)=\Xi _{22}^{\text{e}} .%
		\end{cases}%
	\end{equation}%
	This expression defines the solution for any initial moment of time $\tau _{%
		\mathrm{i}}$. Furthermore, the expressions for $\Xi ^{\text{e}}$ are simplified
	for $-\tau _{\mathrm{i}}=\tau _{\mathrm{a}}\gg 1$. Then, we obtain the
	expression for the occupation of the upper diabatic energy level: 
	\begin{equation}
		P_{\text{d}}(z)=|\alpha (z)|^{2}=\delta e^{-\frac{\pi \delta }{2}}\left\vert D_{-1-i\delta
		}(-z)\right\vert ^{2}.  \label{LZPaPdi}
	\end{equation}%
	Taking into account the relationship between the diabatic and adiabatic
	bases, for the respective upper-level occupation probability in the
	adiabatic basis, we have%
	\begin{equation}
		P_{\text{a}}(z)=|\beta (z)\gamma _{+}-\alpha (z)\gamma _{-}|^{2},
	\end{equation}%
	and this results in the following time dependence%
	\begin{equation}
		P_{\text{a}}(z)=e^{-\frac{\pi \delta }{2}}\left\vert D_{-i\delta }(-z)\gamma
		_{+}-\sqrt{\delta }e^{-\frac{i\pi }{4}}D_{-1-i\delta }(-z)\gamma
		_{-}\right\vert ^{2}.  \label{LZPaPd}
	\end{equation}%
	These expressions define the \textit{time dependence of the upper-level occupation
	probabilities}. They are further simplified if we are interested in the
	asymptotic solution with $\tau \gg 1$.
	
	\paragraph*{Extending Zener's 1932 approach}\mbox{}\\
	
	Consider now another, more conventional, way to solve this problem of
	developing Zener's approach. Below, we will extend Zener's 1932 approach. We will study the asymptotics of $%
	\alpha $ and $\beta $ and find the coefficients $A_{\pm }^{\prime }$ after
	obtaining asymptotics from the initial conditions. Consider now now the asymptotic
	solution with Eq.~\eqref{InfAsympt}, here with the asymptotics of the parabolic
	cylinder functions Eq.~\eqref{asympt2}. 
	
	We find the asymptotic values of $\alpha 
	$ and $\beta $: 
	\begin{subequations}\begin{eqnarray}
		\alpha (-\tau _{\mathrm{a}}) &\approx &A_{+}^{\prime }\Theta _{1}\exp {%
			\left( i\Phi \left( \tau _{\mathrm{a}}\right) \right) },
		\label{ZenWebFuncApprox} \\
		\beta (-\tau _{\mathrm{a}}) &\approx &\left( -e^{-\delta \pi
			/2}A_{+}^{\prime }+e^{\delta \pi /2}A_{-}^{\prime }\right) \Theta _{2}\exp {%
			\left( -i\Phi \left( \tau _{\mathrm{a}}\right) \right) },   \\
		\alpha (\tau _{\mathrm{a}}) &\approx &A_{-}^{\prime }\Theta _{1}\exp {\left(
			i\Phi \left( \tau _{\mathrm{a}}\right) \right) },   \\
		\beta (\tau _{\mathrm{a}}) &\approx &\left( -e^{\delta \pi /2}A_{+}^{\prime
		}+e^{-\delta \pi /2}A_{-}^{\prime }\right) \Theta _{2}\exp {\left( -i\Phi
			\left( \tau _{\mathrm{a}}\right) \right) },  
	\end{eqnarray} 
	where 
	\begin{eqnarray}
		\Theta _{1} &=&\frac{\sqrt{2\pi }}{\Gamma \left( 1+i\delta \right) }\exp
		\left( -\frac{\pi }{4}\delta \right),  \\
		\Theta _{2} &=&\frac{1}{\sqrt{\delta }}\exp \left( -i\frac{\pi }{4}-\frac{%
			\pi }{4}\delta \right),     \label{ZenWebFuncSubst}
	\end{eqnarray}\end{subequations} 
	and $A_{\pm }^{\prime }$ are the asymptotic coefficients, which play the
	same role as $A_{\pm }$ for the exact solution. Let us match this asymptotic
	solution with the transition from a diabatic to adiabatic basis, Eq.~%
	\eqref{eigenstates}, at $\tau _{\mathrm{a}}\gg 1$. Then, we obtain 
	\begin{subequations}\begin{eqnarray}
		\begin{pmatrix}
			c_{+}(-\tau _{\mathrm{a}}) \\ 
			c_{-}(-\tau _{\mathrm{a}})%
		\end{pmatrix}
		&=&%
		\begin{pmatrix}
			\alpha (-\tau _{\mathrm{a}}) \\ 
			\beta (-\tau _{\mathrm{a}})%
		\end{pmatrix}%
		, \\
		\begin{pmatrix}
			c_{+}(\tau _{\mathrm{a}}) \\ 
			c_{-}(\tau _{\mathrm{a}})%
		\end{pmatrix}
		&=&%
		\begin{pmatrix}
			\beta (\tau _{\mathrm{a}}) \\ 
			-\alpha (\tau _{\mathrm{a}})%
		\end{pmatrix}%
		,
	\end{eqnarray}\end{subequations}
	where $c_{\pm }$ are the coefficients of the decomposition for a wave
	function $\left\vert \psi \right\rangle $ with the adiabatic wave functions $%
	\left\vert \psi _{\pm }\right\rangle $, 
	\begin{equation}
		\left\vert \psi \right\rangle =c_{+}\left\vert \psi _{+}\right\rangle
		+c_{-}\left\vert \psi _{-}\right\rangle .
	\end{equation}%
	Then, we use the initial conditions~\eqref{ZInitCond}, and we would like to
	describe the probability of finding the TLS in the upper energy level, which
	is $|c_{+}(\tau _{\mathrm{f}})|^{2}\approx |\beta (\tau _{\mathrm{f}})|^{2}$.
	For this, we need to find an absolute value of the coefficients $A_{\pm
	}^{\prime }$. We describe the coefficients from the first initial condition~\eqref{ZInitCond} and the first equation in the system~\eqref{ZenSystEq};
	hence, we obtain $i\hbar \dot{\alpha}=-\frac{\Delta }{2}$. Then, before using
	the ansatz~\eqref{ZenerAnsatz}, we need to rewrite the initial condition in
	terms of the variable $z$ introduced in Eq.~\eqref{ZenerZprime} 
	\begin{equation}
		\frac{e^{-\frac{i\pi }{4}}}{\sqrt{\delta }}\frac{d\alpha }{dz}=1.
	\end{equation}%
	For the next step, we need to define the derivative for the parabolic
	cylinder function at infinity, as shown in Eq.~(\ref{asympt2}), which reads \cite%
	{GradshteynRyzhik}: 
	\begin{equation}
		\lim\limits_{\left\vert z\right\vert \rightarrow \infty }\frac{d}{dz}%
		D_{-i\delta -1}\left( -z\right) \approx -\frac{1}{2}e^{-z^{2}/4}z^{-i\delta
		}.  \label{derivative}
	\end{equation}%
	Now, we determine the coefficients $A_{\pm }^{\prime }$, for which we use the
	ansatz~\eqref{ZenerAnsatz} and asymptotics of the parabolic cylinder
	functions~\eqref{derivative}: 
	\begin{equation}
		\frac{e^{-\frac{i\pi }{4}}}{2\sqrt{\delta }}A_{\pm }^{\prime }e^{-i\frac{%
				\tau ^{2}}{2}}\left( \sqrt{2}\tau \right) ^{-i\delta }e^{\frac{\pi \delta }{4%
		}}=1.
	\end{equation}%
	It follows that
	\begin{equation}
		A_{+}^{\prime }=A_{-}^{\prime }=2\sqrt{\delta }e^{-\delta \pi /4}e^{i\frac{%
				\pi }{4}}\exp\left[i\frac{\tau ^{2}}{2}\right]\exp\left[i\delta \ln (\sqrt{2}\tau )\right].
		\label{Apm^prime}
	\end{equation}%
	Finally, we obtain the \textit{probability of the lower-energy-level occupation
	after the transition}:
	\begin{equation}
		|\alpha (z_{\text{f}})|^{2}\approx \frac{2\pi \delta }{|\Gamma \left(
			1+i\delta \right) |^{2}}e^{-\pi \delta }=\frac{2\pi \delta e^{-\pi \delta }}{%
			\delta ^{2}|\Gamma \left( i\delta \right) |^{2}}=1-e^{-2\pi \delta }.
	\end{equation}%
	Here, we used Eq.~\eqref{about_Gamma} with the normalization condition $%
	\left\vert \alpha \right\vert ^{2}+\left\vert \beta \right\vert ^{2}=1$.
	Then, we obtain the probability of the upper-level occupation: 
	\begin{equation}
		|\beta (z_{f})|^{2}=\exp\left[-2\pi \delta \right]=\mathcal{P}.
	\end{equation}
	
	In addition to finding the final upper-level occupation probability, after defining
	the coefficients $A_{\pm }^{\prime }$, this approach gives us the asymptotics
	for the coefficients $\alpha $ and $\beta $ of the wave function $\left\vert
	\psi (t)\right\rangle =\alpha (t)\left\vert 0\right\rangle +\beta
	(t)\left\vert 1\right\rangle $ after passing the avoided-level region. If we
	consider the phase of $\alpha (z_{f})$, using Eqs.~\eqref{about_Gamma},~%
	\eqref{Xi_12} and~\eqref{Apm^prime}, we can find the transition phase change:%
	\begin{equation}
		\text{Arg}[\alpha (z_{\text{f}})]= \frac{\pi }{4}+\text{Arg}\left[ \Gamma (1-i\delta )\right]+2\Phi (\tau _{\mathrm{a}}).
		\label{Zener_approximate_result}
	\end{equation}%
	(Note that Eq.~\eqref{Zener_approximate_result} is not in Zener's paper.)
	Hence, we obtained Eq.~(\ref{P}) twice using Zener's method with the full single-transition evolution phase Eq.~\eqref{Xi_12}. Then we can do the same transformation as in Eqs.~(\ref{Zener_Tot_single_trans_ev_matrix}, \ref{Zener_transfer_matrix}), and we obtain the same transition matrix with the Stokes phase Eq.~\eqref{StocesPhase}. 
	We have $\alpha (\tau _{\text{a}})=\Xi _{12}^{\text{e}}$
	because of the initial conditions~\eqref{ZInitCond}. Consequently, we obtain the
	same final result for the probability of excitation and the phase, as in Eq.~%
	\eqref{Xi_12}.
	
	To summarize, in this section we have presented two approaches, generalizing
	the works by \cite{Zener1932} and \cite{Vitanov1996}.\ We obtained the exact
	solution in terms of the parabolic cylinder special functions for a single-passage problem in the general case in terms of the exact transfer matrix $\Xi ^{%
		\text{e}}$, Eq.~\eqref{LZSM_ExactSol}. Then, we obtained a simpler limit
	variant of the transfer matrix, which describes the evolution from $z_{\mathrm{i}%
	}$ to $z_{\mathrm{f}}$, Eq.~\eqref{TMDiabatic}. As a particular case, we
	have the single-passage problem, starting from the ground state and
	ending far from the avoided-level crossing; then, the probability of
	the transition is given by the LZSM formula, while the phase difference is
	defined by the Stokes phase. Zener obtained only the probability of the LZSM
	transition; apparently, he was not interested in the phase change of the wave function.
	However, it happened to be straightforward to define it using Zener's approach.
	
	\subsection{With contour integrals (Majorana)}
	
	\label{App_Majorana} Ettore Majorana studied the spin
	orientation in a dynamic magnetic field \cite{Majorana1932} with
	components $H_{x}=-\Delta $, $H_{y}=0$ and $H_{z}=-vt$. The English version of
	Majorana's paper is available in the book \cite{Bassani2006} and in the
	second edition in \cite{Cifarelli2020}, which is commented on by M.~Inguscio \cite%
	{Inguscio2020}; see also in \cite{Wilczek2014}. We note again that we use
	 convenient and uniform notations for all the four LZSM approaches, which
	here differ from the ones used by Majorana by only changing notation.
	
	Note that Majorana studied first Eq.~\eqref{P} and later Zener studied the same equation.
	In his approach, Majorana used the Laplace transform instead of using
	the special parabolic functions in Zener's method. Given the importance of
	Majorana's approach, details can be found elsewhere \cite{Kofman2022}, while
	here we present its key aspects.
	
	Following Majorana, the system of equations~\eqref{ZenSystEqSolve} can be
	transformed by the substitutions 
	\begin{equation}
		\alpha =f_{1}\exp {\left( \frac{i}{2}\tau ^{2}\right) },\text{ }\beta
		=f_{2}\exp {\left( -\frac{i}{2}\tau ^{2}\right) },  \label{MajSubst}
	\end{equation}from which we obtain the system of two equations
	\begin{equation}
	\begin{cases}
		\dot{f_1}  =i\sqrt{2\delta }f_2\exp {(-i\tau ^{2}),} \\ 
		\dot{f_2}  =i\sqrt{2\delta }f_1\exp {(i\tau ^{2}),}%
	\end{cases}
	\label{initialsystem}
	\end{equation}%
	which can be split into two independent equations%
	\begin{equation}
		\frac{d^{2}f_{1,2}}{d\tau ^{2}}\pm 2i\tau \frac{df_{1,2}}{d\tau }+2\delta
		f_{2,1}=0.  \label{MajoranaDiffEq1}
	\end{equation}%
	Note that there is some difference from \cite{Majorana1932} in denoting the time
	variable $\tau =\sqrt{2}\tau ^{\prime }$. Following Majorana, this equation
	can be solved by the two-sided Laplace transform; then, for the Laplace
	transforms of the functions $f_{1,2}(\tau )$, one obtains 
	\begin{equation}
		F_{1,2}(s)=C_{\delta }\exp({\mp is^{2}/4})s^{\mp 1-i\delta }.
	\end{equation}%
	Here, the constant of integration $C_{\delta }$ should be defined from an
	initial condition, by assuming the system being initially in the ground
	state. Then, we can find $f_{1}(\tau )$ and $f_{2}(\tau )$: 
	\begin{equation}
		f_{1,2}(\tau )=\oint_{C_{1,2}^{\pm }}e^{s\tau }F_{1,2}(s)\;ds,
	\end{equation}%
	where $C_{1,2}^{\pm }$ are the steepest descent contours. The integrals have
	two contributions: the first one is from the saddle-point and the second one
	is from the vicinity of zero. One can derive the contribution from the
	quasi-intersection points by using the saddle-point method in complex
	space. We then obtain the approximate solutions of Eq.~\eqref{MajoranaDiffEq1} for two cases, $\tau <0$ and $\tau >0$, here in their
	general form. 
	Then using
	the substitutions~\eqref{MajSubst}, these can be written for the asymptotes $\alpha (\tau )$ and $%
	\beta (\tau )$ at large $\tau $%
	\begin{subequations}
	\begin{align}
		& \tau <0:
		\begin{cases}
		&\alpha (\tau )=C_{\delta }\sqrt{4\pi}(-2i\tau )^{-i\delta -1}\exp \left( -i%
		\frac{\tau ^{2}}{2}-i\frac{\pi}{4}\right),  \label{MajApproxSollution} \\
		& \beta (\tau )=C_\delta\sqrt{\frac{2\pi i}{\delta}}(-2i\tau)^{-i\delta}\exp\left(-i\frac{\tau^2}{2}\right),
		\end{cases} \\
		& \tau >0:
		\begin{cases}
		&\alpha (\tau )=C_\delta\sqrt{4\pi}(-2i\tau)^{-i\delta-1} \exp\left(-i\frac{\tau^2}{2}-i\frac{\pi}{4} \right)+C_\delta\frac{2\pi i}{\Gamma(i\delta+1)}\tau^{i\delta}\exp\left(i\frac{\tau^2}{2}\right), \\
		& \beta (\tau )=C_\delta\sqrt{\frac{2\pi i}{\delta}}(-2i\tau)^{-i\delta}\exp\left(-i\frac{\tau^2}{2}\right)+C_\delta\sqrt{\frac{\delta}{2}}\frac{2\pi i}{\Gamma(i\delta+1)}\tau^{i\delta-1}\exp\left(i\frac{\tau^2}{2}\right)
		,\end{cases}
	\end{align}
	\end{subequations} where
	\begin{equation}
		C_{\delta }=\sqrt{\frac{\delta }{2\pi }}\exp{\left(-\frac{\pi\delta}{2}\right)}.
	\end{equation}
	From Eq.~\eqref{MajApproxSollution}, we can obtain the absolute values for
	the upper- and ground-diabatic-level occupations after the LZSM transition:
	\begin{subequations} \begin{align}
		& |\alpha (\tau \rightarrow \infty )|^{2}=1-e^{-2\pi \delta }, \\
		& |\beta (\tau \rightarrow \infty )|^{2}=e^{-2\pi \delta }.
	\end{align}\end{subequations}
	In addition, we can find the phase shift, which is accumulated during the transition
	process: 
	\begin{equation}
		\text{Arg}[\alpha (\tau \rightarrow \infty )]-\text{Arg}[\alpha (-\tau
		\rightarrow \infty )]=
		\frac{\pi }{4}+\text{Arg}\left[ \Gamma (1-i\delta )\right]+2\Phi (\tau _{\mathrm{a}}),
	\end{equation}
	We realize that this equation fully coincides with Eq.~\eqref{Zener_approximate_result}, so we can do the same transformations to obtain the Stokes phase from it.
	As a result of this approach,
	we obtain the asymptotic formulas for the amplitude 
	of the TLS wave function after a single passage with the linear excitation in an elegant
	mathematical way, by using Laplace transformations and contour integration.
	
	\subsection{Using the WKB approximation and the phase integral method (St\"{u}%
		ckelberg)}
	
	\label{App_Stuckelberg}
	
	In his 54-page-long paper, Ref.~\cite{Stueckelberg1932} (with the English
	translation in \cite{Stueckelberg_EN}), E.C.G.~St\"{u}ckelberg studied
	the transitions between two energy states. In brief, he considered the problem
	of inelastic collisions and reduced the Schr\"{o}dinger equation to a system
	of two coupled second-order differential equations, which he solved
	developing the WKB
	approximation and phase integral methods. Here, the difficulty consists
	of analytic continuation of the WKB solution through the so-called Stokes
	lines \cite{DiGiacomo2005}. In what follows, we give the most important
	aspects of this, while the details can be found in Ref.~\cite{Child1974a}
	and references therein.
	
	Following \cite{Child1974a}, we start from the two coupled wave equations
	\begin{subequations}\begin{equation}
		\left. 
		\begin{array}{c}
			\left[ \frac{d^{2}}{dr^{2}}+k_{1}^{2}(r)\right] u_{1}(r)=\alpha (r)u_{2}(r),
			\\ 
			\left[ \frac{d^{2}}{dr^{2}}+k_{2}^{2}(r)\right] u_{2}(r)=\alpha (r)u_{1}(r),%
		\end{array}%
		\right\}  \label{two_eqs}
	\end{equation}%
	where 
	\begin{eqnarray}
		k_{i}^{2}(r) &=&2\mu \left( E-V_{i}(r)\right) /\hbar ^{2},  \label{ki} \\
		\alpha (r) &=&2\mu V_{12}(r)/\hbar ^{2}.  \label{alpha}
	\end{eqnarray}\end{subequations} 
	We now reduce the Schr\"{o}dinger equation, which describes the
	collision of atoms and molecules, by expanding the full wave function with
	the ones of the electrons in the atoms and spherical functions \cite%
	{Kotova1969, Eu1970}. Assuming that the motion of the electrons has been
	solved and that the motion of the nuclei is slow with respect to that of the
	electrons, this problem is reduced to two equations~(\ref{two_eqs}), for the nuclear wave functions $u_{i}(r)$, with $r$ being the distance between the
	nuclei, $\mu $ being their reduced mass, and the potential curves for the
	two states being $V_{1}(r)$ and $V_{2}(r)$, which quasicross at $r=R$ and that have a coupling $V_{12}(r)$ between the states.
	
	The approach by St\"{u}ckelberg is done to eliminate $u_{2}(r)$, resulting in a
	fourth-order equation for $u_{1}(r)$. Developing the standard WKB technique 
	\cite{LandauLishitz_QM}, this equation can be solved by means of an
	expansion in powers of $\hbar $,%
	\begin{equation}
		u_{1}(r)=\exp \left[ \frac{i}{\hbar }\left( S_{0}+\hbar S_{1}+...\right) %
		\right] .
	\end{equation}%
	The zero-order terms give $S_{0}/\hbar =\pm \int k_{\pm }(r)dr$, where $%
	k_{\pm }$ are determined by means of Eq.~(\ref{ki}) in terms of the
	adiabatic potential functions $V_{\pm }(r)$, which are defined as%
	\begin{equation}
		V_{\pm }=\frac{1}{2}\left( V_{1}-V_{2}\right) \pm \frac{1}{2}\sqrt{\left(
			V_{1}-V_{2}\right) ^{2}+4V_{12}^{2}}.
	\end{equation}%
	Analysis of the first-iteration terms shows that this description becomes
	invalid near the classical turning points $r=a_{\pm }$, where $V_{\pm
	}(a_{\pm })=E$ and near the transition points $r=r_{\pm }$, where $%
	k_{+}(r_{\pm })=k_{-}(r_{\pm })$. The latter takes place when the adiabatic
	terms intersect, $V_{+}(r_{\pm })=V_{-}(r_{\pm })$; here, $r_{\pm }$ are the
	complex values with both having the same real part $R=\frac{1}{2}\left( r_{+}+r_{-}\right) $.
	The vicinities of these points define non-semi-classical regions that
	should be bridged by changes in the asymptotic solutions along
	contours suitably defined in the complex $r$ plane. Far from the
	turning-point regions, the solution becomes 
	\begin{equation}
		u_{1}(r)\approx C\sin \left( \int_{a_{\pm }}^{r}k_{\pm }(r)dr+\frac{\pi }{4}%
		\right) .  \label{quasiclass1}
	\end{equation}%
	Around the transition zone, we obtain, to the left, at $r\ll R$,%
	\begin{subequations}\begin{equation}
		\left[ 
		\begin{array}{c}
			u_{1}(r) \\ 
			u_{2}(r)%
		\end{array}%
		\right] \approx \left[ 
		\begin{array}{c}
			\frac{A_{1}^{\left( \pm \right) }}{\sqrt{k_{-}}}\exp \left( \pm
			i\int_{R}^{r}k_{-}dr\right) \\ 
			\frac{A_{2}^{\left( \pm \right) }}{\sqrt{k_{+}}}\exp \left( \pm
			i\int_{R}^{r}k_{+}dr\right)%
		\end{array}%
		\right]  \label{quasiclass2left}
	\end{equation}%
	and to the right, at $r\gg R$,%
	\begin{equation}
		\left[ 
		\begin{array}{c}
			u_{1}(r) \\ 
			u_{2}(r)%
		\end{array}%
		\right] \approx \left[ 
		\begin{array}{c}
			\frac{B_{1}^{\left( \pm \right) }}{\sqrt{k_{+}}}\exp \left( \pm
			i\int_{R}^{r}k_{+}dr\right) \\ 
			\frac{B_{2}^{\left( \pm \right) }}{\sqrt{k_{-}}}\exp \left( \pm
			i\int_{R}^{r}k_{-}dr\right)%
		\end{array}%
		\right] .  \label{quasiclass2right}
	\end{equation}\end{subequations} 
	Here, the upper and lower signs refer to positive- and negative-momentum
	solutions, respectively. Then, the left-hand and right-hand coefficients become related%
	\begin{equation}
		\left[ 
		\begin{array}{c}
			B_{1}^{\left( \pm \right) } \\ 
			B_{2}^{\left( \pm \right) }%
		\end{array}%
		\right] =\left[ 
		\begin{array}{cc}
			\sqrt{\mathcal{P}} & -\sqrt{1-\mathcal{P}}e^{\mp i{\phi }_{\text{S}%
			}} \\ 
			\sqrt{1-\mathcal{P}}e^{\pm i{\phi }_{\text{S}}} & \sqrt{\mathcal{P}%
			}%
		\end{array}%
		\right] \left[ 
		\begin{array}{c}
			A_{1}^{\left( \pm \right) } \\ 
			A_{2}^{\left( \pm \right) }%
		\end{array}%
		\right]
	\end{equation}%
	with $\mathcal{P}=e^{-2\pi \delta }$, 
	\begin{equation}
		\delta =\frac{1}{\pi }\text{Im}\left\{ \int_{R}^{r_{+}}\left(
		k_{-}-k_{+}\right) dr\right\} \text{,}
	\end{equation}%
	and certain phase factor ${\phi }_{\text{S}}$. We write this as
	\textquotedblleft certain\textquotedblright\ because the phase factor cannot
	be determined by the phase integral method; this phase factor can be
	determined only by one of the differential equation methods \cite{Child1974a}
	(for more details, see the textbook \cite{Child1974}, particularly
	Appendices C and D therein). 
	
	In his paper, St\"{u}ckelberg did not calculate
	the phase factor; later, it was shown that within such an approach, this phase
	shift can only be defined in the \textit{diabatic limit} ($\delta \ll 1$), with the
	result ${\phi }_{\text{S}}=\pi /4$ \cite{Thorson1971, Delos1972}.
	(Note that our $\delta $ differs from the one in \cite{Child1974a} by a
	factor $\pi $, which is only a matter of notation.)
	
	To interrelate with the linear drive problem, we can linearize the potential
	curves in the vicinity of the transition point%
	\begin{equation}
		V_{1}(r)-V_{2}(r)\approx -\left( F_{1}-F_{2}\right) (r-R),\text{ \ \ }%
		F_{i}=-\left. \frac{\partial V_{i}}{\partial r}\right\vert _{r=R}\text{,}
	\end{equation}%
	and with $V_{12}=\Delta /2=\text{constant}$, we obtain the LZSM formula in the form of
	Eq.~(\ref{P}) with 
	\begin{equation}
		\delta =\frac{\Delta ^{2}}{4\hbar v}\text{, \ \ \ }v=\dot{r}\left\vert
		F_{1}-F_{2}\right\vert .
	\end{equation}
	
	For the double-passage evolution, we combine the nonadiabatic
	transitions, which we just considered above, and the adiabatic evolution
	described by the quasiclassical function in Eq.~(\ref{quasiclass1}), hence
	resulting in Eq.~(\ref{P_double}) with
	
	\begin{equation}
		\Phi _{\mathrm{St}}=\int_{a_{-}}^{R}k_{-}dr-\int_{a_{+}}^{R}k_{+}dr+%
		{\phi }_{\text{S}}\text{.}
	\end{equation}%
	Therefore, we can see that this consists of two parts: the one accumulated during
	the adiabatic motion and the other one ${\phi }_{\text{S}}$,
	called the dynamical or Stokes phase, acquired during the single passage of the
	avoided-crossing region.
	
	\subsection{Duration of the LZSM transition}
	
	\label{Time of the LZSM transition}Transition dynamics matters not only for
	describing a single transition, but also for repeated processes.
	Specifically, when describing periodic driving (which is the subject
	of Appendix \ref{Sec:AppendixB}), if the time of a transition becomes
	larger than the time span between subsequent transitions, $t_{\text{LZSM}%
	}>T_{\text{d}}/2$, we cannot use the adiabatic-impulse model. 
	
	We can split the transition time $t_{\text{LZSM}}=t_{\text{jump}}+t_{\text{%
			relax}}\ $by two terms, where the jump time $t_{\text{jump}}$ describes the
	jump of the probability from an initial value, to the vicinity of a final
	value, and where the relaxation time $t_{\text{relax}}$ states when oscillations of
	the occupation probability decay around the final value $P(t\rightarrow
	\infty )$. Importantly, the transition times differ in the adiabatic and
	diabatic bases; therefore, we define the relaxation times in both bases,
	following \cite{Vitanov1999b}. In the diabatic basis, $P_{%
		\text{d}}(\infty )=1-\mathcal{P}$, and in the adiabatic basis, $P_{\text{a}%
	}(\infty )=\mathcal{P}$. In this subsection, by $P$, we mean the upper-level
	occupation probabilities.
	
	It is straightforward to define the jump time as follows: 
	\begin{equation}
		t_{\text{jump}}=\frac{P(\infty )}{P^{\prime }(0)}.  \label{LZJumpTime}
	\end{equation}%
	Here, we should use the analytic solution for the probabilities in the LZSM
	problem, Eqs.~(\ref{LZPaPdi},~\ref{LZPaPd}), expanding it in a series around $%
	\tau =0$. We find the value of this analytic solution in the zero-bias
	point: 
	\begin{subequations}\begin{eqnarray}
		P_{\text{d}}^{\prime }(0) &=&\sqrt{2\delta \left( 1-\mathcal{P}\right) }\cos
		\chi,  \\
		P_{\text{a}}^{\prime }(0) &=&\frac{e^{-\pi \delta }}{\sqrt{8\delta }},
	\end{eqnarray}%
	where 
	\begin{equation}
		\chi (\delta )=\frac{\pi }{4}+\text{Arg}\left[ \Gamma \left( \frac{1}{2}-\frac{i\delta }{%
			2}\right)\right] -\text{Arg}\left[ \Gamma \left( 1-\frac{i\delta }{2}\right)\right] .
	\end{equation}\end{subequations}
	This very convenient definition, Eq.~(\ref{LZJumpTime}), is suitable for
	most cases, including diabatic dynamics and, in some cases, adiabatic
	dynamics. However, as we see below, such definition becomes invalid for
	the adiabatic dynamics in the \textit{adiabatic limit} $\delta \gg 1$; see also Fig.~%
	\ref{Fig:Singletrans in diff basises}. This is why we need to introduce a
	more elaborated definition.
	
	For obtaining the transition times, we can use the expanded expression for $%
	P(\tau >0)$, Eqs.~(\ref{LZPaPdi},~\ref{LZPaPd}), after passing the avoided-crossing region in the strong-coupling regime with $\delta \gg 1$; then, we
	have 
	\begin{subequations} \begin{eqnarray}
		P_{\text{d}}(\tau &>&0)\approx \frac{1}{2}+\left( \frac{1}{2}-\mathcal{P}%
		\right) \frac{\tau }{\sqrt{\tau ^{2}+2\delta }}-\sqrt{\mathcal{P}\left( 1-%
			\mathcal{P}\right) }\sqrt{\frac{2\delta }{\tau ^{2}+2\delta }}\cos \xi (\tau
		),  \label{P(tau>0)} \\
		P_{\text{a}}(\tau &>&0)\approx \mathcal{P}+\left( 1-2\mathcal{P}\right) 
		\frac{\delta }{8(\tau ^{2}+2\delta )^{3}}+\sqrt{\mathcal{P}\left( 1-\mathcal{%
				P}\right) }\sqrt{\frac{\delta }{2\left( \tau ^{2}+2\delta \right) ^{3}}}\sin
		\xi (\tau ),
	\end{eqnarray}%
	where 
	\begin{equation}
		\xi (\tau )=-\delta +2\delta \ln \left[ \frac{1}{\sqrt{2}}\left( \tau +\sqrt{%
			\tau ^{2}+2\delta }\right) \right] +\tau \sqrt{\tau ^{2}+2\delta }+\frac{\pi 
		}{4}+\text{Arg} \left[ \Gamma \left( 1-i\delta \right)\right] .
	\end{equation}\end{subequations}
	These expressions demonstrate that the approximate solution for the
	probability has two contributions. The first one tends to $P(\infty )$ when $%
	\tau $ tends to infinity; this term defines the jump time $t_{\text{jump}}$.
	The second terms, with $\cos \xi (\tau )$ or $\sin \xi (\tau )$, describe
	decaying  oscillations after passing the avoided-crossing region; these define
	the relaxation time $t_{\text{relax}}$. In particular, we can see that $P_{%
		\text{a}}^{\prime }(0)=0$ at $\delta \gg 1$. Hence, the above definition of
	the jump time becomes invalid.
	
	Therefore, in general, the jump time $t_{\text{jump}}$ can be defined as the
	distance between the starting time $t_{\text{jump}}^{(1)}$ and ending
	time $t_{\text{jump}}^{(2)}$. The former can be defined as a moment when the
	probability leaves the vicinity of the initial probability, $P(t_{\text{jump}%
	}^{(1)})=\eta P(\infty )$, and the latter is defined as a moment when
	entering the vicinity of the final probability, $P(t_{\text{jump}%
	}^{(2)})-P(\infty )=\eta P(\infty )$. Here, $\eta \ll 1$ is the small
	parameter that describes the magnitude of the vicinity near the initial and
	final probabilities. Then, from Eqs.~(\ref{LZPaPdi},~\ref{LZJumpTime}), we
	obtain the jump time for the diabatic basis by expanding the
	parabolic cylinder functions into a series \cite{GradshteynRyzhik, Abadir1993}: 
	\begin{subequations}\begin{equation}
		t_{\text{jump}}^{\text{d}}=\frac{\sqrt{1-\mathcal{P}}}{\sqrt{2\delta }\cos
			\chi (\delta )}.  \label{t_jump^d}
	\end{equation}%
	From this, we can obtain the limit expressions: 
	\begin{equation}
		t_{\text{jump}}^{\text{d}}\approx 
		\begin{cases}
			2\sqrt{\hbar \pi /v}\text{, \ \ }\delta \ll 1, \\ 
			4\sqrt{\hbar \delta /v}\text{, \ \ }\delta \gg 1.%
		\end{cases}%
	\end{equation}%
	For the relaxation time in the diabatic basis, from Eq.~\eqref{P(tau>0)}, we
	have 
	\begin{equation}
		t_{\text{relax}}^{\text{d}}\approx 2\sqrt{\frac{\delta }{v}\left( \frac{1}{%
				\eta ^{2}(e^{2\pi \delta }-1)}-1\right) }.  \label{t_relax^d}
	\end{equation}\end{subequations}
	
	\begin{figure}[t]
		\centering{\includegraphics[width=1.0%
			\columnwidth]{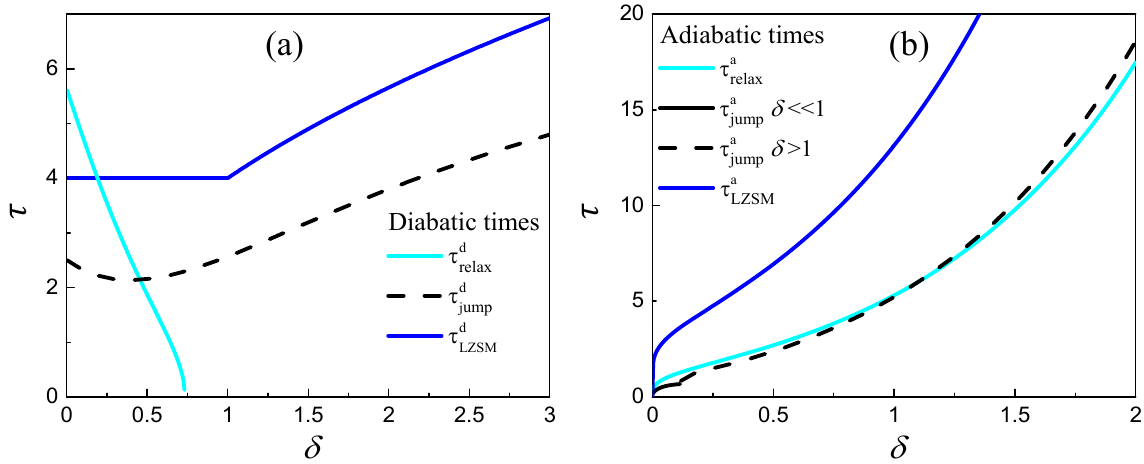}}
		\caption{\textbf{Transition times }$t_{\text{jump}}$\textbf{\ and }$t_{\text{%
					relax}}$\textbf{\ versus the adiabaticity parameter }$\protect\delta $%
			\textbf{. }Here, the dimensionless time is $\protect\tau =t \left[v/2\hbar\right]^{1/2}$. Note that in the adiabatic basis $t_{\text{jump}%
			}^{\text{a}}\approx t_{\text{relax}}^{\text{a}}$, and also that these
			transition times grow exponentially in the \textit{adiabatic limit} ($\protect\delta %
			\gg 1$). In the diabatic basis and the \textit{diabatic limit} ($\protect%
			\delta \ll 1$) $t_{\text{relax}}^{\text{d}}>t_{\text{jump}}^{\text{d}}$, and
			there is no $t_{\text{relax}}^{\text{d}}$ in the adiabatic limit. Note that $\tau_\text{LZSM}$ is plotted using Eq.~\eqref{t_d} and~\eqref{t_a}.}
		\label{Fig:Transition times in diff bases 2}
	\end{figure}
	For the jump and relaxation times in the adiabatic basis, from Eqs.~\eqref{LZPaPd} and~\eqref{LZJumpTime}, it follows 
	\begin{subequations}\begin{equation}
		t_{\text{jump}}^{\text{a}}\approx 
		\begin{cases}
			4\sqrt{\frac{\hbar \delta }{v}}\text{, \ \ }\delta \ll 1, \\ 
			\sqrt{\frac{2}{v}}\left( \frac{1}{\eta }8\delta \exp\left[4\pi \delta\right] \right)
			^{1/6}\text{, \ \ }\delta \gg 1,%
		\end{cases}%
	\end{equation}%
	and 
	\begin{equation}
		t_{\text{relax}}^{\text{a}}\approx \sqrt{\frac{4\delta }{v}\left[ \left( 
			\frac{e^{2\pi \delta }-1}{16\eta ^{2}\delta ^{4}}\right) ^{1/3}-1\right] }.
		\label{t_relax^a}
	\end{equation}%
	In the limiting cases for the relaxation time, we then have 
	\begin{equation}
		t_{\text{relax}}^{\text{a}}\approx 
		\begin{cases}
			\sqrt{\frac{2}{v}}(2\delta /\eta )^{1/3}\text{, \ \ }\delta \ll 1, \\ \\
			\sqrt{\frac{2}{v}}\left( \frac{\exp\left[\pi \delta \right]}{\eta }\right) ^{1/3}(\delta
			/2)^{1/6}\text{, \ \ }\delta \gg 1.%
		\end{cases}
		\label{t_jump^a}
	\end{equation}\end{subequations}
	
	We show the transition times in Fig.~\ref{Fig:Transition times in
		diff bases 2}. To plot these, we have used Eq.~\eqref{t_relax^d} for $t_{\text{%
			relax}}^{\text{d}}$ and Eq.~\eqref{t_jump^d} for $t_{\text{jump}}^{\text{d}}$
	in Fig.~\ref{Fig:Transition times in diff bases 2}(a); then, we used Eq.~\eqref{t_relax^a} for $t_{\text{relax}}^{\text{a}}$ and Eq.~\eqref{t_jump^a}
	for $t_{\text{jump}}^{\text{a}}$ in Fig.~\ref{Fig:Transition times in diff
		bases 2}(b).
	
	Here, for $\delta \ll 1$, we have $t_{\text{relax}}^{\text{d}}\gg
	t_{\text{relax}}^{\text{a}}$. This difference in the relaxation times
	happens because the oscillations of $P_{\text{d}}(t)$ vanish proportionally to 
	$t^{-1}$, but $P_{\text{a}}(t)$ decays proportionally to $t^{-3}$. 
	
	For the case $\delta \gg 1$, we have the opposite situation: $t_{\text{relax}}^{\text{%
			a}}\gg t_{\text{relax}}^{\text{d}}\approx 0$. Note that $t_{\text{relax}}^{%
		\text{a}}\approx t_{\text{jump}}^{\text{a}}$; therefore, we can define $t_{%
		\text{LZSM}}\approx 2t_{\text{relax}}^{\text{a}}$. As a result, the
	equations above can be summarized as Eqs.~(\ref{t_d},~\ref{t_a}) in the main
	text.
	
	\section{Description of a periodically driven two-level system}
	
	\label{Sec:AppendixB}
	
	\subsection{Adiabatic-impulse model (AIM)}
	
	\label{Sec:AIM} In this section, we describe the adiabatic-impulse model in more detail, expanding on Section~\ref{Sec:AIM_Main}.
	
	\subsubsection{Adiabatic evolution}
	
	Consider now the adiabatic evolution when the system passes in one of the adiabatic
	eigenstates $\left\vert E_{\pm }(t)\right\rangle $. This means that the
	adiabatic basis consists of the instantaneous eigenstates of the
	time-dependent Hamiltonian. We can obtain the instantaneous eigenfunctions
	from the Schr\"{o}dinger equation~\eqref{HamEq}, where time $t$ is a
	parameter: 
	\begin{equation}
		H(t)\left\vert E_{\pm }(t)\right\rangle =E_{\pm }(t)\left\vert E_{\pm
		}(t)\right\rangle .  \label{quasiHam}
	\end{equation}
	To describe the system dynamics, we solve the dynamical form of the Schr\"{o}dinger
	equation, Eq.~\eqref{TDSE}, with the initial condition 
	\begin{equation}
		\left\vert E_{\pm }(t_\text{i})\right\rangle =\left\vert E_{\text{i}\pm }\right\rangle .
	\end{equation}%
	We substitute Eq.~\eqref{quasiHam} into the Schr\"{o}dinger equation~\eqref{TDSE}, solve this differential equation, and then obtain the evolution of
	the upper- and lower-energy-level wave functions 
	\begin{subequations}\begin{eqnarray}
		\left\vert E_{\pm }(t)\right\rangle &=&\exp {\left( -\frac{i}{\hbar }%
			\int_{t_\text{i}}^{t}E_{\pm }(t^{\prime })dt^{\prime }\right) }\left\vert E_{\text{i}\pm
		}\right\rangle =\exp {\left( \mp i\zeta \right) }\left\vert E_{\text{i}\pm
		}\right\rangle,   \label{Zevol} \\
		\zeta (t) &=&\frac{1}{2\hbar }\int_{t_\text{i}}^{t}\Delta E(t)\;dt.  \label{AIMzeta}
	\end{eqnarray}\end{subequations} 
	As a result, we obtain the adiabatic evolution matrix, Eq.~%
	\eqref{AdiabaticEvolution}.
	
	\subsubsection{Multiple-passage evolution}
	
	Here we describe some aspects of the multiple-passage evolution in relation
	to Sec.~\ref{Sec:AIM_Main}. To raise the matrix 
	$\Xi $ to the $n$-th power [see Eq.~\eqref{SingleTransitionMatrix}], we first need to diagonalize
	it. For this, we use the unitary matrix 
	\begin{equation}
		A=%
		\begin{pmatrix}
			A_{11} & -A_{21}^{\ast } \\ 
			A_{21} & A_{11}^{\ast }%
		\end{pmatrix}%
		,\;\;\text{    }AA^{\dagger }=1,\;\;\text{    }|A_{11}|^{2}+|A_{21}|^{2}=1,
	\end{equation}%
	such that $A\Xi A^{\dagger }=%
	\begin{pmatrix}
		e^{i\phi } & 0 \\ 
		0 & e^{-i\phi }%
	\end{pmatrix}%
	$, where $\phi $ is the desired value. Consider the equation $\Xi
	=A^{\dagger }\begin{pmatrix}
		e^{i\phi } & 0 \\ 
		0 & e^{-i\phi }%
	\end{pmatrix}A$ and obtain 
	\begin{equation}
		\begin{cases}
			\Xi _{11}=|A_{11}|^{2}e^{i\phi }+|A_{21}|^{2}e^{-i\phi } \\ 
			\Xi _{21}=-2iA_{11}A_{21}\sin \phi .%
		\end{cases}
		\label{AIMOnePerEvolSub}
	\end{equation}%
	From these, we define $\phi $: $\cos \phi =$\textrm{Re}$\Xi _{11}$. In the
	next step, we find the matrix $\Xi _{n}$ of the evolution for $n$ periods 
	\begin{subequations}\begin{equation}
		\Xi _{n}=\Xi ^{n}=A^{\dagger }%
		\begin{pmatrix}
			e^{in\phi } & 0 \\ 
			0 & e^{-in\phi }%
		\end{pmatrix}%
		A=%
		\begin{pmatrix}
			\Xi _{n11} & -\Xi _{n21}^{\ast } \\ 
			\Xi _{n21} & \Xi _{n11}^{\ast }%
		\end{pmatrix}%
		,
	\end{equation}%
	where we simplify the obtained matrix elements, here taking into account Eq.~\eqref{AIMOnePerEvolSub}: 
	\begin{eqnarray}
		\Xi _{n11} &=&\cos (n\phi )+i\text{\textrm{Im}}(\Xi _{11})\frac{\sin {(n\phi
				)}}{\sin \phi }, \\
		\Xi _{n21} &=&\Xi _{21}\frac{\sin {(n\phi )}}{\sin \phi }.  
	\end{eqnarray}\end{subequations} 
	Then, we obtain the probability of the upper-level occupation of the TLS
	during their respective time intervals: 
	\begin{subequations}\begin{eqnarray}
		P_{+}^{(1)}(n) &=&\left\vert \Xi _{n21}\right\vert ^{2}=\left\vert \Xi
		_{21}\right\vert ^{2}\frac{\sin ^{2}{n\phi }}{\sin ^{2}\phi },\;\;\ \ \ \text{ for }%
		(t-nT_{\mathrm{d}})\in (t_{2},T_{\mathrm{d}}+t_{1}), \\
		P_{+}^{(2)}(n) &=&2Q_{1}\frac{\sin ^{2}{n\phi }}{\sin ^{2}\phi }+Q_{2}\frac{%
			\sin {2n\phi }}{\sin \phi }+\mathcal{P}\cos {2n\phi },\;\;\ \ \ \text{ for }(t-nT_{%
			\mathrm{d}})\in (t_{1},t_{2}),
	\end{eqnarray}%
	where 
	\begin{align}
		& Q_1=\mathcal{P}\left[\mathcal{P}\sin^2{\zeta_-}+(1-\mathcal{P})(1+\cos{%
			\zeta_+}\cos{\zeta_-})\right], \\
		& Q_2=2\mathcal{P}(1-\mathcal{P})\cos{(\zeta_1+{\phi}_\text{S})}%
		\cos{(\zeta_2+{\phi}_\text{S})}.  
	\end{align}\end{subequations}
	For the time-averaged value, we need to average over many periods $n\gg 1$,
	so we obtain 
	\begin{subequations}\begin{eqnarray}
		\overline{P_{+}^{(1)}} &=&\frac{|\Xi _{21}|^{2}}{2\sin ^{2}{\phi }}=\frac{1}{%
			2}\frac{|\Xi _{21}|^{2}}{|\Xi _{21}|^{2}+(\text{\textrm{Im}}{\Xi _{11}})^{2}}%
		,  \label{AIMAvPlz} \\
		\overline{P_{+}^{(2)}} &=&\frac{Q_{1}}{\sin ^{2}{\phi }}.
	\end{eqnarray}\end{subequations}
	We can neglect the difference between $\overline{P_{+}^{(1)}}$ and $\overline{
		P_{+}^{(2)}}$ in the case of the slow-passage limit $\mathcal{P}\ll 1$, that is, $%
	\delta \gg 1$. Here, we use $\mathcal{P}$ as a small parameter, and in the
	first approximation, we obtain Eq.~\eqref{ShIF}.
	
	\subsection{Rotating-wave approximation (RWA)}
	
	\label{Sec:RWA}
	
	Consider now the case of strong excitation, $\Delta \ll \sqrt{A\hbar \omega }$
	at $k\hbar \omega \approx \Delta E$, where the latter condition means that
	the energy of $k$ photons approximately equals the energy distance
	between the qubit levels $\Delta E$. Here, we follow Refs.~\cite{Silveri2015,
		Ono2019} presenting the formalism valid not only for the cosine driving but
	rather for a generic excitation $\varepsilon (t)=\varepsilon _{0}+\widetilde{%
		\varepsilon }(t)$ for any function $\widetilde{\varepsilon }(t)$ with the
	period $T_{\mathrm{d}}=2\pi /\omega $.
	
	Let us first split the Hamiltonian into a stationary part $H_{0}$ and a
	time-dependent part $V(t)$:
	\begin{equation}
		H=H_{0}+V(t)=-\frac{\Delta }{2}\sigma _{x}-\frac{\varepsilon _{0}}{2}\sigma
		_{z}-\frac{\widetilde{\varepsilon }(t)}{2}\sigma _{z}.
	\end{equation}%
	To make it convenient to solve the Bloch equations, we perform the unitary
	transformation:
	\begin{equation}
		U=\exp \left[ -i\frac{\eta (t)}{2}\sigma _{z}\right] =\cos {\frac{\eta }{2}}
		-i\sigma _{z}\sin {\frac{\eta }{2}},\;\;\;\;\;\eta (t)=\frac{1}{\hbar }
		\int_{0}^{t}dt^{\prime }\;\widetilde{\varepsilon }\left( t^{\prime }\right) .
	\end{equation}%
	We use the operator $U(t)$ to link the wave function in the rotating and
	stationary coordinate systems, $\psi =U(t)\psi ^{\prime }$. Then, we
	substitute this function into the Schr\"{o}dinger equation and obtain $%
	i\hbar U\dot{\psi}^{\prime }+i\hbar \dot{U}\psi ^{\prime }=HU\psi ^{\prime }$.
	In the rotating coordinate system, we need $i\hbar \dot{\psi}^{\prime
	}=H^{\prime }\psi ^{\prime }$, thus obtaining the new Hamiltonian%
	\begin{equation}
		H^{\prime }=\!U^{\dag }HU-i\hbar U^{\dag }\dot{U}=-\frac{\varepsilon _{0}}{2}
		\sigma _{z}-\frac{\Delta }{2}\left( e^{i\eta }\sigma _{+}+h.c.\right), 
		\label{H2}
	\end{equation}%
	with $\sigma _{+}=\frac{1}{2}\left( \sigma _{x}+i\sigma _{y}\right) $. Then,
	the preparatory stage is finalized by the Fourier series expansion%
	\begin{equation}
		\Delta e^{i\eta }=\sum_{m=-\infty }^{\infty }\Delta _{m}\,e^{im\omega t},
		\label{expansion}
	\end{equation}%
	where the complex-valued amplitude is given by the inverse Fourier transform%
	\begin{equation}
		\Delta _{m}=\Delta \frac{\omega }{2\pi }\int\limits_{0}^{2\pi /\omega
		}dt\;e^{-im\omega t}e^{i\eta (t)}=\Delta \int\limits_{0}^{1}d\tau ^{\prime
		}\exp \left[ i\eta (\tau ^{\prime })-i2\pi m\tau ^{\prime }\right] .
		\label{Delta}
	\end{equation}%
	Then, the Hamiltonian becomes%
	\begin{equation}
		H^{\prime }=\!-\frac{\varepsilon _{0}}{2}\sigma _{z}+\frac{1}{2}
		\sum_{m=-\infty }^{\infty }\left( \Delta _{m}e^{im\omega t}\sigma
		_{+}+h.c.\right)= -\frac{1}{2}\sum_{m=-\infty }^{\infty }
		\begin{pmatrix}
			\varepsilon_{0} & \Delta_me^{im\omega t} \\ 
			\Delta_me^{-im\omega t} & -\varepsilon_{0}
		\end{pmatrix}.
	\end{equation}
	With this Hamiltonian, it is convenient to solve the Bloch equations for the
	density matrix,%
	\begin{equation}
		\dot{\rho}_{ij}=-\frac{i}{\hbar }\left[ H,\rho \right] _{ij}-\frac{\rho
			_{ij}-\rho _{ij}^{(0)}}{\tau _{ij}},
	\end{equation}%
	where $\rho _{ij}$ is the density matrix, $\tau _{ij}$ is the relaxation
	rate, $\rho ^{(0)}$ is the equilibrium density operator with $\rho
	_{01}^{(0)}=\rho _{10}^{(0)}=0$, $Z^{(0)}=~\rho _{00}^{(0)}-~\rho
	_{11}^{(0)}=~\tanh \frac{\Delta E}{2k_{B}T}$, and $k_{B}$ is the Boltzmann
	constant. For this, we parameterize the density matrix
	\begin{equation}
		\rho =\frac{1}{2}\left( 1\sigma _{0}+X\sigma _{x}+Y\sigma _{y}+Z\sigma
		_{z}\right) =\frac{1}{2} 
		\begin{pmatrix}
			1+Z & X-iY \\ 
			X+iY & 1-Z%
		\end{pmatrix}
		\equiv 
		\begin{pmatrix}
			\rho _{00} & \rho _{01} \\ 
			\rho _{10} & \rho _{11}%
		\end{pmatrix}
		.  \label{RWAPodstDens}
	\end{equation}
	
	For the moment, we assume that the system is driven close to a resonance,
	where $\Delta E\approx \left\vert \varepsilon _{0}\right\vert \approx k\hbar
	\omega $. Then, we omit the \textquotedblleft
	fast-rotating\textquotedblright\ terms and leave only terms with $m=k$. We
	can write down the Bloch equations with the Hamiltonian $H^{\prime }$
	component-wise: 
	\begin{subequations}\begin{eqnarray}
		\dot{Z} &=&-\frac{\Delta _{k}}{2\hbar }\mathrm{Im}(e^{-ik\omega
			t})-(Z-Z^{(0)})\Gamma _{1}, \\
		\dot{\rho}_{10} &=&i\frac{\Delta _{k}}{2\hbar }e^{ik\omega t}Z-i\frac{%
			\varepsilon _{0}}{\hbar }\rho _{10}-\rho _{10}\Gamma _{2},  
	\end{eqnarray}%
	where 
	\begin{equation}
		Z^{(0)}=\rho _{00}^{(0)}-\rho _{11}^{(0)}=\tanh \frac{\Delta E}{2k_{\text{B}%
			}T}.  \label{RWAZ0}
	\end{equation}\end{subequations}
	Here $\Delta _{k}$ was defined in Eq.~\eqref{Deltak}. After the substitution $%
	\rho _{10}\exp {\left( -ik\omega t\right) }=\widetilde{X}+i\widetilde{Y}$,
	we obtain the system of equations 
	\begin{subequations}\begin{eqnarray}
		\dot{\widetilde{X}} &=&\left( k\omega +\frac{\varepsilon _{0}}{\hbar }%
		\right) \widetilde{Y}-\widetilde{X}\Gamma _{2}, \\
		\dot{\widetilde{Y}} &=&-\left( k\omega +\frac{\varepsilon _{0}}{\hbar }%
		\right) \widetilde{X}+\frac{\Delta _{k}}{\hbar }Z-\widetilde{Y}\Gamma _{2}, 
		 \\
		\dot{Z} &=&-\frac{\Delta _{k}}{\hbar }\widetilde{Y}-(Z-Z^{(0)})\Gamma _{1}.	
	\end{eqnarray}\end{subequations}
	The stationary solution of these equations can be obtained after substituting $%
	\dot{\widetilde{X}}=\dot{\widetilde{Y}}=\dot{Z}=0$. Consider now the
	low-temperature case, $T\rightarrow 0$, then $Z^{(0)}\approx 1$. Finally, we
	obtain the stationary value for the probability of the upper diabatic state;
	summing all possible resonant terms, we obtain the \textit{qubit upper-level
	occupation probability}
	\begin{subequations}\begin{eqnarray}
		\overline{P}_{\mathrm{up}}^{(k)} &=&\overline{\rho }_{11}^{(k)}=\frac{1}{2}%
		\left( 1-\overline{Z}^{(k)}\right),  \\
		\overline{P}_{\mathrm{up}} &=&\sum_{k}\overline{P}_{\mathrm{up}}^{(k)}=\frac{%
			1}{2}\sum_{k=-\infty }^{\infty }\frac{\left\vert \Delta _{k}\right\vert ^{2}%
		}{\left\vert \Delta _{k}\right\vert ^{2}+\frac{\Gamma _{1}}{\Gamma _{2}}%
			\left( k\hbar \omega -\varepsilon _{0}\right) ^{2}+\hbar ^{2}\Gamma
			_{1}\Gamma _{2}}.    \label{RWAP+}
	\end{eqnarray}\end{subequations}
	Note that for a complex-valued $\Delta _{k}$, what matters is its absolute
	value.
	
	Finally, to obtain the upper-level occupation probability for a
	given bias $\widetilde{\varepsilon }(t)$, we must calculate the functions 
	$\Delta _{k}$. Although more examples can be seen in \cite{Silveri2015, Ono2019},
	for the sinusoidal modulation, with the Jacobi–Anger expansion following
	Eq.~\eqref{Deltak}, it is straightforward to see that%
	\begin{equation}
		\Delta _{m}(x)=\Delta J_{m}(x),\;\;\;\;x=\frac{A}{\hbar \omega }.
		\label{Jacobi-Anger}
	\end{equation}%
	It is useful to recall here the asymptote at $x\gg 1$:
	
	\begin{equation}
		J_{m}(x)\approx \sqrt{\frac{2}{\pi x}}\cos \left[ x-\frac{\pi m}{2}-\frac{
			\pi }{4}\right] .
	\end{equation}%
	This explicitly demonstrates that the occupation probability is
	quasiperiodic in driving the amplitude with the period $\delta A=2\pi \hbar
	\omega $.
	
	\subsection{Floquet theory}
	
	\label{Sec:Floquet}
	
	From Floquet theory \cite{Son2009}, we can obtain a solution $\left\vert
	\psi (t)\right\rangle $ for the Schr\"{o}dinger equation~\eqref{TDSE} with
	any time-periodic Hamiltonian $H(t)=H(t+nT_{\mathrm{d}})$ for any integer $n$%
	\begin{equation}
		\left\vert \psi (t)\right\rangle =e^{-i\epsilon t}\left\vert \Phi
		(t)\right\rangle,   \label{FloqSubst1}
	\end{equation}%
	where $\left\vert \Phi (t)\right\rangle $ denote the time-periodic Floquet
	modes with the same period as the Hamiltonian $H(t)$ and $\epsilon $ is the
	so-called quasienergy. When we substitute Eq.~\eqref{FloqSubst1} into Eq.~\eqref{TDSE}, we can obtain an equation for the quasienergy 
	\begin{equation}
		\left( H(t)-i\hbar \frac{d}{dt}\right) \left\vert \Phi (t)\right\rangle
		=\epsilon \left\vert \Phi (t)\right\rangle .  \label{SchrodEqFloquet}
	\end{equation}%
	Periodic functions can be decomposed into harmonic functions using the
	Fourier series expansion
	\begin{subequations}
		\label{FloqSubst2}
		\begin{eqnarray}
			\left\vert \Phi (t)\right\rangle &=&\sum_{n=-\infty }^{\infty }\left\vert
			\Phi _{n}\right\rangle e^{in\omega t}, \\
			H(t) &=&\sum_{n=-\infty }^{\infty }H^{[n]}e^{in\omega t},
		\end{eqnarray}%
		where the superscript $[n]$ defines the number of the 2x2 part of the
		Hamiltonian matrix. 
		For the next step, we rewrite the Hamiltonian in its exponential form 
	\end{subequations}
	\begin{equation}
		H(t)=-\frac{\Delta }{2}\sigma _{x}-\frac{\varepsilon _{0}}{2}\sigma _{z}-%
		\frac{A}{4}\sigma _{z}\left( e^{i\omega t}+e^{-i\omega t}\right), 
	\end{equation}%
	and we obtain the system of stationary equations for the quasienergies
	using the Schr\"{o}dinger equation~\eqref{TDSE} with the substitutions~\eqref{FloqSubst1} and~\eqref{FloqSubst2}%
	\begin{equation}
		\sum_{i}\sum_{m}\langle 0,n|H_{\mathrm{F}}|i,m\rangle \langle i,m|\epsilon
		_{j}\rangle =\epsilon _{j}\langle 0,n|\epsilon _{j}\rangle,
	\end{equation}%
	where $H_{\mathrm{F}}$ is the Floquet Hamiltonian, $\epsilon _{j}$ is the
	quasienergy eigenvalue, $|\epsilon _{j}\rangle $ is an eigenvector, and $%
	\left\vert i,m\right\rangle $ is the $m$-th Fourier component of the $i$-th
	energy level. We can also write it in terms of matrix elements 
	\begin{equation}
		\epsilon _{j}\Phi _{j,n}=\left( -\frac{\Delta }{2}\sigma _{x}-\frac{%
			\varepsilon _{0}}{2}\sigma _{z}+n\omega \right) \Phi _{n}-\frac{A}{4}\sigma
		_{z}\left( \Phi _{n-1}+\Phi _{n+1}\right), 
	\end{equation}%
	where $j$ defines the number of a state, which is either $0$ or $1$ for a
	TLS after decomposing the wave function, Eq.~\eqref{psi(t)}.
	Then, we obtain the time-independent Floquet Hamiltonian, which is defined by 
	\begin{subequations}
		\begin{equation}
			\left\langle i_{n}\right\vert H_{\mathrm{F}}\left\vert j_{k}\right\rangle
			=H_{0,1}^{[n-k]}+n\hbar \omega \delta _{ij}\delta _{nk},
		\end{equation}%
		where $i$ and $j$ define the number of the state, or 
		\begin{equation}
			H_{\mathrm{F},nk}=H^{[n-k]}+n\hbar\omega  \delta _{nk}I,  \label{HFnk}
		\end{equation}%
		for 2x2 elements ($I$ is 2x2 identity matrix) with three nonvanishing
		Fourier components of the Hamiltonian $H^{\left[ n-k\right] }$, where $n$ and$~k$ are the integer numbers corresponding to different states: 
	\end{subequations}
	\begin{equation}
		H^{[0]}=-\frac{1}{2}%
		\begin{pmatrix}
			\varepsilon _{0} & \Delta \\ 
			\Delta & -\varepsilon _{0}%
		\end{pmatrix}%
		,\text{ \ }H^{[+1]}=H^{[-1]}=-\frac{1}{4}%
		\begin{pmatrix}
			A & 0 \\ 
			0 & -A%
		\end{pmatrix}.
	\end{equation}%
	%
	%
	%
	%
	%
	%
	%
	%
	%
	%
	%
	%
	%
	%
	%
	%
	%
	%
	This consists of the 2x2 submatrices, and Eq.~(\ref{HFnk}) defines its
	element in the $n$-th column and the $k$-th row. Hence, the Floquet states
	matrix can be written as%
	\begin{equation}
		H_{\mathrm{F}}=\left( 
		\begin{array}{c|cc|cc|cc|c}
			\ddots &  &  &  &  &  &  &  \\ \hline
			& \rule{0pt}{4ex} -\frac{\varepsilon _{0}}{2}+(n-1) \hbar \omega & -\frac{\Delta }{2} & -%
			\frac{A}{4} & 0 & 0 & 0 &  \\
			& \rule{0pt}{4ex} -\frac{\Delta }{2} & \frac{\varepsilon _{0}}{2}+(n-1) \hbar \omega & 0 & 
			\frac{A}{4} & 0 & 0 &  \\[2ex]  \hline
			& \rule{0pt}{4ex} -\frac{A}{4} & 0 & -\frac{\varepsilon _{0}}{2}+n\hbar \omega & -\frac{%
				\Delta }{2} & -\frac{A}{4} & 0 &  \\ 
			& \rule{0pt}{4ex} 0 & \frac{A}{4} & -\frac{\Delta }{2} & \frac{\varepsilon _{0}}{2}+n\hbar
			\omega & 0 & \frac{A}{4} &  \\[2ex] \hline
			& \rule{0pt}{4ex} 0 & 0 & -\frac{A}{4} & 0 & -\frac{\varepsilon _{0}}{2}+(n+1)\hbar \omega & 
			-\frac{\Delta }{2} &  \\ 
			& \rule{0pt}{4ex} 0 & 0 & 0 & \frac{A}{4} & -\frac{\Delta }{2} & \frac{\varepsilon _{0}}{2}%
			+(n+1)\hbar \omega &  \\[2ex] \hline
			&  &  &  &  &  &  & \ddots%
		\end{array}%
		\right) .  \label{FloquetHamiltonian}
	\end{equation}%
	The eigenvalues of this matrix can be found numerically if we take a finite number
	of these 2x2 blocks. Then, we can obtain the time-averaged upper-level
	occupation probability 
	\begin{equation}
		\overline{P}_{\mathrm{up}}=\sum_{n}\sum_{j}|\left\langle 1_{n}|\epsilon
		_{j}\right\rangle \left\langle \epsilon _{j}|0_{(n=0)}\right\rangle |^{2}.
	\end{equation}%
	The Schr\"{o}dinger equation with the Hamiltonian $H_{\mathrm{F}}$ in the
	general case cannot be solved analytically; therefore, we first consider the
	eigenvalue problem with $\Delta =0\label{FUnpertCondition}$; next, we can
	use perturbation theory with the small parameter being $\Delta $.
	We denote $H_{0}$ as the unperturbed Hamiltonian. In the situation when $%
	\Delta =0$, there is no coupling between the $\left\vert 0\right\rangle $
	and $\left\vert 1\right\rangle $ states; therefore we can write the same
	Hamiltonian $H_{0}^{\prime }$ for both states: 
	\begin{equation}
		\widetilde{H}_{0}(\text{for state }0\text{ or }1)=%
		\begin{pmatrix}
			\ddots &  &  &  &  \\ 
			& b+(n-1)\hbar \omega & a & 0 &  \\ 
			& a & b+n\hbar \omega & a &  \\ 
			& 0 & a & b+(n+1)\hbar \omega &  \\ 
			&  &  &  & \ddots%
		\end{pmatrix}%
	\end{equation}%
	where $b=-\frac{\varepsilon _{0}}{2}$, $a=-\frac{A}{4}$ for the $\left\vert
	0\right\rangle $ state and $b=\frac{\varepsilon _{0}}{2}$, $a=\frac{A}{4}$
	for the $\left\vert 1\right\rangle $ state. To find the eigenfunctions, we use the
	Schr\"{o}dinger equation~\eqref{SchrodEqFloquet} with Hamiltonian $%
	\widetilde{H}_{0}^{\prime },$ in which $\Delta =0$. As a result, using the
	Jacobi–Anger formula~\eqref{Jacobi-Anger}, we obtain 
	\begin{equation}
		|\widetilde{\Phi }_{n}(t)\rangle =e^{in\omega t}e^{-i(2a/\hbar \omega )\sin
			\omega t}=\sum_{k=-\infty }^{\infty }J_{k}\left( -\frac{2a}{\hbar \omega }%
		\right) e^{i(n+k)\omega t}=\sum_{k=-\infty }^{\infty }J_{k-n}\left( \frac{A}{%
			2\hbar \omega }\right) e^{-ik\omega t}.
	\end{equation}%
	The eigenstates of the unperturbed Hamiltonian $H_{0}$ are the following: 
	\begin{subequations}
		\begin{eqnarray}
			\left\vert \widetilde{0},n\right\rangle &=&\sum_{k=-\infty }^{\infty
			}J_{k-n}\left( \frac{A}{2\hbar \omega }\right) \left\vert 0,k\right\rangle, 
			\\
			\left\vert \widetilde{1},m\right\rangle &=&\sum_{k=-\infty }^{\infty
			}J_{k-m}\left( -\frac{A}{2\hbar \omega }\right) \left\vert 1,m\right\rangle .
		\end{eqnarray}%
		For the Floquet Hamiltonian~\eqref{FloquetHamiltonian}, using the
		addition theorem for Bessel functions 
	\end{subequations}
	\begin{equation}
		J_{n}(2z)=\sum_{m}J_{m}(z)J_{n-m}(z),
	\end{equation}%
	we can obtain the matrix elements for the Floquet Hamiltonian based on this~\eqref{FloquetHamiltonian}: 
	\begin{subequations}
		\label{FloqQuasHam}
		\begin{eqnarray}
			\left\langle 0_{n}^{\prime }\right\vert H_{\mathrm{F}}\left\vert
			1_{m}^{\prime }\right\rangle &=&\sum_{k=-\infty }^{\infty }\sum_{l=-\infty
			}^{\infty }J_{k-n}(z)J_{l-m}(-z)\left\langle 0_{k}\right\vert
			H_{F}\left\vert 1_{l}\right\rangle =-\frac{\Delta }{2}J_{m-n}\left( \frac{A}{%
				\hbar \omega }\right),  \\
			\left\langle 1_{n}^{\prime }\right\vert H_{\mathrm{F}}\left\vert
			0_{m}^{\prime }\right\rangle &=&-\frac{\Delta }{2}J_{n-m}\left( \frac{A}{%
				\hbar \omega }\right),  \\
			\left\langle 0_{n}^{\prime }\right\vert H_{\mathrm{F}}\left\vert
			0_{m}^{\prime }\right\rangle &=&\left( -\frac{\varepsilon _{0}}{2}+n\hbar\omega
			\right) \delta _{nm}, \\
			\left\langle 1_{n}^{\prime }\right\vert H_{\mathrm{F}}\left\vert
			1_{m}^{\prime }\right\rangle &=&\left( \frac{\varepsilon _{0}}{2}+n\hbar\omega
			\right) \delta _{nm}.
		\end{eqnarray}%
		From Eqs.~\eqref{FloqQuasHam}, if we have the condition for
		the multi-photon resonances, $\varepsilon _{0}\approx n\hbar \omega $, then
		we can neglect all other coupling terms, except the one between $\left\vert
		0_{0}^{\prime }\right\rangle $ and $\left\vert 1_{-n}^{\prime }\right\rangle 
		$; therefore, we obtain a 2x2 matrix Hamiltonian: 
	\end{subequations}
	\begin{subequations}\begin{equation}
		H_{\text{RWA}}=-\frac{1}{2}%
		\begin{pmatrix}
			\varepsilon _{0} & \Delta J_{-n}^{\prime } \\ 
			\Delta J_{-n}^{\prime } & -\varepsilon _{0}+\frac{n\hbar \omega }{2}%
		\end{pmatrix}%
	\end{equation}%
	where
	\begin{equation}
		J_{n}^{\prime }=J_{n}\left( \frac{A}{\hbar \omega }\right).
	\end{equation}\end{subequations}
	
	\paragraph*{\textbf{Generalized Van Vleck Perturbation Theory}} \mbox{} \\
	
	The RWA is the first
	approximation in Floquet theory; for the next approximations, we can use
	generalized Van Vleck perturbation theory (GVVPT) \cite{Son2009}. The
	perturbation (small) parameter is 
	\begin{equation}
		\lambda =-\frac{\Delta }{2}.
	\end{equation}
	Here, we can rewrite the Floquet Hamiltonian in the basis of the states $%
	\left\vert 0_{n}^{\prime }\right\rangle $ and $\left\vert 1_{m}^{\prime
	}\right\rangle $:
	\begin{subequations}\begin{equation}
		\widetilde{H}_{\mathrm{F}}=\widetilde{H}_{0}^{\prime }+\lambda \widetilde{V},
	\end{equation}%
	where 
	\begin{eqnarray}
		\widetilde{H}_{0}^{\prime }&=&\left( 
		\begin{array}{c|cc|cc|cc|c}
			\ddots &  &  &  &  &  &  &  \\ \hline
			& \rule{0pt}{1ex} -\frac{\varepsilon _{0}}{2}-\hbar \omega & 0 & 0 & 0 & 0 & 0 &  \\ 
			& 0 & \frac{\varepsilon _{0}}{2}-\hbar \omega & 0 & 0 & 0 & 0 &  \\[1ex] \hline
			& 0 & 0 & -\frac{\varepsilon _{0}}{2} & 0 & 0 & 0 &  \\ 
			& 0 & 0 & 0 & \frac{\varepsilon _{0}}{2} & 0 & 0 &  \\ \hline
			& 0 & 0 & 0 & 0 & -\frac{\varepsilon _{0}}{2}+\hbar \omega & 0 &  \\ 
			& 0 & 0 & 0 & 0 & 0 & \frac{\varepsilon _{0}}{2}+\hbar \omega &  \\ \hline
			&  &  &  &  &  &  & \ddots%
		\end{array}%
		\right), \\ \widetilde{V}&=&%
		\begin{pmatrix}
			\begin{array}{c|cc|cc|cc|c}
				\ddots &  &  &  &  &  &  &  \\ \hline
				& 0 & J_{0}^{\prime } & 0 & J_{1}^{\prime } & 0 & J_{2}^{\prime } &  \\ 
				& J_{0}^{\prime } & 0 & J_{-1}^{\prime } & 0 & J_{-2}^{\prime } & 0 &  \\ 
				\hline
				& 0 & J_{-1}^{\prime } & 0 & J_{0}^{\prime } & 0 & J_{1}^{\prime } &  \\ 
				& J_{1}^{\prime } & 0 & J_{0}^{\prime } & 0 & J_{-1}^{\prime } & 0 &  \\ 
				\hline
				& 0 & J_{-2}^{\prime } & 0 & J_{-1}^{\prime } & 0 & J_{0}^{\prime } &  \\ 
				& J_{2}^{\prime } & 0 & J_{1}^{\prime } & 0 & J_{0}^{\prime } & 0 &  \\ 
				\hline
				&  &  &  &  &  &  & \ddots%
			\end{array}%
		\end{pmatrix}.
		\label{FloquetHamUnperturbed}
	\end{eqnarray}\end{subequations} 
	We can reduce the infinite-dimensional matrix of the Floquet Hamiltonian~\eqref{FloquetHamiltonian} to a 2x2 matrix Hamiltonian $h$ by using the nearly degenerate
	perturbation formalism in the GVVPT \cite{Aravind1984}. Following 
	perturbation theory, the 2x2 matrix of Hamiltonian $h$ and its eigenstates solutions $\phi $
	can be expanded in powers of $\lambda $ 
	\begin{subequations}
		\begin{eqnarray}
			h &=&\sum_{m=0}^{\infty }\lambda ^{m}h^{(m)}, \\
			\phi &=&\sum_{m=0}^{\infty }\lambda ^{m}\phi ^{(m)}.
		\end{eqnarray}%
		For the $n$-photon resonance, the Floquet states are nearly degenerate, so 
	\end{subequations}
	\begin{subequations}\begin{equation}
		\phi _{-}^{(0)}=\left\vert \widetilde{0}_{0}\right\rangle \text{ and }\phi
		_{+}^{(0)}=\left\vert \widetilde{1}_{0}\right\rangle
	\end{equation}%
	and 
	\begin{equation}
		h^{(0)}=\frac{1}{2}%
		\begin{pmatrix}
			-\varepsilon _{0} & 0 \\ 
			0 & \varepsilon _{0}-2n\hbar \omega%
		\end{pmatrix},
	\end{equation}\end{subequations}
	where the two states are nearly degenerate, that is, $\varepsilon _{0}\approx
	n\hbar \omega $. Using GVVPT, a few high-order terms can be obtained: 
	\begin{subequations}
		\begin{eqnarray}
			\phi _{-}^{(1)} &=&\sum\nolimits_{k}^{\prime }\frac{-J_{k}^{\prime }}{%
				\varepsilon _{0}+k\hbar \omega }\left\vert \widetilde{1}_{k}\right\rangle, 
			\\
			\phi _{+}^{(1)} &=&\sum\nolimits_{k}^{\prime }\frac{J_{k}^{\prime }}{%
				\varepsilon _{0}+k\hbar \omega }\left\vert \widetilde{1}_{-n-k}\right\rangle
			, \\
			h^{(1)} &=&\left\langle \phi ^{(0)}|V^{\prime }|\phi ^{(0)}\right\rangle
			=J_{-}^{\prime }%
			\begin{pmatrix}
				0 & 1 \\ 
				1 & 0%
			\end{pmatrix}%
			, \\
			h^{(2)} &=&\left\langle \phi ^{(0)}|V^{\prime }|\phi ^{(1)}\right\rangle
			-h^{(1)}\left\langle \phi ^{(0)}|\phi ^{(1)}\right\rangle
			=\sum\nolimits_{k}^{\prime }\frac{J_{k}^{\prime 2}}{\varepsilon _{0}+k\hbar
				\omega }%
			\begin{pmatrix}
				-1 & 0 \\ 
				0 & 1%
			\end{pmatrix}%
			, \\
			h^{(3)} &=&\left\langle \phi ^{(1)}|V^{\prime }|\phi ^{(1)}\right\rangle
			-\left\langle \phi ^{(1)}|\phi ^{(1)}\right\rangle
			h^{(1)}=-\sum\nolimits_{k}^{\prime }\left( \sum\nolimits_{l}^{\prime }\frac{%
				J_{k}^{\prime }J_{l}^{\prime }J_{k+l+n}^{\prime }}{\left( \varepsilon
				_{0}+k\hbar \omega \right) \left( \varepsilon _{0}+l\hbar \omega \right) }+%
			\frac{J_{k}^{\prime 2}J_{-n}^{\prime }}{\left( \varepsilon _{0}+k\hbar
				\omega \right) ^{2}}\right) 
			\begin{pmatrix}
				0 & 1 \\ 
				1 & 0%
			\end{pmatrix}%
			.
		\end{eqnarray}%
		Here, the summation $\sum\nolimits_{k}^{\prime }$ is from $-\infty $ to $%
		\infty $, with the prime standing for $k\neq -n$. As a result of GVVPT, we
		obtain the Hamiltonian, which includes $n$-photon coupling channels 
	\end{subequations}
\begin{equation}
		H_{\text{GVVPT}}=\frac{1}{2}%
		\begin{pmatrix}
			-\varepsilon _{0}-\delta _{n} & \widetilde{\Delta }_{n} \\ 
			\widetilde{\Delta }_{n} & \varepsilon _{0}+\delta _{n}-2n\hbar \omega%
		\end{pmatrix},
	\label{GVVPTHam}
	\end{equation}%
	\begin{subequations}
	where 
	
		\begin{eqnarray}
			\widetilde{\Delta }_{n} &=&J_{-n}^{\prime }2\lambda
			-\sum\nolimits_{k}^{\prime }\left( \sum\nolimits_{l}^{\prime }\frac{%
				J_{k}^{\prime }J_{l}^{\prime }J_{k+l+n}^{\prime }}{\left( \varepsilon
				_{0}+k\hbar \omega \right) \left( \varepsilon _{0}+l\hbar \omega \right) }+%
			\frac{J_{k}^{\prime 2}J_{-n}^{\prime }}{\left( \varepsilon _{0}+k\hbar
				\omega \right) ^{2}}\right) 2\lambda ^{3}+O(\lambda ^{5}), \\
			\delta _{n} &=&\sum\nolimits_{k}^{\prime }2\frac{J_{k}^{\prime 2}}{%
				\varepsilon _{0}+k\hbar \omega }\lambda ^{2}+O(\lambda ^{4}).
		\end{eqnarray}%
		Then, we find the eigenvalues of the GVVPT Hamiltonian~\eqref{GVVPTHam} 
	\end{subequations}
	\begin{subequations}\begin{equation}
		\epsilon _{0,1}=-\frac{n\hbar \omega }{2}\pm \hbar \widetilde{\Omega}_\text{R},
	\end{equation}%
	where 
	\begin{equation}
		\left( \hbar \widetilde{\Omega}_\text{R}\right) ^{2}=(n\hbar \omega
		-\varepsilon _{0}-\delta _{n})^{2}+\widetilde{\Delta }_{n}^{2}.
	\end{equation}\end{subequations}
	This gives the $n$\textit{-photon time-dependent upper-level} occupation probability 
	
	\begin{equation}
		P_{\text{up}}^{(n)}(t)=\frac{1}{2}\frac{\widetilde{\Delta }_{n}^{2}}{\left(
			\hbar \widetilde{\Omega}_\text{R}\right) ^{2}}\left(1-\cos\widetilde{\Omega}_%
		\text{R} t\right),
	\end{equation}%
	and \textit{averaged} occupation probability 
	\begin{equation}
		\overline{P}_{\text{up}}^{(n)}=\lim_{t\rightarrow \infty }\frac{1}{t}%
		\int_{0}^{t}P_{\text{up}}^{(n)}(t^{\prime })dt^{\prime}=\frac{1}{2}\frac{%
			\widetilde{\Delta }_{n}^{2}}{\widetilde{\Delta }_{n}^{2}+(n\hbar \omega
			-\varepsilon _{0}-\delta _{n})^{2}}.
	\end{equation}
	When we take only the first term of this, we can reach the RWA limit 
	\begin{equation}
		\overline{P}_{\text{up}}^{(n)}=\frac{1}{2}\frac{\Delta _{n}^{2}}{\Delta
			_{n}^{2}+(n\hbar \omega -\varepsilon _{0})^{2}},
	\end{equation}%
	where $\Delta _{n}$ is the same as in the RWA and as was defined in Eq.~%
	\eqref{Jacobi-Anger}.
	
	\subsection{Rate-equation and white-noise approach}
	
	\label{Sec:Rate}
	
	In this subsection, we describe a TLS dynamics modeling the
	decoherence by adding classical noise and then solving the rate equation.
	We demonstrate that such approach gives exactly the same solution as the
	one by solving the Bloch equation in the RWA.
	
	\subsubsection{Transition rate}
	
	Consider a transition rate in a driven TLS, following \cite%
	{Berns2006} [see also \cite{Otxoa2019}]. This is described by the Hamiltonian
	\begin{subequations}\begin{equation}
		H(t)=-\frac{1}{2}\left( 
		\begin{array}{cc}
			\varepsilon (t) & \Delta \\ 
			\Delta & -\varepsilon (t)%
		\end{array}%
		\right),   \label{H}
	\end{equation}
	\begin{equation}
		\varepsilon (t)=\varepsilon _{0}+A\cos \omega t+\delta \varepsilon (t),
		\label{h(t)}
	\end{equation}\end{subequations}
	where $\delta \varepsilon (t)$ stands for classical noise, which models
	decoherence here.
	
	After the unitary transformation%
	\begin{equation}
		U=\exp \left( i\frac{\phi (t)}{2}\sigma _{z}\right), \text{ \ \ }\phi (t)=%
		\frac{1}{\hbar }\int\limits_{0}^{t}dt^{\prime }\varepsilon (t^{\prime }),
		\label{U}
	\end{equation}%
	we obtain%
	\begin{subequations}\begin{equation}
		H^{\prime }(t)=U^{\dag }HU-i\hbar U^{\dag }\dot{U}=\left( 
		\begin{array}{cc}
			0 & \Delta (t) \\ 
			\Delta ^{\ast }(t) & 0%
		\end{array}%
		\right),   \label{H_prime}
	\end{equation}%
	where 
	\begin{equation}
		\Delta (t)=\Delta e^{-i\phi }.  \label{fi}
	\end{equation}\end{subequations}
	
	Perturbation theory gives the transition rate:%
	\begin{equation}
		W=\lim_{\delta t\rightarrow \infty }\frac{\left\vert A_{t,t^{\prime
			}}\right\vert ^{2}}{\delta t},\text{ \ \ }A_{t,t^{\prime }}=\frac{1}{2}%
		\int\limits_{t}^{t^{\prime }}dt^{\prime }\,\Delta (t^{\prime }),\text{ \ \ }%
		\delta t=t^{\prime }-t.  \label{perturb}
	\end{equation}%
	Here, the limit implies $\delta t\gg T_{2}$. Perturbation theory is valid,
	provided that the change of the qubit population is slow on the scale of $%
	T_{2}$, which means that we have $W\ll T_{2}$~\cite{Berns2006, Danon2014}.
	
	We now add the averaging over white noise $\left\langle ...\right\rangle
	_{\delta \varepsilon }$ and rewrite this: 
	\begin{equation}
		W=\frac{1}{4}\lim_{\delta t\rightarrow \infty }\frac{1}{\delta t}%
		\int\limits_{0}^{\tau }d\tau \int\limits_{t}^{t^{\prime }}dt^{\prime
		}\left\langle \Delta (t+\tau )\Delta ^{\ast }(t^{\prime })\right\rangle
		_{\delta \varepsilon }.  \label{with_averaging}
	\end{equation}
	
	Next, we use the Jacobi–Anger expansion, Eq.~(\ref{Jacobi-Anger}), to remove
	the sine from the exponent%
	\begin{eqnarray}
		e^{-i\phi (t)} =\exp \left( -i\frac{\varepsilon _{0}}{\hbar }t-i\frac{A}{%
			\hbar \omega }\sin \omega t-i\delta \phi (t)\right) = 
		e^{-i\frac{\varepsilon _{0}}{\hbar }t}e^{-i\delta \phi
			(t)}\sum_{n=-\infty }^{\infty }J_{n}(x)e^{-in\omega t}.  
	\end{eqnarray}%
	Here, we define%
	\begin{equation}
		\delta \phi (t)=\frac{1}{\hbar }\int\limits_{0}^{t}dt^{\prime }\delta
		\varepsilon (t^{\prime })\text{ \ and\ \ }x=\frac{A}{\hbar \omega }.
	\end{equation}
	
	If the noise is frequency independent (i.e., white) in a low-frequency region, the
	averaging gives \cite{Du2013a}%
	\begin{equation}
		\left\langle e^{i\delta \phi (t)}e^{-i\delta \phi (t^{\prime
			})}\right\rangle _{\delta \varepsilon }=\exp{\left(-\Gamma _{2}\left\vert
			t-t^{\prime }\right\vert \right)}.
	\end{equation}%
	Then, calculating the integral
	
	\begin{equation}
		\int\limits_{-\infty }^{\infty }d\tau e^{-i\left( \frac{\varepsilon _{0}}{%
				\hbar }-n\omega \right) \tau }e^{-\Gamma _{2}\left\vert \tau \right\vert }=%
		\frac{2\Gamma _{2}}{\left( \frac{\varepsilon _{0}}{\hbar }-n\omega \right)
			^{2}+\Gamma _{2}^{2}},
	\end{equation}%
	we obtain the transition rate%
	\begin{subequations}
	\begin{eqnarray}
		W &=&\sum_{n=-\infty }^{\infty }W_{n},  \label{W} \\
		W_{n} &=&\frac{1}{2}\Gamma _{2}\frac{\Delta _{n}^{2}}{\left( \frac{%
				\varepsilon _{0}}{\hbar }-n\omega \right) ^{2}+\Gamma _{2}^{2}},  \label{Wn}
		\\
		\Delta _{n} &=&\Delta J_{n}(x).
	\end{eqnarray}\end{subequations}
	Note that the transition rate $W$ consists of the partial contributions $%
	W_{n}$, each of which is essential in the vicinity of the $n$-th resonance, $%
	\varepsilon _{0}\sim n\hbar \omega $.
	
	\subsubsection{Rate equation}
	
	The rate equation for a TLS reads%
	\begin{equation}
		\frac{dP_{+}}{dt}=\left( W+\Gamma _{1}^{\prime }\right) P_{-}-\left(
		W+\Gamma _{1}\right) P_{+}.  \label{dP+}
	\end{equation}%
	Here, $\Gamma _{1}$ stands for the relaxation rate from the excited to the
	ground state, and $\Gamma _{1}^{\prime }=\Gamma _{1}\exp\left[-\Delta E/k_{\mathrm{B}%
		}T\right]$ stands for the thermally-excited inverse relaxation. From this, it is apparent
	how to take temperature into account. In the following, we assume the
	low-temperature limit ($T=0$) and adopt $\Gamma _{1}^{\prime }=0$.
	
	As the second equation, we can write down the equation for $dP_{-}/dt$,
	analogously to Eq.~(\ref{dP+}), or use the condition $P_{+}+P_{-}=1$. Then,
	in the stationary regime, with the zero left-hand side in Eq.~(\ref{dP+}), we obtain%
	\begin{equation}
		P_{+}=\frac{W}{2W+\Gamma _{1}}\text{.}  \label{P+}
	\end{equation}
	
	In the vicinity of the $k$-th resonance, that is, at $\varepsilon _{0}\sim
	k\hbar \omega $, the probability is given by the $k$-th term $W_{k}$, Eq.~(%
	\ref{Wn}):%
	\begin{equation}
		P_{+}^{(k)}=\frac{W_{k}}{2W_{k}+\Gamma _{1}}\text{.}
	\end{equation}%
	The summation of all possible resonances gives the \textit{full upper-level}
	occupation probability:
	
	\begin{equation}
		P_{+}=\sum_{k=-\infty }^{\infty }P_{+}^{(k)}=\frac{1}{2}\sum_{k=-\infty
		}^{\infty }\frac{\Delta _{k}^{2}}{\Delta _{k}^{2}+\frac{\Gamma _{1}}{\Gamma
				_{2}}\left( \frac{\varepsilon _{0}}{\hbar }-k\omega \right) ^{2}+\Gamma
			_{1}\Gamma _{2}}.  \label{P1}
	\end{equation}%
	This exactly coincides with the solution of the Bloch
	equations in the RWA, Eq.~(\ref{PupRWA}).
	
	\subsubsection{From Bessel to Airy}
	
	It may be instructive to rewrite the Bessel function via the Airy function,
	as in Ref.~\cite{Berns2006}, (see also Ref.~\cite{Malla2019})%
	\begin{equation}
		J_{k}(x)\approx \sqrt[3]{\frac{2}{x}}\mathrm{Ai}\left( \sqrt[3]{\frac{2}{x}}%
		\left( k-x\right) \right),   \label{Bessel-Airy}
	\end{equation}%
	which is valid for $n\gg 1$ and $n>x$. Here, we can obtain Eq.~(\ref%
	{Bessel-Airy}) using Eqs.~(8.455) and (8.433) from Ref.~\cite%
	{GradshteynRyzhik} and the definition of the Airy function of the first
	kind, $\mathrm{Ai}\left( u\right) =\frac{1}{\pi }\int_{0}^{\infty }dt\cos
	\left( \frac{t^{3}}{3}+ut\right) $.
	
	With this remark, we can proceed with the series in Eq.~(\ref{W}). From the
	denominator in Eq.~(\ref{Wn}), it follows that we can make the replacement in
	the nominator: $k\longrightarrow \varepsilon _{0}/\hbar \omega $; then, the
	respective Airy function goes out of the summation. We can execute the
	summation by expanding the fraction into two elementary fractions: 
	\begin{equation}
		\frac{1}{\left( \frac{\varepsilon _{0}}{\hbar }-k\omega \right) ^{2}+\Gamma
			_{2}^{2}}=\frac{1}{\Gamma _{2}\omega }\text{\textrm{Im}}\frac{1}{z-k},\ \ \ \ \text{
			\ }z=\frac{\varepsilon _{0}}{\hbar \omega }-i\frac{\Gamma _{2}}{\omega }.
	\end{equation}%
	To calculate the sum, consider the expansion of the cotangent \cite%
	{GradshteynRyzhik} and its approximation for $\left\vert z\right\vert \gg 1$:
	
	\begin{equation}
		\cot \pi z=\frac{1}{\pi z}+\frac{1}{\pi }\sum_{k=-\infty }^{\infty }\frac{z}{%
			k\left( z-k\right) }\approx \frac{1}{\pi }\sum_{k=-\infty }^{\infty }\frac{1%
		}{z-k}.  \label{exact_series}
	\end{equation}%
	As a result, we obtain%
	\begin{equation}
		W\approx \frac{\pi \Delta ^{2}}{2\hbar \omega }\left( \frac{2\hbar \omega }{A%
		}\right) ^{2/3}\mathrm{Ai}^{2}\left( \sqrt[3]{\frac{2\hbar \omega }{A}}\frac{%
			\varepsilon _{0}-A}{\hbar \omega }\right) \times \text{\textrm{Im}}\cot %
		\left[ \pi \left( \frac{\varepsilon _{0}}{\hbar \omega }-i\frac{\Gamma _{2}}{%
			\omega }\right) \right] .
	\end{equation}%
	This formula, together with Eq.~(\ref{P+}), is very convenient for
	further analysis of the upper-level occupation probability, of which the
	example follows.
	
	\subsubsection{Double-passage regime}
	
	Consider now the case when the decoherence rate is of the order of a driving
	period, $\Gamma _{2}\gtrsim \omega $. Then, $\cot (...)\approx i$ and, making
	use of Airy function asymptotics at a large negative argument (i.e., for
	the relevant $A>\varepsilon _{0}$), we obtain:%
	\begin{eqnarray}
		W \approx \frac{\pi \Delta ^{2}}{2\hbar \omega }\sqrt[3]{\frac{4}{x^{2}}}%
		\mathrm{Ai}^{2}\left( \sqrt[3]{\frac{2}{x}}\frac{\varepsilon _{0}-A}{\hbar
			\omega }\right) \approx  
		 \frac{1}{2}\frac{\Delta ^{2}}{A}\left( 1-\frac{\varepsilon _{0}}{A}%
		\right) ^{-1/2}\cos ^{2}\left[ \frac{2\sqrt{2}}{3}\frac{A}{\hbar \omega }%
		\left( 1-\frac{\varepsilon _{0}}{A}\right) ^{3/2}-\frac{\pi }{4}\right] .
		\label{W_asymptotics}
	\end{eqnarray}
	
	Equation~(\ref{W_asymptotics}) can be simplified to leading order in $%
	\varepsilon _{0}/A$:%
	\begin{equation}
		W\approx \frac{1}{2}\frac{\Delta ^{2}}{A}\cos ^{2}\left[ \frac{2\sqrt{2}}{3}%
		\frac{A}{\hbar \omega }-\sqrt{2}\frac{\varepsilon _{0}}{\hbar \omega }-\frac{%
			\pi }{4}\right] .  \label{W_leading}
	\end{equation}
	
	Importantly, the above result can be obtained by directly considering the
	double-passage regime, as pointed out by \cite{Berns2006}. Indeed, from the
	adiabatic-impulse model, in the fast passage limit $(\Delta ^{2}/A\hbar
	\omega <1)$, for the upper-level occupation probability, it follows see [Eq.~\eqref{with_Fi_St} and also in Refs.~\cite{Shevchenko2010, Chatterjee2018}]:
	\begin{subequations}\begin{eqnarray}
		P_{+}^{\mathrm{double}} &\simeq &2\pi \frac{\Delta ^{2}}{A\hbar \omega }%
		\left( 1-\frac{\varepsilon _{0}^{2}}{A^{2}}\right) ^{-1/2}\sin ^{2}\Phi _{%
			\mathrm{St}},  \label{PII_1} \\
		\Phi _{\mathrm{St}} &=&-\frac{\varepsilon _{0}}{\hbar \omega }\arccos \frac{%
			\varepsilon _{0}}{A}+\frac{A}{\hbar \omega }\left( 1-\frac{\varepsilon
			_{0}^{2}}{A^{2}}\right) ^{1/2}-\frac{\pi }{4}.  \label{FiSt}
	\end{eqnarray}\end{subequations}
	Equation~(\ref{PII_1}) can be simplified to leading order in $\varepsilon
	_{0}/A$:%
	\begin{equation}
		P_{+}^{\mathrm{double}}\simeq \frac{2\pi }{\hbar \omega }\frac{\Delta \ ^{2}%
		}{A}\sin ^{2}\left[ \frac{A}{\hbar \omega }-\frac{\pi }{2}\frac{\varepsilon
			_{0}}{\hbar \omega }-\frac{\pi }{4}\right] .  \label{PII_2}
	\end{equation}%
	Remarkably, the oscillations in this latter formula are very close to the
	ones in Eq.~(\ref{W_leading}). Compare the respective terms in the two
	equations: $\frac{2\sqrt{2}}{3}\simeq 0.94$, which meets $1$, and $\sqrt{2}%
	\simeq 1.4$, which meets $\pi /2\simeq 1.57$. From Eq.~(\ref{P+}), we have $%
	P_{+}\simeq T_{1}W$, and then, we have identical multiplier factors before
	sine in Eq.~(\ref{PII_2})\ and from Eq.~(\ref{W_leading}), provided the
	relaxation time is equal to two driving periods, $T_{1}=2T_{\mathrm{d}}$,
	where $T_{\mathrm{d}}=2\pi /\omega $.
	
	
	
	
	\newpage \nocite{apsrev41Control} 
	\bibliographystyle{apsrmp4-1}
	\bibliography{LZSM2,1}
	
\end{document}